\documentclass[a4paper,11pt,twoside]{ThesisStyle}

\usepackage[english]{babel}
\usepackage{amsmath}
\usepackage{slashed}
\usepackage{amsfonts}
\usepackage{fancyhdr}
\usepackage{amssymb}
\usepackage{color} 
\definecolor{Valentia}{RGB}{233,78,82}
\definecolor{Titleblue}{RGB}{114, 146, 162}
\usepackage{mdframed}
\usepackage{multirow} 
\usepackage{multicol} 
\usepackage{scrextend} 
\usepackage{tikz}
\usepackage[absolute]{textpos} 
\usepackage{colortbl}
\usepackage{array}
\usepackage{comment}
\usepackage{tikz}
\usepackage{tikz-feynman}
\tikzfeynmanset{compat=1.1.0}
\usetikzlibrary{positioning}
\usepackage{appendix}

\usepackage{chngcntr}
\usepackage{etoolbox}
\usepackage{lipsum}

\AtBeginEnvironment{subappendices}{%
\section*{Appendix}
\addcontentsline{toc}{section}{Appendix}
\counterwithin{figure}{section}
\counterwithin{table}{section}
}

\usepackage{dirtytalk}
\usepackage{csquotes}

\newcommand{\be}{\begin{equation}}
\newcommand{\ee}{\end{equation}}
\newcommand{\ba}{\begin{array}}
\newcommand{\ea}{\end{array}}
\newcommand{\into}{\ensuremath{\,\rightarrow\,}}
\newcommand{\eq}[1]{Eq.~(\ref{#1})}
\newcommand{\SO}[1]{\ensuremath{\mathrm{SO}(#1)}}
\newcommand{\SU}[1]{\ensuremath{\mathrm{SU}(#1)}}
\newcommand{\U}[1]{\ensuremath{\mathrm{U}(#1)}}
\newcommand{\tr}{\operatorname{tr}}
\newcommand{\vev}[1]{\langle #1 \rangle}
\newcommand{\diag}{\operatorname{diag}}

\usepackage{amsmath}
\usepackage{amssymb}
\usepackage{cite}
\usepackage{bbm}
\usepackage{color}
\usepackage{url}
\usepackage{authblk}
\usepackage{mathtools}
\usepackage{blindtext}
\usepackage{titlesec}

\newcommand{\Tr}{\operatorname{Tr}}

\newcommand{\MP}{\ensuremath{M_\mathrm{Pl}}}
\newcommand{\GN}{\ensuremath{G_\mathrm{\scriptscriptstyle{N}}}}

\definecolor{deepfuchsia}{rgb}{0.76, 0.33, 0.76}

\usepackage[bottom]{footmisc}

\usepackage{amsmath}
\usepackage{amssymb}
\usepackage{cite}
\usepackage{bbm}
\usepackage{color}
\usepackage{url}
\usepackage{authblk}
\usepackage{tikz}
\usetikzlibrary{tikzmark}
\usepackage{multirow}
\usepackage{multirow}
\usepackage[table,xcdraw]{xcolor}
\usepackage{enumerate}

\usepackage[normalem]{ulem}
\usepackage{comment}

\definecolor{deepfuchsia}{rgb}{0.76, 0.33, 0.76}

\usepackage{amsmath,amssymb}
\usepackage[T1]{fontenc}
\usepackage[utf8]{inputenc}
\usepackage[english]{babel}

\usepackage[left=0in,right=1.3in,top=1.1in,bottom=1.1in,includefoot,includehead,headheight=13.6pt]{geometry}

\usepackage{silence}

\usepackage{aecompl}
\usepackage{url}

\usepackage[printonlyused,withpage]{acronym}

\usepackage{ifpdf}
\usepackage{graphicx}

\graphicspath{{.}{images/}}

\usepackage{color}
\definecolor{linkcol}{rgb}{0,0,0.4}
\definecolor{citecol}{rgb}{0.5,0,0}

\usepackage[nottoc, notlof, notlot]{tocbibind}
\usepackage{minitoc}
\def\diag{\operatorname{Diag}}

\usepackage{rotating}                    
\usepackage{fancyhdr}                    

\let\headruleORIG\headrule
\renewcommand{\headrule}{\color{black} \headruleORIG}

\usepackage{colortbl}
\arrayrulecolor{black}

\fancypagestyle{plain}{
  \fancyhead{}
  \fancyfoot{}
  
}

\usepackage{algorithm}
\usepackage[noend]{algorithmic}
\usepackage{scrextend}

\makeatletter

\def\cleardoublepage{\clearpage\if@twoside \ifodd\c@page\else%
  \hbox{}%
  \thispagestyle{empty}
  \newpage%
  \if@twocolumn\hbox{}\newpage\fi\fi\fi}

\makeatother

{%

\hrulefill
\vspace*{0.5cm}%
\end{minipage}
}

\let\minitocORIG\minitoc
\renewcommand{\minitoc}{\minitocORIG \vspace{2em}}

\usepackage{subfigure}
\usepackage{multirow}
{ \begin{list}%
	{$\bullet$}%
	{\setlength{\labelwidth}{25pt}%
	 \setlength{\leftmargin}{30pt}%
	 \setlength{\itemsep}{\parsep}}}%
{ \end{list} }

\renewcommand{\epsilon}{\varepsilon}

\newenvironment{vcenterpage}
{\newpage\vspace*{\fill}\thispagestyle{empty}}
{\vspace*{\fill}}

\ifpdf
  \usepackage[pagebackref,hyperindex=true]{hyperref}
\else
  \usepackage[dvipdfm,pagebackref,hyperindex=true]{hyperref}
\fi

\renewcommand*{\backref}[1]{}
\renewcommand*{\backrefalt}[4]{%
\ifcase #1 %
(Not cited.)%
\or
(Cited on page~#2.)%
\else
(Cited on pages~#2.)%
\fi}

\hypersetup
{
bookmarksopen=true,
pdftitle="Cosmological consequencies of spontaneous symmetry breaking",
pdfauthor="Giacomo Ferrante",
pdfsubject="Manuscript topic in a few words", 
pdfmenubar=true, 
pdfhighlight=/O, 
colorlinks=true, 
pdfpagemode=UseNone, 
pdfpagelayout=SinglePage, 
pdffitwindow=true, 
linkcolor=linkcol, 
citecolor=citecol, 
urlcolor=linkcol 
}

\begin{document}

\begin{titlepage}

\newgeometry{left=2.5cm, bottom=3cm, top=2cm, right=2.5cm}

\tikz[remember picture,overlay] \node[opacity=1,inner sep=0pt] at (73.6mm, -124.25mm){\includegraphics{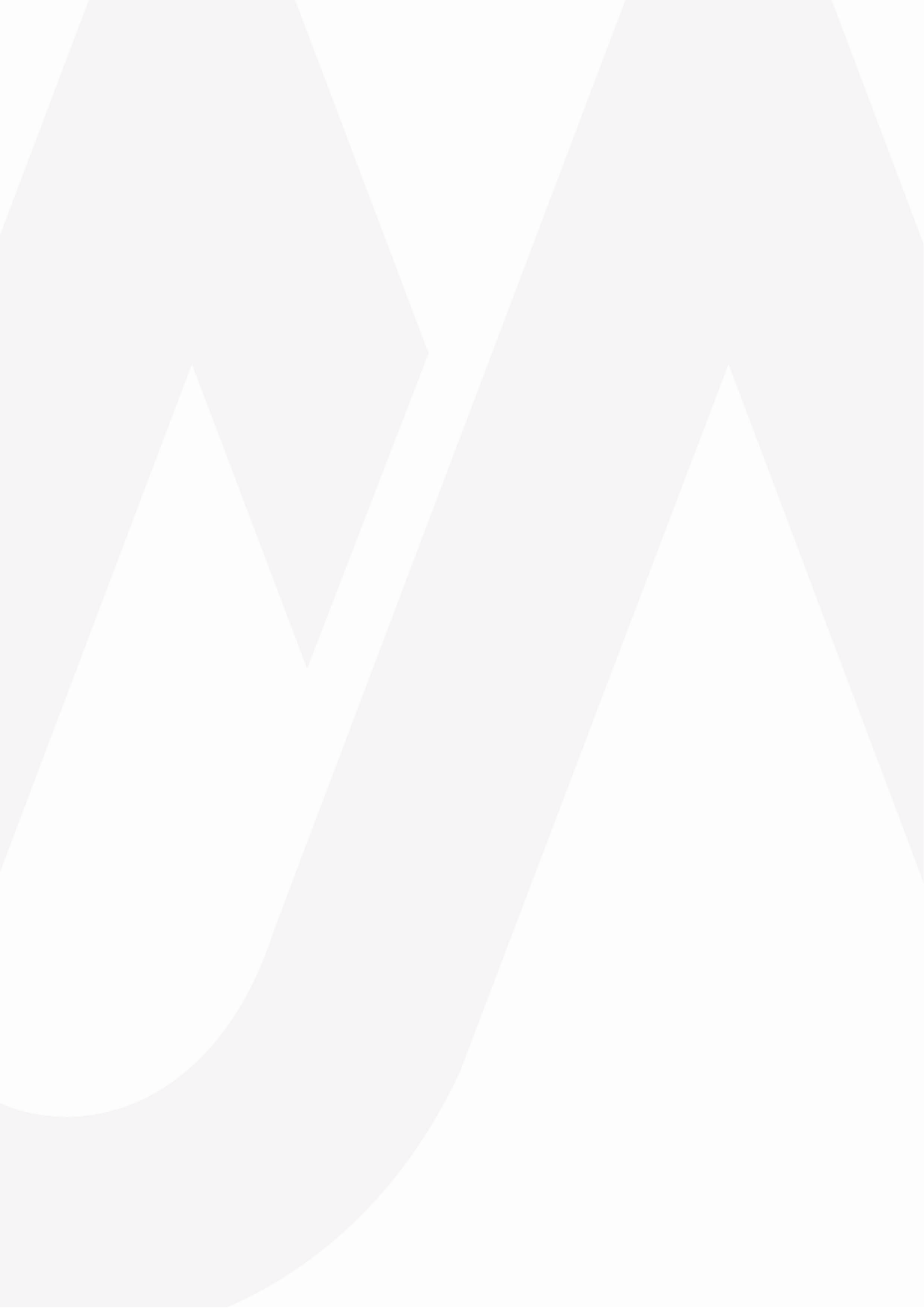}};

{\fontfamily{phv}\fontseries{mc}\selectfont
\centering
\color{Valentia}
\fontsize{18}{13}\selectfont
\textbf{THÈSE POUR OBTENIR LE GRADE DE DOCTEUR\\ DE L’UNIVERSITÉ DE MONTPELLIER}

\normalsize
\color{black}

\bigskip
\textbf{En Physique}

\bigskip
\textbf{École doctorale: Information, Structures, Systèmes}

\bigskip
\textbf{Unité de recherche: Laboratoire Univers et Particules de Montpellier}

\color{Titleblue}
\fontsize{17}{20.4}\selectfont
\vspace{2cm}
\textbf{COSMOLOGICAL CONSEQUENCES\\OF SPONTANEOUS SYMMETRY BREAKING}


\vspace{4cm}
\fontsize{15}{18}\selectfont
\color{black}
\textbf{Présentée par Giacomo FERRANTE\\
le XXVI septembre 2025}

\bigskip
\fontsize{13}{15.6}\selectfont
\textbf{Sous la direction de Felix BRÜMMER}

\vspace{1.5cm}
\normalsize
\textbf{Devant le jury composé de}\\
\bigskip
\fontsize{10}{12}\selectfont
\vspace{1.5mm}
\begin{tabular}{p{14cm}l}
\textbf{Aldo DEANDREA, Professeur, Institut de Physique des deux Infinis de Lyon} & \textbf{Président} \\
\textbf{Valerie DOMCKE, Senior faculty member, Theoretical Physics Department, CERN} & \textbf{Rapporteuse} \\
\textbf{Alberto SALVIO, Associate Professor, University of Rome and INFN Tor Vergata} & \textbf{Rapporteur} \\
\textbf{Felix BR\"UMMER, Maître de conférences, Laboratoire Univers et Particules de Montpellier} & \textbf{Directeur de thèse} \\
\textbf{Yann MAMBRINI, Directeur de Recherche, CNRS and Université Paris Saclay} & \textbf{Examinateur} \\
\textbf{Alberto MARIOTTI, Professor, Vrije Universiteit Brussels} & \textbf{Examinateur} \\

\end{tabular} 
 
\vspace{\fill}
\includegraphics[scale=1]{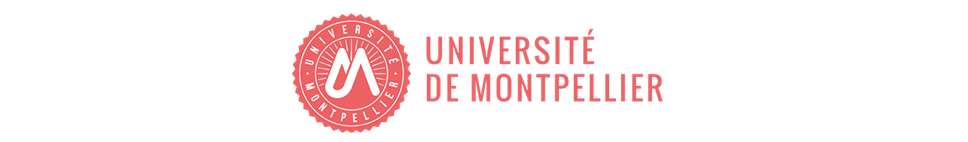}
\vspace{-15mm}}
\end{titlepage}

\newgeometry{top=2cm, bottom=2.5cm, left=2cm, right=2cm}


\pagenumbering{arabic}
\sloppy

\titlepage

\pagenumbering{roman}

\setcounter{page}{0}
\cleardoublepage

\section*{Acknowledgments}

I would like to express my deepest, and most sincere gratitude to my advisor, Felix, for his guidance and support over the past three years. Looking back on the beginning of this journey, I am extremely proud of how much I have learned along the way, and this is largely due to his mentorship and the countless discussions we shared.

I am deeply thankful to Michele for his friendship and advice. Working with him has been a great pleasure and a lot of fun, and I feel that I have learned enormously simply by observing his way of reasoning and approaching problems. Not to mention the tennis matches, the countless “manifs”, and the unconventional dance performances he introduced me to. I sincerely hope that our friendship will continue both within and beyond physics.

My thanks go to Valerie Domcke and Alberto Salvio for agreeing to review this manuscript and for their insightful comments. I am also grateful to Aldo Deandrea, Yann Mambrini, and Alberto Mariotti for their questions and for making the day of my thesis defense both inspiring and instructive.

I feel deeply indebted to “Prof.” Alfredo for his support during my Master’s thesis and for his continued encouragement since then. Working with him during the (very) early years of my career has strongly influenced the kind of researcher I aspire to become. If I look back on those days with nostalgia, it is largely thanks to Antonio, one of the smartest, and at the same time craziest, people I have ever met. Working side by side with him during those hectic days taught me a great deal, and I am grateful for his sincere friendship.

During my journey, I was fortunate to meet many brilliant people who made these three years one of the happiest periods of my life, despite the inevitable hardships of a PhD. I would like to thank Marco for always being a role model and a source of inspiration, both within and beyond physics. I have always thought of him as an academic big brother. I greatly enjoyed our countless discussions about politics and philosophy, not to mention the fact that he hosted me on his couch for a month when I first arrived in Montpellier and was struggling to find an apartment.\footnote{I am also deeply grateful to Théo and Hugo for opening their home to me even though we barely knew each other. This immediately showed me what kind of amazing people they are.}

My deep thanks to Khalil and Théo for being such good friends during these last three years, as well as for being so inspirational. I could always count on them for help and advice in the most critical moments of my PhD.\footnote{I feel I should apologise to Khalil for annoying him with all the questions about the bureaucratic procedures concerning the organisation of the thesis defence. His only fault was to be much more well organised than I was.} I will carry the parties, “manifs'', padel and football games, and karaoke nights as the sweetest memories of these years, with the only exception of the night we went to the Panama club.

A huge shoutout to the “Menton office” (old and new) for making coming to the lab every day a pleasure, for the laughs, the jokes, and all the shared complaints. I really felt like I was part of a small family. A big, big acknowledgment to the “Lemon friends” for the extraordinary trip to the most beautiful city in the world (after Ostia Lido): Menton. My greatest gratitude goes to all the amazing and brilliant people who made leaving Montpellier so sad: Alice, Benjamin, Chadi, Daniel, Elsa, Francesca, Guille, Hugo, Jack, Julien, Juliette, MJ, Natalie, Pierre, Thomas A., Thomas M., Theodore, Théo D. and Tristan.

I would also like to acknowledge all the amazing people I met during conferences and schools, for the interesting discussions and a lot of fun. Among others, I am especially thankful to Enrico, Giorgio, Maya, Martina, Pablo, and Riccardo.

I owe special thanks to my lifetime friends Alessio, Binny, Hilda, Moncio, Riccardo, Soraya, and Vyali for our legendary nights (including and especially the ones we barely survived). The fact that we are still close to each other after so many years, and despite life having taken us down different roads, is proof of how deep our friendship is. I feel truly blessed because of this.

A seat of honor in these acknowledgments is reserved for my parents, Isabella and Giuseppe, for always cheering for me and for always letting me free to explore and question, while gently guiding me through life. I will always be grateful to them for teaching me, through their example, how to be a kind of person to be proud of. If I achieved this goal, it is entirely thanks to the way they raised me. I also thank my grandparents, Ettore and Luciana; my uncles, Cristiana and Roberto; and my “little” cousins, Rocco, Eva, and Bully, for all the moments we shared and for always making home a place I am happy to return to. A special thanks to my brother Germano, whom I deeply admire and with whom I share a childish sense of wonder for the world, for growing up with me.

Dulcis in fundo, I want to thank Benedetta for her unconditional love and support over the past three years, and for always being there despite the many kilometres of distance. I also thank her for all these years of fun, laughter, and adventures around the world. She is one of the people I admire most, as well as my favourite person, and I am truly happy to have shared this journey with her.

\dominitoc
\tableofcontents


\chapter*{List of acronyms}
\mtcaddchapter[List of acronyms]


\begin{acronym}
\renewcommand{\\}{}
\acro{1PI}{one-particle irreducible}
\end{acronym}

\begin{acronym}
\renewcommand{\\}{}
\acro{BBN}{Big-Bang nucleosynthesis}
\end{acronym}

\begin{acronym}
\renewcommand{\\}{}
\acro{BPS}{Bogomol'nyi-Prasad-Sommerfield}
\end{acronym}

\begin{acronym}
\renewcommand{\\}{}
\acro{CDM}{cold dark matter}
\end{acronym}

\begin{acronym}
\renewcommand{\\}{}
\acro{CMB}{cosmic microwave background}
\end{acronym}

\begin{acronym}
\renewcommand{\\}{}
\acro{CS}{cosmic string}
\end{acronym}

\begin{acronym}
\renewcommand{\\}{}
\acro{CW}{Coleman-Weinberg}
\end{acronym}

\begin{acronym}
\renewcommand{\\}{}
\acro{DM}{dark matter}
\end{acronym}

\begin{acronym}
\renewcommand{\\}{}
\acro{DW}{domain wall}
\end{acronym}

\begin{acronym}
\renewcommand{\\}{}
\acro{EFT}{effective field theory}
\end{acronym}

\begin{acronym}
\renewcommand{\\}{}
\acro{ET}{Einstein Telescope}
\end{acronym}

\begin{acronym}
\renewcommand{\\}{}
\acro{FIMP}{feebly-interacting massive particle}
\end{acronym}

\begin{acronym}
\renewcommand{\\}{}
\acro{FLRW}{Friedmann–Lemaître–Robertson–Walker}
\end{acronym}

\begin{acronym}
\renewcommand{\\}{}
\acro{FOPT}{first-order phase transition}
\end{acronym}

\begin{acronym}
\renewcommand{\\}{}
\acro{GUT}{grand-unified theory}
\end{acronym}

\begin{acronym}
\renewcommand{\\}{}
\acro{GW}{gravitational wave}
\end{acronym}

\begin{acronym}
\renewcommand{\\}{}
\acro{IR}{infrared}
\end{acronym}

\begin{acronym}
\renewcommand{\\}{}
\acro{KZ}{Kibble-Zurek}
\end{acronym}

\begin{acronym}
\renewcommand{\\}{}
\acro{LISA}{Laser Interferometer Space Antenna}
\end{acronym}

\begin{acronym}
\renewcommand{\\}{}
\acro{LVK}{LIGO-Virgo-KAGRA}
\end{acronym}

\begin{acronym}
\renewcommand{\\}{}
\acro{MB}{Maxwell-Boltzmann}
\end{acronym}

\begin{acronym}
\renewcommand{\\}{}
\acro{NG}{Nambu-Goto}
\end{acronym} 

\begin{acronym}
\renewcommand{\\}{}
\acro{NGB}{Nambu-Goldstone boson}
\end{acronym}

\begin{acronym}
\renewcommand{\\}{}
\acro{NI}{natural inflation}
\end{acronym} 

\begin{acronym}
\renewcommand{\\}{}
\acro{NP}{new Physics}
\end{acronym}

\begin{acronym}
\renewcommand{\\}{}
\acro{PBH}{primordial black hole}
\end{acronym}

\begin{acronym}
\renewcommand{\\}{}
\acro{pNGB}{pseudo Nambu-Goldstone boson}
\end{acronym}

\begin{acronym}
\renewcommand{\\}{}
\acro{PT}{phase transition}
\end{acronym}

\begin{acronym}
\renewcommand{\\}{}
\acro{PTA}{pulsar timing array}
\end{acronym}

\begin{acronym}
\renewcommand{\\}{}
\acro{QED}{quantum electrodynamics}
\end{acronym}

\begin{acronym}
\renewcommand{\\}{}
\acro{RGE}{renormalisation group equation}
\end{acronym}

\begin{acronym}
\renewcommand{\\}{}
\acro{sFOPT}{strongly first-order phase transition}
\end{acronym}

\begin{acronym}
\renewcommand{\\}{}
\acro{SGWB}{stochastic gravitational wave background}
\end{acronym}

\begin{acronym}
\renewcommand{\\}{}
\acro{SM}{Standard Model}
\end{acronym}

\begin{acronym}
\renewcommand{\\}{}
\acro{SOPT}{second-order phase transition}
\end{acronym}

\begin{acronym}
\renewcommand{\\}{}
\acro{SSB}{spontaneous symmetry breaking}
\end{acronym}

\begin{acronym}
\renewcommand{\\}{}
\acro{UV}{ultraviolet}
\end{acronym}

\begin{acronym}
\renewcommand{\\}{}
\acro{VEV}{vacuum expectation value}
\end{acronym}

\begin{acronym}
\renewcommand{\\}{}
\acro{WIMP}{weakly-interacting massive particle}
\end{acronym}

\begin{acronym}
\renewcommand{\\}{}
\acro{wFOPT}{weakly first-order phase transition}
\end{acronym}

\mainmatter

\chapter{Introduction}
\label{chap:intro}
\minitoc

Our current understanding of Nature is based on two fundamental cornerstones: the Standard Model of particle physics and the $\Lambda$CDM model of cosmology. Together, they provide a robust description of Physics across a wide range of energies and scales. However, in spite of their tremendous experimental success, they both share flaws. The Standard Model fails to explain observed phenomena, including neutrino oscillations, the nature of dark matter, and the asymmetry between matter and antimatter abundances in the Universe. Besides these problems, the Standard Model also presents some naturalness issues or puzzles, namely the strong CP problem, the cosmological constant puzzle and the electroweak hierarchy problem. The latter, in particular, is due to the lack of a convincing explanation for the smallness of the Higgs boson mass. On the other hand, the Standard Model of Cosmology is plagued by tensions, the most infamous one being the Hubble tension, and it does not explain the origin of dark matter nor the one of dark energy. Likewise, it does not provide an explanation for the origin of the initial conditions for our Universe. 

At the core of the Standard Model lies the concept of spontaneous symmetry breaking which, through the Higgs mechanism, accounts for the origin of the masses of the elementary particles we have observed so far.\footnote{Neutrinos are the only exception. The origin of their masses still remains unexplained.} In particular, it explains the reason why the $W^\pm$ and $Z$ gauge bosons are massive, while the photon is not. When the Higgs boson acquires a non-zero vacuum expectation value, it breaks the electroweak gauge symmetry, $\SU{2}_w \times\U{1}_Y$, down to the electromagnetic subgroup, $\U{1}_{\rm em}$. For each broken generator, a Nambu-Goldstone boson appears as a flat direction in the vacuum manifold emerging after the breaking. The Nambu-Goldstone bosons are ``eaten'' by the gauge bosons, which become massive. The photon is associated with the unbroken symmetry and, therefore, remains massless. The discovery of the Higgs boson in 2012 provided a striking confirmation of the Standard Model, highlighting the central role played by spontaneous symmetry breaking in fundamental Physics. Beyond the electroweak sector, spontaneous symmetry breaking could occur in new gauge theories with potentially interesting phenomenology.

The phenomenon of spontaneous symmetry breaking is intimately connected to Nambu-Goldstone bosons and pseudo-Nambu-Goldstone bosons, which are particularly relevant, for they are among the few types of elementary scalar fields whose masses are naturally small. In fact, due to their notorious sensitivity to the ultraviolet-scale, it is in general difficult to obtain light scalar fields in a natural way, with the Higgs boson being no exception. This is the aforementioned electroweak hierarchy problem. Nambu-Goldstone bosons and pseudo-Nambu-Goldstone bosons are related to (approximately) flat directions in the vacuum manifold emerging after spontaneous symmetry breaking, and their masses are ensured to be small by an (approximate) shift symmetry. New classes of naturally light scalars are compelling as they could shed light on the reason why the measured Higgs boson mass is small, possibly solving the electroweak hierarchy problem. Moreover, they play a fundamental role in cosmology as they can provide viable dark matter candidates, or serve as the inflaton driving a stage of exponential expansion of the Universe. 

In the Hot Big Bang scenario, the Universe began its evolution in an extremely hot and dense state. At sufficiently high temperatures, the electroweak symmetry was restored and the Higgs boson had a vanishing vacuum expectation value. In this symmetric phase, the early Universe was quite different from what we observe today, as all the known particles were massless. As the Universe expanded and cooled, a  cosmological phase transition, in which the Higgs field acquired a non-zero VEV and the electroweak symmetry was spontaneously broken, occurred at $T\simeq 100\,{\rm GeV}$. This picture illustrates the cosmological relevance of spontaneous symmetry breaking. In fact, if other gauge sectors existed beyond the Standard Model, they may have undergone a phase transition in the early Universe as well. The observable signatures related to cosmological phase transitions, including gravitational wave signals, dark matter production, and baryogenesis, make them an interesting way of probing New Physics at energy scales that are far beyond the reach of current colliders. 

In this thesis, we investigate the cosmological consequences of spontaneous symmetry breaking, with particular emphasis on scenarios in which the vacuum manifold has a non-trivial structure. Our aim is to examine the role that spontaneous symmetry breaking may play in tackling some of the issues that we have outlined above in both the Standard Model and the $\Lambda$CDM model.

Firstly, we focus on models in which symmetries are broken by scalar fields in large irreducible representations, giving rise to accidentally flat directions in the vacuum manifold. These directions are not associated with any Nambu-Goldstone boson or broken generator, and are lifted by radiative corrections since there is no symmetry protecting them. If the theory is assumed perturbative, the resulting fields, which we dub ``accidents'' to stress the fact that their presence is not directly related to the symmetries of the model, represent a new class of light elementary scalars. Since they are naturally the lightest charged particles, accidents are compelling candidates for dark matter, whose stability follows from the gauge structure of the model, without the need to impose any ad-hoc symmetry. Moreover, the potential of such accidentally light scalars is naturally flat and, hence, suitable to drive slow-roll inflation. We explicitly construct a model of hybrid inflation in which the inflaton is an accidentally light scalar. We also discuss the possibility of generating detectable gravitational waves after the end of inflation. 

Furthermore, when the vacuum manifold has a non-trivial topology, it can give rise to stable topological defects. Such objects consist of finite-energy, extended field configurations whose stability is ensured by the topology of the vacuum. We will study cosmic strings and domain walls forming at the end of inflation in the scenario of hybrid accident inflation discussed above. Such topological defects are expected to produce a stochastic background of gravitational waves during their evolution. Moreover, we focus on the production of dark monopoles during cosmological phase transitions triggered by thermal fluctuations. Our interest is in investigating their possible contribution to the present-day dark matter relic abundance. 

This thesis is organised as follows. In chapter \ref{chap:EffPot}, we review the effect of radiative corrections on scalar field potentials. We begin with an introduction to the problem of naturalness, showing in a concrete example that the masses of scalar fields are quadratically sensitive to the ultraviolet-scale. This is the origin of the electroweak hierarchy problem. We then discuss the tool that allows us to investigate spontaneous symmetry breaking at an arbitrary order in perturbation theory: the effective potential. This is particularly useful in classically scale-invariant theories, such as the Coleman-Weinberg model. This formalism can be extended to include the effect of thermal corrections, which has a central role in the study of cosmological phase transitions. Chapter \ref{chap:DM} is devoted to an introduction to  the Standard Model of Cosmology. After motivating the need for an extra non-baryonic matter component in the Universe, we discuss two of the main paradigms aiming at explaining the nature of dark matter: freeze-out and freeze-in. The last part of this chapter deals with cosmic inflation, showing how it solves the horizon problem, generating, at the same time, the initial conditions from which our Universe evolved. The theory of preheating is also discussed. In chapter \ref{chap:TopoDef}, we review topological defects: extended field configurations that are stable due to the non-trivial topology of the vacuum manifold emerging in some scenarios of spontaneous symmetry breaking. We show the existence of monopole, cosmic string, and domain wall solutions in simple models, and we comment on the cosmological impact such objects may have if they were produced during cosmic history. This amounts to providing an extra matter component in the case of monopoles (topic to which we devote chapter \ref{chap:MonoDM}), or gravitational wave generation for cosmic strings and domain walls. This ends the first, introductory, part of the thesis. Chapter \ref{chap:Accidents} is based on the publication
\begin{itemize}
\item \cite{Brummer:2023znr}: F. Brümmer, G. Ferrante, M. Frigerio, and T. Hambye, “Accidentally light scalars from large
representations,” JHEP, vol. 01, p. 075, 2024, 2307.10092.
\end{itemize}
We investigate a novel type of scalar fields, the accidents, which emerge from spontaneous symmetry breaking by large representations. Accidents are massless at the tree-level, even though they are not Nambu-Goldstone bosons nor pseudo-Nambu-Goldstone bosons. They appear as extra flat directions in the vacuum manifold, which are not associated with any symmetry. Radiative corrections generate a mass, lifting the accidentally flat direction. In chapter \ref{chap:AccInf}, we make use of accidents to construct a successful model of hybrid inflation. This discussion is based on the publication
\begin{itemize}
\item \cite{Brummer:2024ejc}: F. Brümmer, G. Ferrante, and M. Frigerio, “Hybrid inflation and gravitational waves from accidentally light scalars,” Phys. Rev. D, vol. 110, no. 10, p. 103506, 2024, 2406.02531.
\end{itemize}
In our model, the flatness of the potential required to drive slow-roll inflation is readily explained by the accident nature of the inflaton field, and, most importantly, it is protected against radiative corrections. At the end of inflation, a signal of gravitational waves is produced during the process of tachyonic preheating. We also study minimal modifications of the model, which feature the production of cosmic strings and domain walls. The latter produce a stochastic gravitational wave background which could explain the detection recently reported by the PTA collaborations. Chapter \ref{chap:MonoDM} is based on the works 
\begin{itemize}
\item \cite{smallpaper}: F. Brümmer, G. Ferrante, T. Fischer, and M. Frigerio, “No room for monopole dark matter,” 9 2025, 2509.21924.
\item \cite{bigpaper}: F. Brümmer, G. Ferrante, T. Fischer, and M. Frigerio, “The price for monopole dark matter,” to appear.
\end{itemize}
We investigate the production of monopoles from cosmological phase transition in a dark $\SO{3}$ model broken to $\SO{2}$ by a real scalar triplet. This model features the presence of two stable relics, the massive gauge bosons associated with the broken generators, and magnetically charged dark monopoles. We thoroughly explore the parameter space, and study different realisations of the phase transition, assessing whether a scenario in which the relic abundance observed in the Universe today is mostly composed of monopoles, with massive vector bosons being under-abundant, is possible. Finally, in chapter \ref{chap:Concl} we summarise our discussion and we outline future prospects.  

\chapter{Radiative corrections: naturalness and the effective potential}
\label{chap:EffPot}
\minitoc

The first part of this chapter introduces the concept of naturalness in particle physics, with a focus on the electroweak hierarchy problem. After giving a precise definition of what it means for a parameter to be naturally small, we scrutinise the masses of scalar, fermion, and gauge boson fields, thus framing the hierarchy problem. We thoroughly show its implications in a minimal toy-model, exposing the quadratic sensitivity of scalar field masses to the \ac{UV} scale. The interesting conclusions we draw from this toy-model apply to the \ac{SM} as well, and lay the groundwork for a discussion on the electroweak hierarchy problem and some notable attempts at solving it.

In the second part we discuss an extremely powerful tool to compute radiative corrections to scalar field potentials: the effective action. We show two different ways in which the one-loop effective potential can be practically computed: by summing an infinite series of one-loop diagrams, and through the background field method. For the former we also study a concrete example, showing the explicit computation of the effective potential in a scale-invariant version of scalar electrodynamics, also known as the Coleman-Weinberg potential.

Finally, in Sec.~\ref{sec:TQFT}, we see how to extend the effective potential computation to include thermal effects. This has interesting implications in the cosmological \ac{PT} context. An overview of the tunneling process in \ac{PT}s of the first order is also provided.

\section{Naturalness}

The notion of naturalness is deeply related to that of fine-tuning. In general, we say that a parameter of our theory is natural if its numerical value is not the result of high-precision cancellation between some other more fundamental, unrelated parameters. While many quantities in the \ac{SM} are natural, with few of them being predicted by naturalness-oriented reasoning, there are three major naturalness puzzles that are still open today: the cosmological constant problem \cite{Bousso:2007gp}, the electroweak hierarchy problem and the strong CP problem \cite{Hook:2018dlk}. Such puzzles all share the inconvenience of a parameter (the cosmological constant, the Higgs mass, and the QCD $\theta$-term, respectively) that is observed to be smaller than what one would expect according to naturalness.  

The history of particle physics has validated naturalness reasoning as a powerful tool for discovering \ac{NP}. In fact, the apparent presence of accidental cancellations between unrelated parameters has often hidden the fact that our understanding of Nature was incomplete and had to be extended.\footnote{The history of the past 100 years of particle physics is studded with successful postdictions, and some predictions, of the naturalness principle. For a detailed historical review we refer the interested reader to Refs.~\cite{Murayama:2000dw,Giudice:2008bi,Craig:2022eqo}} Borrowing the words of Craig \cite{Craig:2022eqo}:   
\begin{displayquote}
    Throughout its history, naturalness has been variously framed as a
pragmatic strategy, a bedrock principle, an aesthetic criterion, and a catastrophic folly. In truth, it is a bit of each.
\end{displayquote}
In fact, despite its success so far, we are not guaranteed that naturalness keeps working all the way to the top in our climbing up the ladder of energy scales. However, on the one hand, we still need a discerning principle to orient future experimental efforts and, on the other hand, if naturalness had to fail at some point, it would be a big lesson to take home on how Nature works.

In this section, we will deal only with the electroweak hierarchy problem, applying the notion of naturalness to the mass terms of our theory. There is a number of excellent reviews on naturalness with a focus on the electroweak hierarchy problem; see, among the others, Refs.~\cite{Giudice:2008bi, Craig:2022eqo} and references therein.

\subsection{Technical naturalness}\label{sec:TechNat}

Naturalness was firstly brought to the attention of particle physicists by Dirac \cite{Dirac:1938mt}. His idea was that dimensionless parameters in a theory should be of order one, in appropriate units. In what could be regarded as one of the first attempts to use naturalness as a guiding principle in particle Physics, Dirac's aim was to explain the huge hierarchy between the scale of gravity, $m_{\rm Pl} \sim 10^{19}\,{\rm GeV}$, and the proton mass $m_p \sim 1\,{\rm GeV}$. With the idea that these two scales have to be related by some fundamental theory containing only order-one dimensionless constants, he proceeded to elaborate a model of cosmology where the Planck mass changes with time; a model which, we now know, is not realised in Nature. 

This formulation of the naturalness principle, known today as \textit{“Dirac naturalness''}, was refined few decades later by 't Hooft \cite{tHooft:1979rat}. If an operator $O$ in the Lagrangian explicitly breaks a symmetry of the theory, we say that it is \textit{“technically natural''} for its coefficient $c_O$ to be small. In fact, as in the limit $c_O \rightarrow 0$, the theory recovers a larger symmetry, $O$ cannot be radiatively generated by other operators, and loop corrections to $c_O$ will always be proportional to $c_O$ itself. Therefore, if $c_O \ll 1$, so will be its radiative corrections, making it natural for it to be small without the need for fine-tuning. In this case, the problem (if it can be called that) of a small $c_O$ is, then, relegated to the \ac{UV}, where we can postulate that some unknown mechanism predicts it to be small. This contrasts with a symmetry-preserving coupling, which, even if it is small in the \ac{UV}, receives large radiative corrections as we run it down in energy, thus requiring some fine-tuning to explain a small measured value in the \ac{IR}.  The concept of technical naturalness will be made more clear in a moment as we discuss mass terms and spurion analysis.

\subsection{Masses and naturalness}\label{sec:MassNat}

Equipped with the notion of technical naturalness, we can already frame the source of the hierarchy problem, simply using the concept of symmetries and spurions. This provides a very solid, model-independent, statement of the problem. The concept of spurions is very powerful as it allows us to study the selection rules of broken symmetries. Let us consider, for concreteness, an explicit example. Consider a version of \ac{QED} where the photon is massive. As far as we are concerned, we can think of it as the theory of an Abelian Higgs model endowed with a charged, massive Dirac fermion $\Psi$, in the low-energy limit in which the heavy scalar field has been integrated out. The Lagrangian density reads
\be
    \mathcal{L} = i \overline{\Psi} \slashed{D}\Psi + M_{\Psi,0} \overline{\Psi}\Psi + \frac{1}{4}F_{\mu\nu}F^{\mu\nu} + \frac{m_A^2}{2}A_\mu A^\mu\,.
\ee
The Dirac fermion can be written in terms of its Weyl spinor components as $\Psi=(\psi,\xi^\dagger)$. In the massless limit, in which $M_{\Psi,0}\rightarrow 0$, $\Psi$ possesses the typical chiral symmetry of fermions 
\be
    \U{1}_\chi:\quad\psi \rightarrow e^{i\alpha}\psi,\quad\xi^\dagger \rightarrow e^{-i\alpha}\xi^\dagger\,.
\ee
Such a symmetry is explicitly broken by the fermion mass term, which pairs up Weyl components with the same chirality. According to the definition of technical naturalness we have given above, we can, then, conclude that it is technically natural for $M_{\Psi,0}$ to be smaller than the \ac{UV} cutoff of the theory, which we take to be the mass of the Abelian Higgs. To see practically what this means, we promote $M_{\Psi,0}$ to a spurion field of (the explicitly broken) chiral symmetry. This amounts to promoting the fermion mass parameter to a fictitious field that transforms under the chiral symmetry in a way such that the mass term is now invariant under $\U{1}_\chi$
\be\label{eq:MPsi0}
    M_{\Psi,0} \rightarrow e^{-2 i\alpha}M_{\Psi,0}\,.
\ee
Now, in the theory in which the mass parameter has been promoted to a spurion field, chiral symmetry is restored and we expect physical observables to obey its selection rules. In particular, we expect  the renormalised fermion mass (which we here take to be also its pole mass by properly choosing the renormalisation conditions) to transform as $M_{\Psi}\rightarrow e^{-i2\alpha}M_{\Psi}$. However, as the only spurion with such a transformation property is the tree-level, bare mass of $\Psi$ we know that
\be
    M_{\Psi} = M_{\Psi,0}\times (\cdots)\,.
\ee
An explicit computation will tell us what the exact couplings and numerical factors are (the dots in the above expression) but for what concerns us, we have already found that it is natural for the fermion to be light. In particular, even if we had another scale in the theory (e.g., the Abelian Higgs \ac{VEV}), such scale would affect the fermion mass at most logarithmically, for it does not possess the correct transformation property given in Eq.~\eqref{eq:MPsi0}. What we have seen here, in a concrete example, is technical naturalness at work. By following the same line of reasoning for the gauge boson mass we find that, since $m_A$ breaks gauge symmetry, it is technically natural, and quantum corrections to $m_A$ will be proportional to $m_A$ itself. It is worth stressing the fact that both chiral symmetry and gauge symmetry are inherent symmetries of fermion and gauge boson fields, respectively, and they do not depend on the model in question. Therefore, we can conclude that fermion and gauge boson masses are, in general, technically natural.

If we now try to extend the above argument to elementary scalar fields we realise that there is no such a thing as an intrinsic symmetry that is explicitly broken by their mass terms. This is the culprit of the hierarchy problem: the mass of an elementary scalar field is not, in general, technically natural as there is no symmetry protecting it; therefore, it is not natural for it to be smaller than the cutoff scale up to which the theory is defined. As we will see in a concrete example below, this translates to the statement that scalar field masses are quadratically sensitive to the \ac{UV} scale or, analogously, that their mass renormalisation is not multiplicative but, rather, additive. 

Another, more heuristic way of seeing why scalar field masses are not technically natural only takes into account the number of degrees of freedom of a given representation of the Lorentz group \cite{Giudice:2008bi}. A massless spin-1 boson has two degrees of freedom while a massive one has three. Analogously, a massless chiral fermion has two degrees of freedom while a massive Dirac fermion has four. In fact, while a massless Dirac fermion possesses a definite chirality, a massive one does not. Since radiative corrections cannot generate new degrees of freedom out of nowhere, massless vector bosons and massless fermions are protected by Lorentz symmetry. On the contrary, a massless scalar field has the same amount of degrees of freedom as massive one, implying no symmetry protection from the Lorentz group.

\subsection{The hierarchy problem in a toy-model}\label{sec:ToyModHier}
We will now put flesh on the bones of the above discussion by considering the simplest toy model where the hierarchy problem makes its appearance: a theory of two real interacting scalar fields, $\varphi$ and $\phi$, whose masses are hierarchical, $m \ll M$. This model, in spite of its simplicity, will allow us to draw some conclusions that hold for the \ac{SM} as well. To simplify computations we impose a $\mathbb{Z}_2$ symmetry, under which both fields are charged. In the full theory, the renormalised Lagrangian density expressed in terms of renormalised quantities and counterterms is \footnote{In this simple toy model, the wavefunction renormalisation of both the scalar fields is trivial.}
\be
\begin{aligned}\label{eq:Lfull}
    \mathcal{L} \supset & +\frac{1}{2}(\partial_\mu \varphi)^2-\frac{m^2}{2}\varphi^2 - \frac{\lambda_{\varphi}}{4!}\varphi^4 + \frac{1}{2}(\partial_\mu \phi)^2 - \frac{M^2}{2}\phi^2 - \frac{\lambda_{\phi}}{4!}\phi^4 - \frac{\lambda_{\varphi\phi}}{4}\varphi^2\phi^2\\
    & -\left(Z_m -1\right) \frac{m^2}{2} \varphi^2 - \left(Z_{\lambda_{\varphi}}-1\right)\frac{\lambda_\varphi}{4!}\varphi^4 - \left(Z_M -1\right) \frac{M^2}{2} \phi^2 - \cdots\,,
\end{aligned}
\ee
where the dots contain the remainder of the counterterms, which we have not written explicitly for the sake of brevity. Terms proportional to counterterms give rise to Feynman diagrams we need to take into account when computing loop corrections to our theory.

Suppose that experiments allow us to access low energies, of the order of $m$. Since at those energies the heavy scalar field cannot be produced as a propagating degree of freedom, we can simplify our life and integrate it out. This amounts to working with an \ac{EFT}, which is the low-energy limit of the theory described by Eq.~\eqref{eq:Lfull}, where operators involving $\phi$ do not appear. The process of integrating out the heavy mode from Eq.~\eqref{eq:Lfull} modifies the couplings:
\be
\begin{aligned}
    \mathcal{L}_{\rm EFT} \supset &+\frac{1}{2}(\partial_\mu \varphi)^2-\frac{m_E^2}{2}\varphi^2 - \frac{\lambda_E}{4!}\varphi^4\\
    & -(Z_{m_E}-1)\frac{m_E^2}{2}\varphi^2 - (Z_{\lambda_E}-1)\frac{\lambda_E}{4!}\varphi^4\,.
\end{aligned}
\ee
The subscript $``E"$ denotes the parameters of the \ac{EFT}, which in general are different from those of the full theory. The correct way of dealing with theories with separation of scales is by matching the computations of the full theory and the ones of the  \ac{EFT} at some matching scale, and then running the couplings in the  \ac{EFT} down to low energies, in order to extract physical predictions in the \ac{IR}. This procedure is called \textit{``matching and running''} and is nicely reviewed in Ref.~\cite{Cohen:2019wxr}. Since we are after exposing the hierarchy problem, the quantity we shall focus on is the light scalar mass.

\begin{figure} \centering
 \includegraphics[width=.9\textwidth]{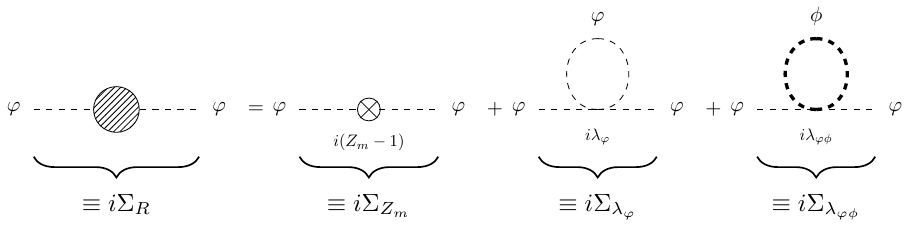}
 \caption{One-loop contributions to the light scalar two-point function. We include, among the radiative corrections, also the diagram arising from the counterterm.}
\label{fig:m1L}
\end{figure}

We begin by computing the mass of $\varphi$ in the full theory, where, at the one-loop level, the light scalar two-point function receives contributions from the diagrams in Fig.~\ref{fig:m1L}. 
The correction coming from the counterterm can be read straightforwardly from the Lagrangian
\be
    i \Sigma_{Z_m} = - i m^2(Z_m -1)\,.
\ee
A light-scalar loop contributes as
\be
    i\Sigma_{\lambda_{\varphi}} = - i\frac{\lambda_\varphi}{2} \int \frac{d^4p}{(2\pi)^4}\frac{i}{p^2-m^2}\,.
\ee
By naive power counting, this integral is quadratically divergent. We decide to make use of dimensional regularisation \cite{Bollini:1972ui, tHooft:1972tcz}, and analytically continue the number of spacetime dimensions to $D=4-2\epsilon$, with $\epsilon\ll1$ a small, positive number. The integral then becomes
\be\label{eq:mvarphi1l}
    i\Sigma_{\lambda_\varphi} =  -i\frac{\lambda_\varphi}{2}\tilde\mu^{2\epsilon}\int \frac{d^D p}{(2\pi)^D}\frac{i}{p^2 - m^2} = i m^2 \frac{\lambda_\varphi}{32\pi^2}\left(1 + \frac{1}{\epsilon} +\log\frac{\mu^2}{m^2} + \mathcal{O}(\epsilon)\right)\,,
\ee
where the term $\tilde\mu^{2\epsilon}$ has been added to ensure that the action remains dimensionless even in $D$ dimensions and, in the second equality, we have defined $\mu^2 \equiv 4\pi e^{-\gamma_E}\tilde\mu^2$. Here, $\gamma_E \simeq 0.577$ is the Euler-Mascheroni constant. To solve the above integral, we have used the general identity \cite{Schwartz:2014sze}
\be
    \int \frac{d^D k}{(2\pi)^D}\frac{k^{2a}}{(k^2 - \mathcal{M}^2)^b} = i (-1)^{a-b}\left(\mathcal{M}^2\right)^{\frac{D}{2}+a-b}\frac{\Gamma(a+\frac{D}{2})\Gamma(b-a-\frac{D}{2})}{\Gamma(b)\Gamma(\frac{D}{2})}\,,
\ee
and we have expanded the result for small $\epsilon$, remembering that $\mu^\epsilon = 1+ \epsilon \log \mu + \mathcal{O}(\epsilon^2)$ and $\Gamma(\epsilon) = 1/\epsilon + \gamma_E + \mathcal{O}(\epsilon)$.

Unsurprisingly, the result diverges in the limit $\epsilon \rightarrow 0$. However, all the divergent terms will be absorbed by properly choosing the counterterm $Z_m$. Such a choice for the counterterms defines the $\overline{\rm MS}$ renormalisation scheme. Before writing explicitly $Z_m$, we first need to evaluate the other radiative contribution to $m^2$, coming from a loop of the heavy scalar. The integral can be computed straightforwardly from the above one just by replacing $\lambda_\varphi \rightarrow \lambda_{\varphi \phi}$ and $m^2 \rightarrow M^2$:
\be
    i\Sigma_{\lambda_{\varphi\phi}} =  i M^2 \frac{\lambda_{\varphi\phi}}{32\pi^2}\left(1 + \frac{1}{\epsilon} +\log\frac{\mu^2}{M^2} + \mathcal{O}(\epsilon)\right)\,.
\ee
In order to remove the divergent terms in the one-loop contributions to the $\varphi$ self-energy, we define
\be
    Z_m = 1+ \frac{1}{\epsilon} \left(\frac{\lambda_\varphi}{32\pi^2}+\frac{\lambda_{\varphi\phi}}{32\pi^2}\frac{M^2}{m^2}\right)\,.
\ee
The renormalised self-energy of the light scalar field is, then,
\be
    \Sigma_R \equiv \Sigma_{\lambda_\varphi}+\Sigma_{\lambda_{\varphi\phi}}+\Sigma_{Z_m} = \frac{\lambda_{\varphi}}{32\pi^2} m^2 \left(1 + \log \frac{\mu^2}{m^2}\right) +\frac{\lambda_{\varphi\phi}}{32\pi^2} M^2 \left(1 + \log \frac{\mu^2}{M^2}\right)\,.
\ee
Notice that in the $\overline{\rm MS}$ scheme, the choice of the counterterms is merely based on convenience, and it is not, by any means, informed by physical measurements. As a result, the renormalised mass $m^2$ appearing in the Lagrangian is, in general, different from the physical mass, which is, instead, the pole of the propagator. For this reason, we will refer to the renormalised mass as the $\overline{\rm MS}$ mass. After including radiative corrections, the inverse $\varphi$ propagator is
\be
\Delta^{-1}_\varphi(p^2) = p^2 - m^2 + \Sigma_R\,.
\ee
Therefore, the physical mass $m_{\rm phys}^2$ is
\be\label{eq:mphys}
    m_{\rm phys}^2 = m^2 - \Sigma_R = m^2 - \frac{\lambda_{\varphi}}{32\pi^2} m^2 \left(1 + \log \frac{\mu^2}{m^2}\right) -\frac{\lambda_{\varphi\phi}}{32\pi^2} M^2 \left(1 + \log \frac{\mu^2}{M^2}\right)\,.
\ee
This is our physical input in the renormalisation procedure. At first sight, it may seem worrisome that a physical parameter, $m_{\rm phys}^2$, depends on the choice of $\mu$, which is an arbitrary scale. However, this inconsistency is only apparent, and it is due to the fact that we have not specified what the $\overline{\rm MS}$ mass is, yet. Indeed, we choose it in a way such that it cancels the unphysical $\mu$-dependence in Eq.~\eqref{eq:mphys}. Practically speaking, this is done by inverting Eq.~\eqref{eq:mphys}, using the fact that $m_{\rm phys}^2 = m^2$ at the tree level:
\be
    m^2(\mu) = m_{\rm phys}^2 + \frac{\lambda_{\varphi}}{32\pi^2} m_{\rm phys}^2 \left(1 + \log \frac{\mu^2}{m_{\rm phys}^2}\right) +\frac{\lambda_{\varphi\phi}}{32\pi^2} M^2 \left(1 + \log \frac{\mu^2}{M^2}\right)
\ee

We repeat the same exercise in the low-energy \ac{EFT}, where $\varphi$ is the only propagating degree of freedom, getting
\be
    i\Sigma_{{\lambda}_E} =  i m_E^2 \frac{\lambda_E}{32\pi^2}\left(1 + \frac{1}{\epsilon} +\log\frac{\mu^2}{m_E^2}\right)\,.
\ee
Therefore, choosing the \ac{EFT} mass counterterm as
\be
    Z_{m_E} = 1+ \frac{1}{\epsilon} \frac{\lambda_E}{32\pi^2}\,,
\ee
gives the $\overline{\rm MS}$ low-energy mass 
\be\label{eq:mEMS}
    m_{E}^2(\mu) = m_{\rm phys}^2 + \frac{\lambda_{E}}{32\pi^2} m_{\rm phys}^2 \left(1 + \log \frac{\mu^2}{m_{\rm phys}^2}\right)\,.
\ee

To make sure that the \ac{EFT} reproduces, in its regime of validity, the same predictions as the full theory, we now impose that the $\overline{\rm MS}$ masses agree with each other at some matching scale $\mu_M$: $ m_{E}^2 (\mu_M) = m^2 (\mu_M)$. Intuitively, a reasonable choice for such matching scale is the physical mass of the heavy particle, $M_{\rm phys}$, as it denotes the threshold above which $\phi$ becomes dynamical and the \ac{EFT} breaks down. The matching condition enforces the following relation 
\be\label{eq:mMatch}
    m_E^2(M_{\rm phys}) = m^2(M_{\rm phys}) -\frac{\lambda_{\varphi\phi}}{32\pi^2}M_{\rm phys}^2\,,
\ee
where we have imposed $\lambda_{\varphi} = \lambda_E$ at tree-level.\footnote{In principle, also the quartic couplings run with energy. However, in our discussion this turns out to be a two-loop effect and, hence, we neglect it, ignoring any scale dependence of dimensionless parameters.}  Notice, also, that $M = M_{\rm phys}$ at the lowest order in perturbation theory. We already see where the problem is: by matching the \ac{EFT} onto the full theory, we get a threshold correction to the light mass which is proportional to the heavy one. This is precisely the source of the (electroweak) hierarchy problem.

Our ultimate goal is to derive a prediction for the light mass in the \ac{EFT}, at some low energy which can be fixed to be of the order of $m_{\rm phys}$. To this end, we need to compute the running of $m_E$. The \ac{RGE} for $m_E$ is given by
\be
    \frac{d m_E^2}{d \log \mu^2} = \gamma_{m_E} m_E^2\qquad{\rm with}\qquad \gamma_{m_E} = \lim_{\epsilon \rightarrow 0} \epsilon \lambda_E \frac{\partial Z_{m_E}}{\partial \lambda_E} = \frac{\lambda_E}{32\pi^2}\,.
\ee
We can solve the \ac{RGE} assuming that $m_E^2$ appearing on the right-hand side of the above equation is constant, as its running is a two-loop effect, and then run $m_E^2$ from the matching scale $M_{\rm phys}$ down to $\mu$. Then, in the approximation in which we only retain the leading log contribution,\footnote{Due to the large hierarchy between $m$ and $M$, the leading log approximation is not a good one. Nonetheless, here we assume that coupling constants are small enough so that we can expand the resummed expression for $m_E^2(\mu)$.} we have
\be
    m_E^2(\mu) = m^2(M_{\rm phys})+\frac{\lambda_E}{32\pi^2}m^2(M_{\rm phys})\log\left(\frac{\mu^2}{M_{\rm phys}^2}\right)-\frac{\lambda_{\varphi\phi}}{32\pi^2}M_{\rm phys}^2\,,
\ee
where we have used Eq.~\eqref{eq:mMatch} as the boundary condition for the \ac{RGE}. The physical mass $m^2_{\rm phys}$ is related to the $\overline{\rm MS}$ mass above by $m_{\rm phys}^2 -  m_{E,\overline{\rm MS}}^2(\mu=m_{\rm phys})=0$, which can be inverted to give
\be\label{eq:mphysEFT}
     m_{\rm phys}^2 = m^2(M_{\rm phys}) + \frac{\lambda_E}{32\pi^2}m^2(M_{\rm phys})\left(1+\log \frac{M_{\rm phys}^2}{m_{\rm phys}^2}\right) - \frac{\lambda_{\varphi\phi}}{32\pi^2}M_{\rm phys}^2\,,
\ee
up to corrections which are of higher order in the couplings.

After this lengthy calculation, let us recap briefly what we have achieved and comment on the result. We have started with a theory containing a light elementary scalar field and a heavy one. Since we wanted to extract predictions valid at energies of the order $m$, we decided to integrate out the heavy field and work with an \ac{EFT} containing only $\varphi$ as a propagating degree of freedom. After computing the one-loop corrected mass of the light field in both the full theory and the \ac{EFT}, we have matched the two computations at the scale $M$. The matching condition ensured us that the prediction of the \ac{EFT} is the same as the one of the full theory, in the regime where the former is defined. The matching condition, however, introduced a large correction to $m_{\rm phys}^2$, proportional to the other scale in the game, $M_{\rm phys}$. Now, by looking at Eq.~\eqref{eq:mphysEFT}, we see that, if we insist on having a light scalar field in the spectrum, we need to fine-tune the high-energy parameter $m^2(M_{\rm phys})$ against the threshold correction coming from the heavy field, to a degree which is as large as the hierarchy between $m_{\rm phys}^2$ and $M_{\rm phys}^2$. Or, seen from a different perspective, a small change in the numerical value of the \ac{UV} parameter $m^2(M_{\rm phys})$ will have a huge impact on our low-energy prediction for $m^2_{\rm phys}$. As a final comment, let us consider the limit in which $\lambda_{\varphi\phi} \rightarrow 0$, and the two fields decouple. From Eq.~\eqref{eq:mphysEFT} we can see that, in this decoupling limit, there is no fine-tuning whatsoever: $m_{\rm phys}^2$ is proportional to the only scale, that is $m^2$. This may appear as obvious but will be relevant in the next section when addressing the electroweak hierarchy problem.

This example has shown explicitly what we have introduced in a somewhat more abstract, but at the same time more general, sense in Sec.~\ref{sec:MassNat}. The absence of symmetry protection causes the masses of scalar fields to be sensitivity to the \ac{UV} scale of the theory, in this case $M_{\rm phys}$. Therefore, requiring a scalar field to be lighter than the \ac{UV} scale always brings some degree of fine-tuning in the game. It is important to stress the fact that this fine-tuning is physical only if the scalar field physical mass is calculable, i.e., if it is a prediction in terms of physical parameters of a more fundamental theory. In this sense, the fine-tuning required by imposing that $m_{\rm phys}^2 \ll M_{\rm phys}$ in Eq.~\eqref{eq:mphysEFT} is not physical, since what we can measure is only $m_{\rm phys}^2$ as a whole, while, for example, $m^2(M_{\rm phys})$ is not a measurable quantity. On the contrary, the parameter $m^2$ is fixed once a physical measurement of $m_{\rm phys}^2$ is carried out. To promote this fine-tuning to a physical effect, we need to embed our model into some finite theory in which there are no infinities and parameters can be used to make predictions. Such a theory can be, e.g., string theory which is, indeed, finite, and which we assume completes our understanding of Nature by incorporating gravity.

\subsection{The electroweak hierarchy problem}

With all the above in mind, we can easily understand the usual statement that ``there is no hierarchy problem within the \ac{SM}''. The reason to this is twofold. First of all, due to its chiral nature, the \ac{SM} is a one-scale theory: there is only one dimensional parameter, that is the Higgs mass or, analogously, its \ac{VEV}. In fact, if we compute radiative corrections to the Higgs mass within the \ac{SM} (with emphasis on ``within''), we realise that there is no need to tune anything. In particular, one-loop corrections will be all proportional to the only scale in the game which is, again, the Higgs mass itself, analogously to what happens in the toy-model in Sec.~\ref{sec:ToyModHier} for vanishing $\lambda_{\varphi\phi}$. Otherwise stated, the Higgs mass is technically natural in the \ac{SM}.\footnote{We are intentionally neglecting the presence of the hypercharge Landau pole, pretending for a moment that the \ac{SM} is a \ac{UV}-complete theory that needs no extension.} The second reason why within the \ac{SM} there is no electroweak hierarchy problem is that the Higgs mass cannot be computed in terms of other observables and is, therefore, just an input parameter that is provided by measurements. 

Sometimes, loop corrections to the Higgs mass are computed in the \ac{SM} using a cutoff regulator
\be\label{eq:WrongHierProb}
    \delta m_H^2=\frac{\Lambda^2}{16 \pi^2}\left(-6 y_t^2+\frac{9}{4} g^2+\frac{3}{4} g^{\prime 2}+6 \lambda\right) ,
\ee
thus giving the impression that the Higgs mass receives quadratic corrections proportional to some large scale $\Lambda$. We now know that this is not true. In fact, the $\Lambda$ appearing above is just an unphysical cutoff used to regularise the otherwise divergent integral. Once physical renormalisation conditions dictated by measurements are imposed, any dependence on $\Lambda$ will disappear, leaving no hierarchy problem behind. This is seen even more clearly by using dimensional regularisation. If we focus only on the one-loop contribution coming from the top quark (which is the largest one), we see that in dimensional regularisation there is no quadratic divergence and $\delta m_H^2 \propto m_t^2$, which is the only scale appearing in the loop integral. 

\begin{figure} \centering
 \includegraphics[width=.25\textwidth]{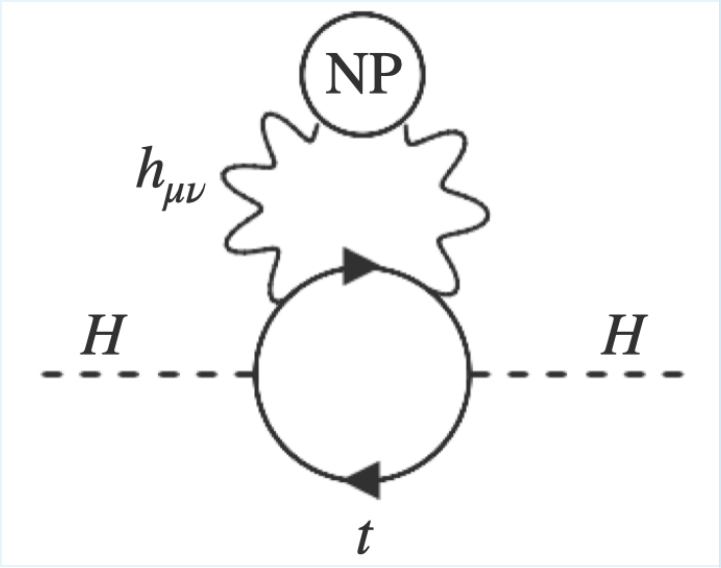}
 \caption{Even if the Higgs does not couple directly to \ac{NP}, gravity will mediate the interaction between the new sector and the \ac{SM}, via a top quark loop.}
\label{fig:HiggsGravNP}
\end{figure}

An actual problem arises as soon as we try to embed the \ac{SM} in some sort of \ac{UV} completion, in which the Higgs mass becomes a computable parameter. Now, the \ac{SM} should be seen as an \ac{EFT} which is valid up to some \ac{NP} scale $\Lambda_{\rm NP}$, which, for example, may coincide with the mass of a new particle. By trying to extend the \ac{SM} we have bought ourselves two problems: on the one hand, we have introduced in the game a second scale, making fine-tuning necessary to explain why $m_H^2 \ll \Lambda_{\rm NP}^2$ and, on the other hand, we have made the Higgs mass a computable quantity in terms of some fundamental parameters, promoting fine-tuning to a physical problem. At this point, one may ask the reason behind the urge to look for an extension of the \ac{SM} and get ourselves into trouble. First of all, if we try to extrapolate the \ac{SM} up to arbitrary energies, we realise that its validity breaks down at energies of the order of $10^{41}\,{\rm GeV}$, where the hypercharge gauge coupling hits a Landau pole, confirming the interpretation of the \ac{SM} as an \ac{EFT} which is valid only up to some scale. Moreover, there are many physical phenomena that have been observed and that the \ac{SM} cannot account for; among the others, the microscopic nature of dark matter, neutrino masses, the observed baryon asymmetry. Each of this problems calls for an extension of the \ac{SM}. Even assuming that all these observations are explained by light Physics, or by Physics that is not coupled to the \ac{SM}, we expect gravity to be completed into a quantum theory at scales of the order of the Planck mass $m_{\rm Pl}$. In this case, even if whatever \ac{NP} is out there does not interact directly with the \ac{SM}, loops mediated by gravity, like the one shown in Fig.~\ref{fig:HiggsGravNP}, will feed into the Higgs mass, pulling its value towards $\Lambda_{\rm NP}\sim m_{\rm Pl}$.

The fact that the \ac{SM} suffers from a problem of naturalness when \ac{UV}-completed was realised by Georgi et Al.~\cite{Georgi:1974yf} first, and by Susskind \cite{Susskind:1978ms} few years later, in the context of grand unification (where the Higgs mass parameter is, indeed, computable). Since then, a number of solutions have been put forward, which we can broadly divide into symmetry-related mechanisms and cosmological mechanisms. In the former case, one tries to introduce new symmetries that protect the mass of the Higgs. The most notable examples are supersymmetry \cite{Martin:1997ns}, where the Higgs boson mass is protected by the chiral symmetry of its fermionic superpartner, and composite Higgs models \cite{Contino:2010rs, Panico:2015jxa}, where the Higgs is a \ac{pNGB} (pretty much like pions in the \ac{SM}) and, therefore, its mass is protected by (approximate) shift-symmetry.\footnote{Notice, however, that no fully satisfactory, asymptotically free \ac{UV}-completion of composite Higgs models has been constructed as of yet \cite{Panico:2015jxa}. } Due to the absence of any hint of \ac{NP} up to few TeV, however, these solutions have lost most of their original appeal \cite{Dine:2015xga}, for they all predict new states close to the electroweak scale (save requiring some fine-tuning themselves). The apparent failure of such strategies has motivated interest in a different approach to the problem. Instead of trying to mitigate the amount of fine-tuning suffered by the Higgs, we may accept the fact that a small Higgs mass is, indeed, unnatural in the sense of the definition of naturalness we have given in Sec.~\ref{sec:TechNat}. However, some early-Universe conspiracy may have chosen such a small value for $m_h^2$. The idea is that there is a landscape of different possible values for the Higgs mass, and some mechanism operating in the early Universe dynamically selects the value we measure today. This reasoning can be implemented in different ways, depending on how the landscape is realised, and on the selection mechanism. For example, in \textit{``relaxion''} models \cite{Graham:2015cka,Fonseca:2019lmc}, the Higgs is coupled to an axion-like particle. The cosmological rolling of the relaxion down its potential, and, in particular, the field value it takes at the end of it, determines the electroweak scale today. In its most minimal version, the relaxion is halted at the right place in field space, i.e.~at the field value needed for a small Higgs mass, by its coupling to $\Tr G \tilde{G}$, whose value depends on $m_h^2$.  In this case, the landscape is entirely contained inside our observable Universe. Another notable example is \textit{``$N$Naturalness''} \cite{Arkani-Hamed:2016rle}, in which our Universe contains $N$ copies of the \ac{SM}. Statistically, the Higgs mass value will vary in a very wide range across the different copies, thus providing the landscape. The selection mechanism is implemented by reheating. Even if the inflaton is coupled with equal strength to all the $N$ copies of the \ac{SM}, if the branching rate of its decay to each individual sector scales as the inverse of the Higgs mass in that sector, copies with a low electroweak scale will be predominantly reheated. Yet another possibility goes under the name of \textit{``sliding naturalness''} \cite{TitoDAgnolo:2021nhd,TitoDAgnolo:2021pjo}. Here, two light scalar fields are coupled to some trigger operator (an operator whose value depends on the Higgs \ac{VEV}, like the aforementioned $\Tr G\tilde{G}$). The landscape is provided by the multiverse, i.e.~causally disconnected patches each with a different value for the Higgs mass. The combined cosmological evolution of the two light scalar fields is engineered to crunch all those Universes where either $m_h^2 =0$ or $m_h^2 \gg 125\,{\rm GeV}$.

Far from being an exhaustive record of all the possibilities that have been explored so far, the aim of this subsection is more that of sketching what the main ideas are. We refer the interested reader to Refs.~\cite{Cohen:2019wxr,Koren:2020pio,Craig:2022eqo} and references therein for a more detailed account of the existing (vast) literature on the electroweak hierarchy problem.

\section{The effective potential}
\label{sec:EffPot}
In classical field theories, the problem of finding the vacuum state amounts to minimising a function of the fields, i.e.~the classical potential. This generally allows us  to express the  \ac{VEV} of the field(s) in terms of the parameters describing the theory. Quantum corrections, however, introduce new field-dependent terms that are not manifest in the Lagrangian, and that modify the functional shape of the classical potential, possibly shifting the \ac{VEV} away from its classical value. This is intimately related to the electroweak problem, as we have seen in the first part of this chapter and, in particular, in Sec.~\ref{sec:ToyModHier}, where radiative corrections destabilised the tree-level mass of scalar fields. There, we have carried out the computation of the one-loop corrected mass for the light scalar field, $\varphi$, by computing the relevant Feynman diagrams and renormalising the theory. However, this procedure can become quickly cumbersome and unpractical as the number of degrees of freedom coupled to the scalar field increases. Moreover, to be able to reconstruct the whole one-loop-corrected potential for $\varphi$, we should have repeated the same procedure for the self-interaction coupling, $\lambda_E$. Only then, we would have been able to estimate the \ac{VEV} at the one-loop order.

In this section, we will derive the \textit{effective potential} $V_{\rm eff}$: a function whose minimisation gives the \ac{VEV} of the full theory, in terms of the renormalised couplings, at a given order in perturbation theory. The problem of finding the \ac{VEV} in a quantum theory reduces to that of minimising $V_{\rm eff}$, analogously to the classical case. The effective potential proves to be a powerful tool especially in (classically) scale-invariant theories, in which the tree-level potential lacks a quadratic term. We regard as belonging to this category both classically flat directions in field space, such as \ac{pNGB}s \cite{Weinberg:1972fn} or accidentally light scalars \cite{Brummer:2023znr}, and models in which the second derivative of the potential is chosen to vanish at the origin \cite{Coleman:1973jx}. In general, the sign of the quadratic term at the symmetry-preserving point will decide about the latter being a local maximum or minimum, and, hence, about whether the symmetry is broken or not. However, in scale-invariant cases, where the quadratic term is absent, the study of symmetry breaking requires the inclusion of quantum corrections. The investigation of radiatively-broken symmetries, as opposed to spontaneously broken symmetries where the potential features a negative second derivative at the origin already at the tree level,  was initiated by Coleman and E.~Weinberg in Ref.~\cite{Coleman:1973jx}. 

This section is inspired by the discussion in Ref.~\cite{Miransky:1994vk}.

\subsection{The effective action}\label{subsec:EffActGenFunc}
Let us start by considering a generic theory of an arbitrary scalar field $\varphi(x)$, described by the action $S\left[\varphi\right] = \int d^4x \mathcal{L}[\varphi]$. The generating functional $Z$, i.e.~the vacuum-to-vacuum transition amplitude, in the presence of an external source $J$ is defined as \footnote{Since the external source $J$ is, in general, a function of time, the in-vacuum and out-vacuum, defined respectively as the vacuum states at $t \rightarrow + \infty$ and $t\rightarrow -\infty$, are different.}
\be\label{eq:Z}
    Z\left[J\right] \equiv \langle 0_{\rm out} \rvert 0_{\rm in} \rangle = \int \mathcal{D}\varphi \exp \bigg\{i S\left[\varphi\right] + i \int d^4 x J(x)\varphi(x)\bigg\}\,.
\ee
The $n$-point Green's function can be obtained from $Z$ as
\be\label{eq:Gn}
    G^{(n)}(x_1, \dots, x_n)\equiv \langle 0 \rvert \varphi(x_1)\dots \varphi(x_n)\rvert 0 \rangle = \frac{1}{Z[0]}(-i)^n\left[\frac{\delta^n}{\delta J(x_1)\dots \delta J(x_n)}Z[J]\right]_{J=0}\,,
\ee
where the normalisation $Z[0]$ removes all the vacuum diagrams at $J=0$. Therefore, $Z$ can be expanded in a power series
\be
    Z[J] = \sum_{n=0}^{\infty}\frac{i^n}{n!}\int d^4 x_1 \dots d^4 x_n\,G^{(n)}\left(x_1, \dots, x_n\right)J(x_1)\dots J(x_n)\,.
\ee
It is easy to prove that, analogously, the functional $W \equiv -i \log Z[J]$ generates only the connected Green's functions $G_{\rm conn}^{(n)}(x_1,\dots,x_n)$. As an example, we can compute the lowest order $n$-point functions but, first, we define the \textit{classical field}, $\varphi_c(x)$ as
\be\label{eq:varphic}
   \varphi_c(x) \equiv\frac{\delta W[J]}{\delta J(x)} 
\ee
which represents the average of all the quantum fluctuations in the presence of the external source $J$, and should be regarded as a functional of $J$. For $n=1$,
\be
     \frac{\delta W[J]}{\delta J(x)}\bigg \rvert_{J=0} = -i \frac{1}{Z[0]}\frac{\delta Z[J]}{\delta J(x)}\bigg \rvert_{J=0} = \langle 0 \rvert \varphi(x) \lvert 0 \rangle = \langle \varphi(x) \rangle\,,
\ee
which is the 1-point function and is connected by definition. From the above equation we see that $\varphi_c(x)\rvert_{J=0} = \langle \varphi(x) \rangle$, making it more clear why we called $\varphi_c$ the classical field. For the ease of notation, in what follows we define $\varphi_i \equiv \varphi (x_i)$ and $J_i \equiv J(x_i)$, with $i=1,\dots,n$ a positive integer. For $n=2$, 
\be
\begin{aligned}
   -i \frac{\delta^2 W[J]}{\delta J_1 \delta J_2}\bigg|_{J=0} &= (-i)^2 \frac{\delta}{\delta J_1}\left[\frac{1}{Z[J]}\frac{\delta Z[J]}{\delta J_2}\right]_{J=0} \\
   &=  \big[\langle 0 \rvert \varphi_1\varphi_2 \lvert 0 \rangle - \langle 0 \rvert \varphi_1 \lvert 0 \rangle  \langle 0 \rvert \varphi_2 \lvert 0 \rangle\big]\,,
\end{aligned}
\ee
where the first term in the second line represents the entire 2-point function $G^{(2)}(x_1,x_2)$, as defined in Eq.~\eqref{eq:Gn}, while the second term are the disconnected pieces. This argument can be easily generalised to an arbitrary $n$. Therefore, we can expand $W$ as 
\be
     i W[J] = \sum_{n=0}^{\infty}\frac{i^{n-1}}{n!}\int d^4 x_1 \dots d^4 x_n\,G_{\rm conn}^{(n)}\left(x_1, \dots, x_n\right)J_1\dots J_n\,.
\ee

Let us define the \textit{effective action} as the Legendre transform of $W$
\be\label{eq:EffAct}
    \Gamma[\varphi_c] = W[J] - \int d^4 x\, J(x) \varphi_{c}(x)\,.
\ee
We will now proceed to prove that the effective action is the generating functional of amputated, \ac{1PI} diagrams, where \ac{1PI} diagrams are those that cannot be separated into two non-trivial diagrams by cutting an internal line. Amputated $n$-point functions are defined as 
\be\label{eq:Gamp}
    G^{(n)}_{\rm amp}(x_1,\dots,x_n) \equiv i^{n-1}\prod_{k=1}^n \int d^4 y_k \left[G^{(2)}(x_k,y_k)\right]^{-1}G^{(n)}(y_1,\dots,y_n)\,,
\ee
or, in other words, Green's functions from which the external propagators have been removed. On a more rigorous footing, what we would like to show is that, if we write the effective action as the series expansion
\be\label{eq:Gamman}
    \Gamma[\varphi_c]=\sum_{n=2}^{\infty}\frac{1}{n!}\int d^4 x_1 \dots d^4 x_n \Gamma^{(n)}(x_1,\dots\,x_n)\varphi_{c,1}\dots \varphi_{c,n}\,,
\ee
the functions $\Gamma^{(n)}$ are the \ac{1PI}, amputated Green's functions. From the definition of $\varphi_c(x)$ in Eq.~\eqref{eq:varphic}, it follows that
\be\label{eq:A}
  \frac{\delta \varphi_{c,1}}{\delta J_2}\bigg\rvert_{J=0}=\frac{\delta^2 W}{\delta J_1\,\delta J_2}\bigg\rvert_{J=0} = i G_{\rm conn}^{(2)}(x_1,x_2),
\ee
where we have used the fact that $W$ is the generator of the connected Green's functions. Moreover, Eq.~\eqref{eq:EffAct} gives
\be\label{eq:B}
    \frac{\delta \Gamma[\varphi_c]}{\delta \varphi_{c,1}} = -J_1\quad\Longrightarrow\quad\frac{\delta}{\delta J_2}\frac{\delta \Gamma[\varphi_c]}{\delta \varphi_{c,1}} = - \delta_D (x_1 - x_2)\,,
\ee
with $\delta_D(x-y)$ the Dirac delta. Eq.~\eqref{eq:A} and Eq.~\eqref{eq:B} allow us to express the Dirac delta as
\be\label{eq:Gamma2}
    \delta_D(x_1 - x_2) = -i \int d^4 x_3\, \Gamma^{(2)}(x_2,x_3)\,G_{\rm conn}^{(2)}(x_3,x_1)\,,
\ee
which proves that $\Gamma^{(2)}$ is amputated, cfr.~Eq.~\eqref{eq:Gamp}. The same holds for the 3-point function. In fact, we can compute the third functional derivative of $W$ as
\be
\begin{aligned}
    \frac{\delta^3 W}{\delta J_1\,\delta J_2\,\delta J_3} &= i \int d^4 x_4 \, G^{(2)}_{\rm conn}(x_3,x_4)\,\frac{\delta}{\delta \varphi_{c,4}}\frac{\delta^2 W[J]}{\delta J_1\,\delta J_2}\\
    &= -i \int d^4 x_4 \,G^{(2)}_{\rm conn}(x_3,x_4)\, \frac{\delta}{\delta \varphi_{c,4}}\left[\frac{\delta^2 \Gamma[\varphi_c]}{\delta \varphi_{c,1}\,\delta \varphi_{c,2}}\right]^{-1},
\end{aligned}
\ee
where, in the first equality, we made use of Eq.~\eqref{eq:A}, while in the second equality we used the fact that the second derivative of $W$ is related to the second derivative of $\Gamma$ through Eq.~\eqref{eq:Gamma2}. By computing the derivative with respect to $\varphi_c$, we finally get
\be
    \frac{\delta^3 W[J]}{\delta J_1\,\delta J_2\,\delta J_3} = -i \int d^4 x_4\,d^4 x_5\,d^4 x_6\,G^{(2)}_{\rm conn}(x_3,x_4)\,G^{(2)}_{\rm conn}(x_1,x_5)\,G^{(2)}_{\rm conn}(x_2,x_6)\,\frac{\delta^3 \Gamma[\varphi_c]}{\delta \varphi_{c,4}\,\delta \varphi_{c,5}\,\delta \varphi_{c,6}}\,,
\ee
which, once inverted, tells us exactly that $\Gamma^{(3)}$ represents the \ac{1PI}, amputated 3-point function. The argument can be extended to arbitrary $n$ \cite{Itzykson:1980rh}. 

It is now clear why $\Gamma$ is called the \textit{effective action}. In fact, starting from its definition in Eq.~\eqref{eq:EffAct}, we can easily prove that
\be
    \frac{\delta \Gamma[\varphi_c]}{\delta \varphi_c(x)}\left(\langle\varphi(x)\rangle\right)=0\,,
\ee
showing that the \ac{VEV} $\langle \varphi \rangle$ is a solution to the equations of motions derived from $\Gamma$. Moreover, at the tree level, where the theory is free, the only \ac{1PI}, amputated vertex is $\Gamma^{(2)}(x_1,x_2) = i \left[ G_{\rm conn}^{(2)}(x_1,x_2)\right]^{-1}$ and the effective action, as defined in Eq.~\eqref{eq:Gamman}, is
\be\label{eq:PhysInterpEffAct}
    \Gamma [\varphi_c] = \frac{1}{2}\int d^4x_1\,d^4x_2 \,\Gamma^{(2)}(x_1, x_2)\,\varphi_{c,1}\,\varphi_{c,2}=\frac{1}{2}\int d^4x \left(\partial_\mu\varphi_c\partial^\mu\varphi_c - m^2 \varphi_c^2\right)\,,
\ee
which is the classical action. Going beyond the tree-level approximation, $\Gamma[\varphi_c]$ contains quantum corrections at a given order in perturbation theory.

In analogy with the classical case, it can be shown that $\Gamma[\varphi_c]$ is related to the average energy of the ground state of the system, represented by $\varphi_c$ \cite{Miransky:1994vk}. With this interpretation in mind, we can extract the effective potential, $V_{\rm eff}(\varphi_c)$, from $\Gamma[\varphi_c]$ in the following way: we  take a momentum expansion of the effective action and define the zeroth-order term in such an expansion as the effective potential. This  also provides us with an operative definition of the effective potential, that allows us to obtain $V_{\rm eff}$ at a given order in perturbation theory, once the Lagrangian of the theory is specified.

Eq.~\eqref{eq:Gamman} can be written in terms of \ac{1PI}, amputated vertices in momentum space
\be
    \Gamma^{(n)}(x_1,\dots,x_n) = \prod_{i=1}^n \left[\int \frac{d^4 k_i}{(2\pi)^4}\right](2\pi)^4 \delta_D\left(\sum_i k_i\right)\tilde{\Gamma}^{(n)}(k_1,\dots,k_n)e^{-i \sum_{i} k_i x_i}\,,
\ee
where the $\tilde{\Gamma}^{(n)}(k_1,\dots,k_n)$ so defined are the Fourier transforms of $\Gamma^{(n)}(x_1,\dots,x_n)$. Expanding $\tilde{\Gamma}^{(n)}$ around $p_i = 0$ we can write the effective action as 
\be
    \Gamma[\varphi_c]= \int d^4 x\sum_{n=0}^\infty \frac{1}{n!}\left[\tilde{\Gamma}^{(n)}(0,\dots,0)\varphi_c^n(x)+\cdots\right]
\ee
where the dots represent higher orders in momentum expansion, which are not of interest to us. We now define the effective potential as the effective action computed for a translationally-invariant classical field $\varphi_c(x) = \varphi_c$:
\be\label{eq:VeffDefDiag}
    \Gamma(\varphi_c) = -\int d^4 x V_{\rm eff}(\varphi_c), \qquad {\rm with}\qquad V_{\rm eff}(\varphi_c) = - \sum_{n=0}^{\infty} \frac{1}{n!}\tilde{\Gamma}^{(n)}(0,\dots,0)\varphi_c^n\,,
\ee
i.e.~we have shown that the effective potential is given by the sum of all the \ac{1PI}, amputated diagrams computed at vanishing external momenta. Such diagrams can be computed at a given order in perturbation theory. Moreover, we see, with the definition above, that the effective potential is minimised by the vacuum state of the theory. In fact, Eq.~\eqref{eq:B} implies that, at vanishing external current $J=0$
\be
    \frac{\delta \Gamma(\varphi_c)}{\delta \varphi_c}\left(\langle \varphi \rangle\right) = 0 \qquad\Longrightarrow \qquad\frac{\delta V_{\rm eff}(\varphi_c)}{\delta \varphi_c} \left(\langle \varphi \rangle\right)= 0.
\ee
In other words, the \ac{VEV} of the full theory containing also quantum corrections, can be obtained by minimising the effective potential.

\subsection{Computing the one-loop effective potential}
We will now compute the effective potential at the one-loop level for theories involving scalar, fermion and vector fields, as done in the seminal work by Coleman and E.~Weinberg \cite{Coleman:1973jx}. We will finally get a formula that applies to general theories and that is extremely useful for practical computations.

\subsubsection{Scalar fields}\label{sec:VeffScal}
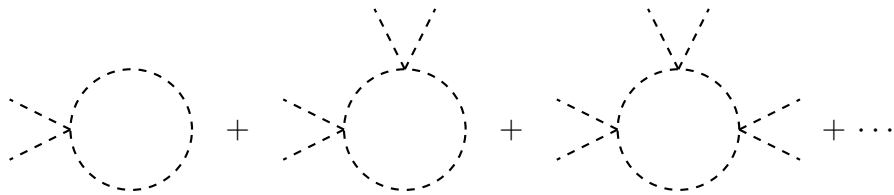
\begin{figure}
\centering
\begin{tikzpicture}[scale=0.8]

  \begin{scope}[shift={(-5,0)}]
    \draw[thick, dashed] (0,0) circle (1cm);

    \draw[thick, dashed] (-1,0) -- (-2,0.5);
    \draw[thick, dashed] (-1,0) -- (-2,-0.5);
    
  \end{scope}

\node at (-3.25,0) {\large $+$};

  \begin{scope}[shift={(-0.5,0)}]
    \draw[thick, dashed] (0,0) circle (1cm);

    \draw[thick, dashed] (-1,0) -- (-2,0.5);
    \draw[thick, dashed] (-1,0) -- (-2,-0.5);

    \draw[thick, dashed] (0,1) -- (-0.5,2);
    \draw[thick, dashed] (0,1) -- (0.5,2);

  \end{scope}

\node at (1.25,0) {\large $+$};

  \begin{scope}[shift={(4,0)}]
    \draw[thick, dashed] (0,0) circle (1cm);

    \draw[thick, dashed] (-1,0) -- (-2,0.5);
    \draw[thick, dashed] (-1,0) -- (-2,-0.5);

    \draw[thick, dashed] (0,1) -- (-0.5,2);
    \draw[thick, dashed] (0,1) -- (0.5,2);

    \draw[thick, dashed] (1,0) -- (2,0.5);
    \draw[thick, dashed] (1,0) -- (2,-0.5);

  \end{scope}

\node at (7,0) {\large $+\,\cdots$};

\end{tikzpicture}
\caption{Sum of \ac{1PI}, connected diagrams contributing the one-loop effective potential in the theory defined in Eq.~\eqref{eq:RealScalar}.}
\label{fig:1PIScalar}
\end{figure}

The renormalisable action for a real scalar field, $\phi$, charged under a $\mathbb{Z}_2$ symmetry reads
\be\label{eq:RealScalar}
    S[\phi] = \int d^4 x\, \left[\frac{1}{2}\partial_\mu\phi\partial^\mu\phi - V_0(\phi)\right]\,,\qquad {\rm with}\qquad V_0 = \frac{\mu^2}{2}\phi^2 + \frac{\lambda}{4!}\phi^4\,.
\ee
According to the definition in Eq.~\eqref{eq:VeffDefDiag}, quantum corrections to the tree-level potential $V_0$ are given by the sum of all the \ac{1PI} diagrams with vanishing external momentum, $\tilde{\Gamma}^{(n)}(0,\dots,0)$. At the one-loop level, the quartic interaction in Eq.~\eqref{eq:RealScalar} gives rise to the \ac{1PI} diagrams in Fig.~\ref{fig:1PIScalar}. 

The presence of even Green's functions only is due to the $\mathbb{Z}_2$ symmetry of the model. By inspecting such diagrams, we can be easily persuaded that the $2n$-th diagram $\tilde{\Gamma}^{(2n)}$ contains $n$ internal propagators and $n$ vertices:
\be
    \tilde{\Gamma}^{(2n)} = \frac{1}{2n} \left(\frac{-i\lambda}{2}\right)^n \int \frac{d^4 p}{(2\pi)^4}\left(\frac{i}{p^2 - \mu^2}\right)^n\, 
\ee
where the overall $2n$ at the denominator is due to the symmetries of the diagrams, while the factor of 2 dividing $\lambda$ comes from the invariance of each vertex under the exchange of the two external lines. Then, the effective potential at the one-loop order is
\be
    V_1(\phi_c) = i \sum_{n=1}^\infty \int \frac{d^4 p}{(2\pi)^4} \frac{1}{2n}\left(\frac{\lambda}{2}\frac{\phi_c^2}{p^2 - \mu^2}\right)^n = -\frac{i}{2} \int \frac{d^4 p}{(2\pi)^4}\log\left(1- \frac{\lambda}{2}\frac{\phi_c^2}{p^2 - \mu^2}\right)\,.
\ee
We now perform a Wick rotation to imaginary time, $p^0_E = -i p^0$, moving to Euclidean space where $p_E^2 = -p^2$, and we define $m^2(\phi_c) \equiv \mu^2 + \lambda \phi_c^2/2$ so that
\be\label{eq:V1ReScalar}
    V_1(\phi_c) = \frac{1}{2}\int \frac{d^4 p_E}{(2\pi)^4} \log\left(p_E^2 + m^2(\phi_c)\right)\,
\ee
after discarding a field-independent term. The above integral is \ac{UV}-divergent and has to be regularised. We delay the discussion about regularisation and renormalisation of the effective potential to the next section, once it becomes clear that we can treat it once and for all for any fields to which the scalar field is coupled.

We can easily generalise the above computation to the theory of $N_s$ interacting, complex scalar fields
\be
    \mathcal{L}= \partial_\mu \phi^i \partial^\mu \phi_i^\dag - V_0 \left(\phi^i,\phi_i^\dag\right)\,. 
\ee
Extending the definition above Eq.~\eqref{eq:V1ReScalar} to the multi-field case, we define
\be
    \left(\mathcal{M}_s^2\right)^i_j \equiv \frac{\partial^2 V_0 }{\partial \phi_i^\dag \partial \phi^j}\Bigg|_{\phi_c,\phi_c^\dag}\,
\ee
which is the factor inserted at any vertex. The indices $i$ and $j$ above run over the field components $i,j = 1,\dots,N_s$. Analogously to the case with a single, real scalar field, $\tilde{\Gamma}^{(n)}$ will have $n$ vertices and $n$ internal propagators and will carry an overall factor $1/2n$, accounting for symmetry under discrete rotations by $2\pi/n$ and reflection. Therefore, after summing all the one-loop \ac{1PI} contributions we obtain
\be
    V_1^{(s)} = \frac{1}{2}\Tr \int \frac{d^4p_E}{(2\pi)^4}\log\left(p_E^2 + \mathcal{M}_s^2\right)\,,
\ee
where the trace is over the different field components. 

\subsubsection{Fermions}\label{sec:VeffFerm}
Let us move to a theory of $N_f$ fermions described by
\be
    \mathcal{L}=i\, {\overline{\psi}}_i \slashed{\partial} \psi^i - {\overline{\psi}}_i \left(\mathcal{M}_f\right)^i_j \psi^j\,,
\ee
where the mass matrix $\mathcal{M}_f$ contains both Dirac and Majorana masses and is a linear function of $\phi_c$. As the trace of an odd number of Dirac matrices vanishes, only diagrams with an even number of vertices contribute to $V_1$. $\Gamma^{(2n)}$ contains $2n$ vertices, each contributing with $\Tr_j\left[(-i)^{2n} \mathcal{M}_f^{2n}\right]$, with the trace running over the different fermionic fields $j=1,\dots,N_f$. The $2n$ vertices are connected by $2n$ fermionic propagators that form the loop giving a factor $\Tr_\sigma \left[i^{2n} \slashed{p}^{2n}/(p^2)^{2n}\right]$, where, now, the trace is over the spinor components. Hence, summing over all the one-loop contributions with an even number of vertices, we get
\be
    V_1^{(f)} = -i 2\Tr \sum_{n=1}^\infty \int \frac{d^4 p}{(2\pi)^4} \frac{1}{2n} \left(\frac{\slashed{p}}{p^2}\mathcal{M}_f\right)^{2n}\,.
\ee
Here, the trace is both on spinor and field components. Before summing over all the \ac{1PI} diagrams, let us comment on the prefactors in the above equation. The factor $1/2n$ is, again, due to symmetry under discrete rotations, and the minus sign in front is the usual sign coming from fermionic loops. To compute the sum above we express $V_1$ as
\be\label{eq:V1Ferm}
    V_1^{(f)} = -i\Tr \sum_{n=1}^\infty \int \frac{d^4 p}{(2\pi)^4}\frac{1}{2n}\left(\frac{\slashed{p}}{p^2}\mathcal{M}_f\frac{\slashed{p}}{p^2}\mathcal{M}_f\right)^n\,,
\ee
pairing the vertices. We can now write the fermion mass matrix as
\be 
\mathcal{M}_f = A + i B \gamma^5\,,
\ee
with $A$ and $B$ both Hermitian matrices. This allows us to recast the term inside parenthesis in Eq.~\eqref{eq:V1Ferm} as
\be
    \frac{\slashed{p}}{p^2}\mathcal{M}_f\frac{\slashed{p}}{p^2}\mathcal{M}_f= \frac{1}{p^2}\mathcal{M}_f\mathcal{M}_f^\dag\,,
\ee
by virtue of $\{\gamma^5,\gamma^\mu\}=0$. This leads to
\be
    V_1^{(f)} = - i\Tr \sum_{n=1}^\infty \int \frac{d^4 p}{(2\pi)^4}\frac{1}{2n}\left(\frac{\mathcal{M}_f\mathcal{M}_f^\dag}{p^2}\right)^n = -\frac{1}{2}\Tr \int \frac{d^4 p_E}{(2\pi)^4)}\log\left(p_E^2 + \mathcal{M}_f\mathcal{M}_f^\dag\right)\,,
\ee
where, we remind, we are summing over both spinor and field components. 

\subsubsection{Gauge bosons}\label{sec:VeffGauge}
Consider now the theory of a scalar field $\phi$ in some representation of a gauge group $G$. Since we are interested in the one-loop corrections to the potential for $\phi$ coming from gauge interactions, we focus on the following part of the Lagrangian
\be
    \mathcal{L} \supset -\frac{1}{4} F_{\mu \nu}^A F^{\mu\nu,A} + \left(D_\mu^A \phi^a\right)^\dagger D^{\mu,A} \phi^a\,
\ee
with $F_{\mu \nu}^a$ the gauge field strength tensor and $D_\mu^A$ the covariant derivative. Uppercase latin indices are in the adjoint of $G$, while lowercase ones are in the representation of $\phi$. In the Landau gauge, the gauge boson propagator is
\be
    \Pi^{\mu}{}_{\nu} = -i\,\frac{\eta^\mu{}_\nu - \frac{p^\mu p_\nu}{p^2}}{p^2}\,,
\ee
with $\eta^{\mu}{}_{\nu}$ the Minkowski metric, and diagrams with both scalar and vector lines in the loops vanish. The only term in the Lagrangian contributing to $V_{\rm eff}$ is
\be\label{eq:LintGauge}
    \mathcal{L} \supset \frac{1}{2}\left(\mathcal{M}_g^2\right)_{AB} A_\mu^A A^{\mu,B}\,, \qquad {\rm where} \qquad \left(\mathcal{M}_g^2\right)_{AB} \equiv g^2\Tr\Big\{\left[\left(T_A\right)^a{}_b\, \phi_a\right]^\dagger \left(T_B\right)^b{}_c\,\phi^c\Big\}\,.
\ee
In the above, $A_\mu$ are the gauge fields, $g$ is the gauge coupling and $T_A$ are the generators of $G$ in the representation of $\phi$, while the trace is meant over the representation indices. Each vertex, together with the insertion of a pair of external scalar legs, carries a factor $-i \mathcal{M}_g$. Therefore,
\be
    V_1^{(g)} = -i\Tr \sum_{n=1}^\infty \int \frac{d^4 p}{(2\pi)^4}\frac{3}{2n}\left(\frac{\mathcal{M}_g}{p^2}\right)^n = \frac{3}{2}\Tr \int \frac{d^4 p_E}{(2\pi)^4} \log \left(p_E^2 + \mathcal{M}_g^2\right)\,,
\ee
where the factor of 3 comes from the trace over the gauge boson propagator and represents the 3 physical degrees of freedom of a massive vector field. Here, the trace runs over the different gauge bosons.

\subsubsection{Renormalising the effective potential}
In a generic theory, the Lagrangian will contain scalar, fermion and vector degrees of freedom. Given that, in the Landau gauge, the contributions coming from fields with different spins factorise as $V_1 = V_1^{(s)} + V_1^{(f)}+V_1^{(g)}$, the most general expression for the one-loop effective potential can be summarised as\footnote{$\mathcal{M}_f^2$ is understood as a short-hand notation for $\mathcal{M}_f \mathcal{M}_f^\dagger$.}
\be
    V_1(\phi_c) = \sum_{i=\{s,f,g\}}\frac{n_i}{2}\Tr \int \frac{d^4 p_E}{(2\pi)^4}\log\left(p_E^2 + \mathcal{M}_i^2(\phi_c)\right)\,,
\ee
where the trace runs over the different field components, and $n_i = \{+1, -4, +3 \}$.\footnote{Here, we are considering Dirac fermions. If one prefers to count in terms of Weyl spinors instead, then $n_f = -2$.} We change to polar coordinates and integrate over the solid angle 
\be
    V_1(\phi_c) = \sum_{i=\{s,f,g\}} \frac{n_i}{16\pi^2}\Tr\int_0^{\Lambda} p_E^3\,d p_E\, \log\left(p_E^2 + \mathcal{M}_i^2(\phi_c)\right)\,,
\ee
where we have introduced a hard cutoff to regularise the otherwise \ac{UV}-divergent integral. $\Lambda$ is in general arbitrary and very large, so we  work under the assumption that $\Lambda^2 \gg \mathcal{M}_i^2(\phi_c)$. Forgetting about both the uninteresting $\phi$-independent terms, and those terms vanishing as $\Lambda \rightarrow \infty$, we have
\be\label{eq:VeffGeneral}
V_1(\phi_c) =\sum_{i=\{s,f,g\}} \frac{n_i}{64\pi^2}\Tr\Bigg\{2\Lambda^2 \mathcal{M}_i^2(\phi_c) + \mathcal{M}_i^4(\phi_c)\left[\log\left(\frac{\mathcal{M}_i^2(\phi_c)}{\Lambda^2}\right)-\frac{1}{2} \right]\Bigg\}\,.
\ee
This expression for the effective potential is extremely general and easy to apply to the specific model under exam. The dependence on the un-physical cutoff $\Lambda$ is absorbed by the counterterms once the renormalisation conditions are specified. If the tree-level potential is renormalisable, i.e.~it is a polynomial of degree 4 in $\phi$, the non-physical dependence on $\Lambda$ can be removed by adding counterterms to the Lagrangian, which are themselves at most quartic polynomials. However, if the theory is non-renormalisable and the tree-level potential contains terms of degree 5 or higher, we see that the effective potential will introduce even higher order terms, and renormalisation will require the introduction of a never-ending tower of terms of higher and higher orders. In the next section we will see a concrete example of how to compute and renormalise the effective potential.

\subsection{The Coleman-Weinberg potential}

As a concrete example, we study the theory of a complex scalar field charged under $\U{1}$, by further imposing, as a renormalisation condition, that
\be\label{eq:CWRenCond}
    \frac{d^2 V_{\rm eff}}{d \phi_c^2}\bigg\rvert_{\phi_c=0} = \mu^2 = 0.
\ee
This model, which at the classical level is simply the theory of massless scalar \ac{QED}, was studied by Coleman and Weinberg to show that radiative corrections can induce symmetry breaking \cite{Coleman:1973jx}. Due to the above renormalisation condition, we would conclude, just by staring at the tree-level potential, that the minimum of the potential is at the origin, where $\U{1}$ is preserved. However, we will see that radiative corrections will turn this conclusion upside down.\footnote{Radiative corrections are still “small'', in the sense that perturbativity is ensured and higher-loop corrections are smaller than lower-order ones. However, here we are in a somewhat pathological case, due to the choice in Eq.~\eqref{eq:CWRenCond}, where quantum corrections to the potential second derivative are larger than the tree-level value.} 

The Lagrangian of the model, in terms of renormalised quantities, is
\be
    \mathcal{L} = \frac{1}{2} \lvert D_\mu \phi\rvert^2 - \frac{\lambda}{4!}\left(\lvert \phi \rvert^2\right)^2 - \frac{1}{4}F_{\mu \nu}F^{\mu \nu}\;(+\;{\rm counterterms})\,,
\ee
where the covariant derivative is $D_\mu \equiv \partial_\mu - i e A_\mu$. We now expand the complex scalar field in its real components $\phi = \phi_1 + i\phi_2$:\footnote{The normalisation of the complex field is chosen in order for the real fields to be canonically normalised.}
\be
    \mathcal{L}=\frac{1}{2}\left(\partial_\mu \phi_1 + e A_\mu \phi_2\right)^2 + \frac{1}{2}\left(\partial_\mu \phi_2 - e A_\mu \phi_1\right)^2 - \frac{\lambda}{4!}\left(\phi_1^2 + \phi_2^2\right)^2 -\frac{1}{4}F_{\mu \nu}F^{\mu \nu}\;(+\;{\rm counterterms})\,.
\ee
To apply the formula derived in Eq.~\eqref{eq:VeffGeneral}, we need to compute the scalar mass matrix
\be
    \left(\mathcal{M}_s^2\right)_{ij} = \frac{\partial^2 V}{\partial \phi_i \partial \phi_j}\bigg\rvert_{\phi_c} = \frac{\lambda}{6}\begin{pmatrix} 3 \phi_{1c}^2 + \phi_{2c} & 2\phi_{1c}\phi_{2c} \\ 2 \phi_{1c} \phi_{2c} & \phi_{1c}^2 + 3 \phi_{2c}^2\end{pmatrix}\,.
\ee
We can now make use of the $\U{1}$ invariance of the theory to set $\phi_{2c} = 0$, so that the mass matrix simplifies to $\mathcal{M}_s^2 = \lambda/6\,{\rm diag} \left(3\phi_1^2\,,\phi_1^2\right)$. The effective potential is
\be
\begin{aligned}
    V_{\rm eff}(\phi_c) = & \; \frac{\lambda}{4!} \phi_c^4 + \frac{A}{2}\phi_c^2 + \frac{B}{4!}\phi_c^4\\    
    &+\frac{1}{64\pi^2} \tr\Bigg\{2 \Lambda^2 \mathcal{M}_s^2(\phi_c) + \mathcal{M}_s^4(\phi_c)\left[\log\left(\frac{\mathcal{M}_s^2(\phi_c)}{\Lambda^2}\right)-\frac{1}{2}\right]\Bigg\}\\
    & + \frac{3}{64\pi^2}\Bigg\{2\Lambda^2(e\phi_c)^2 + (e \phi_c)^4 \left[\log\left(\frac{(e\phi_c)^2}{\Lambda^2}\right)-\frac{1}{2}\right]\Bigg\}\,,
\end{aligned}
\ee
where $\phi_c^2 \equiv \phi_{1c}^2+\phi_{2c}^2$ and, in the first line, we have written the counterterms explicitly, neglecting, once again, field-independent pieces. The second line shows the contribution from the scalar field self-interaction, while in the third line we find the piece coming from the gauge interaction. To find the counterterms, we need to specify the renormalisation conditions; one is given in Eq.~\eqref{eq:CWRenCond} and will fix $A$, while to fix $B$ we impose that the fourth derivative of the effective potential at some arbitrary, non-vanishing scale $M$ is the renormalised quartic coupling $\lambda$
\be\label{eq:CountB}
    \frac{d^4 V_{\rm eff}}{d \phi_c^4}\bigg\rvert_{\phi_c = M} = \lambda\,.
\ee
Solving the renormalisation condition for $A$ and $B$, and plugging them back inside the expression for the effective potential we finally find
\be
    V_{\rm eff} = \frac{\lambda}{4!}\phi_c^4 + \frac{1}{64\pi^2}\left(\frac{5\lambda^2}{18}+3 e^4\right)\phi_c^4\left[\log\left(\frac{\phi_c^2}{M^2}\right)-\frac{25}{6}\right]\,.
\ee
We now make the assumption that $\lambda \ll e^2$, neglecting the effect of scalar loops. This assumption will prove to be consistent with the presence of a non-trivial minimum in the effective potential, a posteriori. The minimum of the potential has to obey the following equation
\be 
    \frac{d V_{\rm eff}}{d\phi_c}\bigg\rvert_{\phi_c = \langle \phi \rangle} = \frac{\lambda}{6}\langle\phi\rangle^3 + \frac{3 e^4}{16\pi^2}\langle\phi\rangle^3 \left[\log\left(\frac{\langle \phi \rangle^2}{M^2}\right) -\frac{25}{6} \right] + \frac{3 e^4}{32\pi^2}\langle \phi \rangle^3 = 0\,.
\ee
Since the renormalisation scale $M$ is arbitrary, we can fix $M = \langle \phi \rangle$ to simplify our calculation. After this choice has been made, the above equation tells us that the effective potential will admit a minimum at $\langle \phi \rangle \neq 0$ only if
\be\label{eq:lambdaCW}
    \lambda = \frac{33}{8 \pi^2}e^4\,,
\ee
thus justifying our initial assumption that $\lambda \ll e^2$. With such a choice for the scalar self-coupling, our effective potential now looks like
\be
    V_{\rm eff} = \frac{3 e^4}{64\pi^2}\phi_c^4 \left[\log\left(\frac{\phi_c^2}{\langle \phi \rangle^2}\right)-\frac{1}{2}\right]\,.
\ee
We see that, starting at the tree-level from a scale-invariant potential (remember that we have chosen that the second derivative of the potential vanishes at the origin as a renormalisation condition), gauge interactions, that only contribute to the potential radiatively, have induced the spontaneous breaking of $\U{1}$. Therefore, the gauge boson $A_\mu$ picks up a mass $m_A^2 = e^2 \langle \phi \rangle^2$ while the mass of the Abelian Higgs field, at the minimum of the potential is 
\be
    m_s^2 = \frac{d^2 V_{\rm eff}}{d \phi_c^2}\bigg\rvert_{\phi_c = \langle \phi \rangle} = \frac{3 e^4}{8\pi^2}\langle \phi \rangle^2\,.
\ee

\subsection{Background field method}
In this section we provide an alternative way to compute the effective potential: the \textit{background field method}. Such a a method, due to Jackiw \cite{Jackiw:1974cv}, makes it easier to go beyond the one-loop approximation, where summing all the \ac{1PI} graphs, as done so far, is not a trivial task. Even though one-loop corrections will suffice us for the rest this thesis, this approach tells us something more about the effective action: the effective action can also be thought of as the action obtained after integrating out quantum fluctuations around the classical solution to the equations of motion. 

In what follows, we will focus on the theory of one real scalar field, $\phi$, even if the proof can be extended to other fields as well. Given the generating functional in Eq.~\eqref{eq:Z}, we look for field configurations that minimise the exponent, i.e.~the saddle point configurations $\phi_s(x)$. The action for a generic real scalar field reads
\be
    S[\phi] = \int d^4 x\, \left[\frac{1}{2}\partial_\mu\phi\partial^\mu\phi - V_0(\phi)\right]
\ee
and, hence, $\phi_s(x)$ is defined by
\be\label{eq:SaddleSol}
\partial^2\phi_s(x) + V_0'(\phi_s) = J(x),
\ee
where a prime denotes derivative with respect to the field. Consider small fluctuations around the saddle point solution
\be
    \phi_s(x) \rightarrow \phi_s(x) + \eta(x).
\ee
Up to quadratic order in $\eta$, the generating functional is\footnote{Terms that are higher order in $\eta$ will not contribute to the one-loop effective potential.}
\be
    Z[J] = e^{i S[\phi_s] + i \int d^4 x J(x) \phi_s(x)} \int \mathcal{D}\eta \exp\Bigg\{\frac{i}{2}\left[\partial_\mu \eta \partial^\mu \eta - \eta^2 V_0''(\phi_s)\right]\Bigg\}\,.
\ee
By integrating by parts, the integral over $\eta$ can be recast as a Gaussian integral
\be
\begin{aligned}
    \int\mathcal{D}\eta \exp\Bigg\{\frac{i}{2}\left[\partial_\mu \eta \partial^\mu \eta - \eta^2 V_0''(\phi_s)\right]\Bigg\} &= \int\mathcal{D}\eta \exp\Bigg\{\frac{i}{2}\eta \left[\partial^2 V_0''(\phi_s)\right]\eta\Bigg\}\\ &= \left[\det \left(\partial^2 + V_0''(\phi_s)\right)\right]^{-1/2}\,,
\end{aligned}
\ee
where, in the last equality, we have made use of the result derived in App.~\ref{app:GaussInt}. Hence, remembering that for a generic hermitian operator $\mathcal{O}$, we have $\det \mathcal{O} = \exp \tr \log \mathcal{O}$, the functional $W$ is
\be
    W[J] = S[\phi_s] + \int d^4 x J(x)\phi_s(x) + \frac{i}{2} \tr \log \left(\partial^2 + V_0''(\phi_s)\right),
\ee
with the last term being the result of having integrated out quantum fluctuations around the classical, saddle solution. Indeed, by computing $\phi_c$, according to Eq.~\eqref{eq:varphic}, we get\footnote{The saddle point solution is a function of the external source $J$, as per Eq.~\eqref{eq:SaddleSol}.}
\be
    \phi_c(x) = \phi_s(x) + \underbrace{\frac{i}{2}\frac{\delta}{\delta J} \tr \log \left(\partial^2 + V_0''(\phi_s)\right)}_{\rm quantum\;corrections}\,.
\ee
Moreover, by virtue of the definition in Eq.~\eqref{eq:EffAct}, we see that the effective action is
\be
    \Gamma[\phi_s] = S[\phi_s] + \underbrace{\frac{i}{2} \tr \log \left(\partial^2 + V_0''(\phi_s)\right)}_{\rm quantum\;corrections}\,,
\ee
that matches the interpretation of the effective action we gave below Eq.~\eqref{eq:PhysInterpEffAct}. According to our definition of the effective potential, given at the end of Sec.~\ref{subsec:EffActGenFunc}, we now consider translationally invariant saddle solutions. This allows us to write the impractical trace  appearing above in a more useful fashion:
\be
    \tr \log \left(\partial^2 + V_0''(\phi_s)\right) = \int d^4 x \int \frac{d^4 p }{(2\pi)^4}\log\left(-p^2 + V_0''(\phi_s)\right).
\ee
Finally, we can write down the effective potential as
\be\label{eq:V1}
    V_{\rm eff}(\phi) = V_0(\phi) - \frac{i}{2}\int \frac{d^4 p}{(2\pi)^4}\log\left(-p^2 + V_0''(\phi)\right)\equiv V_0(\phi) + V_1(\phi)\,.
\ee

\section{Thermal field theory}\label{sec:TQFT}

The formalism developed in the previous section can be generalised to include thermal effects. While quantum field theory successfully describes the interaction among particles in vacuum, it is not suited for describing what happens when we are in the presence of many particles in thermal equilibrium at some temperature $T\neq 0$. In this section, we will extend the effective potential with the effects of a non-zero temperature. We will see that this is particularly relevant to study \ac{PT}s taking place in the early Universe, where thermal effects were sizeable and absolutely non-negligible.

Among the many reviews and textbooks written on the topic we mention Refs.~\cite{Quiros:1999jp,Kapusta:2006pm,Salvio:2024upo}.

\subsection{The thermal effective potential}
We derive here the one-loop effective potential including thermal corrections for scalar, fermion, and gauge bosons field, analogously to what we did in Sec.~\ref{sec:EffPot} for zero-temperature field theory. We will make use of the \textit{imaginary time formalism} described in Ref.~\cite{Quiros:1999jp}. The basic idea is that, if $\beta \equiv 1/T$ is different from zero, the Euclidean time $\tau$ is periodic (anti-periodic) with period $\beta$ for bosons (fermions). This (anti-)periodicity implies that the frequencies of a field propagator will be discretised 
\be
    \Delta (x) = \frac{1}{\beta} \sum_{n = -\infty}^\infty \int \frac{d^3 p}{(2\pi)^3} e^{- i \omega_n \tau + i \vec p \cdot \vec x} \frac{i}{p^2 - m^2 + \omega_n}\,,
\ee
where $\omega_n$ are the Matsubara frequencies
\be
    \omega_n = \begin{cases}
        \frac{2 n\pi}{\beta},\quad {\rm bosons}\\
    \frac{2(n+1)\pi}{\beta},\quad {\rm fermions}
    \end{cases} \qquad {\rm for}\quad n\in \mathbb{Z}\,.
\ee
From the above argument we can extract the Feynman rules in momentum space in the imaginary-time formalism 
\be\label{eq:ThermFeynRules}
\begin{aligned}
    &{\rm boson\;propagator:}\qquad \frac{i}{p^2 - m^2},\qquad{\rm where}\quad p^\mu = \left(2in\pi\beta^{-1},\vec p\right),\\
    &{\rm fermion\;propagator:}\qquad \frac{i}{\slashed{p} - m},\qquad{\rm where}\quad p^\mu = \left(2(n+1)i\pi\beta^{-1},\vec p\right),\\
    &{\rm  loop\; integral:}\qquad  \frac{i}{\beta}\sum_{n=-\infty}^\infty \int \frac{d^3p}{(2\pi)^3}\,,\\
    &{\rm vertex:}\qquad -i\beta(2\pi)^3 \delta \left(\sum_i \omega_i\right)\delta\left(\sum_i \vec{p}_i\right)\,.
\end{aligned}
\ee
Computing the thermal effective potential amounts to resumming the same class of loops as for the zero-temperature scenario, i.e.~the \ac{1PI} diagrams, evaluating them with the Feynman rules given in Eq.~\eqref{eq:ThermFeynRules}.

\subsubsection{Scalar field}\label{sec:VThermScal}
Consider, again, the scalar field of Sec.~\ref{sec:VeffScal}.  To take into account thermal  corrections, we resum the one-loop diagrams shown in Fig.~\ref{fig:1PIScalar}, using the Feynman rules given in Eq.~\eqref{eq:ThermFeynRules}. We find
\be
    V_1(\beta,\phi_c)=\frac{1}{2\beta}\sum_{n=-\infty}^\infty \int \frac{d^3p}{(2\pi)^3}\log\left[\omega_n + \omega^2(\phi_c)\right]\,,
\ee
where we have defined $\omega^2(\phi_c)\equiv \lvert\vec{p}\rvert^2 + m^2(\phi_c)$. The scalar field mass in a background of $\phi_c$ is given above Eq.~\eqref{eq:V1ReScalar}. It is clear that the sum over all the Matsubara frequencies diverges. However, such divergent part does not depend on $\phi_c$ and, hence,  it will contribute with a constant to the potential. To isolate the $\phi_c$-dependent part in the above expression, which is finite, we proceed in the following way. We define 
\be
    v(\omega) \equiv \sum_{n=-\infty}^\infty \log\left(\omega_n^2 + \omega^2\right)\,,
\ee
so that its derivative with respect to $\omega$ gives
\be\label{eq:dvdomega}
    \frac{\partial v(\omega)}{\partial \omega} = \sum_{n=-\infty}^\infty \frac{2\omega}{\omega_n^2 + \omega^2} = 2\beta \left[\frac{1}{2}+ \frac{e^{-\beta\omega}}{1- e^{-\beta\omega}}\right]\,.
\ee
In the last equality, we used the relation \cite{Quiros:1999jp} 
\be
    \sum_{n=-\infty}^\infty \frac{y^2}{y^2 + n^2} = -\frac{1}{2y} + \frac{\pi}{2} + \pi \frac{e^{-2\pi y}}{1 - e^{-2\pi y}}\,.
\ee
By integrating Eq.~\eqref{eq:dvdomega}, we get $v(\omega)$ and, therefore,
\be
    V_1(\beta, \phi_c) = \int \frac{d^3 p}{(2\pi)^3}\left[\frac{\omega}{2} +\frac{1}{\beta} \log\left(1 - e^{-\beta\omega}\right)\right]\,,
\ee
where we have dropped field-independent terms. It can be shown that the first term in the square bracket is nothing but $V_1(\phi_c)$: the zero-temperature, one-loop effective potential in Eq.~\eqref{eq:V1ReScalar}. All in all, the thermal one-loop effective potential reads
\be
\begin{aligned}\label{eq:ThermEffPotScal}
    V_1(T,\phi_c) &= V_1(\phi_c) + \frac{T^4}{2\pi^2} \int_0^\infty dq\, q^2 \log\left[1 - \exp\left(-\sqrt{q^2 +\frac{m^2(\phi_c)}{T^2}}\right)\right]\\
    &\equiv V_1(\phi_c) + \frac{T^4}{2\pi^2} J_B\left(\frac{m^2(\phi_c)}{T^2}\right)\,.
\end{aligned}
\ee
In the second line, we have defined the thermal bosonic function $J_B$. Whenever the field values and the temperatures of interest are such that $m^2(\phi_c) \ll T^2$, $J_B$ admits an expansion which drastically simplifies the expression for the thermal effective potential in Eq.~\eqref{eq:ThermEffPotScal}, allowing for an analytic study:
\be\label{eq:JBexpansion}
    J_B(x) = -\frac{\pi^4}{45} + \frac{\pi^2}{12}x - \frac{\pi}{6} x^{3/2}  - \frac{1}{32}x^2 \log \frac{x}{a_B} + \mathcal{O}\left(x^3\right)\,,
\ee
with $a_B = 16\pi^2 \exp(3/2 - 2\gamma_E)$.
This expansion is known as the \textit{high-temperature expansion}, since it holds for temperatures much larger than the field-dependent mass.

\subsubsection{Fermion fields}\label{sec:VThermFerm}
Analogously to what we did for scalar fields, we start from the theory described in Sec.~\ref{sec:VeffFerm} and we compute the relevant \ac{1PI} diagrams using the imaginary-time formalism
\be
    V_1^{(f)}(\beta,\phi_c)=-\frac{1}{\beta}\Tr \sum_{n=-\infty}^\infty \int \frac{d^3p}{(2\pi)^3}\log\left[\omega_n + \omega^2(\phi_c)\right]\,,
\ee
where, now, $\omega_n$ are the fermionic Matsubara frequencies, and $\omega^2 = p^2 + \mathcal{M}_f\mathcal{M}_f^\dagger$. The trace sums over the spinor and field components. Following the same steps outlined in Sec.~\ref{sec:VThermScal}, we get
\be
\begin{aligned}
    V_1^{(f)}(T,\phi_c) &= V_1^{(f)}(\phi_c) -\frac{T^4}{2\pi^2} \tr \int_0^\infty dq\, q^2 \log\left[1 + \exp\left(-\sqrt{q^2 +\frac{\mathcal{M}_f\mathcal{M}_f^\dagger}{T^2}}\right)\right]\\
    &\equiv V_1^{(f)}(\phi_c) - \frac{T^4}{2\pi^2} \tr J_F\left(\frac{\mathcal{M}_f\mathcal{M}_f^\dagger}{T^2}\right)\,,
\end{aligned}
\ee
which has to be summed over the different spinor components. Notice the difference in sign in the integral compared to the bosonic case. The  thermal fermionic function $J_F$ can be expanded in the high-temperature limit as
\be\label{eq:JFexpansion}
    J_F(x) = \frac{7\pi^4}{360} - \frac{\pi^2}{24}x - \frac{1}{32}x^2 \log \frac{x}{a_F} \mathcal{O}\left(x^3\right)\,,
\ee
where $a_F = \pi^2 \exp(3/2 - 2\gamma_E)$.

\subsubsection{Gauge boson field}

Finally, for gauge boson fields we just give the result
\be
\begin{aligned}
    V_1^{(g)}(T,\phi_c) &= V_1^{(g)}(\phi_c) + \frac{T^4}{2\pi^2} \int_0^\infty dq\, q^2 \log\left[1 - \exp\left(-\sqrt{q^2 +\frac{\mathcal{M}_g^2}{T^2}}\right)\right]\\
    &\equiv V_1^{(g)}(\phi_c) + \frac{T^4}{2\pi^2} J_B\left(\frac{\mathcal{M}_g^2}{T^2}\right)\,.
\end{aligned}
\ee

\subsection{Phase transitions}
Considering a generic model with an arbitrary number of bosonic and fermionic fields, the potential in the presence of a thermal bath of particles can be written as
\be
    V(\phi,T) = V_{\rm tree}(\phi) +  V_1(\phi) + V_1(T,\phi)\,,
\ee
where $V_{\rm tree}(\phi)$ is the tree-level potential, $V_1(\phi)$ is the one-loop, zero-temperature effective potential given in Eq.~\eqref{eq:VeffGeneral}, and
\be\label{eq:VTgen}
    V_1(T,\phi) = \frac{T^4}{2\pi^2}\left[\sum_{b} n_b J_B\left(\frac{m_b^2(\phi)}{T^2}\right) - \sum_{f} (2 n_f) J_F\left(\frac{m_f^2(\phi)}{T^2}\right)\right]\,,
\ee
with the subscripts $b$ and $f$ respectively referring to the bosons and fermions present in the model. $n_b$ counts the number of bosonic field components while $n_f$ counts the number of Weyl spinors (hence the factor of $2$). For temperatures $T$, and field values $\phi$ such that $m_{b,f}^2(\phi) \ll T^2$, we can perform the high-temperature expansion of the otherwise complicated thermal functions $J_{B,F}$, cfr.~Eqs.~\eqref{eq:JBexpansion} and \eqref{eq:JFexpansion}. Truncating such an expansion at the first non-trivial order, we write the thermal correction to the potential as
\be\label{eq:ThermQuadCorr}
    V_1(T,\phi) \simeq \frac{T^2}{24}\left(\sum_b n_b m_b^2(\phi) + \sum_f n_f m_f^2(\phi)\right) \equiv \frac{g^2}{24}T^2\phi^2\,.
\ee
In the last equality, we have defined, following the approach of Ref.~\cite{Salvio:2023qgb}, a collective coupling $g$. This allows us to expose a remarkable fact: thermal corrections introduce a large positive quadratic coefficient in the effective potential. To see what the implications of this conclusion are, let us consider a theory that, at $T=0$, is described by a tree-level potential featuring a minimum at $\langle \phi\rangle = \eta$, where spontaneous symmetry breaking occurs: $G\rightarrow H$. As long as temperature is high enough, say $T \gg \eta$ for couplings of order one, the large quadratic thermal correction in Eq.~\eqref{eq:ThermQuadCorr} deforms the potential, stabilising the origin $\phi=0$, where the symmetry $G$ is preserved, as shown in Fig.~\ref{fig:ThermEffPot}. This phenomenon goes under the name of \textit{symmetry restoration} and implies, for example, that if the Universe was reheated to a temperature larger than $\eta$, $G$ was restored in the very early Universe. However, the Universe cools down as it expands, and the temperature decreases, until at some point, when $T\sim \eta$, a \textit{cosmological phase transition} takes place. This can happen smoothly, as in a \ac{SOPT}, or via thermal tunneling, as in a \ac{FOPT}, depending on the shape of the potential.

\subsubsection{Second-order phase transition}
As an example of a potential giving a \ac{SOPT}, let us consider the theory of a real scalar field, charged under a $\mathbb{Z}_2$ symmetry, and with some unspecified interactions. The high-temperature potential can be written as
\be\label{eq:VSOPT}
    V(T,\phi) \simeq D\left(T^2 - T_0^2\right)\phi^2 + \frac{\lambda(T)}{4}\phi^4\,,
\ee
where $D$ and $\lambda(T)$ are generic coefficients whose explicit form depends on the specific model under consideration. The approximate equality is there to remind ourselves that this is just an expansion of the exact thermal effective potential in Eq.~\eqref{eq:VTgen}. At $T=0$, the potential features a minimum at $\phi^2 = 2 DT_0^2/\lambda$, where the $\mathbb Z_2$ symmetry gets spontaneously broken. For $T\gg T_0$, the quadratic coefficient in the potential is positive and the symmetry is restored. A \ac{PT} happens at $T\simeq T_0$, when the origin turns from a minimum to a maximum of the potential, and the scalar field smoothly adjusts to the new minimum
\be\label{eq:SOPTPot}
    \phi(T) = \sqrt{\frac{2D\left(T_0^2-T^2\right)}{\lambda(T)}}\,.
\ee
The potential in Eq.~\eqref{eq:VSOPT} is shown in the left-hand panel in Fig.~\ref{fig:ThermEffPot}, for different values of the temperature. A smooth transition from the symmetric phase to the broken one, at temperatures around $T\simeq T_0$, is the trademark of a \ac{SOPT}.

\subsubsection{First-order phase transition}\label{sec:FOPTIntro}

\begin{figure}[h!]
\begin{center}
\includegraphics[width=.45\textwidth]{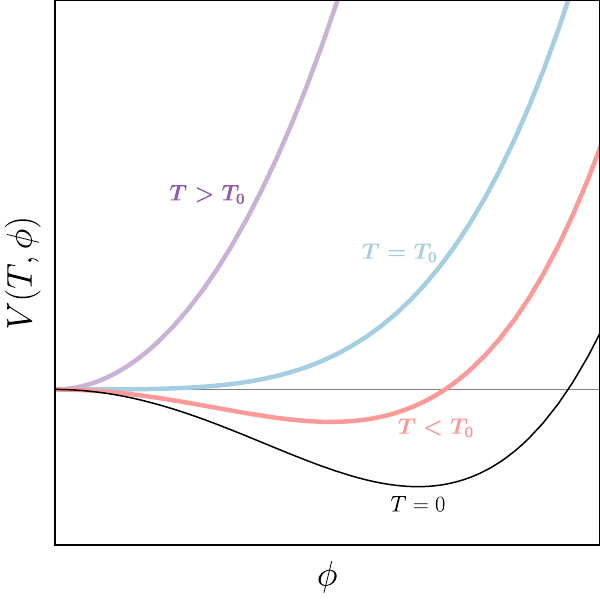}\quad\includegraphics[width=.45\textwidth]{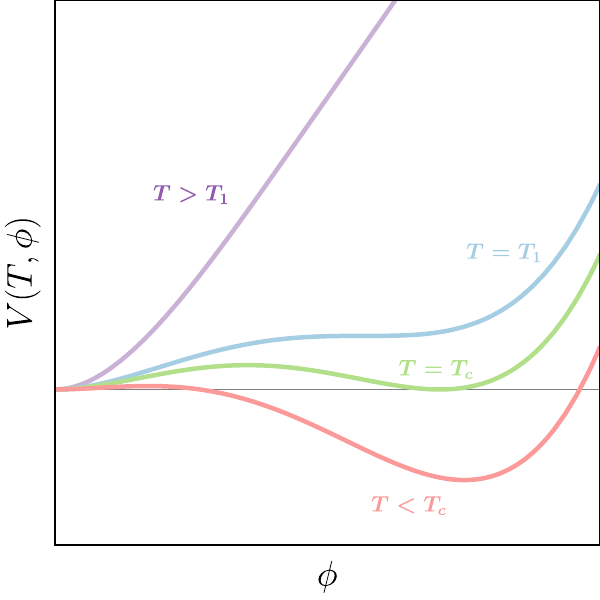}
    \caption{\textit{Left-hand panel.} Sketch of the evolution of the thermal effective potential in Eq.~\eqref{eq:VSOPT} for different values of $T$. At large temperatures, the symmetry-preserving point $\phi=0$ is stabilised by thermal corrections. At $T=T_0$, the origin turns into a maximum of the potential. \textit{Right-hand panel.} Same as in the left-hand panel, but for the potential in Eq.~\eqref{eq:VFOPT}, giving a \ac{FOPT}. The main difference with respect to the \ac{SOPT} scenario presented in the left-hand plot is the presence of a barrier separating two minima. For $T<T_c$, the minimum at the origin becomes metastable and the \ac{PT} occurs via bubble nucleation.}
\label{fig:ThermEffPot}  
\end{center} 
\end{figure}

Consider now a different scenario where the thermal effective potential features a cubic term
\be\label{eq:VFOPT}
    V(T,\phi) = D\left(T^2 - T_0^2\right)\phi^2 - ET\phi^3 + \frac{\lambda(T)}{4}\phi^4\,.
\ee
This term arises, for example, from bosonic interactions, as it can be seen from Eq.~\eqref{eq:JBexpansion}. The presence of a cubic term, changes the evolution of the potential with respect to the one in Eq.~\eqref{eq:VSOPT}, as depicted in the right-hand panel in Fig.~\ref{fig:ThermEffPot}. For large temperatures the potential features only one minimum, at the origin: symmetry is restored by thermal effects. At a temperature
\be
    T_1^2 = \frac{8\lambda(T_1) D}{8\lambda(T_1)D-9E^2} T_0^2
\ee
a second extremum appears, as a flex point.  Below $T_1$ we see the presence of a second local minimum, separated from the origin by a barrier. Such a barrier is induced by the cubic term. At the critical temperature
\be\label{eq:Tc}
    T_c^2 = \frac{\lambda(T_c)D }{\lambda(T_c)D - E^2}T_0^2
\ee
the two minima become degenerate in energy and, for $T<T_c$, the origin becomes metastable. Even if the two minima are still separated by a barrier, the field can tunnel from the local minimum to the global one. The \ac{PT} proceeds via nucleation of bubbles of true vacuum in a false-vacuum background. Finally, at $T=T_0$, the barrier vanishes.

It is worth mentioning that Eq.~\eqref{eq:VFOPT} is just an example of a potential leading to a \ac{FOPT}. In fact, the presence of a cubic term is not a necessary condition for the \ac{PT} to proceed via tunneling. For instance, a model featuring a classically scale-invariant scalar potential with the only addition of fermionic degrees of freedom is known to undergo a \ac{FOPT}. The decay rate of the \ac{SM} electroweak vacuum is studied in a similar setup, where, at large field values, the Higgs potential is dominated by the quartic term, and the relevant coupling is that to the top quark; see e.g.~Ref.~\cite{Devoto:2022qen} and references therein.  

The phenomenology of \ac{FOPT}s is very rich and intriguing. Among the many possible implications we mention: \ac{PBH} formation \cite{Kodama:1982sf,Liu:2021svg,Hashino:2021qoq,Kawana:2022olo,Gouttenoire:2023naa,Lewicki:2023ioy,Baldes:2023rqv,Banerjee:2024cwv} (see, however, the recent work in Ref.~\cite{Franciolini:2025ztf}), \ac{GW} production \cite{Caprini:2018mtu,Caprini:2019egz,Maggiore:2018sht}, heavy \ac{DM} production in bubble collision \cite{Watkins:1991zt,Falkowski:2012fb,Giudice:2024tcp}.

\subsection{Tunneling rate and bubble nucleation}\label{sec:Tunnel}

Due to the presence of a barrier separating the two minima of the thermal effective potential, a \ac{FOPT} proceeds via tunneling from the metastable vacuum to the stable one. This translates into the nucleation of bubbles of the broken phase in a symmetric background. Such bubbles, once nucleated, expand driven by the difference in potential energy between the inside and the outside.  The key quantity to study the progress of the \ac{PT} is the \textit{tunneling rate}, $\Gamma$, which represents the probability per unit time, per unit volume of one bubble nucleating in a homogeneous, metastable background. For \ac{PT}s driven by thermal fluctuations this is given by \cite{Linde:1980tt,Linde:1981zj}
\be
    \Gamma \simeq T^4 e^{-S_3/T}\,,
\ee
where $S_3$ is the so called \textit{bounce action}, i.e.~the Euclidean action of the ${\rm O}(3)$-symmetric solution to the equations of motion that interpolates between the metastable and the stable minima:
\be\label{eq:bounceS3}
    S_3 = 4\pi \int_0^\infty dr r^2 \left[\frac{1}{2}\left(\frac{d \phi_b(r)}{d r}\right)^2 + V\left(\phi_b(r), T\right)\right]\,.
\ee
Here, $V(\phi,T)$ is a generic thermal effective potential, for example the one given in Eq.~\eqref{eq:VFOPT}. The bounce solution $\phi_b(r)$ is the solution to the Euclidean equation of motion
\be\label{eq:bounceEoM}
    \frac{d^2\phi_b}{dr^2}+\frac{2}{r}\frac{d\phi_b}{dr} = \frac{\partial V(\phi_b, T)}{\partial \phi_b}\,,
\ee
with boundary conditions 
\be
    \lim_{r\rightarrow \infty} \phi_b(r) = 0\,,\qquad \frac{d\phi_b}{dr}\Big\rvert_{r=0}=0\,,
\ee
where we have chosen the origin as the false vacuum. Following Coleman \cite{Coleman:1977py}, we can interpret the radial coordinate $r$ as a time coordinate, and the field $\phi_b$ as a spatial coordinate. Then, the above equation of motion describes the classical problem of a particle in the inverted potential $- V$, experiencing a friction term that gets smaller and smaller with time. The particle is initially at rest somewhere between the global maximum (which corresponds to the global minimum of $V$), and the minimum (which corresponds to the barrier in Lorentzian time). The bounce solution is the trajectory of the particle that rolls down the upturned potential and arrives at rest at the second maximum (the metastable minimum of $V$). Finding the bounce solution amounts to finding the correct initial position, $\phi_{b,0}\equiv\phi_b(r = 0)$, from which the particle has to start rolling to come at rest at the second maximum. Once the bounce solution, $\phi_b(r)$, has been found, the action $S_3$ is known and we can compute the tunneling rate $\Gamma$. 

Usually, solving Eq.~\eqref{eq:bounceEoM}  for generic potentials is an arduous task which has to be carried out numerically. Coleman's interpretation provides us with a practical way of computing the bounce solution. We start by picking one value of $\phi_{b,0}$, close to the global maximum, and let the particle roll down. If the particle overshoots (undershoots) the second maximum we modify our initial guess increasing (decreasing) the value of $\phi_{b,0}$. By iterating this procedure until we find the correct $\phi_{b,0}$, we can compute the bounce solution and, hence, the bounce action $S_3$. In concrete, this procedure requires to specify the precise values of the parameters in $V(\phi,T)$. Alternatively, a semi-analytical approach can be adopted by taking advantage of the reparametrisation invariance of  Eq.~\eqref{eq:bounceEoM} \cite{Adams:1992bn,Levi:2022bzt}. For potentials of the form of the one in Eq.~\eqref{eq:VFOPT}, all the parameter-dependence in Eqs.~\eqref{eq:bounceEoM} and \eqref{eq:bounceS3} can be absorbed in one single parameter $\kappa$, which is given by a combination of $D$, $E$ and $\lambda$. The numerical procedure outlined above can then be used to compute the bounce solution for a generic value of $\kappa$, obtaining a semi-analytic expression for $S_3$ as a function of $\kappa$ and, hence, of the potential parameters. A specific example of this will be given in Ch.~\ref{chap:MonoDM}.

An analytic expression for the bounce action is known for a handful of potentials. Here, we will briefly review the \textit{thin-wall limit} and the \textit{thick-wall limit}.\footnote{Another notable example is the Fubini-Lipatov potential $V(\phi) = -\lambda \phi^4$, which is usually taken as a proxy to compute the decay rate of the Higgs vacuum in the \ac{SM}. See, e.g., Ref.~\cite{Devoto:2022qen} for a review.}

\subsubsection{Thin-wall limit}

The thin-wall limit applies whenever the energy difference between the two minima is small \cite{Coleman:1977py}. If this is the case, the potential can be generically written as
\be
    V(\phi) = V_+(\phi) + \frac{\epsilon}{2 a} (\phi -a)\,.
\ee
$V_+$ is a symmetric potential with minima at $\phi = \pm a$, while the term proportional to $\epsilon$ breaks the degeneracy between the two minima, rendering $\phi=-a$ the global minimum. We talk about thin-wall approximation if $\epsilon \ll 1$. This limit applies to weak \ac{FOPT}s, where the tunneling happens right after the critical temperature has been reached, at $T\simeq T_c$. If the energy difference between the two minima is small, the particle in Coleman's interpretation has to start close to the first maximum, which here is $\phi = -a$, and wait a long time until the friction term in Eq.~\eqref{eq:bounceEoM} becomes negligible. At this point it quickly rolls down, due to the absence of friction, and comes at rest at $\phi = + a$. The equation of motion for the thin-wall bounce is
\be\label{eq:EoMtW}
    \frac{d^2 \phi_b}{dr^2}\simeq \frac{\partial V}{\partial \phi_b}\,, 
\ee
where we have neglected the friction term. Going back to the Lorentzian picture, this bounce action describes a field profile that closely resembles a step function: inside the nucleated bubble the field is constant $\phi_b(r) \simeq -a$; at $r \simeq R$, with $R$ the radius of the bubble, the field profile sharply interpolates between $-a$ and $+a$; for $r \gg R$, $\phi_b(r)\simeq +a$. In other words, the field tunnels directly to the stable minimum (this is not, in general, the case, as we will see below for the thick-wall scenario). With this approximate solution for the bounce, we can write Eq.~\eqref{eq:bounceS3} as \cite{Coleman:1977py}
\be\label{eq:S3tW}
    S_3 \simeq 4\pi R^2 S_1 - \frac{4\pi}{3}R^3 \epsilon\,,
\ee
where
\be
    S_1 = \int_0^\infty dr \left[\frac{1}{2}\left(\frac{d \phi_b(r)}{d r}\right)^2 + V\left(\phi_b(r)\right)\right] \simeq \int_{-a}^{+a} d\phi \sqrt{2 V(\phi)}\,.
\ee
In the last equality, we have made use of the equation of motion in Eq.~\eqref{eq:EoMtW} to change the integration variable. Intuitively, Eq.~\eqref{eq:S3tW} receives two competing contributions: a positive surface term, proportional to $S_1$, and a negative volume term, proportional to the energy difference between the two vacua $\epsilon$. By extremising Eq.~\eqref{eq:S3tW} with respect to $R$, we find the critical radius $R_c$ of a nucleated bubble, i.e.~the radius that minimises the total energy of the bubble. All in all, we find, for a thin-walled bubble,
\be
    R_c^{\rm tw} = \frac{2S_1}{\epsilon}\,,\qquad S_3^{\rm tw} = \frac{16\pi}{3}\frac{S_1^3}{\epsilon^2}\,. 
\ee
In the limit $\epsilon \rightarrow 0$, both the critical radius and the action diverge to infinity. In fact, in this limit, the two minima are exactly degenerate and tunneling is forbidden.

\subsubsection{Thick-wall limit}\label{sec:thick-wall}

In the thick-wall scenario, the tunneling process happens between two widely separated minima, whose energy difference is large compared to the size of the barrier. This is what typically happens in supercooled \ac{PT}s (see Ch.~\ref{chap:MonoDM}), in which bubbles start nucleating only at $T\ll T_c$. Due to the large energy difference between the minima, the particle in the upturned potential starts rolling almost immediately (for small values of $r$), when the friction term is still relevant. The initial position has to be chosen far away from the global maximum in order not to overshoot. This translates into a field profile for the bounce solution that is spread over a distance $\delta R \simeq R$, hence the name “thick-wall''. The tunneling process drives the field right across the barrier, to a field value $\phi_*$, which is far from the stable minimum. After tunneling, the field classically rolls down the potential, from $\phi_*$ to the global minimum. In the thick-wall approximation, the bounce action reads \cite{Nardini:2007me}
\be
    S_3 \simeq \frac{4\pi}{3}R^3 \langle V\rangle + 2\pi R^2 \left(\frac{\delta \phi}{\delta R}\right)^2 \delta R\,,
\ee
where $\delta \phi$ is the field excursion between the metastable vacuum and the tunneling point, in this case $\delta \phi = \phi_*$, and $\langle V \rangle$ is the average potential energy density inside the bubble, which we take to be $\langle V\rangle = V(\phi_*)$. Remembering that, in the thick-wall approximation, $\delta R = R$, we find
\be
    S_3 = \frac{4\pi}{3}R^3 V(\phi_*) + 2\pi R \phi_*^2\,.
\ee
As in the thin-wall limit, we compute the critical radius by minimising the action and then we plug it back into $S_3$ to find
\be
    \left(R_c^{\rm Tw}\right)^2 = \frac{\phi_*^2}{-2 V(\phi_*)}\,,\qquad S_3^{\rm Tw} = \frac{4\pi}{3}\frac{\phi_*^3}{\sqrt{-2V(\phi_*)}}\,. 
\ee
For a given potential $V$, the expression for the tunneling point $\phi_*$ can be found by extremising $S_3^{\rm Tw}$ with respect to $\phi_*$.
\newpage

\begin{subappendices}
\section{Gaussian integrals}
\label{app:GaussInt}
In this appendix, we review the method to calculate integrals of the form
\be
    \mathcal{I} = \int_{-\infty}^{\infty} dx \exp\left[-\frac{1}{2}a x^2 + J x\right]\,,
\ee
which are very common in path integral computations. The first step consists in completing the square at the exponent, by summing and subtracting $J^2/(2a)$, so that the above integral can be written as
\be
    \mathcal{I} = \int_{-\infty}^{\infty} dx \exp\left[-\frac{a}{2}\left(x - \frac{J}{a}\right)^2 + \frac{J^2}{2 a}\right]\,.
\ee
After performing the change of variable $x' = \sqrt{a}\left(x + J/a\right)$, we get
\be
    \mathcal{I}= \frac{e^{\frac{J^2}{2a}}}{\sqrt{a}}\int_{-\infty}^{\infty}dx'\,e^{-\frac{{x'}^2}{2}}\,, 
\ee
which shows that $\mathcal{I}$ is a Gaussian integral. Definite Gaussian integrals are best computed in two dimensions, so we make use of the following trick 
\be
    \left(\int_{-\infty}^{\infty}dx\,e^{-\frac{x^2}{2}}\right)^2 = \int_{-\infty}^{\infty}dx\,dy\,e^{-\frac{x^2}{2}}e^{-\frac{y^2}{2}}= 2\pi \int_0^{\infty}r\,dr\, e^{-\frac{r^2}{2}}
=\pi \int_0^{\infty} dr^2 e^{-\frac{r^2}{2}} = 2\pi\,.
\ee
Therefore, we can easily apply the above result to our integral to show that
\be
    \mathcal{I} = \sqrt{\frac{2\pi}{a}} e^{\frac{J^2}{2a}}\,.
\ee
Such a reasoning is easily generalised to multi-dimensional integrals, where the integration variable $x$ is rather an $n$-dimensional vector $\vec{x}$, that, in general, can be complex: $ax^2 \rightarrow \vec{x}^\dag A \vec{x}$, where $A$ is an $n\times n$ matrix. After diagonalising $A$, the $n$-dimensional integral becomes a product of $n$ one-dimensional Gaussian integrals:
\be
    \int_{-\infty}^{\infty} d \vec{x} \exp \left[\frac{1}{2} \vec{x}^\dag A \vec{x} + \vec{J}^\dag\vec{x}\right] = \sqrt{\frac{\left(2\pi\right)^n}{\det A}}\exp\left[\frac{1}{2}\vec{J}^\dag A \vec{J}\right]\,.
\ee
The determinant at the denominator comes from the product of the eigenvalues of $A$.

\end{subappendices}

\chapter{The Standard Model of Cosmology}
\label{chap:DM}
\minitoc

Our current understanding of the Universe is based on two foundation stones. The first one is \textit{the cosmological principle}, which assumes that the Universe is isotropic and homogeneous on scales larger than $\sim100\,{\rm Mpc}$. Isotropy implies that the Universe appears the same in all directions from any given point, i.e.~there is no preferred orientation. Homogeneity suggests that the energy distribution in the Universe is uniform. Even though such assumptions are supported by the \ac{CMB}, it is important to stress the fact that the cosmological principle has not been experimentally proven, since we have only observed the \ac{CMB} from our location in the Universe. We cannot be sure that the Universe remains isotropic and homogeneous when observed from different points in space and time. Nonetheless, we have been known for quite a long time that we do not possess the status of special observers in the solar system, let alone the Universe. Therefore, we can reasonably assume that any other observer would see a Universe as smooth as we do on large scales. The second pillar of modern cosmology is the experimental measurement that our Universe is expanding with time, as proved for the first time by Hubble in 1929 \cite{Hubble:1929ig}.

These foundations imply that the geometry of our Universe, on large scales, is described by the \ac{FLRW} metric, which represents a spatially homogeneous and isotropic space, expanding with time. Focusing on the scenario in which spatial curvature vanishes, the \ac{FLRW} metric takes the form
\be\label{eq:FLRW}
ds^2 = -dt^2 + a^2(t) \left[ dr^2 + r^2 \left( d\theta^2 + \sin^2\theta\, d\phi^2 \right) \right],
\ee
where $a(t)$ is the scale factor.

By plugging the \ac{FLRW} metric into the $(0,0)$ component of the Einstein’s equations, we get the Friedmann equation
\be\label{eq:FriedEq}
H^2(t) \equiv \left( \frac{\dot{a}(t)}{a(t)} \right)^2 = \frac{\rho(t)}{3 \MP^2},
\ee
where $H(t)$ is the Hubble parameter, $ \rho(t)$ is the total energy density of the Universe, and $\MP = 2.4 \times 10^{18}\,{\rm GeV}$ is the reduced Planck mass. This equation is of pivotal importance in cosmology as it relates the rate of the expansion of the Universe to its energy content. For example, in the Hot Big-Bang picture, the Universe  started in a very hot and dense state. Expanding, it cooled down, as temperature evolves as $T\propto a(t)^{-1}$, and became more and more sparse, as volume increases as $V(t) \propto a(t)^{3}$.

Our knowledge of the early Universe is limited to two milestone moments in its evolution: \ac{BBN} and the \ac{CMB}. \ac{BBN} occurred at a temperature $T_{\rm BBN} \sim \mathcal O({\rm MeV})$, when light nuclei such as hydrogen, helium, and lithium were first able to form. The excellent agreement between theoretical predictions and observed abundances of such light elements is considered a strong evidence in support of the Standard Model of Cosmology. In particular, the observed abundances of light elements provide stringent bounds on any extra energy component that may have been present in the Universe at the time of \ac{BBN} \cite{Fields:2019pfx}. At a temperature $T_{\rm CMB} \sim \mathcal{O}(0.1\,{\rm eV})$, the Universe became cool enough for photons to decouple from matter, emitting the \ac{CMB}. This relic radiation provides a snapshot of the Universe at the epoch of recombination. The striking fact is that \ac{CMB} shows an extremely homogeneous Universe, in accordance with the cosmological principle. However, the most interesting piece of information is encoded in the small anisotropies observed in the \ac{CMB}. Such anisotropies tell us about the energy content of the Universe at the time, as well as about its initial conditions. Such early-time measurements have to be matched with our observations of the Universe at late times. In fact, today we observe a Universe which is extremely inhomogeneous at small scales (we are a living proof of this fact), and whose expansion has recently started to accelerate \cite{SupernovaSearchTeam:1998fmf}. 

Our best shot at describing the Universe throughout its whole history, in a way that complies with both the theoretical and observational insights introduced above, is the Standard Model of Cosmology, also known as the \textit{$\Lambda$CDM model}. It describes a spatially flat Universe composed primarily of \ac{CDM} and dark energy, the latter being modelled as a cosmological constant $\Lambda$. In this framework, \ac{CDM} is a non-relativistic, non-luminous form of matter that clusters gravitationally, while dark energy drives then accelerated expansion of the Universe at late times.

Despite its conceptual simplicity, $\Lambda$CDM remarkably succeeds in explaining a wide range of observations, including the \ac{CMB} anisotropy spectrum, large-scale structure formation, gravitational lensing, and the accelerating cosmic expansion. It is described by six parameters, which in the convention adopted by the Planck satellite collaboration are \cite{Planck:2018vyg}:
\begin{itemize}
    \item \textbf{Baryon density} ($h^2\Omega_b $): the fraction of the critical density (the total energy density in the Universe today) in ordinary matter;
    \item \textbf{CDM density} ($h^2\Omega_{\rm CDM}$): the corresponding fraction in \ac{CDM};
    \item \textbf{Sound horizon} $(\theta_{\rm MC})$: the angular size of the sound horizon at recombination;
    \item \textbf{Optical depth to reionisation} ($\tau_{\rm reio}$): the mean free path of photons travelling from \ac{CMB} till the moment in which the first stars reionised the Universe;
    \item \textbf{Scalar spectral index} ($ n_s$): describing the scale dependence of the curvature power spectrum;
    \item \textbf{Amplitude of curvature perturbations} ($ A_s$): describing their amplitude. 
\end{itemize}
These parameters are tightly constrained by current cosmological observations, and particularly by high-precision data from Planck \cite{Planck:2018vyg}. While $\Lambda$CDM offers an elegant and powerful framework, it is challenged by the Hubble tension, a discrepancy between the value of the Hubble rate measured at late times and the one inferred from \ac{CMB} (see Ref.~\cite{Kamionkowski:2022pkx} for a review), and by the recent DESI collaboration results, that seem to point towards an evolving dark energy paradigm to explain the expansion of the Universe, rather than a cosmological constant one \cite{DESI:2024mwx,DESI:2025zgx}. Moreover, despite its success, $\Lambda$CDM is a ``heuristic'' model that does not explain the origin of \ac{DM} and dark energy, nor that of the Universe initial conditions. In the rest of this chapter we shall focus on the main paradigms aiming at tackling the problem of the microscopic nature of \ac{DM} (Sec.~\ref{sec:DM}), and on the inflationary paradigm as a possible origin for the primordial fluctuations observed in the \ac{CMB} (Sec.~\ref{sec:Infla}). Before that, we will introduce the thermodynamics of an expanding Universe in Sec.~\ref{sec:Thermo}, setting the stage for our subsequent discussion.

\section{Thermodynamics of the early Universe}\label{sec:Thermo}
\subsection{Thermal equilibrium}\label{sec:ThermEq}
Assuming \ac{SM} particle content and interactions, the Universe was in thermal equilibrium at temperatures $T\gtrsim 100\,{\rm GeV}$. Moreover, the black-body spectrum of the \ac{CMB}, suggests that  equilibrium was maintained till much later times. These observations allow us to describe the thermodynamics of the early Universe using statistical mechanics. 

The thermodynamical properties of an ensemble of particles depend on their phase-space distribution $f(\vec{p},t)$. Starting from $f$, we can compute the particle number density, energy density and pressure as
\be\label{eq:ThermoQuant}
\begin{aligned}
     n(t)&= g \int \frac{d^3p}{(2\pi)^3} f(\vec{p},t)\,,\\
     \rho(t) &= g \int \frac{d^3p}{(2\pi)^3} E(\vec{p})f(\vec{p},t)\,,\\
     P(t) &= g \int \frac{d^3p}{(2\pi)^3} \frac{\lvert\vec{p}\rvert^2}{3 E(\vec{p})}f(\vec{p},t)\,,
\end{aligned}
\ee
respectively, where $g$ are the internal degrees of freedom (spin, gauge multiplicity etc.). Here, we have neglected any possible interaction among particles, so that the energy of a particle of mass $m$ is $E^2(\vec{p}) = \lvert \vec{p}\rvert^2 + m^2$. Since most of our discussion will take place in the early Universe, which was most likely homogeneous and isotropic, the phase-space distribution effectively depends only on the magnitude of momenta $p \equiv \lvert\vec{p}\rvert$, and not on their orientation. As long as kinetic equilibrium is established, i.e.~as long as particles efficiently exchange momentum through elastic scatterings, their phase-space distribution is 
\be\label{eq:feq}
    f_{\rm eq}(p) = \frac{1}{e^{E/T}\pm 1}\,,
\ee
where the $``-"$ sign is for bosons, Bose-Einstein distribution, while the $``+"$ sign is for fermions, Fermi-Dirac distribution.\footnote{We have neglected, here, the chemical potential, assuming that the number of particles is equal to that of antiparticles, at least as long as thermal equilibrium holds.} Assuming an equilibrium distribution, we can explicitly compute the number density, defined in Eq.~\eqref{eq:ThermoQuant}, in the limits of ultra-relativistic ($T \gg m$) and non-relativistic ($T \ll m$) particles
\be\label{eq:neq}
    n(T) \simeq \Bigg\{\begin{array}{lr} 
            c_1\frac{g\zeta(3)}{\pi^2} T^3\quad &{\rm for}\quad T\gg m\,,\\
            g \left(\frac{m T}{2\pi}\right)^{3/2}e^{-m/T}\quad &{\rm for}\quad T\ll m\,,
            \end{array}
\ee
where $c_1 = 1, 3/4$ for bosons and fermions, respectively. In what follows, we will often approximate the equilibrium distribution in Eq.~\eqref{eq:feq} with the  \ac{MB} one $f_{\rm eq}(p)\simeq e^{-E/T}$, neglecting quantum effects. In fact, taking a \ac{MB} distribution simplifies computations, allowing us to get the number density of particles for any temperature 
\be
    n_{\rm MB}(T) \simeq g \frac{m^2 T}{2\pi^2}K_2(m/T) \simeq \Bigg\{\begin{array}{lr} 
            \frac{g}{\pi^2} T^3\quad &{\rm for}\quad T\gg m\,,\\
            g \left(\frac{m T}{2\pi}\right)^{3/2}e^{-m/T}\quad &{\rm for}\quad T\ll m\,,
            \end{array}
\ee
where $K_2(x)$ is the modified Bessel function of the second kind.
We see that, in the non-relativistic limit, the result obtained with the \ac{MB} distribution agrees with the exact one in Eq.~\eqref{eq:neq}. The reason is that, on dimensional grounds, we expect $p \sim T$ and, therefore, $T \ll m$ implies $T \ll E$, in which limit the full distribution in Eq.~\eqref{eq:feq} reduces to the \ac{MB} one. Moreover, we see that in the ultra-relativistic regime, the error we are making is of $17\%$ for bosons and of $10\%$ for fermions. This justifies our use of the \ac{MB} distribution in the following sections, even for particles that are not in the non-relativistic regime.

Analogously, we can compute the energy density $\rho$ using the exact distribution in Eq.~\eqref{eq:feq}, in the two limiting cases
\be
    \rho(T) \simeq \bigg\{\begin{array}{lr} 
            c_2\frac{g\,\pi^2}{30} T^4\quad &{\rm for}\quad T\gg m\,,\\
           m\,n(T)\quad &{\rm for}\quad T\ll m\,,
            \end{array}
\ee
where $c_2 = 1$ ($c_2 = 7/8$) for bosons (fermions). In the presence of multiple relativistic species, potentially each with a temperature $T_i$, the total energy density in radiation is typically written as $\rho_r(T) = g_*(T)\pi^2 T^4/30$, where $g_*(T)$ is the \textit{effective number of relativistic degrees of freedom}
\be
    g_*(T) = \sum_{i={\rm bos}} g_i \left(\frac{T_i}{T}\right)^4+ \frac{7}{8}\sum_{i = {\rm fer}} g_i \left(\frac{T_i}{T}\right)^4\,.
\ee
In the above definition of $g_*$, $T$ represents the temperature of the Universe thermal bath, customarily chosen as the \ac{SM} photon temperature. Particles that are both in \textit{kinetic} and in \textit{chemical} equilibrium with the thermal bath are said to be in \textit{thermal} equilibrium, and share a common temperature $T_i = T$.\footnote{Roughly speaking, particles are in kinetic equilibrium as long as elastic processes are efficient. Chemical equilibrium implies that creation and annihilation of a particle species happen at the same rate. Therefore, the number density of particles in chemical equilibrium would be  constant, if it were not for the expansion of the Universe.} Adopting a \ac{MB} distribution, the energy density is
\be
    \rho_{\rm MB}(T) = \frac{g m^2 T}{2\pi^2}\left[m K_1\left(m/T\right) + 3 T K_2\left(m/T\right)\right]\,,
\ee
with $K_1(x)$ the modified Bessel function of the first kind.

Finally, the pressure is $P \simeq \rho/3$ for relativistic particles, as expected from radiation fluid, and $P \simeq 0$ for non-relativistic particles, i.e.~a matter fluid.  

With the above thermodynamical quantities at hand, we can compute the entropy density of a radiation fluid
\be\label{eq:s}
    s = \frac{\rho + P}{T} = \frac{2\pi^2 g_s(T)}{45}T^3\,,
\ee
where we have introduced the \textit{effective number of entropy density degrees of freedom}
\be
    g_s(T) = \sum_{i={\rm bos}} g_i \left(\frac{T_i}{T}\right)^3+ \frac{7}{8}\sum_{i = {\rm fer}} g_i \left(\frac{T_i}{T}\right)^3\,.
\ee
Notice that $g_*(T) = g_s(T)$ for particles in thermal equilibrium. We will see below, when computing the abundance of \ac{DM}, that the entropy density is a particularly useful quantity. In fact, as long as the Universe evolution is adiabatic, the total entropy density is conserved
\be
    s\,a^3 = \frac{2 \pi^2 g_s(T)}{45} T^3 a^3 = {\rm const.}
\ee

\subsection{The Boltzmann equation}\label{sec:BoltzEq}
In an expanding Universe, thermal equilibrium is not always established. The question whether a particle species is in thermal equilibrium depends on the nature of the interaction it possesses with the thermal bath and, in general, with other particles. In this section we want to study how the number density of particles that are not in equilibrium evolves with time. The formalism we develop here will be used in Sec.~\ref{sec:DM} to compute the relic abundance of \ac{DM}. The equation describing the evolution of the distribution $f$, and, hence, of $n$, in the presence of interactions and beyond thermal equilibrium is the Boltzmann equation
\be
    \hat{L}[f(p,t)] = \hat{C}[f(p,t)]\,,
\ee
where $\hat{L}$ and $\hat{C}$ are the Liouville operator and the collision operator, respectively. In a \ac{FLRW} Universe, the Liouville operator reads \cite{Gondolo:1990dk}
\be
    \hat{L}[f(E,t)] = \left(E \frac{\partial}{\partial t} - H p^2 \frac{\partial}{\partial E}\right) f(E,t)\,,  
\ee
where we have freely traded momentum dependence for a dependence on the energy. Since, at the end of the day, the quantity we are interested in is $n(t)$, we integrate both sides of the Boltzmann equation and we get to
\be
    \dot{n}(t) + 3H n(t) = \frac{g}{(2\pi)^3}\int \frac{d^3 p}{E}\hat{C}[f(E,t)]\,,
\ee
where the dot denotes derivative with respect to cosmic time. The right-hand side of the above equation encodes the effects of creation and destruction of particles due to inelastic interactions. In the absence of inelastic interactions the integral of $\hat{C}$ vanishes and the Boltzmann equation is trivially satisfied by $n(t) \propto a(t)^{-3}$: particle number density is merely depleted by the effect of the expansion of the Universe. However, in more realistic (and interesting) scenarios, particles will undergo a number of interactions that change their number density. In the rest of this section, we will focus on processes of the form $1\,2\leftrightarrow 3\,4$, where integer numbers label the different particle species, and solve the Boltzmann equation to find how the number density of the species $1$, $n_1$, evolves with time. Our discussion can be easily generalised to processes including an arbitrary number of particles in the initial and final states $1\,2\,\cdots \leftrightarrow i\,j\,\cdots$.

In the case of a two-to-two scattering like the one given above, the collision term entering the Boltzmann equation for $n_1$ reads
\be \label{eq:CollTerm}
\begin{aligned}
    \frac{g_1}{(2\pi)^3}\int\frac{d^3 p_1}{E_1} C[f_1] =  &\int d \Pi_1 d\Pi_2 d\Pi_3 d\Pi_4 (2\pi)^4 \delta^4 (p_1+p_2-p_3-p_4)\\& \times \left[-\lvert\mathcal{M}_{12\rightarrow 34}\rvert^2f_1 f_2 (1 \pm f_3)(1 \pm f_4) + \lvert\mathcal{M}_{34\rightarrow 12}\rvert^2 f_3 f_4 (1 \pm f_1)(1 \pm f_2)\right]\,
\end{aligned}
\ee
where $d\Pi_i\equiv g_i d^3 p_i /(2\pi^3)2 E_i$ is the Lorentz-invariant phase space. The above expression is quite intuitive. The first term in square brackets reduces the number of 1-particles due to  $1\,2 \rightarrow 3\,4$, described by the transition amplitude $\mathcal{M}_{12\rightarrow 34}$. At the same time, the inverse process $3\,4\rightarrow 1\,2$ increases $n_1$. Initial particle momenta are weighted by their distributions $f_i$. For final particles we have to take into account the effect of Bose enhancement or Pauli blocking, therefore momenta are weighted by $1\pm f_i$, where the upper sign is for bosons while the lower one is for fermions. The transition amplitude $\lvert\mathcal{M}\rvert^2$ is averaged over initial and final states and includes symmetry factors for both initial and final particles. In practice, in the following we will only consider scenarios where CP is conserved and, therefore, $\lvert\mathcal{M}_{12\rightarrow 23}\rvert^2 = \lvert\mathcal{M}_{34\rightarrow 12}\rvert^2 \equiv \lvert\mathcal{M}\rvert^2$.

In principle, to get the evolution of the number density for each species, $n_i(t)$, we are faced with the arduous task of solving a set of coupled equations for the different $f_i$. However, we can make a number of simplifying assumption, which will hold in any scenario of interest to us. First of all, we assume that particles 3 and 4 are in chemical and kinetic equilibrium, so that their number densities are only affected by the expansion of the Universe and their distributions are $f_i = f^{\rm eq}_{i} \simeq e^{-E_i/T}$ for $i=3,4$. Notice that we have approximated the equilibrium distributions with the \ac{MB} ones. Moreover, we assume that the effect of Pauli blocking and Bose enhancement can be neglected $1 \pm f_i \simeq 1$.\footnote{The error one makes in neglecting Pauli blocking and Bose enhancement is approximately of 10\%, compatible with that coming from approximating the equilibrium distribution with the \ac{MB} one.}  The Dirac delta in the right-hand side of Eq.~\eqref{eq:CollTerm} enforces energy conservation, $E_1 + E_2 = E_3 + E_4$, leading to $f^{\rm eq}_{3} f^{\rm eq}_{4} - f_1 f_2 = f^{\rm eq}_{1} f^{\rm eq}_{2} - f_1 f_2$, where $f_{1,2}$ are the actual distributions of species 1 and 2 while $f^{\rm eq}_{1,2}$ are their distributions if they were in equilibrium.
This allows us to write the Boltzmann equation as
\be
    \dot{n}_1(t) + 3 H n_1(t) = \int d\Pi_1 d\Pi_2 d\Pi_3 d\Pi_4\,(2\pi)^4\delta^4(p_1 + p_2 - p_3 - p_4)\lvert\mathcal{M}\rvert^2 \left(f^{\rm eq}_{1} f^{\rm eq}_{2} - f_1 f_2\right)\,.
\ee
To evaluate the right-hand side, we take our approximations one step further and assume that particles $1$ and $2$ are in kinetic equilibrium (whether they are also in chemical equilibrium or not is what we are trying to understand by solving the Boltzmann equations). The assumption of kinetic equilibrium usually holds since elastic scatterings are typically more efficient than inelastic ones. Furthermore, in concrete models, particles $1$ and $2$ also have interactions with the thermal bath  that are much stronger than their interaction with species $3$ and $4$. Hence, we can make the ansatz that $f_i = f_i^{\rm eq}n_i/n_i^{\rm eq}$ for $i=1,2$ \cite{Gondolo:1990dk}. The Boltzmann equation, then, takes the familiar form
\be\label{eq:BoltzmannEqn}
    \dot{n}_1(t) + 3 H n_1(t) = \langle \sigma_{12\rightarrow 34} v\rangle (n_1^{\rm eq}n_2^{\rm eq} - n_1 n_2)\,
\ee
where $v = \left[(p_1 \cdot p_2)^2-m_1^2 m_2^2\right]^{1/2}/(E_1 E_2)$ is called the \textit{M\o ller velocity}. In the above expression we have also defined the thermally averaged cross section
\be
    \langle \sigma_{12\rightarrow 34} v\rangle \equiv (n_1^{\rm eq}n_2^{\rm eq})^{-1}\int d\Pi_1 d\Pi_2 d\Pi_3 d\Pi_4 (2\pi)^4 \delta^4(p_1 + p_2 - p_3 - p_4) \lvert \mathcal{M} \rvert^2 e^{-E_1/T} e^{-E_2/T},
\ee
which can be thought as the ``zero-temperature'' cross section where the initial states have been weighted with their equilibrium thermal distribution:\footnote{Here, we define the zero-temperature cross section for $1\,2\rightarrow 3\,4$ as
\be
    \left[\sigma_{12\rightarrow 34}\right]_{`T=0'} \equiv \int d\Pi_3 d\Pi_4 (2\pi)^4 \delta^4(p_1 + p_2 - p_3 - p_4) \lvert\mathcal{M}\rvert^2\,.
\ee
According to our definitions of $d\Pi$ and $\lvert \mathcal{M} \rvert^2$ below Eq.~\eqref{eq:CollTerm}, this represents the cross-section summed over particles both in the initial and final states, and including symmetry factors. Notice that this definition of $\sigma$ is slightly different from the one usually adopted to describe, for example, processes at colliders.
}
\be
    \langle \sigma_{12\rightarrow 34} v\rangle = \int \frac{d^3 p_1}{(2\pi)^3}\frac{d^3 p_2}{(2\pi)^3} e^{-E_1/T}e^{-E_2/T}  v \,\left[\sigma_{12\rightarrow 34}\right]_{{\rm`}T=0{\text{'}}}\,.
\ee
Staring at Eq.~\eqref{eq:BoltzmannEqn}, in the simplified scenario in which particles 1 and 2 and particles 3 and 4 are equal pairwise, we can already grasp the physics by considering two limiting cases. If $\langle \sigma_{11\rightarrow 33} v\rangle n^{\rm eq} \gg H$, the Boltzmann equation tells us that $n_1\simeq n_1^{\rm eq}$. Intuitively, the annihilation of a pair of 1-particles into a pair of 3-particles is efficient and so it is the inverse process. Chemical equilibrium is established and the number density $n_1$ tracks its equilibrium value. In the opposite limit, $\langle \sigma_{11\rightarrow 33} v\rangle n^{\rm eq} \ll H$, the annihilation process is not able to keep up with the Hubble expansion and $n_1 \propto a(t)^{-3}$, as expected in the absence of any interaction. Notice that the transition between the two limiting regimes will take place at some point due to the cooling-down of the Universe. In fact, the quantities $n_{\rm eq}$, $H$ and $\langle \sigma_{12\rightarrow 34} v\rangle$ all depend on the temperature of the thermal bath $T$.

Eq.~\eqref{eq:BoltzmannEqn} is best expressed in terms of the comoving number density (sometimes also referred to as the \textit{``yield''}) $Y\equiv n/s$, where the expansion of the Universe has been factored out by dividing by the entropy density $s$, defined in Eq.~\eqref{eq:s}. Moreover, since the thermally-averaged cross section depends on $T$, it will prove convenient to use temperature instead of time as a variable in the Boltzmann equation. And, since it is good practice to work with dimensionless quantities, we define $z\equiv m_1/T$. After this cosmetic refinements, our Boltzmann equation becomes
\be
    \frac{d Y_1}{d z} = \frac{s N_1 \langle \sigma_{12\rightarrow 34}v\rangle Y^{\rm eq}_1 Y^{\rm eq}_2}{z H}\left(1-\frac{Y_1 Y_2}{Y_1^{\rm eq} Y_2^{\rm eq}}\right)\,,
\ee
where we have assumed that entropy is conserved. The factor $N_1$ above represents the number of $1$-particles produced/destroyed by each scattering process. If the species 1 and 2 are equal to each other, as is the case for Majorana fermions or real scalars, then $N_1 = 2$. This concludes, for the moment, our derivation of the Boltzmann equation. We will come back at it later, when we will solve it to get the relic abundance of \ac{DM} obtained through the freeze-out and freeze-in mechanisms.

\section{Dark matter}\label{sec:DM}
    The presence of a non-baryonic matter component in the Universe is hinted at by a plethora of observations and measurements.\footnote{Here, baryonic is to be understood according to the wicked notation adopted by cosmologists, to whom protons, neutron and electrons are all ``baryons''.} Despite the presence of \ac{DM} being widely accepted by the scientific community, and in spite of the tremendous experimental effort put forward so far, little is known about its nature. This is chiefly due to the fact that all we know about \ac{DM} comes from its gravitational interaction with the surrounding visible matter. This allows us to only access its properties on a macroscopic level. The most precise measurement of the \ac{DM} energy fraction in the Universe comes from the \ac{CMB} \cite{Planck:2018vyg}
\be
    h^2\Omega_{\rm DM} = 0.1200 \pm 0.0012
\ee
where $\Omega_{\rm DM} = \rho_{\rm DM}/\rho_{\rm cr}$ is the energy density of \ac{DM} over the critical energy density of the Universe $\rho_{\rm cr} = 3 \MP^2 H_0^2$. By rescaling the Hubble parameter today as $H_0 = h\, 100\,{\rm km}/{\rm s}/{\rm Mpc}$, with $h\simeq 68$, we see that the above value translates into a fraction of around 25\% the total energy density. On the contrary, the energy fraction of baryonic matter is only $h^2 \Omega_b \simeq 0.02$ \cite{Planck:2018vyg}. Observations also tell us that: \ac{DM} had to be non-relativistic already at the time of structure formation, it has to be stable (or with a lifetime that is larger than the age of the Universe), and it has to interact at most weakly, both with itself and with the \ac{SM} particles.

Even though it seems like we know quite a few things about \ac{DM}, the possible explanations behind it are copious, with many plausible paradigms (see Ref.~\cite{Cirelli:2024ssz} for a vast review, and references therein). In the absence of a detection, be it direct or indirect, current experimental bounds do not allow us to narrow down the different possibilities. Quite on the contrary, the possible \ac{DM} masses span a huge range of values that notably goes from ultra-light \ac{DM}, with a mass that can be as small as $m_{\rm DM}\simeq 10^{-21}\,{\rm eV}$, to \ac{PBH} \ac{DM} that can be as heavy as $m_{\rm DM} \simeq 10^{-12} M_{\odot}$. 

In this thesis, we will focus on the particle \ac{DM} paradigm, in particular studying the \ac{WIMP} as well as the \ac{FIMP} paradigms. We will also explore the possibility of \ac{DM} being composed of dark monopoles. After resuming the main experimental motivations that lead to the entrance of \ac{DM} in the scientific debate, we will review the freeze-out and the freeze-in mechanism as two complementary ways of obtaining the correct abundance of \ac{DM}. We will, then, briefly sketch the idea behind \ac{DM} in the form of point-like topological defects, to whose detailed discussion we devote Ch.~\ref{chap:MonoDM}.

\subsection{Motivations}\label{sec:DMMotivations}

\begin{itemize}
    \item \textbf{Spiral-galaxies rotation curves.} The smallest-scale evidence for \ac{DM} comes from spiral galaxies. Newtonian gravitation allows us to compute the rotation curve of a spiral galaxy, i.e.~the circular velocity of a test mass $m$ as a function of its distance from the centre of the galaxy, $r$. By equating the gravitational force that a mass $M(r)$ contained inside a sphere of radius $r$ exerts on $m$, to the centrifugal force experienced by $m$ while moving along a circular orbit or radius $r$, we get 
    \be
        v(r) = \sqrt{\frac{ \GN M(r)}{r}}\,,
    \ee
    under the assumption of spherical symmetry. At large $r$, most of the visible mass of the galaxy is enclosed by the orbit of the test mass, and $M$ can be approximated as constant. 
    Then, the circular velocity decreases as $v(r) \propto r^{-1/2}$. Measurements of rotation curves, carried out by measuring the Doppler effect of luminous matter, however, suggest a quite different behaviour with a velocity curve that flattens at large $r$. To explain the constant behaviour of $v(r)$ we can, then, postulate the presence of some invisible (dark) matter with a radial density profile $\rho_{\rm DM} \propto r^{-2}$ so that $M(r) \propto r$. 
    
    \item \textbf{Velocity dispersion of the Coma cluster.} On larger scales, we find the presence of galaxy clusters, i.e.~systems of hundreds of thousands of gravitationally bound galaxies. The need for extra, non-luminous matter was first put forward by Zwicky in his work on the Coma cluster \cite{Zwicky:1933gu}. If we model the cluster as a collection of $N \gg 1$ objects of equal mass $m$, we can estimate its total mass, $M = N m$, from the virial theorem
    \be
        N  \frac{1}{2}m v^2 = \frac{1}{2}\frac{N^2}{2}\frac{ \GN m^2}{R}\qquad \Rightarrow\qquad m N = \frac{2 R v^2}{\GN}\,.
    \ee
    By measuring the typical size of the cluster, $R$, and the velocity of each individual mass $m$, we can then infer the virial mass of the system. Zwicky found that the the virial mass so computed is actually much larger than the one measured from luminous matter. Therefore, he argued that the very existence of the cluster was suggesting the presence of extra matter that we are not able to see.
    
    \item \textbf{The bullet cluster.} What is considered today as the most convincing evidence of the existence of \ac{DM} is the observation of two colliding galaxy clusters, known as the bullet cluster, carried out for the first time in 2006 \cite{Clowe:2006eq}. By measuring the distribution of visible matter through X-ray emissions, and that of \ac{DM} through weak lensing, it became clear that the two components are extremely well spatially separated. This happens because, during the collision, the baryonic matter present in the two clusters, that possesses a quite strong self-interaction, experiences a shock wave, while the two \ac{DM} distributions pass through each other. As a result, \ac{DM} gets ``filtered'' and appears today separated from the baryonic component that remains close in space to the collision point.  
    
    \item \textbf{Large-scale structure formation.} Our Universe today is extremely inhomogeneous, as one can infer, for example, from galaxy surveys \cite{BOSS:2016wmc, DESI:2024mwx}. At the same time, \ac{CMB} observations suggest that the primordial inhomogeneities produced during inflation are as small as $\delta \simeq 10^{-5}$ \cite{Planck:2018vyg}. If only baryonic matter were present in the Universe, primordial density fluctuations would start growing under the effect of gravity only after recombination. In fact, before that time, baryons and photons are tightly coupled due to electromagnetic interactions, and form a relativistic fluid with a speed of sound as large as $c_s^2 \simeq 1/3$. Such a large $c_s$ prevents the growth of overdensities, which, instead, oscillate due to the strong pressure exerted by photons. It is only after baryons decouple that they can start forming structures. This, however, gives overdensities too little time to grow and produce the large inhomogeneities we observe today. A non-relativistic, pressureless, matter component that does not couple to photons, on the contrary, is able to cluster already at the time of matter-radiation equality, due to the absence of any pressure effect preventing its overdensities from growing. After baryons decouple they are free to fall into the gravitational wells of \ac{DM} clumps that ``speed-up'' structure formation. The above hand-waving argument can be made more rigorous by solving the linearised Euler equations for the different fluids. Here, however, we content ourselves with a pictorial description and we refer the interested reader to any textbook on cosmology, e.g.~Ref.~\cite{Dodelson:2003ft}, for a more thorough discussion. It is important to stress the fact that the success of \ac{DM} in explaining structure formation hinges upon its being non-relativistic at the time of matter-radiation equality. This sets a lower bound on the \ac{DM} mass, $m_{\rm DM} \gtrsim {\rm KeV}$.
    
    \item \textbf{Acoustic peaks in the \ac{CMB}.} As already mentioned at the beginning of this section, the most precise measurement of \ac{DM} abundance so far is provided by the \ac{CMB}. Its power spectrum presents a series of peaks at different scales, which are due to the acoustic oscillations in the photo-baryonic fluid. Such oscillations are the result of the competition between gravitational collapse and radiation pressure prior to recombination. Their positions and relative amplitudes are sensitive to the total matter content as well as to the relative contributions of baryonic and non-baryonic matter. In particular, \ac{DM} provides the gravitational potential wells in which the photon-baryon fluid oscillates. In a universe without \ac{DM}, baryons would experience a much weaker gravitational potential. The result would be a \ac{CMB} power spectrum with reduced amplitude of the first peak and nearly equal heights of the odd and even peaks. Such a spectrum would be inconsistent with observations from \ac{CMB} experiments, such as Planck \cite{Planck:2018vyg}. Fitting the \ac{CMB} power spectrum, therefore, tightly constraints the relative abundance of dark and baryonic matter components. Once again, the interested reader is referred to Ref.~\cite{Dodelson:2003ft} for a pedagogical introduction to the as intricate as fascinating topic of \ac{CMB} physics.
\end{itemize}

\subsection{Thermal freeze-out and the WIMP paradigm}\label{sec:Freeze-out}
The freeze-out mechanism provides a compelling way of generating the relative energy density of \ac{DM} we observe in the Universe today. Let us consider a stable particle $\chi$ with mass $m_\chi$, and some interaction with the \ac{SM} which is efficient enough for the two sectors to be in thermal equilibrium at large temperatures. For concreteness, and since we have already worked out the Boltzmann equation for this scenario in Sec.~\ref{sec:BoltzEq}, we consider that the process keeping the dark sector and the visible one thermalised is $\chi \chi \leftrightarrow {\rm SM\, SM}$, described at zero temperature by the cross section $\sigma$. We assume this process to be efficient at temperatures $T \gg m_\chi$. At such large temperatures, then, $\chi$ is relativistic and in thermal equilibrium with the \ac{SM} bath. \ac{DM} and \ac{SM} particles share a common temperature and, therefore, a common number density $n\propto T^3$.\footnote{Here, we are implicitly assuming that \ac{DM} is heavier than any \ac{SM} particle. While this does not have to be the case, it slightly simplifies our discussion, since we do not have to keep track of entropy injection from \ac{SM} particles falling out of equilibrium as $T$ drops.} At temperatures $T \simeq m_\chi$, \ac{DM} becomes non-relativistic and, from this moment on, its relative number density gets suppressed as $n_\chi\propto e^{-m_\chi / T}$. This exponential suppression lasts as long as thermal equilibrium is maintained. However, we can suppose that at some temperature $T= T_{\rm fo}$, the interaction keeping $\chi$ in thermal equilibrium falls below the Hubble rate. \ac{DM} is not in equilibrium anymore and its comoving number density, $Y_\chi$, remains frozen to a value that is smaller than the photon one today, as imposed by observations. The question whether the \ac{DM} relic abundance so obtained matches the observed one is, then, a model-dependent one. The appeal of this paradigm is that its prediction in terms of the \ac{DM} relic abundance is insensitive to the Universe evolution prior to the freeze-out moment $T=T_{\rm fo}$.

The evolution of the comoving number density for $\chi$ across the different steps we have sketched above is described by the Boltzmann equation
\be
    \frac{d Y_\chi}{d z} = \frac{s\, 2\langle \sigma v\rangle (Y_\chi^{\rm eq})^2}{z H}\left[1-\frac{Y_\chi^2}{(Y_\chi^{\rm eq})^2}\right]\,,
\ee
as derived in Sec.~\ref{sec:BoltzEq}.
Once $m_\chi$ and $\langle \sigma v\rangle$ are specified, the above differential equation can be solved numerically to give the relic abundance of $ \chi$. Here, however, we will make some approximations to try to get analytic insights. We will focus on the range of temperatures for which $\chi$ is non-relativistic, $z\equiv m_\chi/T \gg 1$, since at larger temperatures the evolution of $n_\chi$ is trivial. At such temperatures the thermally averaged cross section can be expanded in small velocity $v$
\be\label{eq:sigmaExp}
    \langle \sigma v \rangle = \sigma_0 + \frac{6 T}{m_\chi}\sigma_1 + \cdots\,,
\ee
where $\sigma_0$ and $\sigma_1$ are known as the $s$-wave and $p$-wave contributions, respectively. In some specific models, $\sigma_0 
$ vanishes and the $p$-wave term represents the leading contribution. However, for the sake of simplicity, we will keep only the $s$-wave term in the above expansion
\be
\label{eq:BoltzmannChiApprox}
    \frac{dY_\chi}{dz} = \frac{\lambda}{z^2}\left[(Y_\chi^{\rm eq})^2 - Y_\chi^2\right]\,,\qquad {\rm where} \qquad \lambda \equiv \MP m_\chi \sigma_0 \sqrt{\frac{\pi g_s^2}{45 g_*}}\,.
\ee
The above definition for $\lambda$ holds in a radiation-dominated Universe. In the non-relativistic regime, the equilibrium comoving number density of $\chi$ is
\be
    Y_\chi^{\rm eq} = \frac{45 g_\chi}{2^{5/2}\pi^{7/2}g_s}z^{3/2}e^{-z}\,,
\ee
as given by Eq.~\eqref{eq:neq}.
Eq.~\eqref{eq:BoltzmannChiApprox} can be solved numerically for any $z$, while analytic solutions can be derived much before freeze-out, $1 \ll z \ll z_{\rm fo}$, and much after freeze-out, $z \gg z_{\rm fo}$. The latter is the limit we are interested in to compute the abundance of $\chi$ today. 

\underline{$ \boldsymbol{z \ll z_{\rm FO}}$.} At times earlier than freeze-out, \ac{DM} is in thermal equilibrium with the thermal bath and $Y_\chi \simeq Y_\chi^{\rm eq}$. Therefore, we can expand Eq.~\eqref{eq:BoltzmannChiApprox} to the linear order in $\delta Y_{\rm \chi} \equiv Y_\chi - Y_\chi^{\rm eq} \ll Y_\chi^{\rm eq}$
\be
    Y_\chi \overset{z \ll z_{\rm fo}}{\simeq} Y_\chi^{\rm eq} + \frac{z^2}{2\lambda }\,,
\ee
where we have used the fact that $d Y_\chi^{\rm eq}/dz \simeq - Y_\chi^{\rm eq}$ for $z \gg 1$. Our approximation breaks down when $\delta Y_\chi$ becomes of the order of $Y_\chi^{\rm eq}$. This provides a good estimate for the moment of freeze-out $z_{\rm fo}$
\be
    z_{\rm fo} = \log\left( \frac{45}{2^{3/2}\pi^{7/2}}\frac{g_\chi}{g_s}\frac{\lambda}{z_{\rm fo}^{1/2}}\right)\,.
\ee
By solving it iteratively we find
\be
    z_{\rm fo} = \log\left(\frac{45}{2^{3/2}\pi^{7/2}}\frac{g_\chi}{g_s}\lambda\right)-\frac{1}{2}\log\log\left(\frac{45}{2^{3/2}\pi^{7/2}}\frac{g_\chi}{g_s}\lambda\right) + \cdots
\ee

\underline{$ \boldsymbol{z \gg z_{\rm FO}}$.} After freeze-out, the distribution of $\chi$-particles stops tracking the equilibrium one. Hence, for $z \gg z_{\rm fo}$, we can neglect the terms proportional to $Y_\chi^{\rm eq}$ in Eq.~\eqref{eq:BoltzmannChiApprox}:
\be
    \frac{d Y_\chi}{d z} \overset{z \gg z_{\rm fo}}{\simeq} = -\frac{\lambda}{z^2}Y_{\chi}^2\,.
\ee
 The above differential equation can be easily solved in the interval $z_{\rm fo} < z < \infty$, to get today's comoving number density of $\chi$
 \be\label{eq:YChiFO}
    Y_{\chi,0}^{\rm fo} \simeq \sqrt{\frac{45 g_*}{\pi^2 g_s^2}}\frac{z_{\rm fo}}{\MP m_\chi \sigma_0}\,,
 \ee
where $Y_{\chi,0} \equiv Y_{\chi}(z=\infty)$ and we have neglected the contribution at the time of freeze-out since $Y_{\chi}(z_{\rm fo})\gg Y_{\chi,0}$.

\begin{figure}[h!]
\begin{center}
\includegraphics[width=.47\textwidth]{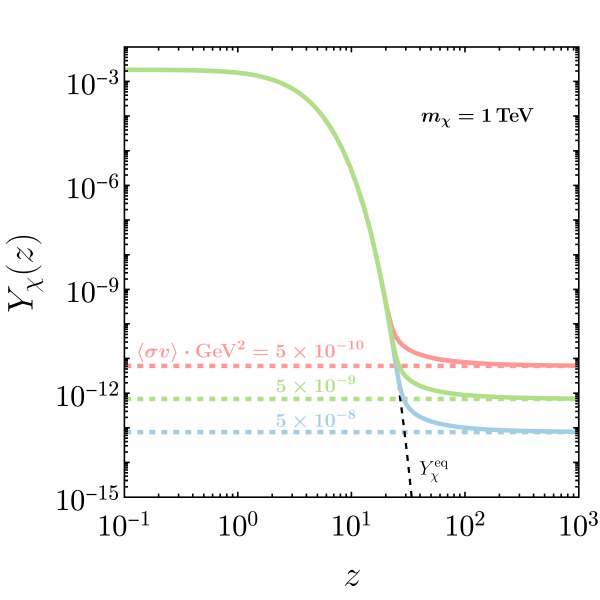}\qquad\includegraphics[width=.47 \textwidth]{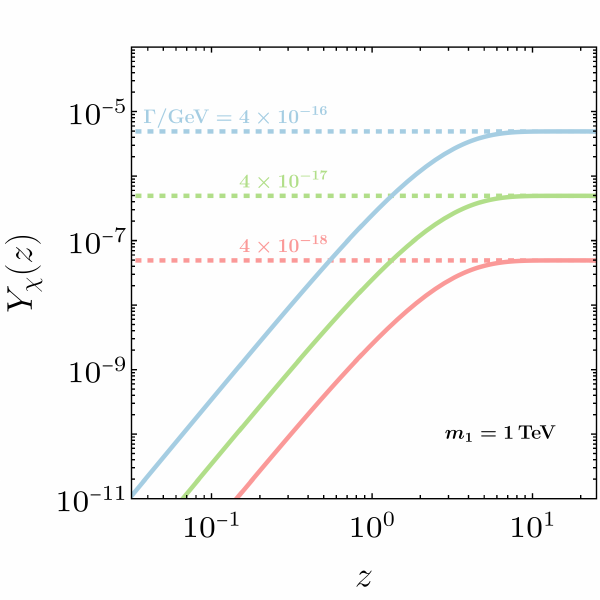}
    \caption{\textit{Left-hand panel.} Comoving number density of a thermal population of $\chi$ for different values of the annihilation cross-section, obtained by numerically solving Eq.~\eqref{eq:BoltzmannChiApprox} (solid lines). For $z\sim 1$, $\chi$ becomes non-relativistic and its number density gets exponentially suppressed. Around $z\sim 20 - 30$ the process keeping $\chi$ in thermal equilibrium freezes-out: the comoving number density stops tracking the equilibrium one (black dashed line), and quickly approaches the analytic estimate given in Eq.~\eqref{eq:YChiFO} (colored dashed lines). The green lines give $h^2 \Omega_\chi \simeq 0.12$ today. \textit{Right-hand panel.} Comoving number density of $\chi$ frozen in by the process $1 \rightarrow \chi\,2$, for $m_1 = 1\,{\rm TeV}$. The solid lines represent the numerical solution to Eq.~\eqref{eq:YChiFI}, while the dashed ones show the asymptotic value for $z\rightarrow \infty$, given in Eq.~\eqref{eq:YChiFI0}. Along the green lines the relic abundance of $\chi$ matches the measured one for $m_\chi = 1\,{\rm MeV}$.}
\label{fig:YChi}  
\end{center} 
\end{figure}

A numerical solution of Eq.~\eqref{eq:BoltzmannChiApprox} is shown in the right-hand plot in Fig.~\ref{fig:YChi}, alongside our analytic approximation at large $z$, cfr.~Eq.~\eqref{eq:YChiFO}.

 Today's \ac{DM} energy density fraction is given by $\Omega_\chi = \rho_\chi / \rho_0$, with $\rho_0 = 3 H_0^2 \MP^2$ the total energy density of the Universe today and $\rho_\chi = m_\chi n_{\chi,0}$ the \ac{DM} one. The Hubble constant today can be expressed as $H_0 = h \times 100\,{\rm km}/{\rm s}/{\rm Mpc}$ so that
 \be\label{eq:OmegaDM}
    h^2\Omega_\chi = \frac{m_\chi n_{\chi,0}}{3 \MP^2 (100\,{\rm km}/s/{\rm Mpc})^2} = \frac{2\pi^2 g_{s,0} T_0^3 Y_{\chi,0} m_\chi}{135 \MP^2 (100\,{\rm km}/s/{\rm Mpc})^2}=0.12\left( \frac{Y_{\chi,0}}{4.4\times 10^{-13}}\right)\left(\frac{m_\chi}{{\rm TeV}}\right)\,,
 \ee
 where we have used $g_{s,0} = 43/11$ and $T_0 = 2.7\,{\rm K}$. Now, using our estimate for the relic comoving number density in Eq.~\eqref{eq:YChiFO} (where we assume $g_s = g_* \simeq 100$) we find
 \be\label{eq:hOmFO}
    h^2 \Omega_\chi^{\rm fo} = 0.12\,\left(\frac{z_{\rm fo}}{23}\right) \left(\frac{4.6\times 10^{-9}\,{\rm GeV^{-2}}}{\sigma_0}\right)\,. 
 \ee
Any dependence on the \ac{DM} mass is only contained into $z_{\rm fo}$ and it is logarithmically mild. Moreover, as also shown in the left-hand panel in Fig.~\ref{fig:YChi}, the relic abundance scales with the inverse of the annihilation cross section: for a larger cross section, $\chi$ remains in equilibrium with the thermal bath for a longer time after it has become non-relativistic, thus experiencing more Boltzmann suppression. We now see, by plugging numbers in the above expression, that the correct relic abundance is obtained for $m_\chi \sim 1\,{\rm TeV}$ and $\sigma_0 \sim 5\times 10^{-9}\,{\rm GeV}^{-2}$, that is, very roughly, for a particle with a mass (almost) at the electroweak scale and with an annihilation cross section that is typical of weak interactions. This (almost) striking coincidence between the weak scale and the relic density of \ac{DM} goes under the name of \textit{``\ac{WIMP} miracle''}.

\subsubsection{The unitarity bound}
Some order-of-magnitude bounds on the mass of the \ac{DM} particle can be readily extracted from Eq.~\eqref{eq:hOmFO}. The $s$-wave contribution  to the cross section describing the annihilation of non-relativistic \ac{DM} into light particles is, by dimensional analysis, $\sigma_0\sim g^2/(4\pi m_\chi^2)$, with $g$ some dark coupling parametrising the interaction between the \ac{SM} and $\chi$. Therefore, the perturbativity requirement $g < 4\pi$, translates into $m_\chi \lesssim 100\,{\rm TeV}$. Particles heavier than $100\,{\rm TeV}$ will have too large a relic abundance, no matter what the value of the perturbative coupling $g$ is. This back-of-the-envelope estimate neglects the fact that $\sigma_0$ is only one term in the expansion of $\langle \sigma v \rangle$, see Eq.~\eqref{eq:sigmaExp}. The thermally averaged cross section receives contributions from infinitely many other terms ($p$-wave, $d$-wave etc.), which, despite being subleading, prevent us from giving a rigorous bound on $m_\chi$. 

A lower bound on the \ac{DM} mass can also be obtained if we assume that \ac{DM} annihilates via weak interactions $\sigma_0 \sim G_{\rm F}^2 m_\chi^2/2\pi$, with $G_{\rm F} = 1.2 \times 10^{-5}\,{\rm GeV^{-2}}$ the Fermi constant. In this case, in order to avoid producing $h^2\Omega_\chi > 0.12$, $\chi$ has to be heavier than $m_\chi \gtrsim {\rm few}\,{\rm GeV}$. This is known as the Lee-Weinberg bound \cite{Lee:1977ua}. Such a constraint can be, however, relaxed if we assume that \ac{DM} annihilates via other channels. Another possibility is that annihilation does not happen into \ac{SM} particles but rather into some other particle populating the dark sector. However, even in these scenarios, \ac{DM} mass cannot be smaller than few KeV. In fact, such a light \ac{DM} would be relativistic around the time of matter-radiation equality, thus spoiling structure formation, as also discussed in Sec.~\ref{sec:DMMotivations}.

\subsection{Non-thermal relics}
Thermal \ac{DM} is particularly attractive since its relic density mostly depends on one parameter only: the annihilation cross section. This makes freeze-out the most predictive scenario for \ac{DM} production. However, there is a plethora of viable scenarios in which \ac{DM} is successfully produced non-thermally. In the freeze-in scenario, particles with small interactions with the thermal bath can still be produced efficiently enough to reach sizeable relic densities \cite{Hall:2009bx}. The extremely large energies achieved during inflation make it a plausible framework to produce heavy particles. This can happen either during inflation, as a consequence to the expansion of the Universe \cite{Chung:1998zb}, or right after it, thanks to the onset of parametric resonance during the preheating stage (see Sec.~\ref{sec:Preh}). \ac{PBH}s are also viable \ac{DM} candidates \cite{Byrnes:2025tji}, even though the available parameter space is more and more constrained by astrophysical and cosmological observations \cite{Carr:2020gox}. The standard scenario leading to \ac{PBH} formation is that of collapsing large curvature perturbation produced during inflation. The large amplitude of the power spectrum that is needed to form \ac{PBH}s can be obtained in the context of ultra-slow-roll inflation \cite{Ivanov:1994pa}, where the potential features an inflection point, or thanks to an auxiliary field dubbed “curvaton'' \cite{Kawasaki:2012wr,Ando:2017veq,Inomata:2020xad,Ferrante:2023bgz}.\footnote{Typically, obtaining a curvature power spectrum large enough to lead to \ac{PBH} formation, in ultra-slow-roll scenarios of inflation, requires a large amount of fine-tuning at the level of the inflationary potential \cite{Cole:2023wyx}.} \ac{PBH}s can also be produced after supercooled \ac{FOPT}s \cite{Kodama:1982sf,Liu:2021svg,Hashino:2021qoq,Kawana:2022olo,Gouttenoire:2023naa,Lewicki:2023ioy,Baldes:2023rqv,Banerjee:2024cwv}. However, see Ref.~\cite{Franciolini:2025ztf} for a recent analysis questioning previous results. Extremely strong \ac{PT}s can lead to the dilution of previously thermalised relics \cite{Hambye:2018qjv,Baker:2019ndr}. Moreover, bubble collisions can reach energies high enough to produce heavy particles \cite{Watkins:1991zt,Falkowski:2012fb,Giudice:2024tcp}. Monopoles, which are stable due to the non-trivial topology of the vacuum manifold, may also form after a cosmological phase transition. Far from being an exhausting review of all the possible scenarios, here we just mentioned some of the main ideas for \ac{DM} production beyond the thermal paradigm. A more comprehensive and detailed list of the all the possibilities can be found in Ref.~\cite{Cirelli:2024ssz}. Below, we will discuss in more details the freeze-in mechanism and the monopole \ac{DM} scenario. The latter will be the main focus of Ch.~\ref{chap:MonoDM}.

\subsubsection{Freeze-in and \ac{FIMP} dark matter}
In Sec.~\ref{sec:Freeze-out}, we have assumed that \ac{DM} possesses interactions strong enough to have reached thermal equilibrium at some point during cosmic history. Another possibility is that \ac{DM} is a \ac{FIMP}, i.e.~its interactions with the thermal bath are so small that a thermal population cannot be produced from the plasma \cite{Hall:2009bx}. Such interactions, however, are efficient enough to produce few particles now and then, so that the integrated number density of \ac{DM} reaches a sizeable value during cosmic history. The process, which is never efficient compared to the Hubble rate, eventually stops when the bath particles producing \ac{DM} become non-relativistic and their number density is exponentially suppressed. Since \ac{FIMP}s are never thermalised in the first place, and since their number density is negligible for most of the Universe evolution, their mass can be lighter than MeV, without altering the rate of the expansion of the Universe at the time of \ac{BBN}. In other words, \ac{BBN} constraints do not apply to \ac{FIMP}s.

Depending on the interaction \ac{DM} has with the visible sector particles, we can identify two scenarios:
\begin{itemize}
    \item \ac{IR} dominated freeze-in: \ac{DM} possesses renormalisable interactions with other particles and the bulk of its relic abundance is produced at ``late'' times, right before it gets frozen;
    \item \ac{UV} dominated freeze-in: the interaction between \ac{DM} and the thermal bath is non-renormalisable and it shuts off at ``early'' times, close to the reheating temperature.
\end{itemize}
While the latter scenario is sensitive to the Universe initial conditions, i.e.~to the reheating dynamics, the former is analogous to freeze-out, in the sense that any dependence of \ac{DM} relic abundance on early-time physics is erased. Here, we will focus on the \ac{IR}-dominated scenario, considering, for simplicity a trilinear interaction leading to a $1 \rightarrow  \chi\,2$ decay, under the further assumption that $m_1 > m_2 + m_\chi$.

The Boltzmann equation describing the evolution of $n_\chi$ reads
\be
    \dot{n}_\chi + 3H n_\chi \simeq \int d\Pi_1 d\Pi_2 d\Pi_\chi(2\pi)^4 \delta^4(p_1 - p_2 -p_\chi)  \lvert \mathcal{M} \rvert^2 \left(f_1 - f_2 f_\chi\right)\,,
\ee
where $\mathcal{M}$ is the matrix element of the decay process, which we have assumed to be equal to the one of the inverse decay. We have also neglected the effect of Bose enhancement and Pauli blocking. Since $\chi$ always possesses a negligible number density, till the moment when the decay shuts down, we take $f_\chi \simeq 0$. Moreover, we assume that particle 1 interacts in a sizeable way with the thermal bath and, therefore, has an equilibrium distribution $f_1 \simeq e^{-E_1/T}$
\be
\begin{aligned}
    \dot{n}_\chi + 3H n_\chi &\simeq \int d\Pi_1 d\Pi_2 d\Pi_\chi (2\pi)^4 \delta^4(p_1 - p_2 - p_\chi) \lvert \mathcal{M}\rvert^2 e^{-E_1/T}\\
    & = \frac{g_1}{2\pi^2} m_1^2\,\Gamma\,T\, K_1(m_1/T)\,,
\end{aligned}
\ee
where, in the second equality, we have integrated the matrix element over final particle phase space to obtain the decay rate $\Gamma$, and we have carried out the integral over $p_1$. Expressing the Boltzmann equation in terms of the yield $Y_\chi = n_\chi/s$ and performing the change of variables $T\rightarrow z=m_1/T$ we get to
\be\label{eq:YChiFI}
    \frac{d Y_\chi(z)}{d z} \simeq -\frac{45}{4\pi^4}\sqrt{\frac{90 g_1^2}{g_* g_s^2}}\frac{\MP\Gamma}{m_1^2}z^3 K_1(z)\,.
\ee
The numerical solution to the above Boltzmann equation, $Y_{\chi}(z)$, is shown in the right-hand panel in Fig.~\ref{fig:YChi}, for benchmark values of $m_1$ and $\Gamma$. The relic abundance corresponds to $\lim_{z\rightarrow \infty}Y_\chi(z)$, that has the analytic form
\be\label{eq:YChiFI0}
    Y_\chi^{\rm fi} \simeq \frac{135}{8\pi^3}\sqrt{\frac{90 g_1^2}{g_* g_s^2}}\frac{\MP\Gamma}{m_1^2}\,.
\ee
The abundance of relics can be, then, obtained via Eq.~\eqref{eq:OmegaDM}
\be\label{eq:hOmFI}
    h^2\Omega_\chi^{\rm fi} = 0.12\, g_1\left(\frac{m_\chi}{1\,{\rm MeV}}\right)\left(\frac{1\,{\rm TeV}}{m_1}\right)^2\left(\frac{ \Gamma}{3.5\times 10^{-17}\,{\rm GeV}}\right)\,,
\ee
where we have assumed $g_* = g_s=100$.
By comparing Eq.~\eqref{eq:YChiFI0} with Eq.~\eqref{eq:YChiFO}, we notice one major difference, which is also evident from the two plots in Fig.~\ref{fig:YChi}: while a larger interaction rate suppresses the abundance of frozen-out relics (see discussion below Eq.~\eqref{eq:hOmFO}), for freeze-in \ac{DM} it has the opposite effect. Moreover, since in this scenario \ac{DM} is never in thermal equilibrium and it never has a cosmologically-relevant energy density, the constraint coming from \ac{BBN} observation is not present and \ac{DM} can be pretty light. Nonetheless, \ac{DM} has to be non-relativistic by the time of structure formation and this poses a lower bound on the \ac{DM} mass of $m_\chi \gtrsim 1\,{\rm KeV}$. 

\subsubsection{Monopole dark matter}
As detailed in Sec.~\ref{sec:Mono}, monopoles are stable, extended field configurations that arise in theories where the vacuum manifold emerging after spontaneous symmetry breaking is non-trivial. Their stability is ensured by the fact that they are the lightest states possessing a magnetic/topological charge. In gauge theories, the monopole mass is typically given by $m_M \sim v/g$, where $v$ is the \ac{VEV} of the scalar field responsible for spontaneous symmetry breaking, and $g$ is the gauge coupling related to the broken local symmetry. The relic abundance of monopoles strongly depends on the details of the cosmological \ac{PT} from which they formed, as well as on their evolution, with monopole-antimonopole annihilation possibly reducing their number density after formation. To set notation, we will refer to monopoles carrying magnetic charge under the \ac{SM} $\U{1}_{\rm em}$ as ``magnetic monopoles''. Monopoles arising from dark gauge sectors that do not mix with the \ac{SM} will be dubbed ``dark monopoles''.

The possible production of magnetic monopoles during a cosmological \ac{PT} has been known for quite a long time in the \ac{GUT} framework \cite{Preskill:1979zi}. In unified scenarios the presence of magnetic monopoles is considered a curse. The reason is that the \ac{GUT} scale needs to be as large as $v_{\rm GUT}\sim 10^{16}\,{\rm GeV}$ to keep the proton long-lived. This results in extremely massive monopoles, with $m_M \gtrsim v_{\rm GUT}$, that overclose the Universe. Moreover, the abundance of magnetic monopoles is constrained by astrophysical observations of the galactic magnetic field \cite{Turner:1982ag}, as well as by direct search experiments \cite{MACRO:2002jdv}, to be much smaller than $h^2\Omega_{\rm DM}$. Therefore, any scenario where monopoles are charged under the \ac{SM} magnetic field, and have a sizeable relic density is almost ruled out.

The possibility that monopoles emerging from a dark gauge sector could comprise a sizeable fraction of \ac{DM} has received relatively little attention \cite{Murayama:2009nj,GomezSanchez:2011orv,Baek:2013dwa,Khoze:2014woa,Kawasaki:2015lpf,Sato:2018nqy,Daido:2019tbm,Graesser:2020hiv}. In such a scenario, astrophysical constraints are evaded altogether, and the symmetry-breaking scale $v$ is not bounded by the proton decay, giving us more freedom to explore a larger portion of parameter space. A careful assessment of whether \ac{DM} can be entirely made of dark monopoles emerging from a $\SO{3}$ dark gauge group is the focus of Ch.~\ref{chap:MonoDM}.

\section{Inflation}\label{sec:Infla}

The inflationary paradigm, i.e.~an hypothetical stage of exponential expansion experienced by the Universe, was first introduced by Alan Guth in 1980 to address the homogeneity and flatness problems that plagued the Hot Big Bang scenario \cite{Guth:1980zm}. Soon after, it was realised that inflation also provides a compelling framework to generate the initial conditions for our Universe \cite{Starobinsky:1982ee, Guth:1982ec, Hawking:1982my}. However, it became clear at the same time that the original version of inflation, in which the Universe undergoes a supercooled \ac{FOPT}, also leads to unacceptably large inhomogeneities that are at odds with \ac{CMB} observations. Moreover, as already pointed out by Guth in his pioneering paper \cite{Guth:1980zm} and confirmed two years later by Guth and E.~Weinberg \cite{Guth:1982pn}, bubble nucleation in an exponentially-expanding Universe generally does not lead to successful percolation, hence raising questions about how to exit from the inflationary stage to recover the Universe we observe. 

For these reasons the slow-roll paradigm \cite{Linde:1981mu, Albrecht:1982wi}, in which a scalar field, the \textit{inflaton}, slowly rolls down an almost-flat potential, quickly became the most popular one in the inflationary community. The more and more precise data coming from \ac{CMB} experiments, culminating with the Planck satellite mission \cite{Planck:2018vyg}, put stringent bounds on the infinitely many possible incarnations of the slow-roll paradigm, constraining the shape of the inflaton potential \cite{Planck:2018jri}.

The rest of this section is organised as follows: in Sec.~\ref{sec:HorProb}, we will review the horizon problem, showing how a period of cosmic inflation can solve it. In Sec.~\ref{sec:SR}, we will introduce slow-roll inflation and its predictions in terms of \ac{CMB} observables. Sec.~\ref{sec:Preh} will be devoted to discussing (p)reheating and how to recover the Universe we observe after a stage of inflation. Much of the discussion contained in this section comes from the beautiful lectures notes on the physics of inflation by Daniel Baumann \cite{Baumann:2011}.  

\subsection{The horizon problem}\label{sec:HorProb}
The horizon problem can be summarised in the following way. By looking at the \ac{CMB}, we observe that the Universe at the time was extremely homogeneous, with temperature fluctuations as small as $\delta T/T\simeq 10^{-5}$. At the same time, if we  take two regions in the \ac{CMB} map that are separated by more than $\theta \simeq 2^{\circ}$ in the sky, and we evolve them backwards in time assuming ``standard'' cosmology, we realise that they have never been in causal contact with each other.\footnote{Here, by ``standard'' cosmology, we mean a Universe that did not experience a stage of inflation.} Therefore, one would rather expect temperature fluctuations among uncorrelated regions to be $\delta T/T \sim \mathcal{O}(1)$. The horizon problem is the lack of an explanation for the homogeneity of the \ac{CMB} and, for this reason, it is sometimes also called the \textit{homogeneity problem}.

Already from its formulation, it is clear that this puzzle has much to do with the causal structure of our Universe, and, specifically, with how light rays propagate in a Universe described by the \ac{FLRW} metric, given in Eq.~\eqref{eq:FLRW}. To study causality in an evolving spacetime, it is convenient to define the \textit{conformal time} $\tau$ as $d\tau \equiv d t/a(t)$, so that the \ac{FLRW} metric now becomes, in polar coordinates, 
\begin{equation}
ds^2 = a^2(\tau)\left[- d\tau^2 + dr^2 + r^2 d\Omega\right]\,,
\end{equation}
which is conformally equivalent to the Minkowski metric, hence the name ``conformal time''. Conformally equivalent metrics are the same up to an overall rescaling (in this case $a(\tau)$) and, therefore, they possess the same causal structure. For simplicity, we consider the radial propagation of a ray of light, whose geodesic is given by $ds^2 = 0$ and, therefore, $r(\tau) = \pm \tau + {\rm const}$. Upon choosing $t=0$ as the origin of the Universe, the maximal distance travelled by a photon, and therefore by any particle, at a generic instant $t$ is
\begin{equation}\label{eq:ComHor}
    d_h(t) = \tau(t) - \tau(0) = \int_0^t \frac{dt'}{a(t')}\,,
\end{equation}
which is also known as the \textit{comoving particle horizon} as it represents the maximal distance over which causal connection can be established. In other words, two points that are outside each other's particle horizon have never been in causal contact. According to the usual convention, we set $a(t=0) = 0$. Two space-time points separated by a distance larger than $d_h(t)$ at a given time $t$, have never been in causal contact with each other. Conformal time is also often expressed in terms of the \textit{comoving Hubble radius} $(a H)^{-1}$, i.e.~the distance beyond which objects recede super-luminally, as
\begin{equation}\label{eq:tauaH}
    \tau = \int (a H)^{-1} d \log a\,.
\end{equation}
The comoving Hubble radius is related to the energy content of the Universe via the Friedman equation (Eq.~\eqref{eq:FriedEq}):
\begin{equation}\label{eq:HubHor}
    (a H)^{-1} \propto a ^{\frac{1}{2}(1+3 w)}\,,
\end{equation}
where $w$ is the equation of state parameter of the Universe $w\equiv P/\rho$. Accordingly, we also find  
\begin{equation}\label{eq:tauw}
    \tau \propto \frac{2}{1+3 w}a^{\frac{1}{2}(1+3 w)}\,,
\end{equation}
as per Eq.~\eqref{eq:tauaH}. Since a matter fluid is characterised by $w = 0$, while $w=1/3$ for radiation, we see that the Hubble radius, and $\tau$ likewise, grows with time for ordinary energy density (remember that $\dot a >0$ according to the Friedman equation). Moreover, we see that $\tau_0 \equiv \tau(t=0) = 0$. We can conclude that the amount of conformal time elapsed between the beginning of the Universe and the moment of recombination (when \ac{CMB} was emitted) is finite, and so it is the particle horizon at that time $d_h (t_{\rm rec}) < \infty$. The finiteness of the particle horizon at recombination implies that there is a finite distance over which causal contact could have been established up to that moment. It is important to stress that our conclusions hinge upon the fact that $(aH)^{-1}(t)$ grows with time, since this leads to $d_h(t_{\rm rec})<\infty$.

\subsubsection{A decreasing Hubble horizon}
\begin{figure}[h!]
\begin{center}
\includegraphics[width=.7\textwidth]{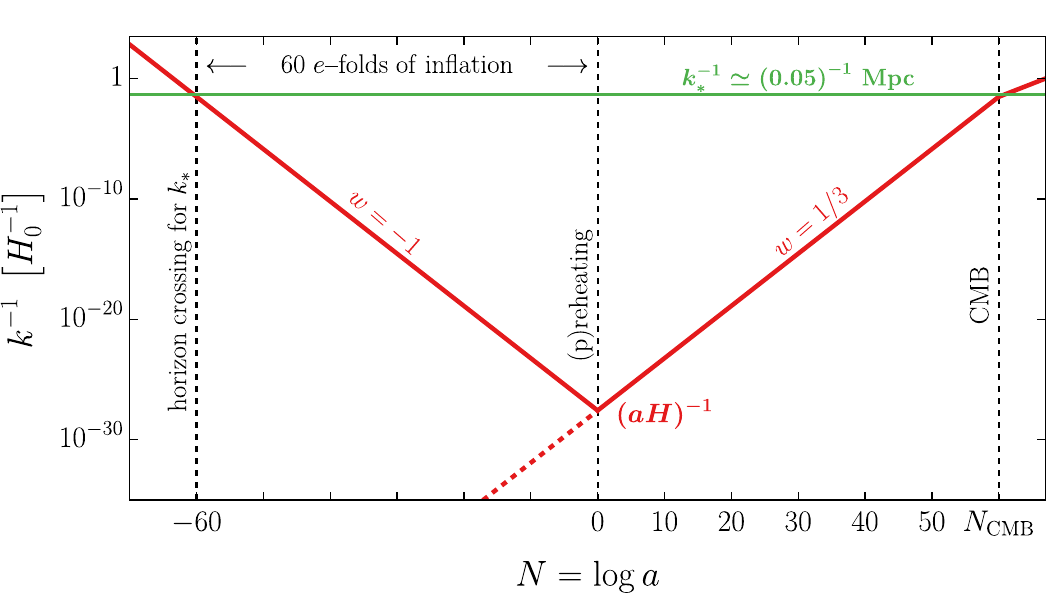}
    \caption{In red, the behaviour of the comoving Hubble horizon $(aH)^{-1}$ as a function of the number of $e$-folds. We set $N=0$ as the moment of reheating, that here is assumed to occur instantaneously.  The green line represents the pivot scale. It crosses the horizon, $k_* = aH$, for the first time 60 $e$-folds before the end of inflation, and  it remains super-horizon till the time of recombination $N_{\rm CMB}$. The Universe is assumed to be radiation dominated, $w=1/3$, from reheating till the moment of \ac{CMB}. For the sake of clarity, we also approximate the moment of matter-radiation equality to coincide with $N_{\rm CMB}$.  The dashed red line shows $(aH)^{-1}$ in the absence of cosmic inflation.}

\label{fig:scales}  
\end{center} 
\end{figure}

We have seen that the horizon problem is linked to the fact that, assuming a ``standard'' cosmological evolution, a finite amount of conformal time elapsed from the origin of the Universe till the \ac{CMB} time. This, in turn, is due to the fact that the Hubble radius grows with the expansion of the Universe, as it is evident from Eq.~\eqref{eq:tauaH}. Therefore, the solution may be as simple as postulating a stage of cosmic history during which the Hubble radius decreased with time instead:
\begin{equation}\label{eq:HubHorDec}
    \frac{d}{dt}(a H)^{-1}<0\,.
\end{equation}
We call such a period \textit{cosmic inflation}. Looking at Eq.~\eqref{eq:HubHor}, we see that, in order for the above inequality to be satisfied, the equation of state of the Universe has to be $w < -1/3$, which describes a fluid with negative pressure. At this stage of the discussion, let us just assume that this is the case for some values of $t$, to prove that this actually solves the horizon problem. We will postpone the investigation of the microphysics leading to $w<-1/3$ to Sec.~\ref{sec:SR}. Now, according to Eq.~\eqref{eq:tauw}, the initial time of the Universe $t=0$ is associated to $\tau_0 = -\infty$, rather than $\tau_0 = 0$. This means that the comoving horizon $d_h(t) = \tau(t) - \tau_0$ is infinite, and causal connection in the \ac{CMB} can be established over arbitrarily large distances. In other words, by simply assuming that there was a period during the Universe evolution in which the comoving Hubble radius decreased, we have given enough conformal time to the \ac{CMB} to homogenise. Yet another way of seeing it is to reason in terms of scales. Any length-scale $l$ is associated to a wave-number $k\sim 1/l$. Typically, when talking about “scales'' in this context we refer to $k$. We say that a scale is sub-horizon if $k\gg aH$. Sub-horizon scales are in causal contact since they are entirely contained inside the Hubble horizon, $l\ll (aH)^{-1}$. On the contrary, super-horizon scales, for which $k\ll a H$, are causally disconnected. Since the Hubble horizon evolves with time, at any time $t$ new scales enter or leave the horizon, with the horizon-crossing condition being $k = (aH)(t)$. As shown in Fig.~\ref{fig:scales}, scales that enter the horizon at a given time, say at recombination, after a period of cosmic inflation do not establish causal contact for the first time. Thanks to the shrinking Hubble horizon, they had already been sub-horizon, and, hence, causally connected once. In order to solve the homogeneity problem, we want cosmic inflation to last long enough for the \ac{CMB} scales to have been sub-horizon at some point in the past. We take as a reference scale the pivot scale used by the Planck collaboration, $k_*\simeq 0.05\,{\rm Mpc^{-1}}$ \cite{Planck:2018vyg}. The duration of inflation is usually measured in terms of the \textit{number of $e$-folds} $N(t) \equiv \log a(t)$. In order to successfully account for the \ac{CMB} homogeneity, we need $N\simeq 50-60$.\footnote{To be precise, this number depends on the cosmic history from the end of inflation to recombination. Details on the reheating model and on the subsequent evolution of the Universe energy density will impact the required number of $e$-folds of inflation \cite{Liddle:2003as}.}

So far, we have defined cosmic inflation as an epoch of cosmic history during which the Hubble radius shrunk, cfr.~Eq.~\eqref{eq:HubHorDec}, since it makes clear how this solves the horizon problem. However, inflation is often described as a stage of accelerated expansion of the Universe. This alternative definition is actually equivalent to Eq.~\eqref{eq:HubHorDec}
\begin{equation}\label{eq:AccExp}
    \frac{d}{dt} (a H)^{-1}\overset{H = \dot a/a}{=} \frac{d}{dt}(\dot a)^{-1} = - \frac{\ddot a}{(\dot a)^2}<0 \qquad \Rightarrow \qquad \ddot a>0\,.
\end{equation}
Another widely used definition of inflation is an epoch during which the Hubble rate varies slowly. Again, this is simply another way of stating that the Hubble radius decreases with time
\begin{equation}\label{eq:epsilon}
    \frac{d}{dt}(a H)^{-1} = - \frac{1}{(a H)^2}\left(\dot a H+ a \dot H\right)\equiv -\frac{1}{a} \left(1 - \epsilon\right) <0 \quad \Rightarrow \quad \epsilon <1\,, \quad {\rm with} \quad \epsilon \equiv -\frac{\dot H}{H^2}\,.
\end{equation}
The parameter defined above is called the \textit{slow-roll parameter} and, as we will see below, it is of fundamental importance. An $\epsilon<1$ implies that the rate of change of the Hubble rate (in proper units) is small.

\subsection{Slow-roll inflation}\label{sec:SR}
In the previous section, we proved that in order to solve the horizon problem we need a shrinking Hubble radius. We have also seen that this can be achieved if the energy density of the Universe is dominated by a fluid with $w <- 1/3$, which, therefore, cannot be matter nor radiation. We will show in this section that obtaining such an apparently-weird equation of state parameter is rather simple.

Consider a scalar field, $\phi$, the inflaton, which, for the sake of simplicity, we take to be real, with a potential $V(\phi)$. Assuming that any non-minimal coupling to gravity vanishes, the action for $\phi$ in a curved space-time with metric $g_{\mu \nu}$ is
\begin{equation}\label{eq:SInfl}
        S = \int d^4 x \sqrt{- g} \left[\frac{1}{2}M_{\rm Pl}^2R + \frac{1}{2}g^{\mu\nu}\partial_\mu \phi \, \partial_\nu \phi - V(\phi)\right]\,,
\end{equation}
where $R$ is the Ricci scalar. For the time being, we are interested in the background evolution of the field. The equation of motion for the background component of $\phi$ is simply the Klein-Gordon equation in a curved space-time\footnote{For a textbook introduction to the topic of QFT in curved space-time see Ref.~\cite{Birrell:1982ix}.}
\begin{equation}\label{eq:EoM}
    \ddot \phi + 3 H \dot\phi = - V'(\phi)\,,
\end{equation}
where a prime denotes derivative with respect to the field. The effect of the space-time geometry is entirely encoded in a friction term due to the expansion of the Universe. Assuming that the Universe energy density is dominated by the inflaton, the Friedman equation reads
\begin{equation}
    H^2 = \frac{1}{3 M_{\rm Pl}^2}\left[\frac{1}{2}(\dot\phi)^2+V(\phi)\right]\,,
\end{equation}
and the equation of state parameter is
\be\label{eq:w}
    w = \frac{P}{\rho}= \frac{\frac{1}{2}(\dot\phi)^2-V}{\frac{1}{2}(\dot\phi)^2+V}\,.
\ee
By deriving both sides of the Friedman equation with respect to time, we get 
\begin{equation}
    \dot H = -\frac{1}{2}\frac{(\dot \phi)^2}{M_{\rm Pl}^2}\,,
\end{equation}
which can be plugged into the definition of $\epsilon$ in Eq.~\eqref{eq:epsilon} to yield
\begin{equation}
    \epsilon = 3 \frac{\frac{1}{2}(\dot\phi)^2}{\frac{1}{2}(\dot\phi)^2 + V(\phi)}\,.
\end{equation}
Therefore, in order for a period of slowly-varying $H$ ($\epsilon \ll 1$) to take place, the potential energy has to be larger than the kinetic one $V(\phi) \gg (\dot \phi)^2$. By taking this limit in Eq.~\eqref{eq:w}, we get $w \simeq -1$. We conclude that cosmic inflation can be driven by a scalar field slowly rolling down its potential so that its potential energy density dominates the Universe. Moreover, in order to solve the homogeneity problem, we would also like this stage of inflation to persist long enough. For this to be case, the field acceleration has to be smaller compared to the friction term in Eq.~\eqref{eq:EoM}
\begin{equation}
    \delta \equiv -\frac{\ddot\phi}{H\dot\phi}\ll 1\,.
\end{equation}
The condition for the duration of inflation can also be expressed in terms of variation of $\epsilon$ with time
\begin{equation}
    \eta \equiv \frac{\lvert \dot\epsilon\rvert}{H \epsilon}\ll 1\,.
\end{equation}
Summing up, in order to get a successful period of inflation we need $\epsilon,\delta \ll1$ or, analogously $\epsilon,\eta\ll1$, which go under the name of \textit{slow-roll conditions}.

We can now make use of the slow-roll condition to simplify our discussion. On the one hand, the smallness of $\epsilon$ implies that the Hubble rate is well approximated by $H^2 \simeq V/(3 M_{\rm Pl}^2)$ while, on the other hand, $\delta\ll1$ allows us to simplify Eq.~\eqref{eq:EoM} as $3 H \dot\phi \simeq \partial V/\partial\phi$, leading to the following expressions for the slow-roll parameters
\begin{equation}\label{eq:SRParV}
    \epsilon = \frac{M_{\rm Pl}^2}{2}\left(\frac{V'}{V}\right)^2\,,\qquad \eta = M_{\rm Pl}^2\frac{V''}{V}\,.
\end{equation}
The above expressions for the slow-roll parameters are particularly useful since they are related only to the shape of the inflationary potential. Their smallness translates into the requirement of either a rather flat inflationary potential, which from the point of view of naturalness is quite worrying, or of a transplanckian field excursion, which raises doubts on the legitimacy of integrating out physics above $\MP$. We recall that the number of $e$-folds of inflation, already introduced above Eq.~\eqref{eq:AccExp}, are
\begin{equation}
    N \equiv \int_{a(t_i)}^{a(t_f)}d \log a = \int_{t_i}^{t_f}H(t) dt\,.
\end{equation}
The extrema of integration are the moment when inflation starts, $t_i$, and the moment when inflation ends, $t_f$, which can be chosen as $\epsilon(t_i) = \epsilon (t_f) = 1$ or $\eta(t_i) = \eta (t_f) = 1$, whichever occur first. Notice that, in the slow-roll approximation, the integral over time can be traded for an integral over field space, according to the approximated equation of motion for the inflaton
\begin{equation}\label{eq:N}
    N = \frac{1}{\sqrt{2\epsilon}}\int_{\phi_i}^{\phi_f} \frac{d\phi}{H}\,.
\end{equation}
By writing the number of $e$-folds in terms of field excursion, it becomes clear that the inflaton plays the role of a clock, determining when inflation starts and, most importantly, when it ends. This is of crucial importance when considering also field fluctuations around the background value.

\subsubsection{CMB observables}
While inflation was firstly proposed in order to solve the horizon problem \cite{Guth:1980zm}, it quickly became clear that it also contained a mechanism to generate the initial conditions from which the Universe started evolving \cite{Starobinsky:1982ee, Guth:1982ec, Hawking:1982my}. In other words,  it explains why the Universe was so homogeneous at the epoch of \ac{CMB} and, at the same time, where the initial seeds that led to the growth of inhomogeneities came from. In a nutshell,  our discussion has been completely classical so far, involving only the background value of the inflaton $\phi$. However, the inflaton has some (small) quantum fluctuations, in accordance with the uncertainty principle
\be
    \phi(\vec{x},t) = \bar{\phi}(t) + \delta \phi(\vec{x},t)\,,
\ee
where the bar denotes the background, homogeneous value, while $\delta \phi \ll \bar{\phi}$ are the field inhomogeneities. Such quantum fluctuation, once stretched by the Universe exponential expansion, become of cosmological relevance. We have concluded the previous section by commenting on the role of the inflaton as a clock for the duration of inflation. Now, due to the presence of small, uncorrelated fluctuations around the homogeneous field value, the precise number of $e$-folds of inflation will vary in different points of the Universe, as per Eq.~\eqref{eq:N}. As a result, different regions of the Universe will inflate for a different amount of time, resulting into the density fluctuations we observe in the \ac{CMB}. To be precise, regions where $\delta \phi (\vec{x},t)>0$ inflate for a shorter time and give rise to underdensities. Indeed, their energy density gets converted to radiation and is redshifted by the Universe expansion. On the contrary, in patches that inflate longer the energy density remains constant as long as vacuum domination holds, eventually resulting into overdensities.

Since, by assumption, the inflaton field dominates the energy density of the Universe, its quantum fluctuations cause the stress-energy tensor of the Universe to fluctuate as well, $\delta \phi \Rightarrow \delta T_{\mu\nu}$. Through the Einstein's equations, stress-energy tensor fluctuations affect the metric that, in turns, backreacts onto the equation of motion of $\phi$. The discussion of metric perturbations is very subtle. Here, we will gloss over many details and we invite the interested reader to give a look at Ref.~\cite{Riotto:2002yw}. In a nutshell, a general metric can be decomposed, analogously to a field, into a background part plus perturbations
\be
    g_{\mu\nu} = \bar{g}_{\mu\nu}(t)+ \delta g_{\mu\nu}(\vec{x},t)\,.
\ee
In our case the background spacetime will be the \ac{FLRW} one. When studying metric perturbations, we are faced with the issue of gauge invariance. In fact, the perturbation of a given quantity is given by the difference between the value it takes on the perturbed metric $g_{\mu\nu}$, and the one it takes on the background one $\bar{g}_{\mu\nu}$. Of course, we want to compare these quantities at the same point in space-time but, since space-time itself depends on the choice of the metric, we must define a map that allows us to univocally relate one point on the background manifold to the same point on the perturbed spacetime. This amounts to choosing a coordinate system. There is not one unique way of fixing this map and, therefore, Physics cannot depend on its choice. In other words general relativity is a gauge theory where the gauge transformation is a coordinate transformation from one local frame to another. Choosing a coordinate system implies fixing a gauge. Metric perturbations can be decomposed according to the scalar-vector-tensor decomposition, i.e.~depending on the way they transform under a local spatial rotation, on a constant-time hypersurface. Of the 10 independent degrees of freedom, 4 can be removed by a coordinate transformation and, therefore, we are left with 2 scalar, 2 vector and 2 tensor perturbations. On top of these metric perturbations, we also have the inflaton one $\delta \phi$. As, at the linear level, the different degrees of freedom evolve independently, we can study them separately. During inflation vector modes can be shown to be negligible; on the contrary, scalar and tensor ones will lead to density fluctuations and primordial \ac{GW}s, respectively, and have huge phenomenological implications since they can be detected in the \ac{CMB}. We fix the comoving gauge $\delta T_{0,i}=0$ (with $i=1,2,3$ running over the spatial coordinates), which amounts to following the world-lines of a free-falling observer. In this gauge the inflaton fluctuations vanish and we are left with 2 scalar metric perturbations, of which one vanishes along the Einstein's equations, plus 2 tensor modes
\be\label{eq:deltagmunu}
    \delta g_{ij} = a^2(1-2\zeta) \delta_{ij} + a^2 h_{ij}\,.
\ee
Here, $h_{ij}$ is a transverse, traceless tensor, while $\zeta$ is known as the \textit{comoving curvature perturbation}, since a comoving spatial slice has curvature $R = \frac{4}{a^2}\nabla^2 \zeta$. We will now proceed to study the evolution of the scalar and of the tensor perturbations separately.

\underline{\textbf{Scalar perturbations.}} By plugging the scalar part of the perturbed metric into the action in Eq.~\eqref{eq:SInfl}, we derive the action describing the evolution of $\zeta$, which, at the quadratic order reads
\be
    S[\zeta] = \frac{\MP^2}{2}\int d^4x \,a^3 \left(\frac{\dot \phi}{H}\right)^2\left[(\dot\zeta)^2 -\frac{1}{a^2}(\partial_i\zeta)^2\right] + \cdots\,.
\ee
In order to get a canonically normalised field, we perform the following field redefinition
\be
    v \equiv z\zeta\,,\qquad {\rm with}\qquad z^2 \equiv 2 a^2 \epsilon\MP^2\,,
\ee
where we have written
\be\label{eq:epsilonSRApprox}
    \epsilon = \frac{\MP^2}{2}\left(\frac{V'}{V}\right)^2 \simeq \frac{1}{2\MP^2}\left(\frac{\dot \phi}{H}\right)^2\,,
\ee
in the slow-roll approximation. Switching to conformal time, we get the action for an harmonic oscillator with a time dependent mass
\be\label{eq:Svk}
    S[v] = \frac{1}{2}\int d\tau d^3x\left[(v')^2-(\partial_i v)^2 -m_{\rm eff}^2(\tau) v^2\right]\,,\qquad {\rm where}\qquad m_{\rm eff}^2(\tau) \equiv -\frac{z''}{z}\,.
\ee
The prime here denotes a derivative with respect to conformal time.
The time-varying mass is an effect of the interaction between $\zeta$ and the background geometry. The background metric, in fact, is a function of time for a two-fold reason: firstly because of the expansion of the scale factor $a(t)$, and, secondly, for the evolution of the inflaton field $\dot\phi$. By varying $S[v]$, we obtain the equation of motion for $\zeta$, which is also known as the \textit{Mukanov-Sasaki equation},
\be
    v_k'' + \left(k^2 - \frac{z''}{z}\right)v_k = 0\,,
\ee
in momentum space. The above equation can be solved exactly in a perfect de Sitter spacetime. However, we know that during inflation the background metric is not exactly a de Sitter one. Departures from de Sitter are encoded in the slow-roll parameters. For example, according to our definition of $\epsilon$ in Eq.~\eqref{eq:epsilon}, we have defined it as the rate at which the Hubble parameter changes over time. In slow-roll inflation, $\epsilon \ll 1$ but it is not exactly zero, as one would need to recover a perfect de Sitter spacetime metric. Therefore, we can expand the Mukanov-Sasaki equation in small departure from a perfect de Sitter spacetime, as we did for the inflaton equation of motion. This departure is parametrised by the slow-roll parameters
\be\label{eq:SRPar}
    \epsilon = -\frac{\dot H}{H^2}\,,\qquad\eta = \frac{\dot\epsilon}{\epsilon H}\,,\qquad \xi= \frac{\dot\eta}{H\eta}\,,
\ee
where we have already discussed the first two in the previous section, and we have introduced a third one, measuring the rate of variation of $\eta$. The parameter $\xi$ is related to the third derivative of the inflaton potential. Given that $z^2 = 2a^2\epsilon\MP^2$, we have 
\be
    \frac{z^{\prime \prime}}{z}=(a H)^2\left[2-\varepsilon+\frac{3}{2} \eta-\frac{1}{2} \varepsilon \eta+\frac{1}{4} \eta^2+\frac{1}{2}\eta \xi\right]\,,
\ee
which is an exact expression. From the definition of $\epsilon$, we write the Hubble radius at the first order in the slow-roll expansion
\be\label{eq:aHSRApprox}
    \left(a H\right)^{-1} = - \tau (1-\epsilon) + \mathcal{O}(\epsilon^2)\,,
\ee
which gives
\be
    \frac{z^{\prime \prime}}{z} \simeq \frac{1}{\tau^2}\left[2+3\left(\varepsilon+\frac{1}{2} \eta\right)\right] \equiv \frac{\nu^2-\frac{1}{4}}{\tau^2}\,,\qquad{\rm with}\qquad\nu \equiv \frac{3}{2}+\varepsilon+\frac{1}{2} \eta\,.
\ee
By solving the above differential equation for $z$, we find $z \propto \tau^{\frac{1}{2}-\nu}$. Notice that, in principle, the slow-roll parameters evolve with time during inflation. However, their relative time variation is of the same order, in the slow-roll expansion, as the parameters themselves, cfr.~Eq.~\eqref{eq:SRPar}. Therefore, we can regard $\nu$ in the above equation as a constant and solve the Mukanov-Sasaki equation to find (remember that $\tau <0$ during inflation)
\be
    v_k(\tau)=\sqrt{\frac{\pi}{2}}(-\tau)^{1 / 2} H_\nu^{(1)}(-k \tau),
\ee
where we have taken the Bunch-Davies vacuum as initial condition, and $H_\nu^{(1)}(x)$ is the Hankel function of the first kind. When the mode $k$ is super-horizon, $- k\tau \ll 1$, and, in the limit of perfect de Sitter ($\nu = 3/2$), 
\be
    \lim_{k\tau\rightarrow 0}v_k(\tau) \simeq -i\frac{1}{k^{3/2}\tau}\,,
\ee
showing that, for superhorizon modes, $\zeta_k = v_k/z = {\rm const}$. The fact that the comoving curvature perturbation does not evolve while super-horizon (up to slow-roll-small corrections) is an extremely important result. In fact, this allows us to relate the value it had when it exited the horizon during inflation to the one at later times, when it got back inside the Hubble horizon. In particular, this means that scales re-entering the horizon at the \ac{CMB}, contain information on what the inflating Universe looked like when they became superhorizon, as shown by Fig.~\ref{fig:scales}. 

Now that we have a solution for the modes $v_k$, we can quantise the field $v$, promoting it to a field operator
\be
    \hat v(\tau, \vec x)=\int \frac{d^3 k}{(2 \pi)^{3 / 2}}\left[\hat{a}_{ k}^{-} v_k(\tau) e^{i \vec k \cdot \vec x}+\hat{a}_{k}^{+} v_k^*(\tau) e^{-i \vec k \cdot \vec x}\right]\,,
\ee
where $a^+_k$ and $a^-_k$ are the creation and annihilation operators, respectively, and they obey the usual commutation relations. The magnitude of vacuum fluctuations when they become super-horizon is 
\be
    \langle \hat v_k \hat v_{k'}\rangle = \lvert v_k \rvert^2 \delta (\vec k + \vec k') \simeq \frac{1}{k^3}\frac{(a H)^2}{(1+\epsilon)^2}\delta (\vec k + \vec k')\equiv P_v(k)\delta (\vec k + \vec k') \,,
\ee
where in the second equality we have used Eq.~\eqref{eq:aHSRApprox}, while in the last one we have defined the power spectrum for the $v$ field. The curvature perturbation power spectrum is $P_\zeta = P_v/z^2$. It is useful to define the dimensionless curvature power spectrum $\Delta_s^2 \equiv k^3 P_\zeta/(2\pi^2)$. Given that $z^2= 2 a^2 \epsilon \MP^2$, we have
\be
    \Delta_s^2 = \frac{1}{\epsilon} \left(\frac{H}{2\pi \MP}\right)^2 = \frac{1}{24\pi^2\epsilon}\frac{V}{\MP^4}\,.
\ee
The quantities $H$ and $\epsilon$ are functions of time but, since they vary very slowly in the slow-roll approximation, the above power spectrum is almost scale invariant. Its amplitude at the pivot scale $k_* = 0.05\,{\rm Mpc^{-1}}$ is usually denoted $A_s\equiv \Delta_s^2(k_*)$ and it is tightly constrained by Planck \cite{Planck:2018vyg}
\be
    \log \left(10^{10} A_s\right) = 3.047 \pm 0.014\,.
\ee
The measurement of the amplitude of the curvature power spectrum contains information on the scale of inflation $H$ and on $\epsilon$ at the moment when the pivot scale $k_*$ crossed the Hubble horizon. 

The scale-dependence of $\Delta_s^2$, also known as the \textit{spectral index} is defined as
\be
    n_s-1 \equiv \frac{d \log \Delta_s^2}{d \log k}\,.
\ee
It can be computed in the slow-roll approximation by noticing that 
\be
    \frac{d \log \Delta_s^2}{d\log k} = \frac{d \log \Delta_s^2}{dN} \frac{d N}{d\log k}\,.
\ee
The first of the two derivatives is 
\be
    \frac{d \log \Delta_s^2}{dN}=2 \frac{d\log V}{d N} - \frac{d\log \epsilon}{d N} = -6\epsilon +2\eta\,,
\ee
where we have now used the expression for the slow-roll parameters in terms of the potential, as given in Eq.~\eqref{eq:SRParV}.
The second one has to be evaluated at horizon crossing when $a H = k$ and, therefore $\log k = N + \log H$:
\be
    \frac{dN }{d \log k} \simeq 1 + \epsilon\,.
\ee
All in all,\footnote{Notice that, if we had used the slow-roll parameters in terms of the Hubble rate, defined in Eq.~\eqref{eq:SRPar}, we would have found $n_s-1 = -2\epsilon -\eta$. This is due to the fact that the two parameters are not exactly equivalent for they are the result of two different expansions.} 
\be
    n_s -1 = -6\epsilon + 2\eta \,.
\ee
In a pure de Sitter background, where $\epsilon =\eta = 0$, the spectrum of curvature perturbation would be exactly scale invariant. The slow-roll parameters introduce a small correction to scale invariance (as small as the slow-roll parameters themselves). The \ac{CMB} data currently available are so precise that they are able to detect such small departures from perfect de Sitter, with the Planck best-fit value for the spectral index being \cite{Planck:2018vyg}\footnote{We mention the recent results from the Atacama Cosmology Telescope which, when combined with Planck and DESI DR1, give a larger preferred value for the spectral index, $n_s = 0.9743\pm 0.0034$ \cite{ACT:2025fju}.}
\be
    n_s = 0.9647 \pm 0.0043\,.
\ee
Moreover, since the slow-roll parameters are related to the inflationary potential, as per Eq.~\eqref{eq:SRParV}, departures from scale-invariance in $\Delta_s^2$ contain information about the shape of $V(\phi)$.

\underline{\textbf{Tensor perturbations.}}
Our discussion so far has focused on the scalar part of the perturbed metric in Eq.~\ref{eq:deltagmunu}. However, what is considered to be the smoking gun of the inflationary paradigm is the production of a stochastic background of \ac{GW}s. The reason is two-fold: the amplitude of such primordial \ac{GW}s depends only on the Hubble rate during inflation making their detection an extremely robust measurement of the scale of inflation; moreover, inflationary \ac{GW}s leave an absolutely unique imprint on the \ac{CMB} polarisation. 

The formalism developed above for scalar metric perturbations can be applied to the tensor part of the perturbed metric in Eq.~\eqref{eq:deltagmunu}. The quadratic action for $h_{ij}$ reads
\be
    S[h_{ij}] = \frac{\MP^2}{8}\int d\tau d^3 x a^2\left[(h_{ij}')^2 - (\nabla h_{ij})^2\right]\,,
\ee
which is the same action as the one of a massless scalar field in a \ac{FLRW} background. The Fourier transform of $h_{ij}$ is defined as
\be
    h_{i j}(\tau, \vec{x})=\int \frac{d^3 k}{(2 \pi)^{3 / 2}} \sum_{\gamma=+, \times} \epsilon_{i j}^\gamma(k) h_{\vec{k}, \gamma}(\tau) e^{i \vec{k} \cdot \vec{x}}\,,
\ee
with $\epsilon^\gamma_{ij}$ the polarisation tensor, and $\gamma=  +,\times$ the two polarisation modes of the \ac{GW}s. In momentum space, then, the action reads 
\be
    S=\sum_\gamma \int \mathrm{d} \tau \mathrm{~d}^3 k \frac{a^2}{4} M_{\mathrm{pl}}^2\left[\left(h_{\vec{k}, \gamma}^{\prime}\right)^2-k^2\left(h_{\vec{k}, \gamma}\right)^2\right]\,.
\ee
We proceed as for $\zeta$, and we define a canonically normalised tensor field 
\be
    v_{\vec k,\gamma} \equiv \frac{a}{2}\MP h_{\vec k,\gamma}\,,
\ee
whose action is
\be
    S=\sum_\gamma \frac{1}{2} \int d \tau d^3 d k\left[\left(v_{\vec{k}, \gamma}^{\prime}\right)^2-\left(k^2-\frac{a^{\prime \prime}}{a}\right)\left(v_{\vec{k}, \gamma}\right)^2\right]\,.
\ee
The action for the canonically normalised \ac{GW} field is essentially composed of the sum of two copies of the action for curvature perturbation in Eq.~\eqref{eq:Svk}, one per each polarisation. The calculation of the \ac{GW} power spectrum goes through essentially in the same way as for $P_\zeta$, leading to the dimensionless tensor power spectrum
\be
    \Delta_t^2 = \frac{2}{\pi^2}\frac{H^2}{\MP^2}\,.
\ee
The power spectrum of tensor modes generated during inflation is, in general, normalised to the curvature perturbation one
\be\label{eq:r}
    r \equiv \frac{\Delta_s^2}{\Delta_t^2} = 16\epsilon \overset{\rm Eq.~\eqref{eq:epsilonSRApprox}}{=} \frac{8}{\MP^2}\frac{\dot \phi^2}{H^2}\,,
\ee
which goes under the name of \textit{tensor-to-scalar ratio}. With no measurement of $r$ so far, the most stringent upper bound is given by the BICEP/Keck collaboration \cite{BICEP:2021xfz}
\be
    r < 0.036\,.
\ee
Near-future experiment will improve this bound, with, for example, \ac{CMB}-S4 that should be able to set an upper limit of $r < 0.001$ at 95\% confidence level \cite{CMB-S4:2020lpa}. Eq.~\eqref{eq:r} makes the dependence of $r$ on the distance travelled by the inflaton in field space clear. By inverting it, we find
\be
    \frac{\Delta \phi}{\MP} \simeq N \sqrt{\frac{r}{8}}\,,
\ee
where we have approximated $r$ as scale-invariant. Assuming $N=60$, as required to solve the \ac{CMB} homogeneity problem, we find
\be
    \frac{\Delta \phi}{\MP} \simeq  \sqrt{\frac{r}{0.001}}\,,
\ee
meaning that near-future experiments will be able to probe inflationary scenarios where $\Delta\phi \sim \mathcal{O}(\MP)$. This is known as the \textit{Lyth bound} \cite{Lyth:1996im}. Trans-planckian field excursions, however, are usually not well theoretically-motivated since questions arise whether higher order corrections to the inflationary potential can be regarded as being under control or not. On the flip side, this means that if one insists on having a model where inflation is realised over sub-planckian field scales, the predicted $r$ will be too small to be possibly detected in the next years.

\subsection{(P)reheating}\label{sec:Preh}

Following the end of inflation, the Universe is left in a deserted state, with any form of energy density produced during inflation, or prior to it, having been diluted by the accelerated exponential expansion. The vacuum energy that drove inflation is now mostly converted to the inflaton kinetic energy, as he oscillates around the minimum of its potential. This energy needs to be transferred to a thermal bath of particles, in order to start the epoch of radiation-domination that preceded the \ac{CMB} emission. This process is called \textit{reheating} and we say that, after its conclusion, the Universe is reheated to a temperature $T_{\rm reh}$, which is the temperature of the thermal bath of particles to which the inflaton has ceded its energy. In general, the inflaton will be always coupled to the \ac{SM} Higgs doublet, $h$, via a portal coupling of the form $\lambda_{\phi H}\phi^2 \lvert h \rvert^2$.\footnote{In some case the \ac{SM} Higgs boson itself drives inflation, aided by a non-minimal coupling to gravity \cite{Bezrukov:2007ep}. See also Ref.~\cite{Cheong:2021vdb} for a recent review on the topic.} However, here we remain agnostic on the possible interactions between the inflaton and \ac{SM} particles, and we consider a generic coupling $\mathcal{L}\supset g^2 \phi^2 \chi^2$, where $\chi$ denotes either the Higgs doublet or a dark-sector scalar with a sizeable coupling to the \ac{SM}. The inflaton produces $\chi$ particles via scattering and/or decay, with the latter channel open only if $\langle \phi \rangle \neq 0$ at the end of inflation, and $m_\phi^2 > m_\chi^2$. Assuming that this is the case, the decay $\phi \rightarrow \chi \chi$ becomes effective starting from the moment when $\Gamma \sim H$, with $\Gamma$ the decay rate. In the approximation in which the process of reheating completes instantaneously, i.e.~all the inflaton energy density is instantaneously transferred to a thermalised population of $\chi$ particles, we can derive the reheating temperature as $T_{\rm reh} \sim \sqrt{\Gamma H}$.  This estimate assumes the classical picture of many isolated inflaton particles decaying at the same time, and it entirely relies on a perturbative description. However, this is not quite the picture one should have in mind when thinking about the post-inflationary Universe. On the contrary, we should imagine an inflaton condensate coherently oscillating and decaying in a non-perturbative way, leading to extremely efficient particle production. This was firstly understood by Kofman et Al. \cite{Kofman:1994rk, Kofman:1997yn}, which dubbed  this stage of particle production via parametric resonance \textit{``preheating''}. The name stresses the fact that after this stage, when most of the inflaton energy has been depleted, and the Universe lies in a highly non-homogeneous state, the classical picture of reheating described above applies, with particles decaying and thermalising perturbatively.

\subsubsection{Parametric resonance}
Due to its coupling to the inflaton field, the field $\chi$ acquires an effective mass $m_\chi(\phi)^2 = m_\chi^2 + g^2\phi^2$, where $m_\chi^2$ is the bare, $\phi$-independent mass. We will work under the assumption that the $\chi$-particles are light, $m_\chi^2 \ll g^2\phi^2$. We quantise the field on a \ac{FLRW} geometry, with modes $\chi_k$ satisfying
\begin{equation}\label{eq:Chik}
	\ddot\chi_k + 3 H \chi_k + \left(\frac{k^2}{a^2} + m_\chi^2 + g^2 \phi(t)^2\right)\chi_k = 0\,.
\end{equation}
The equation of motion for $\chi_k$ is the one of an harmonic oscillator with a time-dependent frequency. Time-dependence comes from both the evolving scale factor, $a(t)$, and the oscillating background of the inflaton field $\phi(t)$. As a case study, we focus here on the model of chaotic inflation $V(\phi) = m^2\phi^2/2$, where the evolution of the inflaton zero-mode is described by $\phi(t) = \Phi \sin mt$. In what follows, we will neglect the effect of the expansion of the Universe. We will come back to this point at the end of this section. Eq.~\eqref{eq:Chik} becomes
\begin{equation}\label{eq:omegak}
	\ddot\chi_k + \omega_k^2(t)\chi_k = 0\,, \qquad {\rm with} \qquad \omega_k^2(t)\equiv k^2 +g^2 \Phi^2 \sin^2 m t\,.
\end{equation}
By performing the change of variables $mt = z$, the above equation takes the form of the well-known Mathieu equation
\begin{equation}\label{eq:Mathieu}
	\chi_k'' + \left(A_k - 2 q \cos 2z\right)\chi_k= 0\,,
\end{equation}
where a prime denotes differentiation with respect to $z$, and $A_k \equiv 2q +k^2/m^2 $, $q\equiv g^2\Phi^2/(4m^2)$. The Mathieu equation admits solutions $\chi_k \propto \exp(\mu_k^{(n)}z)$, each with an exponentially unstable behaviour for $k$ in a given interval $\Delta k^{(n)}$. The different solutions are labelled here by an integer $n$. The exponential growth of $\chi_k$ leads to a fast growth of the occupation number $N_k$, defined as
\begin{equation}\label{eq:nk}
	N_k = \frac{\omega_k}{2}\left(\frac{\lvert \dot \chi_k\rvert^2}{\omega_k^2} + \lvert\chi_k\rvert^2\right)-\frac{1}{2}\,.
\end{equation}
We interpret the exponential growth of $N_k$ as efficient production of $\chi$-quanta with momentum $k$.

\begin{figure}[h!]
\begin{center}
\includegraphics[width=.45\textwidth]{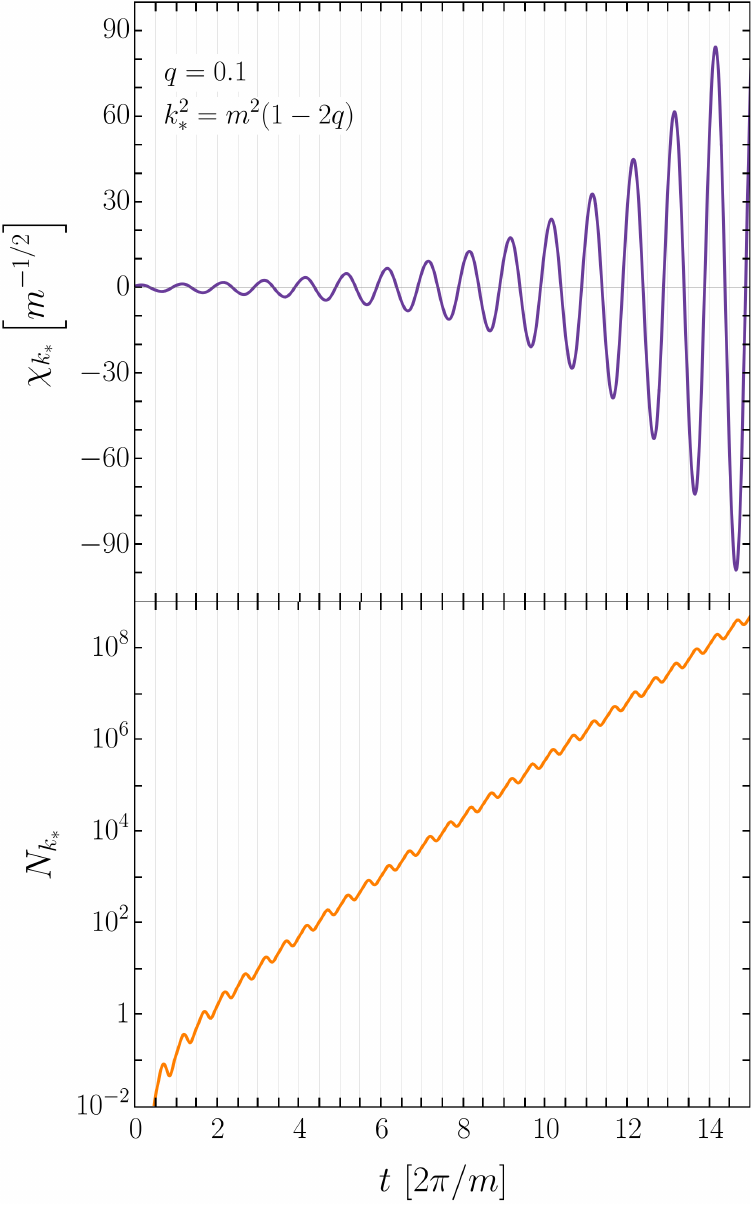}\qquad\includegraphics[width=.45\textwidth]{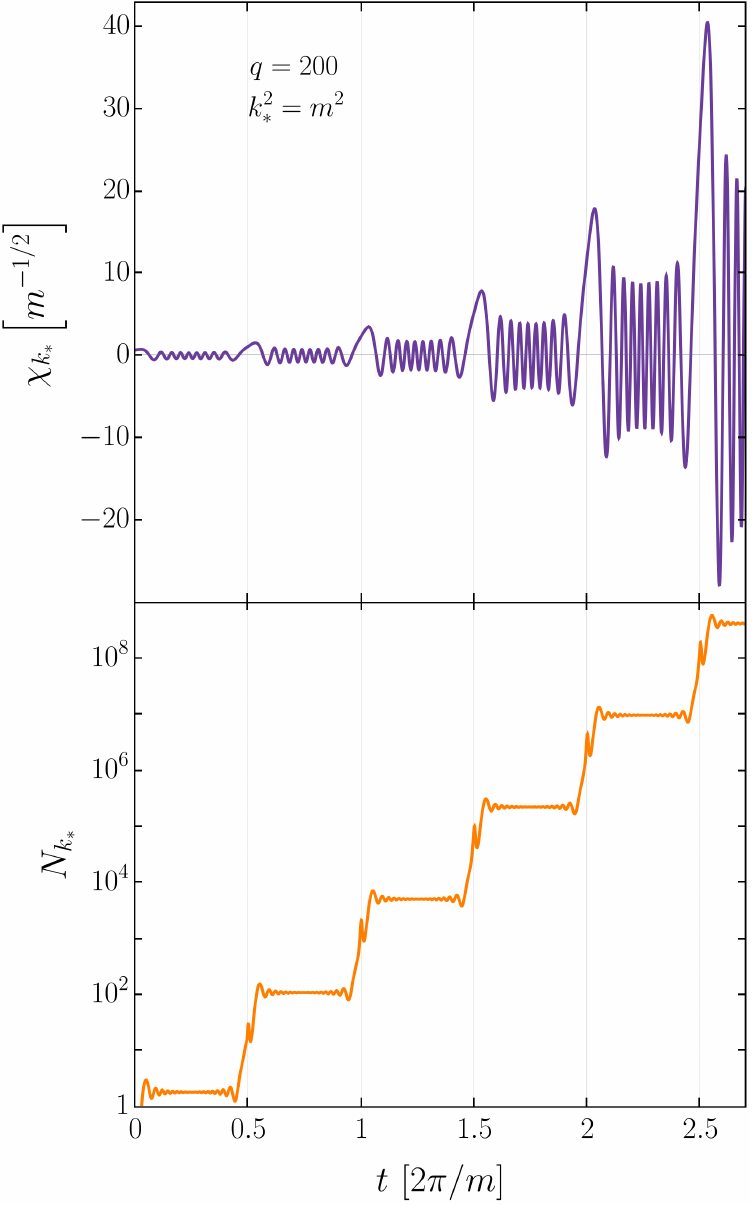}
    \caption{Resonant particle production from an oscillating inflaton field in a flat spacetime. Time is shown in units of $2\pi/m$ so that any integer or half-integer value of $t$ corresponds to $\phi(t)=0$. Purple curves show the evolution of the mode $\chi_{k_*}$ where $k_*$ is the mode for which particle production is the most efficient. In orange, the growth of the occupation number $N_{k_*}$. \textit{Left-hand panel.} In the narrow resonance regime, the mode oscillates once per each inflaton oscillation. The occupation number grow exponentially. \textit{Right-hand panel.} For broad resonance, the evolution of the frequency of $\chi_k$ changes adiabatically for most of the time. Whenever $\phi(t) \simeq 0$, Eq.~\eqref{eq:omegaNonAd} is fulfilled and a burst of particle production occurs. Broad resonances are much more efficient in producing $\chi$ quanta (notice the different time-scale compared to the right-hand plot).}
\label{fig:ParRes}  
\end{center} 
\end{figure}

\underline{\textbf{Narrow resonance.}} We begin our discussion assuming that the amplitude of the inflaton oscillations is smaller than $m$ and that, in particular, $g \Phi \ll m$. In this limit $q \ll 1$ and the resonance bands are known to be quite narrow. They are centred around momenta such that $A_k \simeq l^2$, and they have a width $\Delta k \simeq q^l$, for $l=1,2,\cdots$. Then,  for small $q$ the most relevant band is the first one, $l =1$. The exponent describing the growth in the first band is
\begin{equation}\label{eq:muk}
	\mu_k = \sqrt{\left(\frac{q}{2}\right)^2 - \left(\frac{k}{m}-1\right)^2}.
\end{equation}
Resonance occurs for modes in the range $k\in m\left[1-3q,\,1-q\right]$ where $\mu_k$ is real, with the maximal instability being at the centre of the band, $k_*^2 = m^2(1-2q)\simeq m^2$. There, the exponent takes its largest value $\mu_{k_*} \simeq q/2$. The occupation number of the $\chi$-mode experiencing the largest growth is
\begin{equation}
    N_{k_*} \propto \lvert\chi_{k_*}\rvert^2 \propto e^{qz} \simeq \exp\left(\frac{g^2\Phi^2}{4m}t\right)\,.
\end{equation}
The reason why the effect of parametric resonance is stronger for $k \sim m$ is understood interpreting it as the decay of two $\phi$ particles into two $\chi$ particles, each with momentum $k \simeq m$. In fact, the exponential growth of the occupation number of $k$-modes can also be understood in terms of Bose enhancement \cite{Mukhanov:2005sc}. We plot the numerical solution of Eq.~\eqref{eq:Mathieu}, in the narrow resonance regime and for $k=k_*$, in the left-hand plot in Fig.~\ref{fig:ParRes}. There, we also show how the occupation number $N_{k_*}$ grows with time.

\underline{\textbf{Broad resonance.}}
In the narrow resonance regime, we have assumed that the inflaton performs oscillations that are small compared to its mass, $\Phi\ll m$. However, in many models of inflation one can have $\Phi$ as large as $\MP$, especially when considering the first oscillations. This motivates us to look at the solution of the Mathieu equation in the broad resonance regime, $q \gg 1$. In this regime, the exponential instability of $\chi_k$ occurs for momenta in a much wider range and the instability exponent $\mu_k$, defined in Eq.~\eqref{eq:muk}, is typically larger than in the $q\ll1$ limit discussed above. Particle production is, then, much more efficient. The qualitative difference between broad and narrow resonance can be understood by looking at Fig.~\ref{fig:ParRes}, and comparing the right-hand panel (broad resonance) with the left-hand panel (narrow resonance). We see that, in the broad regime,  the occupation number $N_k$ has a jump any time the inflaton passes through the minimum of the potential $\phi(t)=0$. Moreover, we notice that, contrary to what happens for narrow resonances, the frequency of oscillation of the model $\chi_k$ is much larger than the one of the inflaton. This is understood by noticing that $q\gg1$ implies $g\,\Phi\gg m$; in other words, due to the large amplitude of the inflaton oscillations, the time effective mass of $\chi$ is much larger than the inflaton mass. This means that for most of the time the frequency $\omega_k$ changes adiabatically and the occupation number stays constant. However, for small $\phi(t)$ the rate of change of $\omega(t)$ blows up and the condition for particle production, 
\be\label{eq:omegaNonAd}
\dot \omega_k > \omega^2
\ee
is satisfied. Hence, the sharp increase in $N_k$. Since particle production takes place for $\phi(t)\simeq 0$, we can approximate $\dot\phi \simeq m\Phi$ and write the condition  for particle production in Eq.~\eqref{eq:omegaNonAd} as
\be\label{eq:kEnhance}
    k^2 \lesssim\left(g^2 \phi(t)\,m\,\Phi\right)^{2 / 3}-g^2 \phi(t)^2\,,
\ee
where we treat the momentum $k$ satisfying this condition as a function of $\phi(t)$. By minimizing it,  we find the value of the inflaton field at which particle production occurs for the broadest range of momenta
\be
    \phi_* \simeq \frac{1}{3^{3/4}}\sqrt{\frac{m\Phi}{g}}\,.
\ee
By plugging the above estimate for $\phi_*$ into Eq.~\eqref{eq:kEnhance}, we get the largest momentum experiencing particle production $k_{\rm max}^2 \simeq 2\,g\,m\,\Phi/3^{3/2}$\,. Since, for most of the interval $\lvert\phi(t)\rvert \leq \phi_*$, the value of enhanced momenta is smaller but comparable to $k_{\rm max}$, we can define the typical momentum experiencing particle production as
\be
    k_* = \sqrt{m\,g\,\Phi}\,,
\ee
which implies that $m\ll k_* \ll g\Phi$ in the broad resonance regime.

\underline{\textbf{Stochastic resonance}} We now briefly discuss the effect that the expansion of the Universe, so far neglected, has on parametric resonance. We do so by restoring the dependence on $a(t)$ in the equations of motion for $\phi$ and $\chi_k$. The mode equation is best expressed in terms of $X_k \equiv a^{3/2}(t)\,\chi_k(t)$
\be
    \ddot{X}_k + \omega_k^2 X_k = 0,\qquad {\rm with} \qquad \omega_k^2 \equiv \frac{k^2}{a^2} + g^2\Phi^2(t) \sin^2 mt\,,
\ee
where we have neglected a small term in the definition of the frequency. By comparing the $\omega_k$ defined above with the one defined in Eq.~\eqref{eq:omegak}, we see that the expansion of the Universe has a two-fold effect: it redshifts momenta, and it makes the amplitude of the inflaton oscillations a function of time, typically $\Phi(t)\sim 1/t$ for the quadratic potential we are considering. As a result, the parameter that characterises the instability bands $q$, defined below Eq.~\eqref{eq:Mathieu},  decreases with time. This implies that even if we start in a broad resonance regime, $q\gg1$, after few oscillations $q$ has greatly decreased leading particle productions to a narrow resonance behaviour. Eventually, parametric resonance shuts down. 
Moreover, since now $\Phi$ decreases with time, the effective mass of the $\chi$ modes decreases as well. Therefore, their frequency of oscillation diminishes. Then, the phases that a mode $\chi_k$ possesses at two different times $t_1$ and $t_2$ such that $\phi(t_1)=\phi(t_2)=0$, are uncorrelated from each other. This leads to an interesting stochastic behaviour in the occupation number defined in Eq.~\eqref{eq:nk}, which may either increase or decrease in a random way, whenever $\phi(t) \simeq 0$ \cite{Kofman:1997yn}. Nonetheless, the net effect is that of a large increase over time, with particle production always dominating over particle destruction.

The extremely large number of $\chi$ quanta produced during the resonant regime at some point backreacts onto the motion of the inflaton field. The inflaton acquires a mass $\Delta m^2 = g^2 \langle\chi\rangle^2$ where
\be
    \langle\chi\rangle^2 = \frac{1}{a^3}\int \frac{d^3k}{(2\pi)^3}\lvert X_k(t)\rvert^2\,.
\ee
The backreaction effect becomes relevant when $\Delta m^2 \simeq m^2$ or, analogously, when
\be
    n_\chi(t) \equiv \frac{1}{a^3}\int \frac{d^3k}{(2\pi)^3} n_k(t) \simeq \frac{m^2 \Phi(t)}{g}\,.
\ee
When this happens the resonant behaviour is destroyed and preheating ends. The Universe is left in a highly non-homogeneous, non-thermal state. Large inhomogeneities are also expected to produce a signal of \ac{GW}s appearing to us as a stochastic background. This classical signal is different and complementary to the one discussed in Sec.~\ref{sec:SR}, and it peaks at some characteristic frequency proportional to the inflationary scale \cite{Khlebnikov:1997di, Dufaux:2007pt}.  Scatterings and decays of the different modes will eventually produce a Universe in thermal equilibrium at some temperature $T_{\rm reh}$.

\subsubsection{Tachyonic preheating}
When discussing parametric resonance, we had in mind the picture of an inflaton field oscillating around the minimum of a convex potential. This is strictly true for the chaotic model of inflation we have examined so far, and it is approximately true for small oscillations taking place around the minimum a generic potential. However, more often than not, inflationary potentials feature regions of negative curvature. This is the case, for example, in hybrid inflation potentials \cite{Linde:1993cn}. As explained in greater detail in Ch.~\ref{chap:AccInf}, hybrid inflation models feature two scalar fields. One is the inflaton, whose slow-roll dynamics drives cosmic inflation, while the other is called the \textit{waterfall field} and, as long as its mass is large and positive, it stabilises the inflationary trajectory. The mass of the waterfall field typically decreases as $\phi$ rolls, and eventually turns negative. A minimum away from the inflationary trajectory emerges and inflation ends. As soon as the waterfall field mass turns negative its quantum fluctuations are exponentially enhanced, as it was realised for the first time by Felder et Al.~in Refs.~\cite{Felder:2000hj, Felder:2001kt}. They showed that this exponential enhancement, which is active right at the end of inflation, leads to a kind of particle production that is usually much more efficient even than the broad regime of parametric resonance. Calling our waterfall field $\psi$, its $k$-modes are
\be
    \psi_k = \frac{1}{\sqrt{2\omega_k}}e^{i \omega_k t - i \vec{k}\cdot \vec{x}}\,,\qquad {\rm with} \qquad \omega_k^2 = k^2 + m_\psi(\phi)^2\,.
\ee
Then, as long as the waterfall field mass is positive $m_\psi^2(\phi)>0$, the frequency $\omega_k$ is real and the modes have the usual oscillatory behaviour. However, we have seen that hybrid inflation terminates as soon as $m_\psi^2<0$. When this happens, $\omega_k$ becomes imaginary for $k^2 < \lvert m_\psi^2(\phi)\rvert$  and such modes get exponentially enhanced. In particular, the mode dispersion grows as
\be
    \langle \psi_k^2\rangle =\frac{1}{4\pi^2}\int_0^m k dk e^{2t\sqrt{m^2 - k^2}}\,,
\ee
which is associated with explosive production of $\psi$ quanta. This behaviour holds as long as the background waterfall field is on the concave part of the potential. Close to the global minimum of the potential, the second derivative will change sign shutting down tachyonic resonance. At this point, $\psi$ starts oscillating around the minimum of its potential producing particles in a way described by the theory of parametric resonance.

Analogously to what happens for parametric resonance, tachyonic particle production is an extremely violent process leading to the production of large energy density inhomogeneities. Such inhomogeneities necessarily generate a \ac{GW} background. The analytic treatment of \ac{GW} production during tachyonic preheating was first discussed in Ref.~\cite{Dufaux:2008dn}, which we will follow closely in the discussion below. Such an analytic argument allows us to gain intuition on what the relevant scales involved in the process are.  However, lattice simulations are required for a detailed prediction of the \ac{GW} signal, due to the highly non-linear nature of the preheating process. 

As an exemplary model, we consider the following potential for the waterfall field
\be\label{eq:Vpsi}
    V(\psi) = \frac{\lambda}{4}\left(\psi^2 - v^2\right)^2 + \frac{g^2}{2}\phi^2\psi^2 + V(\phi)\,.
\ee
$V(\phi)$ is the inflationary potential whose detailed form is not relevant here. Inflation ends at the point in field space $\psi_c$ where the waterfall field mass turns negative. For the potential above, 
\be
    \psi_c = \frac{\sqrt{\lambda}}{g}v\,.
\ee
Waterfall field inhomogeneities can be modelled as bubbles of typical radius $R_* \sim 1/k_*$, where $k_*$ is the characteristic momentum enhanced during tachyonic instability. The signal of \ac{GW}s produced by such bubbles is peaked at a frequency $f_*$, where it takes a peak amplitude $h^2\Omega_*$:
\be
    f_* \simeq \frac{4\times 10^{10}\,{\rm Hz}}{R_* \rho_p^{1/4}}\,,\qquad h^2\Omega_* \simeq 10^{-6}\left(R_* H_p\right)^2\,,
\ee
with $\rho_p$ and $H_p$ the energy density and the Hubble rate at the moment when \ac{GW}s are emitted, respectively. We now need to estimate the peak momentum $k_*$. The inflaton arrives at $\phi_c$ with some velocity that depends on the shape of $V(\phi)$. We remain agnostic about the explicit form of the velocity at the critical point and we express it in terms of the dimensionless quantity
\be
    u_c \equiv \frac{g\dot\phi_c}{\lambda v^2}\,.
\ee
Due to its velocity, $\phi$ classically rolls past $\phi_c$ and $\psi$ acquires a negative mass that increases with time $m_\psi^2 = -g^2(\phi_c^2 - \phi^2)$. Close to the critical point, the waterfall field mass grows as $m_\psi^2 \simeq  2 g^2 \phi_c \lvert\dot\phi_c\rvert \Delta t$, where $\Delta t$ is the amount of time elapsed from the moment when $\phi=\phi_c$. Exponential growth of quantum fluctuations becomes efficient when $\Delta t \gtrsim \sqrt{\lvert m_\psi^2\rvert}$. The momentum experiencing the largest amount of enhancement is $k_*^2 \lesssim \lvert m_\psi^2\rvert$, given by
\be
    k_*^3\simeq 2g^2 \phi_c\lvert\dot\phi_c\rvert\,.
\ee
Since, for the model in Eq.~\eqref{eq:Vpsi}, $\rho_p = \lambda v^4/4$ and $H_p^2 = \rho_p/(3\MP^2)$, the \ac{GW} signal will be characterised by
\be
    f_* \simeq \lambda^{1/4}u_c^{1/3}7\times 10^{10}\,{\rm Hz}\,,\qquad h^2\Omega_* \simeq 10^{-6} u_c^{-2/3}\left(\frac{v}{\MP}\right)^2\,.
\ee
We see that decreasing the velocity with which the inflaton approaches the critical point shifts the signal towards smaller frequencies and larger amplitudes, hence bringing it closer to a possibility of detection in the next future. However, the above expressions for $f_*$ and $h^2\Omega_*$ cannot be extrapolated down to arbitrary small values of $u_c$. In fact, for tiny velocities, the effect of quantum fluctuations of the inflaton field has to be taken into account \cite{Dufaux:2008dn}. Another way of decreasing $f_*$ is by taking small values of $\lambda$, which also amounts to decreasing the inflationary scale, as anticipated above. This makes the signal of \ac{GW}s produced during preheating a possible way of probing inflationary scenarios with a scale too low to be possibly detected by \ac{CMB} experiments in the near future. 

\chapter{Topological defects}
\label{chap:TopoDef}
\minitoc

Topological defects are extended field configurations, satisfying the equations of motion, that are stable due to the non-trivial topology of the vacuum manifold $\mathcal{M}$. If the manifold of equivalent vacua $G/H$ presents some non-trivial topological structure, they are expected to be produced during the cosmological \ac{PT} leading to the breaking $G\rightarrow H$\cite{Kibble:1976sj}. This is not the case for the \ac{SM} gauge group whose symmetry breaking pattern $\SU{2}\times \U{1} \rightarrow \U{1}$ is trivial and, therefore, does not lead to the production of finite-energy, stable field configurations. Nonetheless, many extensions of the \ac{SM} predict the presence of topological defects.

Here, we will study monopole configurations (point-like topological defects) in Sec.~\ref{sec:Mono}, \ac{CS} configurations (one-dimensional topological defects) in Sec.~\ref{sec:CS}, and \ac{DW} configurations (sheet-like topological defects) in Sec.~\ref{sec:DW}. We will first prove their existence in some simple models, and, then, we will discuss their interesting cosmological implications in terms of relic abundance, for monopoles, and of \ac{SGWB} production, for \ac{CS}s and \ac{DW}s. In the discussion below, we will mostly follow Refs.~\cite{Vilenkin:2000jqa,Shifman:2012zz}, to which we also refer the reader for a thorough discussion around topological defects and their cosmology. 

\section{Monopoles}\label{sec:Mono}

Monopoles are point-like topological defects arising in models of spontaneous symmetry breaking $G\rightarrow H$ such that the second homotopy group of the vacuum manifold, $\mathcal{M}$, is non-trivial: $\Pi_2(\mathcal{M})\neq I$. This happens if $\mathcal{M}$ contains non-contractible two-spheres $S^2$. A Higgs field profile lying in the vacuum manifold at spatial infinity can be mapped to an $S^2$ sphere in field space. Due to the non-trivial topology of the vacuum, this sphere cannot be contracted to a point and, as a consequence, the Higgs field will take values away from its \ac{VEV} at the core of the solution in physical space. This is a monopole.  If $G$ is gauged, the monopole also carries a magnetic charge under the unbroken generators of the group, with the gauge field having a non-trivial profile at the monopole core. 

\subsection{The 't Hooft-Polyakov monopole}\label{sec:tHooftPolyakov}
The existence of monopoles was pointed out independently by 't Hooft \cite{tHooft:1974kcl} and Polyakov \cite{Polyakov:1974ek}, in the context of an $\SO{3}$ gauge symmetry broken by a triplet Higgs field $\phi$. The most general Lagrangian density at the renormalizable level is
\be
    \mathcal{L} = -\frac{1}{4g^2}G_{\mu\nu}^a G^{\mu\nu,a}+\frac{1}{2}(D_\mu \phi^a)(D^\mu\phi^a) +\frac{\mu^2}{2}\phi^a\phi^a - \frac{\lambda}{4} \left(\phi^a\phi^a\right)^2\,,
\ee
where $a=1,2,3$ is the index running over the field components, while $\mu,\nu=0,\dots, 3$ are Lorentz indices. The gauge field strength tensor and the covariant derivative are defined as
\be
    G_{\mu\nu}^a \equiv \partial_\mu W_\nu^a - \partial_\nu W_\mu^a + \epsilon^{abc}W_\mu^b W_\nu^c\,,\qquad D_\mu \phi^a \equiv \partial_\mu \phi^a + \epsilon^{abc}W_\mu^b\phi^c\,,
\ee
respectively. $\epsilon^{abc}$ is the Levi-Civita tensor. If $\mu^2>0$, $\lambda >0$, the potential energy density is minimised by $\langle \phi^2 \rangle  =  \mu^2/\lambda \equiv \eta^2$, where $\SO{3}$ is broken to $\SO{2}$. Since $\Pi_2[\SO{3}/\SO{2}] =\mathbb{Z}$,\footnote{Pictorially, this can be understood in the following way. We can wrap an $S^2$ sphere on another $S^2$ sphere a number of times  $n = 0,\,\pm 1,\,\pm2,\,\dots$, and each configuration belongs to a distinct homotopy class since it cannot be continuously deformed into another one.} we expect the presence of stable monopole solutions in the spectrum. Such solutions are classified according to the homotopy class they belong to. The one-monopole solution carries one unit of magnetic charge and it corresponds to wrapping once an $S^2$ sphere  on $\mathcal{M}$. This is the least-energy solution with non-vanishing magnetic charge. 

Let us take the limit $\lambda\rightarrow0$, while keeping $\eta$ fixed. This limit is known as the \ac{BPS} limit and leads to simplifications in our discussion, affecting our estimates only by $\mathcal{O}(1)$ factors. The monopole solution is the finite-energy, static (time-independent), non-trivial solution for the Higgs and the gauge fields that minimise the energy. Upon choosing as an ansatz for the gauge field that the only non-vanishing component is the magnetic one, we can write the energy of the monopole configuration as
\be
    E \simeq \int d^3x \left[\frac{1}{2g^2}B_i^a B_i^a + \frac{1}{2}(D_i\phi^a)(D^i \phi^a)\right]\,,\qquad {\rm with}\quad B_i^a = -\frac{1}{2}\epsilon_{ijk}G^a_{jk}\,,
\ee
where the $\simeq$ has to be understood in the sense of the \ac{BPS} limit, in which the potential energy vanishes.
We rewrite the energy as 
\be
    E \simeq \int d^3 x \left[\frac{1}{2}\left(\frac{1}{g}B_i^a-D_i\phi^a\right)^2 + \frac{1}{g} B_i D_i\phi^a\right]\,.
\ee
Starting from the last term in the sum, we define the magnetic charge $Q_M$ of the monopole as 
\be
     Q_M \equiv \frac{1}{g \eta}\int d^3 x B_i D_i\phi^a = \frac{1}{g \eta}\int d^3 x\partial_i(B_i^a\phi^a) = \frac{1}{g \eta}\int_{S_R} d^2 S_i B_i^a\phi^a\,,
\ee
where, in the second equality, we have integrated by parts and we have used the equation of motion for the magnetic field, $D_i B_i^a = 0$. In the last equality, we used Gauss theorem to express $Q_M$ as a surface integral over a large sphere in coordinate space $S_R$. We showed that $Q_M$ is a topological quantity, defined as the integral over a large sphere of the flux of what can be thought as a gauge invariant definition of the magnetic field
\be
    Q_M = \frac{1}{g}\int_{S_R} d^2 S_i \mathcal{B}_i\,,\qquad {\rm with}\quad \mathcal{B}_i \equiv \frac{1}{\eta}B_i^a\phi^a\,.
\ee
Therefore, the monopole energy is
\be
    E \simeq \eta Q_M + \frac{1}{2}\int d^3 x \left(\frac{1}{g}B_i^a -D_i\phi^a\right)^2\,.
\ee
Since the term inside the integral is positive definite, $E$ is minimised for
\be\label{eq:BogoEq}
    \frac{1}{g}B_i^a - D_i\phi^a = 0\,.
\ee
The above equation is known as the Bogomol'nyi equation, and any solution to it also satisfies the equations of motion \cite{Shifman:2012zz}. To solve it is a non-trivial task but we can already make some guesses about the form of the solutions for the Higgs and the gauge fields. In order for the energy of the monopole configuration to be finite, the scalar field solution has to take values in the vacuum manifold far from the monopole core. Moreover, inspired by the topological argument behind the existence of the monopole solution, we identify field space and physical space 
\be\label{eq:philarger}
    \phi^a \rightarrow \eta \frac{x^a}{r}\,,\qquad {\rm for}\quad r\equiv \sqrt{x^a x^a}\rightarrow \infty\,.
\ee
This is called the \textit{hedgehog solution} since the scalar points in the same direction in field space and in coordinate space. Moreover, energy minimisation also requires the covariant derivative $D_i\phi^a$  to vanish at large values of the radius. This can be achieved by judiciously choosing a gauge field profile that, at $r\rightarrow \infty$, compensates for the derivative $\partial_i\phi^a$ (with the Higgs field given in Eq.~\eqref{eq:philarger})
\be
    W_i^a \rightarrow \epsilon^{aij}\frac{x^j}{r^2}\,,\qquad {\rm for}\quad r\rightarrow \infty\,.
\ee
By virtue of the property of the Levi-Civita tensor, $\epsilon^{abi}\epsilon^{cdi}= \delta^{ac}\delta^{bd} - \delta^{ad}\delta^{bc}$, it is straightforward to check that the above gauge field profile exactly cancels the Higgs spatial derivative at infinity. We now write the ansatz for the field profiles as
\be\label{eq:Profiles}
    \phi^a = \eta \frac{x^a}{r}h(r)\,,\qquad W_i^a=\epsilon^{aij}\frac{x^j}{r^2}K(r)\,,
\ee
where $h$ and $K$ are two functions to be determined by solving the Bogomol'nyi equation with the boundary conditions
\be
\begin{aligned}\label{eq:Boundaries}
    &r\rightarrow \infty:\qquad h(r)\rightarrow 1\,,\quad K(r)\rightarrow 1\,,\\
    &r=0 : \qquad h(r)= 0\,,\quad K(r)=0\,.
\end{aligned}
\ee
The first line enforces the correct behaviour at infinity so that the monopole has finite energy, while the second line ensures that the solution is non-singular at the origin. By plugging Eq.~\eqref{eq:Profiles} into Eq.~\eqref{eq:BogoEq} we find two differential equations \cite{Shifman:2012zz}
\be
\begin{aligned}
    & \frac{d h}{d\rho} = \frac{1}{\rho^2}\left(2K -K^2\right)\,,\\
    &\frac{d K}{d\rho} = h(1-K)\,,
\end{aligned}
\ee
where we defined the dimensionless parameter $\rho \equiv g\eta r$. With the boundary conditions in Eq.~\eqref{eq:Boundaries}, the above system of non-linear differential equations is known to admit the analytic solution
\be
    h(r) = \frac{\cosh{g\eta r}}{\sinh{g\eta r}} - \frac{1}{g\eta r}\,,\qquad K(r) = 1 - \frac{g\eta r}{\sinh{g\eta r}}\,.
\ee
As the above solutions satisfy Eq.~\eqref{eq:BogoEq} by definition, the mass of the monopole is entirely given by its magnetic charge 
\be\label{eq:m_M}
    m_M \simeq \eta Q_M = \frac{1}{g} \int_{S_R} d^2 S_i B_i^a\phi^a = \frac{4\pi\eta}{g}\,,
\ee
where, since $S_R$ is a large sphere enclosing the monopole, we can limit ourselves to considering the asymptotic behaviour of the field profiles.

Away from the \ac{BPS} limit, where $\lambda$ is sizeable, the field profiles cannot be obtained analytically. However, the effect of the scalar potential energy on the monopole mass is very mild. It corrects our \ac{BPS} estimate in Eq.~\eqref{eq:m_M} by a factor $f\left(\lambda/g^2\right)$, with $f(0) = 1$ (BPS limit), and $f(\infty) \simeq 1.8$ \cite{Bai:2020ttp}. Moreover, we can still derive an order-of-magnitude estimate for $m_M$ \cite{Preskill:1986kp}
\be
    m_M \simeq 4\pi\left[\frac{1}{g^2 r_W} + \eta^2 \left(r_W - r_s\right)+\lambda \eta^4 r_s^3\right]\,,
\ee
where $r_W$ and $r_s$ are the characteristic radii of the gauge and scalar field profiles, respectively, defined as the scales over which the fields depart from their asymptotic values. We have assumed $r_W>r_s$. The expression above is quite intuitive: the first term represents the energy of a magnetically charged sphere of radius $r_W$; the second term is the gradient energy of the scalar field, which is only cancelled by the gauge field when the latter takes its asymptotic value; the last terms is the potential energy of the monopole configuration in a sphere of radius $r_s$. Notice that if $r_W < r_s$, the gradient energy of the scalar field is always cancelled by the gauge field and the second term in the above sum would not be present. The scales $r_W$ and $r_s$ are obtained by requiring that they minimise the energy of the monopole: $r_W \simeq (g\eta)^{-1}$ and $r_s \simeq (\sqrt{\lambda}\eta)^{-1}$. The radius of the monopole is $r_M = \min\left[r_W,r_s\right]$. In this case, $r_M \simeq 1/(g\eta)$, and we see that this is always larger than its Compton wavelength $\lambda_M \simeq m_M^{-1}$. Therefore, the monopole can be treated as a classical, extended object.

\subsection{Production}\label{sec:KibbleArg}
Monopoles are expected to be produced in the context of cosmological \ac{PT}s. According to the discussion in Sec.~\ref{sec:TQFT}, the symmetry of the model ($\SO{3}$ in the case of the 't Hooft-Polyakov monopole) is restored at temperatures larger than the critical one, $T_c$. For $T<T_c$, symmetry breaking becomes possible and a \ac{PT} takes place. The way in which this happens depends on the shape of the potential. We delay any further discussion about the different possible scenarios of monopole production to Ch.~\ref{chap:MonoDM}, and we provide here a broad, model-independent picture just to get some physical intuition. 

At the onset of the \ac{PT}, the Higgs field randomly chooses one point in the vacuum manifold $\mathcal{M}$ across a region in physical space given by its correlation length $\xi$. Therefore, after the \ac{PT} completes, the Universe is fragmented into domains of typical size $\xi^3$. Inside each domain, the Higgs field is correlated and has the same orientation in field space. However, its orientations in different domains are uncorrelated. We expect that a monopole forms with some probability $p$ at the intersection among different domains. Therefore, the number density of monopoles at production is
\be
    n_M \simeq p \xi^{-3}\,.
\ee
For $\mathcal{M}\cong S^2$, the probability of forming a monopole has been estimated by Kibble to be $p=1/8$ \cite{Kibble:1976sj}.  The value of the correlation length strongly depends on the details of the \ac{PT}. However, we can already provide a lower bound on the monopole number density at production. In fact, we know that the Hubble horizon provides the largest distance over which information can be exchanged at a given time. Then, causality bounds $\xi \leq H^{-1}$, leading to $n_M \geq p H^3$. Expressing the monopole abundance in terms of the comoving number density $Y_M \equiv n_M/s$ and assuming that the Universe is radiation dominated at the moment when the \ac{PT} completes, we have
\be
    Y_M \geq p \frac{45}{2\pi^2 g_s(T_c)}\left(\frac{T_c}{\MP}\right)^3\,.
\ee
This argument is due to Kibble \cite{Kibble:1976sj} and it is very general since it applies to any class of topological defects (with different values of $p$, depending on the dimension of the topological defect) and any type of cosmological \ac{PT}. However, we will show in Ch.~\ref{chap:MonoDM} that often $\xi \ll H^{-1}$, and monopole abundance at production is much larger than the one estimated by the Kibble argument.

\subsection{Annihilation}\label{sec:MonoAnn}
After monopole production, their number density can be reduced by annihilations of a monopole-antimonopole pair \cite{Zeldovich:1978wj,Preskill:1979zi}. This process is active as long as there is a plasma of charged particles interacting with the monopoles. In general, this is always the case. In fact, the broken gauge group $G$ has to be non-Abelian in order for point-like topological defects to exist. This leads to the unavoidable presence of charged gauge bosons in the plasma. The motion of monopoles in the plasma of relativistic charged particles can be described by a Brownian motion with velocity $u \simeq \sqrt{T/m_M}$. Due to their interaction with the plasma, monopoles experience a drag force that can be estimated as \cite{Vilenkin:2000jqa}
\be
\vec{F} \simeq n \sigma v \Delta\vec{p}\,, 
\ee
where the thermal velocity of particles in the plasma is $v\simeq 1$. The number density of relativistic particles is $n = \mathcal{N}\zeta(3)T^3/\pi^2$, where $\mathcal{N}$ are the internal degrees of freedom of the particles in the plasma. The cross section of the interaction between monopoles and light charged particles is $\sigma \simeq q^2/T^2$, with $q$ the electric charge, in units of the gauge coupling $g$, of the particles constituting the plasma. Notice that $\sigma$ does not depend on the value $g$. This is due to the fact that the monopole magnetic charge scales as $Q_M \propto 1/g$. On dimensional ground, the typical momentum transferred in a collision between a monopole and a charged particle is $\Delta \vec{p} \simeq -\vec{u}\,T$. Therefore, the overall drag force experienced by the monopole is given by 
\be\label{eq:dragF}
    \vec{F} \simeq - \beta\,T^2 \vec{u}\,,
\ee
where we have defined $\beta\equiv q^2\,\mathcal{N}\zeta(3)/\pi^2$. The equation of motion for a monopole subject to the drag force $\vec{F}$ then reads 
\be
    m_M \dot{\vec{u}}=-\beta\,T^2 \vec{u}\,,
\ee
with solution $\vec{u}=\vec{u}_0\, e^{-t/\tau}$ for some initial velocity $\vec{u}_0$. The mean free time $\tau \equiv m_M/\beta T^2$ allows us to compute the monopole mean free path 
\be\label{eq:lfree}
    l_{\rm free} = \tau\,u \simeq \frac{1}{\beta}\sqrt{\frac{m_M}{T^3}}\,\,
\ee
that describes the average length travelled by a monopole between two subsequent collisions with plasma particles.
The presence of a magnetic, long-range interaction between monopoles and antimonopoles makes them move towards each other with a drift terminal velocity that is set by the drag force in Eq.~\eqref{eq:dragF}
\be
    u_{\rm drift} \simeq \frac{4\pi}{g^2\,\beta}\frac{n_M^{2/3}}{T^2}\,.
\ee
The typical distance between a monopole and an antimonopole has been taken to be $r \simeq n_M^{-1/3}$.\\
Interactions with the plasma helps forming monopole-antimonopole bound states by dissipating the energy of the pair. Such a bound state is formed when the distance between a monopole and an antimonopole becomes smaller the capture radius $l_{\rm capt}$, that is given by the distance at which the kinetic energy of the monopole becomes equal to its magnetic potential energy
\be
    l_{\rm capt} = \frac{8\pi}{g^2\,T}\,.
\ee
The characteristic rate of bound-state formation is
\be\label{eq:BoundRate}
    \Gamma_d \simeq \frac{u_{\rm drift}}{r} \simeq \frac{4\pi}{\beta\,g^2}\frac{n_M}{T}\,.
\ee
If we assume that any bound state decays in a negligible amount of time, the Boltzmann equation governing monopole abundance reads
\be
   \dot{n}_M + 3H n_M = -\Gamma_d\,n_M \simeq -\frac{4\pi}{\beta g^2 T^2}n_M^2\,.
\ee
The above equation is most conveniently expressed in terms of the yield $Y_M \equiv n_M/T^3$
\be
    \frac{d Y_M}{dt}=-\frac{4\pi}{g^2\,\beta}T\,Y_M^2\,.
\ee
Considering that in radiation domination, where we assume monopoles to be produced and annihilated, the Hubble rate is
\be
    H^2 = \frac{\pi^2\,g_*}{90}\frac{T^4}{\MP^2} = \frac{1}{4\,t^2}\,,
\ee
the equation governing the evolution of monopole yield can be recast as
\be
    \frac{d Y_M}{dT} = \frac{8\pi}{\beta\,g^2}\sqrt{\frac{45}{2\pi^2\,g_*}}\frac{\MP}{T^2}Y_M^2
\ee
leading to 
\be
    \frac{1}{Y_M^f}-\frac{1}{Y_M^p}= \frac{8\pi}{\beta\,g^2}\sqrt{\frac{45}{2\pi^2\,g_*}}\MP\left(\frac{1}{T_f}-\frac{1}{T_p}\right),
\ee
where $p$ and $f$ denote  quantities evaluated at the time of monopole production and at some later time, respectively. As, in general, $Y_M^f \ll Y_M^p$ and $T_f \ll T_p$ we have
\be\label{eq:YAnn}
    Y_M^f \simeq \frac{\beta g^2}{8\pi}\sqrt{\frac{2\pi^2 g_*}{45}}\frac{T_f}{\MP}\,.
\ee
The final number density of monopoles is given by the smallest quantity between the abundance set by production and the one given in Eq.~\eqref{eq:YAnn}. In other words, Eq.~\eqref{eq:YAnn} gives an upper bound on the monopole relic density. As we will see in Ch.~\ref{chap:MonoDM}, this represents a huge obstacle in realising scenarios where monopoles comprise the totality of the \ac{DM} observed in the Universe. Here, $T_f$ is the temperature at the moment when monopole annihilation is not active anymore and freezes-out. This happens when the particles in the plasma become non-relativistic and their equilibrium number density experiences an exponential suppression. The interaction rate between monopoles and charged particles becomes negligible and the monopole mean free path becomes effectively infinite, so that annihilation stops.

\section{Cosmic strings}\label{sec:CS}
 \ac{CS}s are one-dimensional topological defects. They are present in any model in which the manifold of degenerate vacua is non-simply connected, i.e.~if it contains non-contractible loops $S^1$: $\Pi_1[\mathcal{M}]\neq I$. This happens, for example, anytime an Abelian group $\U{1}$ is spontaneously broken. If the theory is gauged, the string core carries a  $\U{1}$-flux associated with the generator broken in the vacuum. 

\subsection{The Abelian string}
The simplest model featuring the presence of string-like topological defects is the Abelian Higgs model, also known as scalar electrodynamics: the model of a complex scalar field $\phi$ charged under a gauged $\U{1}$ symmetry. The Lagrangian density reads
\be\label{eq:AbelianHL}
\mathcal{L} = -\frac{1}{4 g^2} F_{\mu\nu}F^{\mu\nu} + (D_\mu \phi)(D^\mu \phi)^\dagger - \frac{\lambda}{4}\left(\lvert\phi\rvert^2 - \eta^2\right)^2\,,
\ee
with 
\be
    D_\mu \phi = \partial \phi - i A_\mu \phi\,\qquad{\rm and}\qquad F_{\mu\nu}=\partial_\mu A_\nu - \partial_\nu A_\mu\,.
\ee
The potential energy density is minimised for $\lvert\phi\rvert^2 = \eta^2$, where the $\U{1}$ symmetry is broken. Since $\Pi_1[\U{1}] = \mathbb{Z}$, the spectrum of the theory also contains \ac{CS}s. The energy of a static string is 
\be\label{eq:EStr}
    E = \int d^2 x \left[\lvert\left(\partial_i - i A_i\right)\phi\rvert^2 + \frac{1}{2}\left(E_i E^i + B_i B^i\right) + \frac{\lambda}{4} \left(\lvert\phi\rvert^2 - \eta^2 \right)^2\right]\,,
\ee
where $i = x,y$ is the Lorentz index spanning spatial dimensions in the plane orthogonal to the string. The fields $E_i$ and $B_i$ introduced above are the electric and magnetic fields, respectively, and they are defined as $E_i = F_{0i}$, $2B_i= - \epsilon_{ijk}F^{jk}$. Analogously to what we did for monopoles, we look for field configurations that minimise the energy. Requiring that $E<\infty$ constrains the field profiles far from the string core. The Higgs field has to go to its \ac{VEV}, at $r \equiv \sqrt{x^2 + y^2}\rightarrow \infty$. This condition, however, fixes only the absolute value of $\phi$, with its phase that remains unconstrained:
\be
    \phi(\vec x) \rightarrow \eta e^{i n \alpha(\vec x)}\,,\qquad {\rm for}\quad r\rightarrow \infty\,.
\ee
In order for the Higgs field to be single-valued, we want $\phi(\alpha=2\pi)= \phi (\alpha=0)$. Therefore, $n$ has to be an integer. Since it counts the number of times the Higgs field wraps around the string in field space when completing one closed loop in physical space, $n$ takes the name of \textit{winding number}. With the Ansatz above, the space-derivative of the Higgs field diverges logarithmically as $\int d^2x \lvert \partial_i\phi\rvert^2 \sim \int dr/r$, in the absence of a gauge field. Therefore, the profile for $A_i$ has to be chosen in order to cancel this divergence in the covariant derivative.
Moreover, the gauge field has to be pure gauge at infinity, not to generate any field tensor $F_{ij}$. We can quickly check that these requirements are satisfied by
\be
    A_i \rightarrow n \partial_i\alpha(\vec x) = - n \epsilon_{ij}\frac{x_j}{r^2}\,,\qquad {\rm for}\quad r\rightarrow \infty\,.
\ee
In global theories, there is no gauge field compensating for the Higgs kinetic energy at infinity and the energy of the string in Eq.~\eqref{eq:EStr} diverges logarithmically. This is due to the presence of a long-range interaction among global strings mediated by Nambu-Goldstone boson exchange. The winding number can be written as the integral of the gauge field along a closed path, $\gamma_R$, surrounding the string solution at infinity
\be
    n = \frac{1}{2\pi}\oint_{\gamma_R} dx^i A_i = \frac{1}{2\pi}\int_{\Sigma_R} dS^z (\nabla \times A)_z = \frac{1}{2\pi}\int_{\Sigma_R} d^2 x B_z \equiv \frac{g \Phi_B}{2\pi}\,,
\ee
where, in the last equality, we have used Stoke's theorem to write the winding number as the flux of the magnetic field $B_z = (\nabla \times A)_z = F_{xy}$ over the surface enclosed by the path $\gamma_R$, $\Sigma_R$. By construction, $\Sigma_R$ is traversed by the string, making the winding number a measure of the flux carried by the string core in units of $g/(2\pi)$. 

A drastic simplification of the mathematical description of the field profiles minimising the string energy is obtained in the critical coupling limit, in which $2\lambda = g^2$, and the masses of the Higgs and gauge fields are equal. In this limit, the interaction between two identical parallel strings vanishes, as an effect of the attractive scalar self-interaction being balanced by the repulsive gauge one. Practically speaking, the equations of motion for $A_\mu$ and $\phi$ derived from Eq.~\eqref{eq:AbelianHL} become of the first order and can be easily solved. The string tension $\mu$ or, in other words, its energy, in the critical case is
\be
    \mu = \int d^2 x \left[\frac{1}{4g^2}F_{ij}F^{ij} + \lvert D_i \phi\rvert^2 + \frac{g^2}{2}\left(\lvert \phi\rvert^2-\eta^2\right)^2\right]\,.
\ee
Notice that we have used $2\lambda = g^2$ when writing the potential energy term. After integrating by parts and dropping total derivative terms, the above expression for $\mu$ reduces to \cite{Shifman:2012zz}
\be
    \mu = 2\pi \eta^2 n + \int d^2 x\bigg\{ \lvert\left[ \left(\partial_x - i A_x\right)+\left(\partial_y - i A_y\right)\right]\phi\rvert^2+ \frac{1}{2}\left[\frac{1}{g} B_z + g \left(\lvert\phi\rvert^2 - \eta^2\right)\right]^2\bigg\}\,.
\ee
Since the two terms inside the integral are both positive, the string tension is minimised when they vanish separately
\be\label{eq:BogoEqString}
\begin{aligned}
    \left[ \left(\partial_x - i A_x\right)+\left(\partial_y - i A_y\right)\right]\phi &=0\,,\\
    B_z + g^2 \left(\lvert\phi\rvert^2 - \eta^2\right)&=0\,,
\end{aligned}
\ee
for which the string tension is
\be\label{eq:CritStrTens}
    \mu = 2\pi \eta^2 n\,.
\ee
To solve Eq.~\eqref{eq:BogoEqString}, we start from the ansatz
\be\label{eq:StringAnsatz}
    \phi (r) = \eta \varphi(r)e^{i\alpha}\,,\qquad A_i(r) = -\epsilon_{ij}\frac{x_j}{r^2}\left[1-f(r)\right]\,, 
\ee
where we have fixed $n=1$. Upon defining $\rho \equiv g\eta r$ we are left with the following differential equations for the functions $\varphi$ and $f$
\be
\begin{aligned}
    \rho\frac{d\varphi}{d\rho} - f\varphi &= 0\,,\\
    -\frac{1}{\rho}\frac{df}{d\rho}+\varphi^2 - 1 &=0\,.
\end{aligned}
\ee
The boundary conditions
\be
\begin{aligned}
    &\varphi(\infty) = 1\,,\qquad f(\infty) = 0\,,\\
    &\varphi(0) = 0\,,\qquad f(0) = 1\,,
\end{aligned}
\ee
ensure the correct asymptotic behaviour for the solutions, and smoothness at the string core. Finding $\varphi$ and $f$ for a generic $\rho$ requires solving the above Cauchy problem numerically. However, at large $r$ we get\cite{Shifman:2012zz}
\be
    \varphi(r) \simeq 1 - \exp(-g\eta r)\,,\qquad f(r) \simeq \exp(-g\eta r)\,,
\ee
with the two fields reaching their asymptotic behaviour exponentially fast.

Away from the critical limit, the trick we have performed above of integrating by parts and reducing the equations of motion to a set of first-order differential equations does not work. On the contrary, one needs to solve the Euler-Lagrange equations, derived by minimising the energy in Eq.~\eqref{eq:EStr}, which are now of the second order. The ansatz in Eq.~\eqref{eq:StringAnsatz} is still valid but, now, the differential equations for the functions $\varphi$ and $f$ are significantly more complicated \cite{Shifman:2012zz}. Contrary to the critical case, the string tension has to be computed numerically. Surprisingly, the result is different from the one obtained for a critical string, cfr.~Eq.~\eqref{eq:CritStrTens}, only by a slowly-varying function $h(2\lambda/g^2)$ \cite{Vilenkin:2000jqa}. The function $h$ can be easily found in the limiting cases of $2\lambda\gg g^2$ and $2\lambda\ll g^2$
\be
    \mu = 2\pi \eta h(2\lambda/g^2) \simeq 2\pi \eta^2 \times \begin{cases}
        \log \frac{2\lambda}{g^2},\quad 2\lambda \gg g^2\\
        \left(\log \frac{g^2}{2\lambda}\right)^{-1},\quad 2\lambda \ll g^2
    \end{cases}\,.
\ee

\subsection{Cosmological evolution}\label{sec:StrCosmoEvo}
Likewise monopoles, strings are formed when the cosmological phase transition leading to spontaneous symmetry breaking happens. Their number density can be estimated by the Kibble argument presented in Sec.~\eqref{sec:KibbleArg}. However, as we will show below, the string network evolution has an attractor solution: the \textit{scaling regime}. Soon after its formation, the \ac{CS} network reaches the scaling regime independently on the initial number density of strings, so that most of the \ac{CS} phenomenological consequences do not depend on their number density at the moment of production. 

On cosmological scales, local \ac{CS}s can be regarded as one-dimensional objects. In fact, usually, their thickness is much smaller than the typical distance between two neighbouring strings, that is of the order $H^{-1}$. Therefore, we can describe their evolution by means of an effective action where the core of the string is not resolved: the \textit{\ac{NG} action}. This approach is extremely powerful since it does not require us to know the detailed profile of the Higgs and gauge fields. Moreover, it is mostly model-independent, modulo some phenomenological parameters that need to be input by hand. Numerical studies comparing a \ac{NG} description with a field-theoretical one, in which the field profiles at the string core are properly taken into account, found good agreement between the two approaches, see e.g.~Ref.~\cite{Blanco-Pillado:2023sap}. Notice, however, that other numerical studies seem to come to different conclusions, observing deviations from the \ac{NG} behaviour  in field-theoretical simulations of local strings \cite{Hindmarsh:2017qff}. The reasons behind this discrepancy are still under debate.

Assuming that the curvature of long strings is negligible, the \ac{NG} action is the area of the worldsheet swept by the string motion \cite{Goto:1971ce,Vilenkin:2000jqa}
\be\label{eq:NGAction}
    S = -\mu \int d^2\zeta \sqrt{-\gamma}\,,
\ee
where $\mu$ is the string tension and $\gamma$ is the determinant of the metric induced on the worldsheet
\be
    \gamma_{a b } = g_{\mu\nu} \frac{\partial x^\mu}{\partial\zeta^a}\frac{\partial x^\nu}{\partial \zeta^b}\,,\qquad a,b = 0,1.
\ee
$\zeta_a$ are the worldsheet coordinates: $\zeta_0$ can be chose to be time-like, while $\zeta_1$ is space-like and labels points along the string. The \ac{NG} effective action ceases to be valid when two strings cross each other. When this happens the approximation of infinitely-thin objects breaks down around the contact point. Unless there is some topological obstruction, when strings cross each other they inter-commute with probability close to unity, as shown by numerical simulations \cite{Shellard:1987bv}, and they form loops.

We describe the string network by an ensemble of randomly distributed long strings with length $L$ and energy density 
\be\label{eq:RhoInfty}
    \rho_\infty = \frac{\mu}{L^2}.
\ee
Their energy is, therefore, $E = \rho_\infty V$, with $V$ a large physical volume. Soon after their production, \ac{CS}s experience two competing effects. The Universe expansion stretches strings, making the energy stored in the network increase: the string tension $\mu$ remains constant while their characteristic length grows as $L\propto a$, so that $E\propto a$.  At the same time, strings tend to fragment into loops. Loops oscillate under their own tension and quickly evaporate emitting \ac{GW}s (see Sec.~\ref{sec:StringsGWs}) since there is no topology argument justifying their stability.\footnote{For the local strings discussed here, \ac{GW} emission is the preferred decay channel. However, loops of global and superconducting strings may also emit light particles and electromagnetic radiation efficiently \cite{Vilenkin:2000jqa}.} Therefore, loop production counteracts the effect of the Universe expansion by depleting the network energy. To see which one of these two effects prevails we need to estimate the rate of string fragmentation into loops. A string of length $L$ in a volume $V=L^3$ travels on average a distance equal to $L$ before encountering another string. Assuming that strings move at relativistic speed, and that one loop forms at each encounter, we get a loop chopping rate per unit volume of $L^{-4}$. If loops are created with a typical size equal to $L$, the energy variation of a string can be written as
\be
    \dot E \simeq \frac{\dot a}{a}E-\frac{1}{L^4}\mu L V\,,
\ee
where the first term encodes the effect of the Universe expansion, while the second one describes the energy lost by loop production, with each loops carrying away an energy equal to $\mu L$. Written in terms of $\rho_\infty$, defined in Eq.~\eqref{eq:RhoInfty}, the above equation is
\be\label{eq:RhoInftyEvo}
    \dot\rho_\infty \simeq -2 \frac{\dot a}{a} \rho_\infty -\frac{\rho_\infty}{L}\,.
\ee
We are interested in studying the evolution of $L$ compared to the size of the Hubble horizon, which scales as $H^{-1}\propto t$, for a Universe dominated by an ordinary fluid. Therefore, we introduce the variable
\be
    \gamma(t) \equiv \frac{L}{t}\,,
\ee
which, plugged into Eq.~\eqref{eq:RhoInftyEvo}, yields
\be
    \dot \gamma \simeq -\frac{1}{2t}\left(\gamma-1\right)\,.
\ee
It can be easily checked that the solution $\gamma \simeq 1$ represents an attractor. This is justified by the following heuristic argument. A large density of strings means $\gamma \lesssim 1$, which implies $\dot \gamma > 0$ due to an increase in the rate of fragmentation into loops. On the other hand, for $\gamma \gtrsim 1$, $\gamma$ decreases with time due to a less efficient loop production. This argument shows that the string network is naturally lead, by the competition between Universe expansion and loop production, to a scaling regime in which the characteristic string length $L$ remains a constant fraction $\gamma$ of the Hubble radius $H^{-1}$. The precise value of $\gamma$ has to be computed by numerical simulations.
During the scaling regime, the comoving number density of long strings, i.e.~their number per Hubble volume, remains constant since $L$ and $H^{-1}$ both scale in the same way. Moreover, as long as $L\propto t$, the string network energy density redshifts in the same way as the background one, mimicking the equation of state of the Universe. In fact, during the scaling regime the energy density of long strings scales as $\rho_\infty = \mu/L^2 \propto t^{-2}$. On the other hand, the energy of the Universe goes as $\rho_{\rm bkg} \propto H^2 \propto t^{-2}$. Notice that this scaling does not depend on the equation of state parameter $w$ of the Universe. Since $\rho_\infty$ and $\rho_{\rm bkg}$ scale in the same way with $t$,  \ac{CS}s will occupy a constant fraction of the total energy density. This is a peculiarity of  \ac{CS}s that allows them not to dominate the energy density of the Universe.

\subsection{Gravitational radiation}\label{sec:StringsGWs}
Both long strings and loops radiate energy in the form of \ac{GW}s, due to the presence of small-scale ripples like kinks and cusps. For local strings, as the ones we analyse here, loops are generally long-lived. Moreover, they are continuously produced throughout the network evolution. Indeed, numerical simulations show that the total energy density of the network is mostly contained into small loops \cite{Blanco-Pillado:2013qja}. Therefore, here we will focus on the \ac{GW} signal produced by loops which is, in general, much stronger than the one produced by long strings. 

We can get an order-of-magnitude estimate for the power of gravitational radiation emitted by a loop by means of the quadrupole formula. A string loop of length $\ell$ has mass $M \sim \mu\,\ell$ and, hence, quadrupole moment $D \simeq \mu \,\ell^3$. On dimensional grounds, the typical frequency at which such a loop oscillates is $\omega \sim 1/\ell$, so that
\be\label{eq:PGWapprox}
   P_{\rm GW} \sim  \GN \left(\frac{d^3 D}{dt^3}\right)^2 \sim  \GN\mu^2\,,
\ee
with $ \GN= \left(8\pi \MP^2\right)^{-1}$ the Newton constant. Notice that $P_{\rm GW}$ does not depend on the loop length, since the third derivative of the quadrupole moment is constant. Since a string loop will, in general, move at relativistic speed, the quadrupole formula above is not strictly valid. Nonetheless, it allows us to get a rough order-of-magnitude estimate based only on dimensional analysis. Surprisingly enough, we will see below that a more refined analysis does not lead us far from Eq.~\eqref{eq:PGWapprox}.

The oscillation frequencies emitted by a loop of length of $\ell$ are 
\be
    \bar f_k = \frac{2k}{\ell}\,,
\ee
where $k$ is the wave-number, and the factor of $2$ in the numerator takes into account the periodic boundary conditions of the loop. The bar reminds us that these are the frequencies at the moment of emission. At a time $t>\bar t$, subsequent to emission, the frequency $\bar f$ appears redshifted by the Universe expansion
\be
    f(t) = \frac{a(\bar t)}{a(t)}\bar f\,.
\ee
The total radiated power is the sum over all the modes 
\be
    P_{\rm GW} = \sum_k P_{\rm GW}^{(k)} = \sum_k \Gamma_k  \GN\mu^2 = \Gamma  \GN\mu^2\,,
\ee
where the total efficiency $\Gamma$ is a numerical coefficient that depends on the small-scale structure of loops. \ac{NG} numerical simulations find $\Gamma \simeq 50 - 100$ \cite{Maggiore:2018sht}. The precise form of $\Gamma_k$ depends on what kind of features the loop presents at small scales: either cusps or kinks \cite{Gouttenoire:2022gwi}
\be
    \Gamma_k = \frac{\Gamma k^{-q}}{\zeta(q)}\,,\qquad {\rm with}\;\begin{cases}
        q=4/3\quad {\rm cusps}\,,\\
        q= 5/3\quad {\rm kinks}\,.
    \end{cases}
\ee
We see that $P_k$ decreases quite slowly with $k$. This is due to the fact that cusps and kinks are small-scale features propagating on the loop and, therefore, they contribute to the highest frequencies. The energy radiated via \ac{GW}s is subtracted from the loop initial mass $M \simeq \mu \ell$. We assume that, during the scaling regime, all the loops formed at some time $t_i$ have the same initial length, which is a fraction $\alpha$ of the Hubble horizon: $\ell_i = \alpha\,t_i$. This assumption is not exactly correct. On the contrary, numerical simulations have shown that most of the long-string energy goes into small loops, with only about $10\%$ of the loops having an initial length with $\alpha=0.1$ \cite{Blanco-Pillado:2013qja}. However, larger loops are the ones that contribute the most to the \ac{GW} signal, with smaller ones giving a negligible contribution.\footnote{Other numerical studies have found the presence of a second population of small loops giving a sizeable contribution to the \ac{GW} signal, boosting it at large frequencies \cite{Ringeval:2017eww}.} Therefore, here we focus only on the largest loops, i.e.~those produced with $\alpha=0.1$. After being formed, loops start shrinking due to \ac{GW} emission, so that their length evolves as
\be\label{eq:l(t)}
    \ell(t) = \alpha t_i - \Gamma  \GN\mu (t-t_i)\,, 
\ee
where we have assumed that $\Gamma$ does not depend on the loop size. The typical life-time of a loop is, then, $T = \ell_i/(\Gamma  \GN\mu)$. A string network emitting a detectable signal of \ac{GW}s has $ \GN\mu \simeq 10^{-7}$, leading to $T \simeq 10^5 l_i$: string loops are quite long lived. Loops produced with initial length $\ell_i$ will then emit \ac{GW}s in the range of frequency $f > 2/\ell_i$. Therefore, fixing as $f_{\rm det}$ the frequency at which the sensitivity of an hypothetical experiment peaks, only loops produced above the critical length $\ell_{\rm det} = 2/f_{\rm det}$ will emit, at some point in their life-time, \ac{GW}s with $f\simeq f_{\rm det}$.

\begin{figure}[h!]
\begin{center}
\includegraphics[width=.9\textwidth]{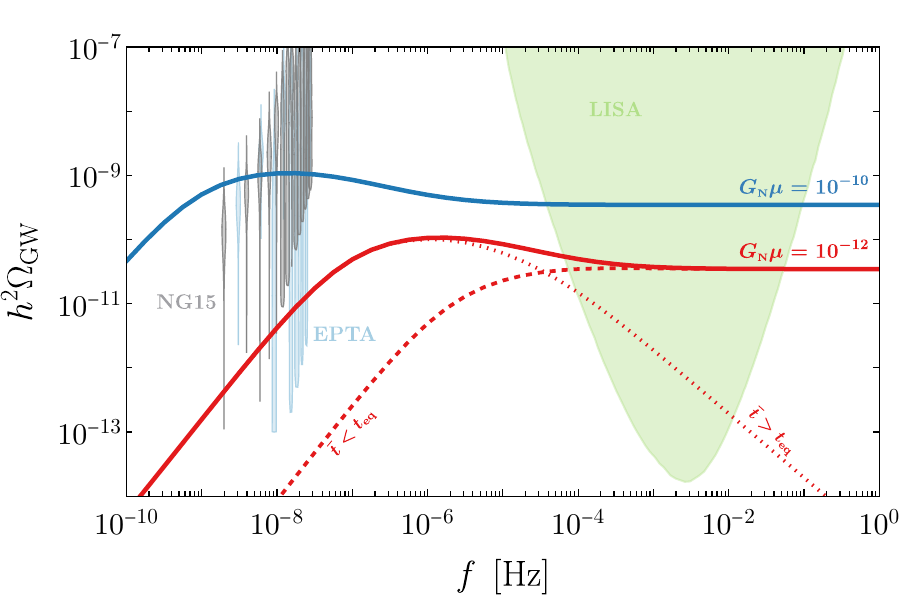}
    \caption{GW spectrum produced by a network of \ac{CS}s with tension $\GN\mu = 10^{-10}$, in blue, and $\GN\mu=10^{-12}$, in red, considering the effect of cusps only ($q=4/3$). We fixed $\alpha=0.1$. The dashed and the dotted lines refer to the signal of \ac{GW}s emitted by loops during radiation domination and matter domination, respectively. The violins represent the signal detected by the NANOGrav (gray) \cite{NANOGrav:2023gor} and EPTA (light blue) \cite{EPTA:2023fyk} collaborations. In green, the projected sensitivity of LISA \cite{LISACosmologyWorkingGroup:2022jok} as derived in Ref.~\cite{NANOGrav:2023hvm}.}
\label{fig:StringGW}  
\end{center} 
\end{figure}

As we have seen, the frequencies of emitted \ac{GW}s depend on the length of the loop. Loops are continuously produced by the string network at different lengths that depend on the time of their production $t_i$. Therefore, we expect at a given time $t$, the presence of a population of loops with different lengths $l$. Stated otherwise, many different loops, created at different time during cosmic history contributes to a given frequency $f$. To accurately predict the \ac{GW} signal emitted by the string network we need to know the loop density distribution as a function of their length and of time. The number density of loops can be estimated analytically \cite{Gouttenoire:2022gwi}
\be\label{eq:nloop}
    \frac{d n(t_i)}{d t_i} = 0.1 \frac{C_{\rm eff}(t_i)}{\alpha t_i^4}
\ee
where $t_i$ is the time of production of a loop that at a later time $t$ has a length $\ell$. The numerical pre-factor $0.1$ accounts for the fact that only $10\%$ of the total energy of long strings goes into loops with $\ell_i \simeq 0.1 t$. By inverting Eq.~\eqref{eq:l(t)}, it can be found
\be\label{eq:ti}
    t_i(\ell,t) = \frac{\ell + \Gamma  \GN\mu t }{\alpha + \Gamma  \GN\mu}\,\,
\ee 
so that Eq.~\eqref{eq:nloop} can be analogously expressed in terms of $\ell$ and $t$.
The parameter $C_{\rm eff}$ describes the efficiency of production of loops. It is \cite{Gouttenoire:2022gwi}
\be
    C_{\rm eff} = \begin{cases}
         5.4, \qquad t_i < t_{\rm eq}\\
        0.39,\qquad t_i>t_{\rm eq}
    \end{cases}\,,
\ee
for loops formed in radiation domination and matter domination, respectively.
Once we know the loop number density and the power radiated by each loop, we can proceed to estimate the total \ac{GW} signal. As usual, we define the energy fraction of \ac{GW}s per logarithmic frequency interval today
\be
    \Omega_{\rm GW}(f) \equiv \frac{f}{\rho_c} \frac{d \rho_{\rm GW}(f,t_0)}{d f} = \frac{G}{3 H_0^2}f \frac{d \rho_{\rm GW}(f,t_0)}{df}\,,
\ee
where $H_0 = 100\,{\rm km/s/Mpc}$ is the Hubble rate today. The energy density of \ac{GW}s that have a frequency $f$ at time $t$, gets contributions from all the loops formed till that moment and that emitted a frequency $\bar f$ at a moment $\bar t$, such that $a(t) f = a(\bar t) \bar f$. Since the frequency emitted by a loop depends on its length $\ell$, $\rho_{\rm GW}(f,t)$ will be given by
\be
    \rho_{\rm GW}(f,t) = P_{\rm GW} \int_{t_F}^{t}  n\left(\frac{a(t)}{a(\bar t)} f,\bar t\right)\left[\frac{a(\bar t)}{a(t)}\right]^4 d\bar t\,,
\ee
with $t_F$ the moment of formation of the network which, here, we take to coincide with the beginning of the scaling regime, neglecting the possible friction exerted on long strings by the plasma. Therefore, 
\be
     \frac{d \rho_{\rm GW}(f,t)}{d f} = P_{\rm GW} \int_{t_F}^{t}  \frac{d n\left(\bar f,\bar t\right)}{d \bar f}\left[\frac{a(\bar t)}{a(t)}\right]^3 d\bar t\,,
\ee
where we have rewritten the integrand in terms of the frequency at emission $\bar f$. The loop number density appearing in the above integral can be found starting from Eq.~\eqref{eq:nloop}:
\be
\begin{aligned}
    \frac{d n(\bar f, \bar t)}{d \bar f} &= \left[\frac{a(t_i)}{a(\bar t)}\right]^3 \frac{dn}{dt_i} \frac{dt_i}{d\ell}\frac{d\ell}{d\bar f} \\
    &=\left[\frac{a(t_i)}{a(\bar t)}\right]^3 \sum_k 0.1\frac{C_{\rm eff}(t_i)}{t_i^4}\frac{1}{\alpha(\alpha+\Gamma  \GN\mu)} \frac{2k}{\bar f^2}\,.
\end{aligned}
\ee
Hence, we have
\be
    h^2\Omega_{\rm GW}(f) = ( \GN\mu)^2 \frac{8\pi h^2}{3H_0^2}\sum_k \frac{2k}{f}\frac{0.1\,\Gamma_k}{\alpha(\alpha+\Gamma  \GN\mu)}\int_{t_F}^{t_0} d\bar t\,\frac{C_{\rm eff}(t_i)}{t_i^4} \left[\frac{a(\bar t)}{a(t_0)}\right]^5 \left[\frac{a(t_i)}{a(\bar t)}\right]^3\,,
\ee
where, we remind, $t_i$ has to be meant as a function of $\ell$, or $f$ analogously, and $t$, as per Eq.~\eqref{eq:ti}. The scale factor $a(t)$ can be found by numerically solving the Friedman equation
\be 
    \left(\frac{\dot a(t)}{a(t)}\right)^2 = H_0^2 \left[\Omega_{r,0} \left(\frac{a(t)}{a(t_0)}\right)^{-4} + \Omega_{m,0} \left(\frac{a(t)}{a(t_0)}\right)^{-3} + \Omega_\Lambda\right]\,,
\ee
for $\Omega_{r,0} \simeq 9.2\times 10^{-5} $, $\Omega_{m,0} \simeq 0.31$, and $\Omega_\Lambda = 1 - \Omega_{r,0} - \Omega_{m,0}$ \cite{Auclair:2019wcv}. We have set the spatial curvature of the Universe to zero. Notice that, for the sake of simplicity, we have neglected an additional factor in the evolution of the radiation energy density given by the annihilation of \ac{SM} species that changes the number of relativistic degrees of freedom over time. The signal of \ac{GW}s emitted by a network of local \ac{CS}s is shown in Fig.~\ref{fig:StringGW}, for different benchmark values of the string tension $\GN\mu$.

Fig.~\ref{fig:StringGW} discloses one of the peculiar features of \ac{GW}s emitted by  \ac{CS} loops. For frequencies emitted deep in radiation domination, the signal presents a characteristic flat shape,  with its amplitude being independent from its frequency. Intuitively, this is the outcome of the combination of two effects. \ac{GW}s produced at earlier times experience a larger redshift suppression. However, at the same time, we see that more loops are formed in the early Universe since $dn/dt_i \propto t_i^{-4}$, as per Eq.~\eqref{eq:nloop}. During radiation domination these two effects are perfectly balanced, leading to the characteristic flat spectrum observed in Fig.~\ref{fig:StringGW} at large frequencies. An intuitive understanding can be derived starting, again, from the quadrupole formula \cite{Gouttenoire:2019kij}. We consider the power emitted in \ac{GW}s, at a time $\bar t$, by a population of $N_l$ loops
\be
    P_{\rm GW}(\bar t) \sim N_{l}(\bar t)  \GN\mu^2\,.
\ee
In the scaling regime, the number density of loops produced at $t_i$ scales as $n \propto t_i^{-3}$, cfr.~Eq.~\eqref{eq:nloop}. Therefore, the number of loops at emission time $\bar t$ is
\be
    N_{l}(\bar t) \sim \left(\frac{\bar t}{t_i}\right)^3 \left(\frac{t_i}{ \bar t}\right)^{3/2}\,,
\ee
where the second parenthesis is the redshift term in a radiation-dominated Universe, $a \propto t^{1/2}$. The scaling of $N_{l}$ with time suggests that the later $\bar t$, the more numerous the loops. Therefore, the dominant contribution to the \ac{SGWB} from a population of loops formed at $t_i$ comes towards the end of their life at a time 
\be
    T= \frac{\alpha\,t_i}{\Gamma  \GN\mu}\,.
\ee
The energy density of radiation is
\be
    \rho_r(\bar t) \sim  \GN^{-1}\bar t^{-2}\,,
\ee
while the one of \ac{GW}s is
\be
    \rho_{\rm GW}(\bar t) \sim P_{\rm GW}(\bar t)\,\bar t/{\bar t}
    ^3 \sim \left(\frac{\bar t}{ t_i}\right)^{3/2}  \GN\mu^2 {\bar t}^{-2}\,.
\ee
It follows that
\be
    \Omega_{\rm GW} \sim \Omega_{r} \frac{\rho_{\rm GW}(\bar t)}{\rho_{r}(\bar t)} \sim \Omega_r (\GN\mu)^2 \left(\frac{\bar t}{ t_i}\right)^{3/2}\,.
\ee
By considering the signal at $\bar t = T$, we find
\be
    \Omega_{\rm GW} \sim \Omega_r \left(\frac{\alpha}{\Gamma}\right)^{3/2}( \GN\mu)^{1/2}\,,
\ee
which is, indeed, independent from $\bar t$ and, therefore, from $\bar f$, and scales as $\Omega_{\rm GW} \propto \sqrt{ \GN \mu}$. The presence of a flat spectrum in radiation domination, makes \ac{CS}s a powerful tool to explore possible variations in the standard cosmic history, that will lead to departures from scale invariance in the \ac{GW} signal \cite{Auclair:2019wcv,Gouttenoire:2019kij}. The fact that the for larger values of the string tension, the spectrum moves towards smaller frequencies is understood as follows. A lower value of $\GN\mu$ suppresses the amount of energy radiated by loops in the form of \ac{GW}s. This implies that loops are more long-lived, according to our estimate for the typical loop life-time below Eq.~\eqref{eq:l(t)}. As a consequence, \ac{GW}s with a frequency $f$ are emitted later in the Universe history, compared to the case of a larger $\GN\mu$, thus experiencing less redshift.

\subsection{Cosmological constraints}\label{sec:StrCosmoConstr}
Due to the fact that they mimic the equation of state of the background, as seen in Sec.~\ref{sec:StrCosmoEvo},  \ac{CS}s never come to dominate the energy budget of the Universe. Nonetheless, their energy density is constrained by cosmological observations.

We have seen that a network of  \ac{CS}s will unavoidably lead to the production of a \ac{SGWB}, with a rather flat signal whose amplitude depends on the parameter $\GN\mu$. The null detection of a background signal by the \ac{LVK} collaboration sets an upper bound on the string tension $\GN\mu \lesssim 10^{-7}$ \cite{KAGRA:2021kbb}. The recent hint of a detection by the \ac{PTA} collaborations could be explained by a network of local strings with $\GN\mu \simeq 10^{-10}$ \cite{NANOGrav:2023hvm, EPTA:2023xxk}. However, due to the flat nature of the signal, \ac{CS}s are not among the favoured candidates to explain the \ac{PTA} signal. This could set the most competitive bound on the string tension so far but it is not clear yet how robust this is, due to the theoretical and numerical uncertainties about the modelling of loop production. Moreover, since \ac{GW}s behave as radiation, they contribute to the expansion rate of the Universe. \ac{BBN} predictions tightly constrain the amount of extra radiation that is allowed in the Universe at the time, resulting into $h^2\Omega_{\rm GW}\lesssim 10^{-6}$ \cite{LISACosmologyWorkingGroup:2022jok}, which is less stringent than the \ac{LVK} one.

GW production aside, the energy density stored in one-dimensional objects of cosmological size produces characteristic density fluctuations in the early Universe. In fact,  \ac{CS}s were originally proposed as a possible alternative to inflation, as an explanation for the generation of inhomogeneities leading to large-scale structure formation \cite{Zeldovich:1980gh}. This is possible because, once produced,  \ac{CS}s keep sourcing sub-horizon perturbations all along the Universe evolution, differently from other sources such as \ac{PT}s. This is the main reason why \ac{CS}s became so popular as an alternative to inflation. In fact, instantaneous sources are only able to produce inhomogeneities within the size of the Hubble horizon at the time in which they are active, which is usually much smaller than the typical scales at which we observe structures today. We can easily estimate the string tension needed to produce sufficient inhomogeneities. For a scaling network, the  overdensity produced by strings on the horizon scale is $\delta \rho \sim \mu t/ t^3 \sim \mu/t^2$, so that
\be
    \frac{\delta \rho}{\rho_r}\sim 30 \GN\mu\,,
\ee
where $\rho_r \sim( 30  \GN t^2)^{-1}$. Since $\delta\rho/\rho \sim 10^{-5}$ is required on galactic scale, we see that a string network with $\GN\mu \sim 10^{-6}$ could provide sufficient inhomogeneities at the correct scales. However, the spectrum of fluctuations so produced is monochromatic. This is in contrast with the acoustic peaks observed in the \ac{CMB}, which then constrain the maximum tension allowed for an hypothetical string network to be $\GN\mu \lesssim 10^{-7}$ \cite{Planck:2013mgr}. This latter bound is considered to be extremely robust since it does not depend on the modelling on the loop production functions which is still under debate, see e.g.~Ref.~\cite{Blanco-Pillado:2019vcs}.

\section{Domain walls}\label{sec:DW}

\ac{DW}s are sheet-like topological defects arising in models with a disconnected vacuum manifold $\Pi_0[\mathcal M]\neq I$. This is the case, for example, whenever a discrete symmetry gets spontaneously broken.

\subsection{$Z_2$ domain walls}

The simplest scenario featuring the presence of \ac{DW}s is the one of a real scalar field $\phi$ charged under a discrete $\mathbb Z_2$ symmetry: $\mathbb Z_2:\phi \rightarrow -\phi$. The Lagrangian density of the model, at the renormalizable level, is 
\begin{equation}\label{eq:LZ2}
    \mathcal{L} = \frac{1}{2} \left(\partial_\mu \phi\right)^2 + \frac{\mu^2}{2}\phi^2 -\frac{\lambda}{4}\phi^4 \equiv \frac{1}{2} \left(\partial_\mu \phi\right)^2 - V(\phi)\,.
\end{equation}
For $\mu^2,\lambda>0$, the potential features a minimum far from the origin, at $\langle \lvert \phi \rvert\rangle = \mu/\sqrt{\lambda} \equiv \eta$, where the $\mathbb{Z}_2$ symmetry is broken. The vacuum manifold is constituted by two disconnected points $\phi = \pm \eta$. Since $\Pi_0[\mathbb Z_2] = \mathbb Z_2$, we expect the presence of a stable field configuration interpolating between the two disconnected minima: a \ac{DW}. The equation of motion for the $\phi$ zero-mode is simply the Klein-Gordon one
\be
    \partial_\mu\partial^\mu \phi = \frac{\partial V(\phi)}{\partial \phi}\,,
\ee
where the potential $V(\phi)$ is defined in Eq.~\eqref{eq:LZ2}. Analogously to what we did for \ac{CS}s and monopoles, we seek static solutions of the equation of motion. In particular, we look for planar solutions orthogonal to the $z$-axis $\phi(z)$. A solution of this kind is known analytically for the quartic potential in Eq.~\eqref{eq:LZ2}
\be\label{eq:phiDW}
    \phi(z) = \eta \tanh \left(\sqrt{\frac{\lambda}{2}} \eta z\right)\,,
\ee
describing a field profile going at $-\eta$ for $z\rightarrow -\infty$, and at $+\eta$ for $z\rightarrow +\infty$. In interpolating between the two disconnected minima, the field passes through $\phi=0$ at $z=0$. This field solution describes a \ac{DW} in the $xy$ plane, centred at $z=0$, and with width $\delta \simeq (\sqrt{\lambda/2}\,\eta)^{-1}$, defined by $\phi(z=\delta/2)\simeq \eta/2$. The thickness of the wall is the result of two competing effects. The surface energy density gets a contribution from the scalar field gradient proportional to $\delta \,\partial_z \phi \sim \eta^2/\delta$, which is larger for thinner walls. The contribution from the potential energy is instead $\delta \lambda \eta^4$. In other words, gradient energy minimisation leads to thicker walls where the field profile changes smoothly, but, at the same time, potential energy is minimised for thinner walls, since the field departs from its minimum value over a smaller region in physical space. The balance between these two effects leads to our estimate for $\delta$.  The energy-momentum tensor of the wall, $T_{\mu\nu}$, can be easily found by plugging Eq.~\eqref{eq:phiDW} into the definition
\be
    T_{\mu\nu} = \partial_\mu \phi(z) \partial_\nu \phi(z) - g_{\mu\nu}\mathcal{L}\,,
\ee
where $g_{\mu\nu}$ is the Minkowski metric. This gives
\be\label{eq:TmunuDW}
    T_\mu^\nu = f(z) \diag (1,1,1,0)\,,\qquad {\rm with} \quad f(z) \equiv \frac{\lambda}{2}\eta^4\left[\cosh\left(\sqrt{\frac{\lambda}{2}}\eta z\right)\right]\,.
\ee
The fact that $T_{x}^x = T_y^y$ follows from Lorentz invariance of the \ac{DW} solution under boosts in the $xy$ plane: the motion of the wall in the plane orthogonal to $z$ cannot be observed. The surface energy density of the \ac{DW} solution is 
\be
    \sigma = \int T_0^0 dz = \frac{2\sqrt{2}}{3}\sqrt\lambda \eta^3\,.
\ee
The symmetry of the energy-momentum tensor in Eq.~\eqref{eq:TmunuDW} suggests that the wall tension in the $x$ and $y$ directions is equal to $\sigma$, meaning that any region of the wall with non-zero curvature will be straightened by the large tension and accelerated to relativistic speed.

\subsection{Cosmological evolution}
The initial number density of \ac{DW}s produced by a phase transition can be estimated using the Kibble argument presented in Sec.~\ref{sec:KibbleArg}, with the initial number density given by $n_{\rm DW} \simeq p \,\xi^{-3}$. By dividing a volume $\xi^3$ into cubic cells, each labelled by $+\eta$ or $-\eta$ according to the \ac{VEV} of $\phi$ in that cell, we can convince ourselves that \ac{DW}s, which are present at the interface between cells with different labels, are formed with probability $p=1/2$. According to percolation theory, we know that, for a cubic lattice, identical cells  percolate if they comprise around $30\%$ of the whole lattice. Since, here, the field chooses randomly between $\langle \phi\rangle=\pm \eta$, we see that our system is above the  percolation threshold. Therefore, we expect the presence of one “infinite'' $+\eta$ cell and one “infinite'' $-\eta$ cell. In other words, the system is dominated by one infinite wall. The presence of finite closed walls of radius $R\sim \xi$ is also to be expected. 

The evolution of a \ac{DW} can be described by an effective action where it is regarded as an infinitely thin, two-dimensional object with no internal structure
\be\label{eq:SDW}
    S = - \sigma \int d^3\zeta \sqrt{-\gamma}\,,
\ee
which is the analogous of the \ac{NG} action used to described the motion of  \ac{CS}s, in Eq.~\eqref{eq:NGAction}. $\sqrt{-\gamma}$ is the determinant of the metric induced on the three-dimensional world-sheet (“world-tube'')
\be
    \gamma_{a,b}= g_{\mu\nu}\frac{\partial x^\mu}{\partial \zeta^a}\frac{\partial x^\nu}{\partial \zeta^b}\,,\qquad \zeta=0,1,2\,.
\ee
The equations of motion can be derived by minimising the action in Eq.~\eqref{eq:SDW}. In other words, a wall evolves in order to minimise the volume of the three-dimensional world-sheet swept by its motion. Solving such equations of motion is quite involved. Luckily, some interesting features of a Universe with \ac{DW}s moving around can be grasped already by the a perfect-gas approximation, in which the \ac{DW} network is regarded as a perfect gas moving with velocity $v$ inside a large box of volume $V\gg\xi^3$. In this approximation we can derive some collective properties of the \ac{DW} network, such as its equation of state parameter \cite{Kolb:1990vq}
\be
    w = v^2 - \frac{2}{3}\,.
\ee
The energy density of a perfect fluid redshifts as $\rho \propto a^{-3(1+w)}$. Therefore, a gas of relativistic walls ($v \simeq 1$)  redshifts as $\rho_{\rm DW} \propto a^{-4}$, as expected from radiation. However, in the opposite limit, in which the walls are non-relativistic ($v\ll1$) we find $\rho_{\rm DW}\propto a^{-1}$. This behaviour is quite intuitive. Indeed, the \ac{DW} network energy density is given by 
\be
    \rho_{\rm DW} \sim \frac{\sigma A}{V}\,,
\ee
where $A$ is the area of a single wall. As $V\propto a^{3}$ and $A \propto a^2$, we find $\rho_{\rm DW} \propto a^{-1}$, as expected. Notice that, since the velocity of \ac{DW}s is expected to be quickly redshifted by the Universe expansion, \ac{DW}s will reach the non-relativistic limit at some point.  This shows that stable \ac{DW}s can quickly dominate the total energy density of the Universe. Therefore, stable \ac{DW}s are cosmologically disfavoured, unless they are formed in \ac{PT}s taking place at extremely low energies.  Moreover, even if \ac{DW}s were not to overclose the Universe, their presence is expected to produce large density fluctuations in the \ac{CMB}, of the order \cite{Vilenkin:2000jqa,Zeldovich:1974uw}
\be
    \frac{\delta \rho}{\rho} \sim  \GN\sigma t_0 \sim  10^{6}\left(\frac{\eta}{\rm GeV}\right)^3\,,
\ee
therefore constraining the symmetry-breaking scale to be $\eta \lesssim {\rm MeV}$.

To avoid the cosmological disaster of \ac{DW}-dominance, one can simply introduce a small term in the Lagrangian density in Eq.~\eqref{eq:LZ2} which explicitly breaks the $\mathbb Z_2$ symmetry. In this way, the degeneracy between the two vacua $\langle\phi\rangle = \pm \eta$ is broken and the \ac{DW}s are not stable. When the explicit breaking term becomes cosmologically relevant, regions of space lying in the false vacuum shrink due to the effect of a pressure proportional to the bias itself. We will discuss an explicit model featuring biased \ac{DW}s in Sec.~\ref{secDW}.

\subsection{Gravitational radiation}
In scenarios in which \ac{DW}s are unstable, a large amount of \ac{GW}s is radiated by the network as a consequence of \ac{DW} collisions and collapse. This leads to the production of a \ac{SGWB}. The detailed modelling of the signal requires numerically simulating the Higgs field evolution \cite{Saikawa:2017hiv}. However, an order-of-magnitude estimate of the predicted relic density can be obtained using the quadrupole formula. A \ac{DW} of typical length $R$ has a mass $M \sim \sigma R^2$. Analogously to what happens for \ac{CS}s we assume that \ac{DW} evolution features a scaling regime where their typical size remains a fixed constant of the Hubble horizon $R \sim t$. The presence of this behaviour has been confirmed by numerical simulation \cite{Saikawa:2017hiv}. The wall quadrupole is, then, $Q\sim M R^2 \sim \sigma t^4$. The quadrupole formula gives
\be
    P_{\rm GW} \sim  \GN\left(\frac{d^3Q}{dt^3}\right)^2 \sim  \GN\sigma^2 t^2\,,
\ee
for the power radiated into \ac{GW}s. The energy density can be found as
\be
    \rho_{\rm GW} \sim \frac{P_{\rm GW} t}{t^3} \sim  \GN \sigma^2\,.
\ee
The above expression is only valid in the far-field regime and, therefore, strictly speaking, it does not apply to \ac{DW}s, which are cosmologically extended objects. Nonetheless, this very naive estimate allows us to see that the energy density of \ac{GW}s scales as $\sigma^2$, a behaviour which is confirmed also by numerical simulations \cite{Saikawa:2017hiv}. A more precise estimate of the energy density of a biased network of \ac{DW}s is given in Sec.~\ref{sec:TopologicalDefects}.

\chapter{Accidentally light scalar fields}
\label{chap:Accidents}
\minitoc

In models with \ac{SSB} by scalar fields in large group representations, we observe that some of the scalar masses can be loop-suppressed with respect to the naive expectation from symmetry selection rules. We present minimal models -- the $\SU{2}$ five-plet and $\SU{3}$ ten-plet -- with such accidentally light scalars, featuring compact tree-level flat directions lifted by radiative corrections. We sketch some potential applications, from stable relics and slow roll in cosmology, to hierarchy and fine-tuning problems in particle physics. 

\section{Introduction}

We consider perturbative, renormalisable models of scalar fields in four-dimensional quantum field theory. When all operators allowed by Lorentz invariance and the internal symmetries of the model are included in the Lagrangian with generic coefficients, all scalar fields are massive, with the only exception of \ac{NGB}s. By Goldstone's theorem, in models with a continuous global symmetry group $G$ spontaneously broken to a subgroup $H$, the \ac{NGB}s, forming the coset $G/H$, are exactly massless to all orders.
As a general rule, the mass of non-Goldstone scalar fields arises at the tree level from the scalar potential. One well-known exception to this rule are \ac{pNGB}s, which appear when the symmetry $F$ of the scalar potential is larger than the symmetry $G$ which defines the model as a whole \cite{Weinberg:1972fn}.
This requires additional fields and interactions, such as gauge or Yukawa couplings, which explicitly break $F$ and induce \ac{pNGB} masses via loops.

It is less well known that there exist models with tree-level massless scalars which are neither \ac{NGB}s nor \ac{pNGB}s in the above sense. In these models, the most general renormalisable potential compatible with the symmetry $G$ is {\it not} invariant under an enhanced continuous symmetry larger than $G$. Still, some non-NGB scalar fields remain massless at the tree level. To distinguish these accidentally tree-level massless fields from \ac{pNGB}s as defined above, we will call them ``accidents'' for short. Accidents were encountered e.g.~in pre-QCD attempts to build renormalisable models of mesons \cite{Bars:1973vp}, and their nature was clearly emphasised in an early precursor \cite{Georgi:1975tz} of the little-Higgs idea (reviewed in Ref.~\cite{Schmaltz:2005ky}).\footnote{One may prefer to extend the definition of \ac{pNGB} to any scalar which is massless at the leading order in the loop expansion. In this case accidents can be dubbed \ac{pNGB}s too, as done e.g.~in Ref.~\cite{Georgi:1975tz}. We retain the name ``accident'', in order to distinguish this peculiar class of light particles. In most of the more recent little-Higgs constructions \cite{Schmaltz:2005ky}, the little Higgs is a conventional \ac{pNGB} rather than an accident.}
Accidents also appear in O'Raifeartaigh-like models of spontaneous supersymmetry breaking \cite{Capper:1976mf}, where they have been dubbed ``pseudo-moduli'' and studied in greater detail more recently \cite{Ray:2006wk,Intriligator:2008fe}; in this case the scalar potential is  additionally constrained by supersymmetry.

In this chapter, we present some models with accidentally light scalars
which are in a sense the most minimal ones. We focus on two examples, with symmetry groups $G=\SU{2}\times\U{1}$ and $G=\SU{3}\times\U{1}$; there are no additional discrete symmetries imposed; the field content is a single scalar multiplet in an irreducible representation. This is to be contrasted with the examples in the literature, which to our knowledge tend to rely on more complicated continuous symmetries (typically multiple copies of the same group), feature additional ad-hoc discrete symmetries, and  contain several scalar fields in various representations. The price to pay to avoid these complications is to take the single scalar field in a large representation of $G$.

The simplest possibilities are the five-plet in the $\SU{2}$ case, 
analysed in Sec.~\ref{sec:su2}, and the ten-plet in the $\SU{3}$ case, analysed in Sec.~\ref{sec:su3}. There are few known results on \ac{SSB} by large representations, see, e.g., Ref.~\cite{Jetzer:1983ij}, and these do not cover our cases of interest, which further motivates our analysis.

The geometry of field space is non-trivial in our models, due to the large field representation involved. There is a compact manifold $M'$ of degenerate tree-level vacua. The continuous symmetry group $G$ is completely broken at a generic point on $M'$. When $G$ is gauged, all points on a single $G$-orbit are identified, but the resulting tree-level vacuum manifold $M$ of physically inequivalent points does not reduce to a single point. Instead, $M$ is parameterised by one or more non-Goldstone flat directions in field space: these correspond to tree-level massless ``accident'' fields. Both the scalar and the vector mass spectrum change when moving along $M$. At special points in $M$, the vacuum symmetry is enhanced (i.e.~a subgroup $H$ of $G$ is unbroken), and additional accidents appear.

In Sec.~\ref{sec:apps} we will briefly discuss a few potential phenomenological applications of accidents in cosmology (\ac{DM}, slow roll) and in particle physics (Higgs, doublet-triplet splitting). A detailed analysis of the phenomenology is left for future work.

\section{The simplest model with accidents: an SU(2) five-plet}\label{sec:su2}

\subsection{Potential and tree-level spectrum}\label{potT}

Our first example of a model with accidentally light scalars is for $G=\SU{2}\times\U{1}$
with a complex scalar in the five-dimensional representation of $\SU{2}$ and with unit $\U{1}$ charge, $\phi\sim {\bf 5}_1$.
The most general $G$-invariant renormalisable potential can be written as
\be
 V=-\mu^2\,S+\frac{1}{2}\left[\lambda \,S^2+ \kappa
 \left(S^2-|S'|^2\right)+ \delta\, A^a A^a\right]\,.
\label{V5}\ee
Here $S$, $S'$ and $A^a$ are bilinears transforming in the singlet and adjoint representation of $\SU{2}$,
\be
S=\phi^\dag\phi\,,\qquad S'=\phi^T \phi\,, \qquad A^a=\phi^\dag T^a\phi\quad(a=1,2,3)\,,
\label{bi5}\ee
where $T^a$ are the $\SU{2}$ five-plet generators, which may be chosen imaginary and antisymmetric and satisfy $\tr\left(T^a T^b \right) = 10\,\delta^{ab}$.
It can be checked that the $\SO{10}$ global symmetry of the free theory 
is broken explicitly by $V$ to $\SU{2}\times\U{1}$: there is no larger continuous accidental symmetry. 

Since our aim is to study \ac{SSB}, we take $\mu^2>0$.
We further take $\lambda$, $\kappa$ and $\delta$ to be positive. While $\lambda$ must be positive for the potential to be bounded from below, negative values for $\kappa$ and $\delta$ are possible, but not of interest here. With our choice, each of the three terms in the quartic potential is positive definite, and
\begin{itemize}
 \item $S$ is non-zero if and only if at least one  \ac{VEV} is non-vanishing,
 \item $S^2-|S'|^2$ vanishes if and only if $\phi$ and $\phi^*$ are aligned in field space, i.e.~$\phi=c\hat\varphi$ for some complex number $c$ and real unit vector $\hat\varphi$,
 \item if $\phi$ and $\phi^*$ are aligned, then $A^a=0$ by antisymmetry; this corresponds to the fact that the adjoint ${\bf 3}$ is contained in the antisymmetric part of the product ${\bf 5}\otimes{\bf 5}$. 
\end{itemize}
A minimum is therefore found by choosing $\vev\phi$ such that $\phi$ and $\phi^*$ are aligned and $\vev{S}=\mu^2/\lambda$: 
\be
 \vev{\phi_j}=\frac{v_j}{\sqrt{2}}\,e^{i\theta}\quad(j=1\ldots 5)\,,\qquad  v_j\in\mathbb{R}\,,\qquad\theta\in[0,\,2\pi)\,,\qquad
v^2\equiv v_j v_j
 =\frac{2\mu^2}{\lambda}\,.
\label{minV}\ee
The remarkable feature of this potential is the existence of one flat direction which is \emph{not} associated to a \ac{NGB}, corresponding to one accidentally massless field. The vacuum manifold is indeed five-dimensional: while the overall scale $v$ of the \ac{VEV} is fixed by the minimisation condition, one is free to rotate the five components $v_j$, and to choose the angle $\theta$. Changes in $\theta$ correspond to the $\U{1}$ Goldstone direction. Of the four directions in $\SO{5}/\SO{4}\simeq S^4$ corresponding to different choices for the orientation of the \ac{VEV},\footnote{More precisely one should divide by a $\mathbb{Z}_2$ since, for any $v$, the points $(v,\theta)$ and $(-v,\theta+\pi)$ are identified.} three are associated to the $\SU{2}$ Goldstone directions, but the fourth one does not correspond to any symmetry generator. When gauging $\SU{2}\times\U{1}$, all points on the Goldstone manifold are identified, but there remains a flat direction of degenerate, physically inequivalent vacua.

\begin{figure}[ht]
\begin{center}
\includegraphics[width=.42\textwidth]{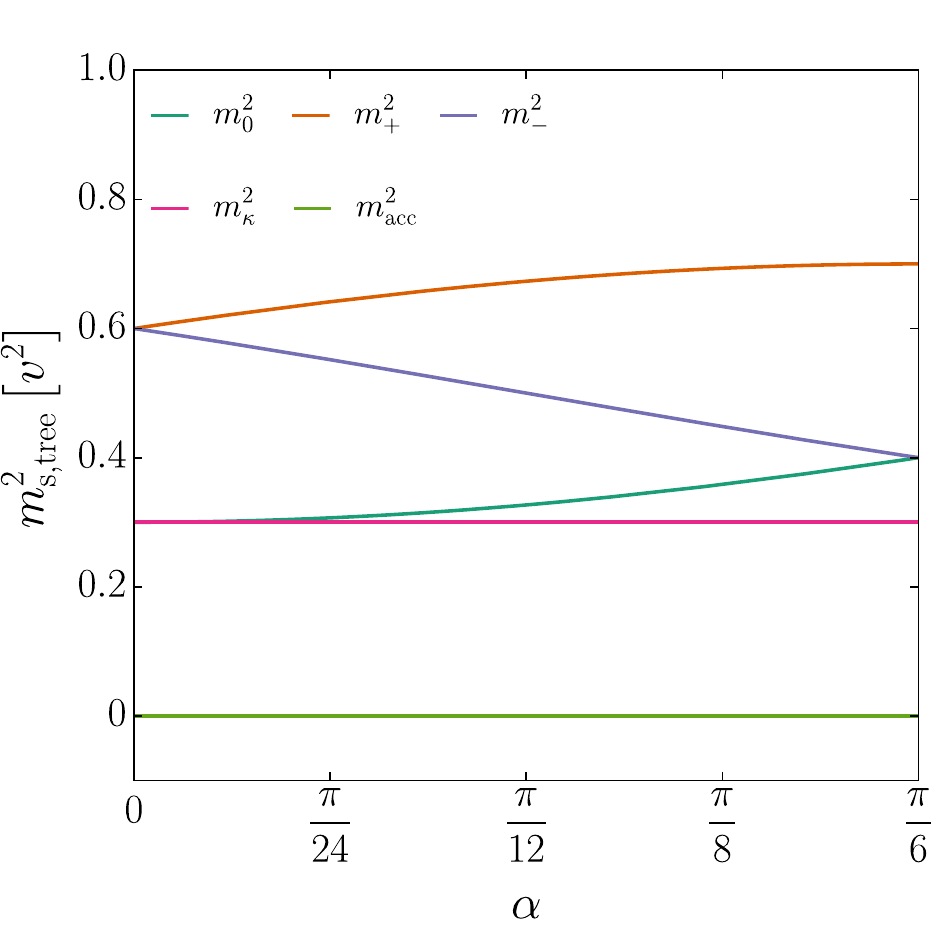}
\qquad\includegraphics[width=.42\textwidth]{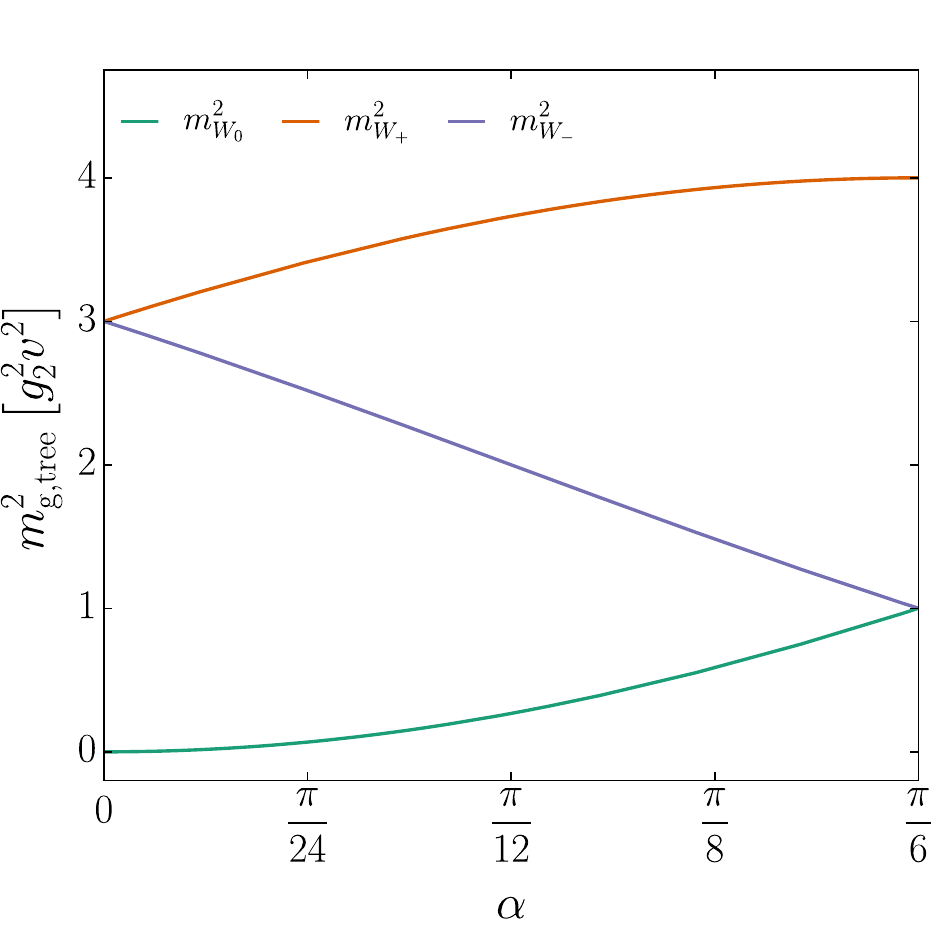}
\caption{\textit{Left-hand panel.} Tree-level masses of the scalar fields moving along the flat direction $\alpha$, for some benchmark values of the couplings, $\kappa = 0.3$ and $\delta = 0.1$. The mass of the radial mode $m_\lambda^2$ is not shown,
as it is controlled by an independent quartic coupling.
\textit{Right-hand panel.} Tree-level masses of the $\SU{2}$ vector bosons in units of $g_2^2 v^2$. At $\alpha = 0$ one of the gauge bosons is massless as $\U{1}'$ is preserved:  at such point the corresponding would-be \ac{NGB} becomes a second accidentally massless scalar.}
\label{fig:TreeMass}  
\end{center}
\end{figure}

Explicitly, an $\SU{2}\simeq\SO{3}$ five-plet can be represented as a traceless symmetric $3\times 3$ matrix, $\Phi \equiv \phi_j \lambda_j$, with a convenient basis given by the symmetric Gell-Mann matrices $\lambda_{1,3,4,6,8}$.  The $\SO{3}$ transformation $O$ acts as $\Phi\to O\Phi O^T$. Using $\SO{3}\times\U{1}$ invariance, the \ac{VEV} can thus be chosen real and diagonal:
\be\label{eq:alpha}
\langle\Phi\rangle=\frac{v}{\sqrt{2}}\left(\lambda_3\sin\alpha+\lambda_8\cos\alpha\right)\,,\qquad\lambda_3=\left(\begin{array}{ccc}1&&\\&-1&\\&&0\end{array}\right)\,,\qquad\lambda_8=\frac{1}{\sqrt{3}}\left(\begin{array}{ccc}1&&\\&1&\\&&-2\end{array}\right)\,.
\ee
The accidentally flat direction is parameterised by the angle $\alpha$.
One can show that $\alpha\sim\alpha+\pi/3$ as well as $\alpha\sim-\alpha$ under $\SU{2}\times\U{1}$, 
hence the fundamental domain of $\alpha$ can be chosen to be $\alpha\in[0,\,\frac{\pi}{6}]$. After diagonalisation, the scalar mass matrix becomes
\be\label{eq:scmass}
M^2(\alpha)=\diag\left[m_\lambda^2,\,m_\kappa^2,\,m_0^2(\alpha),\,m_+^2(\alpha),\,m_-^2(\alpha),\,\,0,\,0,\,0,\,0,\,0\right]\,,
\ee
where
\begin{align}
&m_\lambda^2=\lambda\,v^2,&&m_\kappa^2=\kappa\,v^2,\nonumber\\
&m_0^2(\alpha)=\left(\kappa+4\delta \sin^2 \alpha\right)v^2, &&m_{\pm}^2(\alpha)=\left[\kappa+4\delta\sin^2\left(\alpha\pm\frac{\pi}{3}\right)
\right]v^2\,.
\end{align}
with $m_\lambda$ the mass of the radial mode.
When $G$ is gauged, the gauge boson masses are
\be\label{eq:vmass}
m^2_{W0}(\alpha)= 4\,g_2^2\,\sin^2\alpha\,v^2\,,\qquad
m^2_{W\pm}(\alpha) = 4\,g_2^2\,\sin^2\left(\alpha\pm\frac{\pi}{3}\right)v^2\,,\qquad
m_B^2=g_1^2\,v^2\,,
\ee
where $g_1$ and $g_2$ are the $\U{1}$ and $\SU{2}$ gauge coupling, respectively.
Fig.~\ref{fig:TreeMass} shows the correlations among the tree-level masses of scalars
and gauge bosons, as a function of $\alpha$.
Note that the sum of scalar squared masses remains constant along the flat direction, and the same holds for the sum of vector squared masses.

There exists a distinguished point on the vacuum manifold, corresponding to a null eigenvector of one of the $\SU{2}$ generators, $T^3$ say, 
where a $\U{1}'$ subgroup of $\SU{2}$ is unbroken.
In the above parameterisation, it is given by $\alpha=0$.
At this enhanced-symmetry point, there appears a second accidentally massless scalar, which corresponds to the $\U{1}'$ would-be \ac{NGB}. This should be contrasted with the standard picture in the absence of a flat direction: in that case an enhanced-symmetry vacuum is an isolated minimum of the potential, and an unbroken gauge symmetry implies a massless gauge boson, but no massless scalar. 
There are degeneracies in the massive spectrum too, $m_\kappa=m_0$, $m_+ = m_-$ and
$m_{W+}=m_{W-}$; they are similarly associated with $\U{1}'$-charged states.
The two accidents together form a complex scalar with $\U{1}'$ charge $2$. 

A second distinguished point along the flat direction corresponds to an eigenvector of maximal eigenvalue $2$ of $T^3$, and coincides with the opposite endpoint of the fundamental domain, $\alpha=\pi/6$. 
In this point $G$ is completely broken (as in any generic point), but still there are degeneracies in the mass spectrum, $m_0=m_-$ and $m_{W0}=m_{W-}$.

The extraneous massless degree of freedom which is found at a generic point of the vacuum manifold is not a \ac{pNGB} in Weinberg's strict sense \cite{Weinberg:1972fn}, since the potential admits no accidental symmetry it could correspond to. Nevertheless, in the limit where some of the couplings vanish, the symmetry of the potential is enhanced, and some of the scalar degrees of freedom become \ac{NGB}s.\footnote{In a sense, this is trivially the case for all scalar fields in any model. That is, consider a general model of $N$ real scalars and take the limit where all couplings tend to zero. All scalars then become massless: $N-1$ of them can be regarded as the \ac{NGB}s of the spontaneous breaking of $\SO{N}$ (the symmetry of the kinetic term) to $\SO{N-1}$ (the subgroup preserved by a generic \ac{VEV}), while the radial mode associated to the \ac{VEV} direction can be regarded as the \ac{NGB} of scale invariance.}

In particular,
for $\kappa=\delta=0$ and vanishing gauge couplings, the model has a global $\SO{10}$ symmetry spontaneously broken to $\SO{9}$, and indeed there is only one massive radial mode with non-zero mass $m_\lambda$, plus nine \ac{NGB}s. When $\kappa$ is switched on, the global symmetry is reduced to $\SO{5}\times\U{1}$, cfr.~Eqs.~\eqref{V5}--\eqref{bi5}.
This symmetry is spontaneously broken by the \ac{VEV} to $\SO{4}$, giving rise to five massless \ac{NGB}s, while four scalars acquire a mass $m_\kappa$. When $\delta$ is switched on, the symmetry is explicitly broken to $\SU{2}\times\U{1}$, but five massless modes remain. Four of them are \ac{NGB}s of the exact, spontaneously broken symmetries, while the fifth is the accident. The characteristic feature of this model is that the explicit breaking does \emph{not} induce a tree-level mass proportional to $\delta$ for the accident, which remains light (to the extent that the model is perturbative) even when $\delta$ is of the order of the other quartic couplings.

We conclude by discussing an interesting property of the tree-level mass matrices in the presence of accidents. We will argue that in  models with one or more accidentally-flat directions $\alpha_i$, the quantities $\tr M^2$ and $\tr M^4$ are independent of $\alpha_i$ at the tree level, where $M^2=M^2(\alpha_i)$ stands either for the scalar mass matrix, or the vector boson mass matrix $M_W^2$, or the fermion mass matrix $M_F^2\equiv m_F^\dag m_F$. The argument goes as follows: these traces are quadratic or quartic polynomials in the scalar \ac{VEV} components $v_j$, with all indices contracted in a $G$-invariant way. On the other hand, the scalar potential $V(\phi_j)$ contains, by definition, all independent $G$-invariant polynomials in $\phi_j$ up to degree four. The minimum value of the potential, $V(v_j)$, is constant along the accidentally flat directions $\alpha_i$. This implies that the vacuum is invariant under a symmetry $G_v$ larger than $G$. Let us assume that the action of $G_v$ lifts to a linearly realized action on the initial scalar fields (in the above example, $G_v=\SO{5}\times\U{1})$. Then, the \ac{VEV}s of $G_v$-invariant operators in the potential are constant as the $\alpha_i$ vary. On the other hand, the $G_v$-breaking operators vanish in the vacuum. In conclusion, there is no $\alpha_i$-dependent polynomial that can contribute to the trace of $M^2(\alpha_i)$ or $M^4(\alpha_i)$. One can check that this argument holds for all the models with accidents considered in this chapter.

\subsection{One-loop lifting of the flat direction}
\label{sec:veff}

Let us focus on the special $\U{1}'$ preserving point of the previous section, at $\alpha=0$. At the one-loop level, the tree-level scalar potential is corrected by the Coleman-Weinberg effective potential \cite{Coleman:1973jx, Jackiw:1974cv}, as discussed in details in Sec.~\ref{sec:EffPot}, 
\be
 \Delta V_{\rm CW}=\frac{1}{64\pi^2}\,\text{\rm Str}\,\left({\cal M}^4\log\frac{{\cal M}^2}{\Lambda^2}\right)\,,
\ee
where Str denotes the weighted supertrace, $\Lambda$ is the renormalisation scale, and ${\cal M}^2$ is the scalar-field dependent mass-squared matrix. The one-loop effective potential gives rise to a $\Lambda$-dependent tadpole term for the radial mode of tree-level mass $m_\lambda$. We impose that this tadpole should vanish as a renormalisation condition, i.e.~we define the renormalised \ac{VEV} to be  $v^2=2\mu^2/\lambda$ in terms of the renormalised $\mu$ and $\lambda$. One then finds that all one-loop tadpole terms for the other scalar fields vanish as well, so that the $\U{1}'$ preserving point remains a critical point of the effective potential. Concerning the mass spectrum, the \ac{NGB}s remain massless as they should, while the two modes which were accidentally massless at the tree level pick up a finite and positive one-loop mass,
\be
 m_{\text{acc}}^2=\frac{1}{4\pi^2}\left[3\,g_2^2\, m_{W\pm}^2+\delta\,m_{\pm}^2\;f\left(\frac{m_0^2}{m_\pm^2}\right)\right]\Biggr|_{\alpha=0}\,,
\ee
where $m_{W\pm}^2(0)=3\,g_2^2v^2$, $m_\pm^2(0)=(3\delta+\kappa)v^2$,  $m_0^2(0)=\kappa v^2$, and $f$ is a positive-definite function,
\be
f(x)=1-x+x\log x\,.
\ee
We conclude that the symmetry-enhanced point becomes an isolated minimum of the potential, after including one-loop corrections.

For the second distinguished point on the tree-level vacuum manifold, at its opposite end $\alpha=\pi/6$, one finds that there is similarly no tadpole term induced at one loop for the tree-level massless mode (once the one-loop tadpole for the radial mode has been subtracted). However, its one-loop mass is instead tachyonic, and can be written as
\be
 \tilde m_{\text{acc}}^2=-\frac{1}{4\pi^2}\left[3\,g_2^2\,m_{W-}^2\;f\left(\frac{m_{W+}^2}{m_{W-}^2}\right)+\delta\,m_-^2\;f\left(\frac{m_+^2}{m_-^2}\right)\right]\Biggr|_{\alpha=\frac{\pi}{6}}\,,
\ee
with $m_{W+}^2(\frac{\pi}{6})=4\,g_2^2 v^2$, $m_{W-}^2(\frac{\pi}{6})=g_2^2 v^2$, $m_+^2(\frac{\pi}{6})=(\kappa+4\delta)v^2$, and $m_-^2(\frac{\pi}{6})=(\kappa+\delta)v^2$. Therefore, this second point is a saddle point of the effective potential. 

Note that the accident effective potential does not depend on $g_1$ at one loop. Indeed, gauging $\U{1}$ preserves the $\SO{5}$ symmetry which is recovered for $\delta\into 0$ and $g_2\into 0$. In this limit the accidents become \ac{NGB}s of $\SO{5}/\SO{4}$, and therefore their effective potential must vanish.

\subsection{Coupling to fermions: accident misalignment 
}\label{subsec:abelianhiggs}

The low-energy fluctuations around the $\U{1}'$-preserving point at $\alpha=0$ are described by a simple \ac{EFT}: a $\U{1}'$ gauge theory with one light charged scalar. The scalar mass $m_{\text{acc}}^2$ is loop-suppressed with respect to the masses of the heavier states constituting the \ac{UV} completion. It is interesting to study the question whether the symmetry-enhanced point can be destabilized by loop effects, which would spontaneously break the residual $\U{1}'$. However, we have shown that, with a field content of only scalars and gauge bosons, $m_{\text{acc}}^2$ is always positive. 

\begin{figure}[ht]
 \centering
 \includegraphics[width=.65
\textwidth]{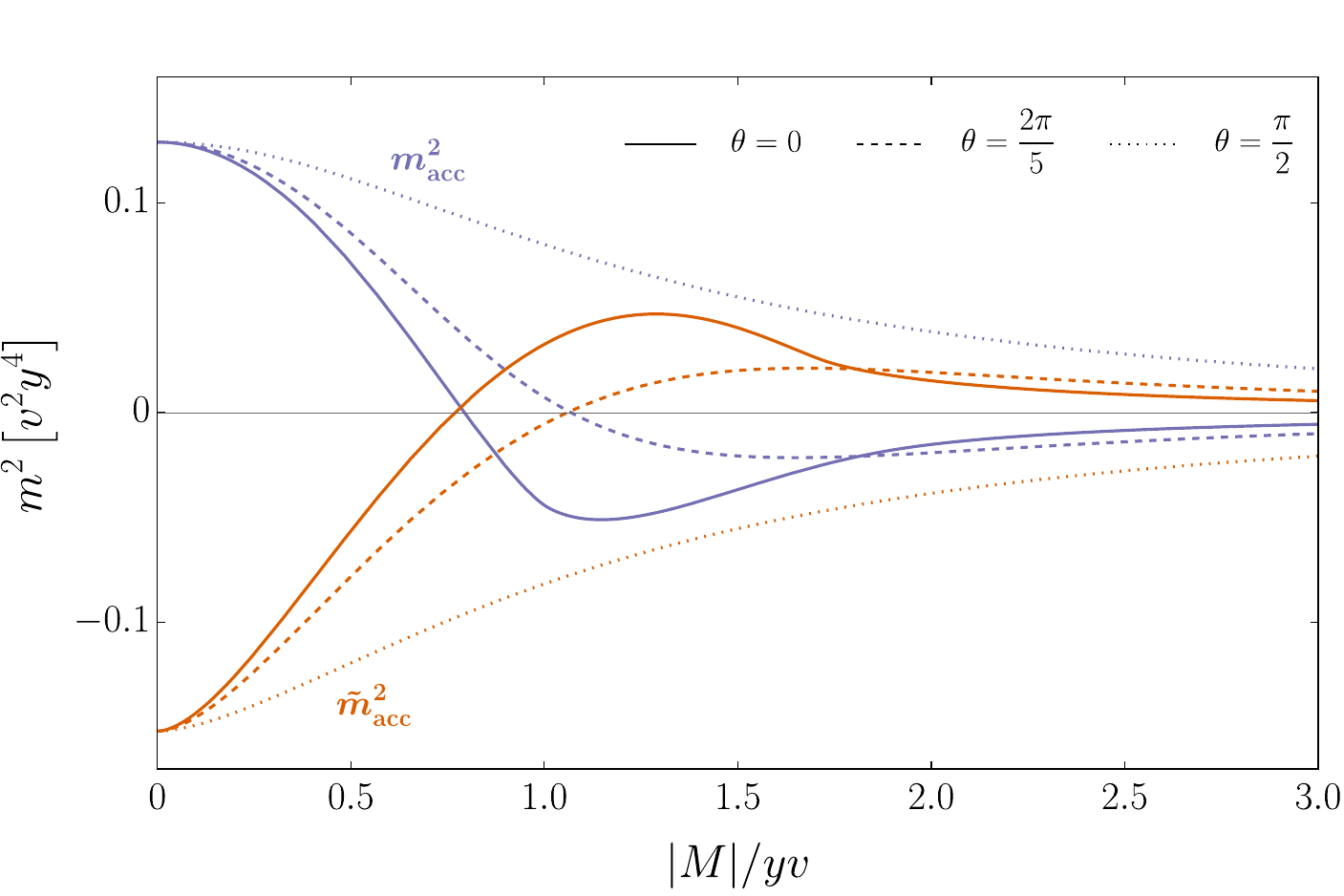}
 \caption{The fermion contribution to the one-loop mass-squared of the accident,  at $\alpha=0$ (purple curves) and $\alpha=\pi/6$ (orange curves), as a function of $|M|/yv$, where $M = \lvert M \rvert e^{i\theta}$ is the Dirac mass and the two Yukawa couplings are assumed to be equal, $y_\chi=y_\psi=y$. For, e.g., $\theta=0$ (solid curves), the $\U{1}'$-preserving point is destabilised for $M\simeq 0.8\, yv$, whereupon the $\U{1}'$-breaking point at $\alpha=\pi/6$ becomes the minimum.}\label{fig:yukmass}
\end{figure}

We therefore add to the model a minimal anomaly-free set of fermion fields coupled to the five-plet $\phi$. We take $\chi$ and $\psi$ to be left-handed Weyl fermions, transforming as $\bf{3_{\pm 1/2}}$ with respect to $\SU{2}\times \U{1}$,\footnote{
Another minimal choice would be $\chi\sim \bf{5_0}$, $\psi\sim \bf{1_{-1}}$, $\bar\psi\sim \bf{1_{1}}$. However in this case the Yukawa couplings are $\SO{5}$ symmetric, therefore they induce no potential for the accidents.} which allows for the terms
\be\label{eq:Lferm}
{\cal L}\supset y_\psi\,\psi^T\Phi \psi
+y_\chi\,\chi^T \Phi^*\chi
+M\, \psi^T\chi\text{+ h.c.}\,,
\ee
where the matrix $\Phi$ was defined above \eq{eq:alpha}. The complex phases of the Yukawa couplings $y_\psi$ and $y_\chi$ can be set to zero without loss of generality, but then the phase of the Dirac mass $M$ becomes physical. In Fig.~\ref{fig:yukmass} we plot the accident mass at $\alpha=0$ induced by fermion loops, in the limit where gauge and scalar self-couplings are negligible, for the special case $y_\chi=y_\psi\equiv y$. For a small $|M|\ll yv$, we find $m_{\rm acc}^2$ is always positive, while for sizeable $|M|$ with a suitable phase, there is a region where the accident becomes tachyonic, so that fermion loops do indeed destabilize the symmetry-enhanced point. In this region, the former saddle point $\alpha=\pi/6$ becomes the minimum of the effective potential, and $\U{1}'$ is spontaneously broken at the loop level.

Note that, even though this spontaneous breaking is induced at one-loop, its scale is the same as for the tree-level breaking of $\SU{2}$, that is to say, the gauge boson masses in the new minimum are all of the order $g_2\,v$. In other words, the \ac{EFT} valid close to the symmetry-enhanced point is not suitable anymore, because the new $\U{1}'$-breaking minimum is at distance of the order of the cutoff $v$ in field space. It would be more interesting to find a configuration where the effective potential has a minimum $\alpha_{\rm min}$ close to (but not exactly at) the $\U{1}'$-preserving point, which would imply a breaking scale $v'\ll v$, where we define $v'/v\equiv\sin(3\alpha_{\rm min})$. In the effective theory well below the scale $v$, one could then identify the accident field with the Higgs boson of $\U{1}'$ breaking. 

To this end, we now discuss the combined effect of scalar self-couplings, gauge couplings and Yukawa couplings. The one-loop effective potential along the accident direction can be evaluated explicitly as a function of $\alpha$ because, on the tree-level vacuum manifold, ${\cal M}^2$ is simply given by the tree-level mass matrix which is easily diagonalised.
Subtracting the radial tadpole and the vacuum energy at the renormalisation point $\alpha=0$, one obtains
\be\begin{split}\label{eq:VCW}
\Delta V_{\rm CW}(\alpha)\Bigr|_{\text{tree-level vacuum}}=\frac{1}{64\pi^2}\Biggl[&\tr\left( M^4(\alpha)\log \frac{M^2(\alpha)}{\Lambda^2}-M^4(0)\log \frac{M^2(0)}{\Lambda^2}\right)\\
+&\,3\,\tr \left( M_W^4(\alpha)\log \frac{M_W^2(\alpha)}{\Lambda^2}- M_W^4(0)\log \frac{M_W^2(0)}{\Lambda^2}\right)\\
-&\,2\,\tr\left( M_F^4
(\alpha)\log \frac{M_F^2(\alpha)}{\Lambda^2}- M_F^4(0)\log \frac{M_F^2(0)}{\Lambda^2}\right)\Biggr]\,.
\end{split}
\ee
Here, the scalar mass matrix is given by \eq{eq:scmass}, and the eigenvalues of the gauge boson mass matrix $M_W^2$ by \eq{eq:vmass}. The $6\times 6$ fermion mass matrix $m_F$ resulting from \eq{eq:Lferm} can likewise be diagonalised analytically, and we denoted $M_F^2\equiv m_F^\dag m_F$. Note that the $\Lambda$-dependence in \eq{eq:VCW} is spurious, since $\tr M^4(\alpha)$, $\tr M_W^4(\alpha)$ and $\tr M_F^4(\alpha)$ are $\alpha$-independent (see Sec.~\ref{potT}): the divergences have already been subtracted, and the resulting effective potential is finite.

\begin{figure}[ht]
\begin{center}
\includegraphics[width=.42\textwidth]{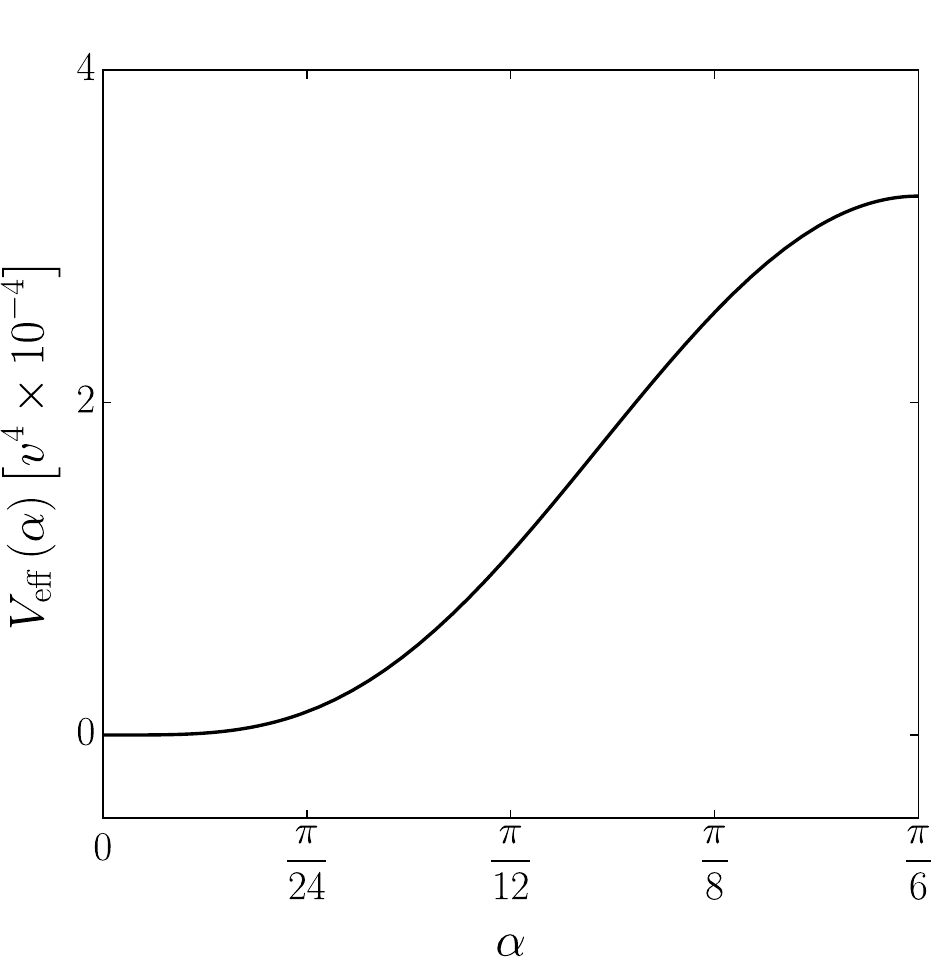}
\qquad\includegraphics[width=.42\textwidth]{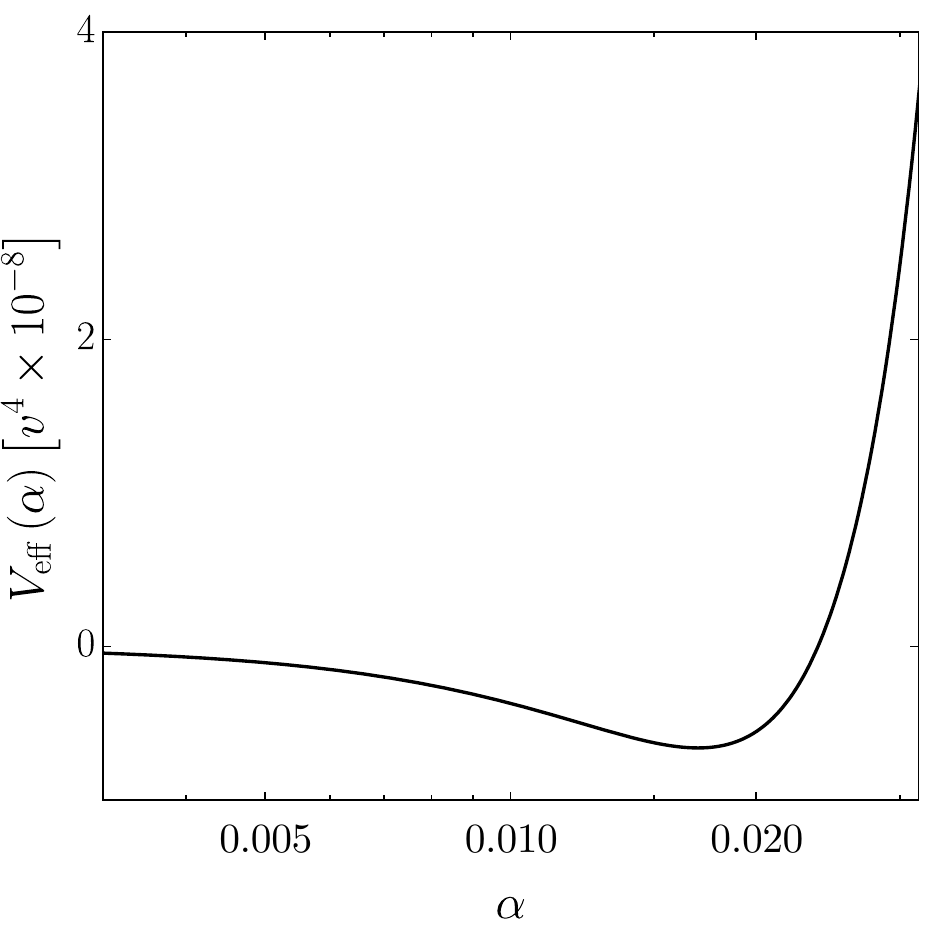}
\caption{\textit{Left-hand panel.} One-loop effective potential along the accident direction $\alpha$, for $\kappa = 1$, $\delta = 0.995$, $\theta = 0$, $\left| M \right|/v= 0.8$, $y_{\chi}=y_{\psi}=1$ and negligible gauge couplings. \textit{Right-hand panel.} A zoom of the left-hand panel in the region close to the origin, showing the minimum at $\alpha
\simeq 0.02$.} 
\label{fig:Veff}  
\end{center}
\end{figure}

The couplings can now be chosen to obtain a minimum of the one-loop corrected potential parametrically close to $\alpha=0$. The potential for an example point in parameter space is shown in Fig.~\ref{fig:Veff}. At this point, the gauge couplings are negligible, and the loop contributions from scalars and fermions largely balance each other, leading to a small misalignment from the $\U{1}'$-preserving direction, with $v'/v \simeq 0.06$.
However, such a minimum can be obtained only at the price of fine-tuning the parameters at the per-mille level. This in turn implies that higher loop corrections could significantly shift the minimum of the potential. To understand the fine-tuning, it is instructive to consider the Fourier expansion of the contributions to the effective potential. One has
\be
\Delta V_{\rm CW}(\alpha)\Bigr|_{\text{tree-level vacuum}} \simeq c_6 \cos \left(6 \alpha \right) + c_{12} \cos \left(12 \alpha \right)\,,
\ee
where $c_6$ and $c_{12}$ are functions of the couplings. Both the fermionic and  bosonic contributions to $c_{12}$ turn out to be numerically suppressed with respect to the respective contributions to $c_6$, by at least two orders of magnitude (and higher harmonics are even more suppressed, which is why they are neglected here).\footnote{To be precise, this statement holds in the fermion sector for all values of the couplings that tend to destabilise the $\alpha=0$ vacuum, see Fig.~\ref{fig:yukmass}.} For generic values of the couplings, the effective potential is therefore dominated by the lowest harmonic, $\Delta V_{\rm CW}\simeq c_6 \cos(6\alpha)$, and the minimum is either at $\alpha=0$ or at $\alpha=\pi/6$ depending on the sign of $c_6$. In order to obtain a small misalignment, the fermionic and bosonic contributions to $c_6$ must cancel for the most part, and moreover $c_6/c_{12}$ must be accurately tuned, in order to achieve $(v'/v)^2 \simeq  1/2 + c_6/(8 c_{12})\ll 1$. The tuning in the coefficients is numerically of the order of higher-loop effects, which we neglected in our analysis. In other words, the exact position of the minimum shown in Fig.~\ref{fig:Veff} is not under theoretical control.

\section{The first in a series of accidents: an SU(3) ten-plet}\label{sec:su3}

Accidentally massless scalars are not specific to the $\SU{2}\times\U{1}$ five-plet model. The same phenomenon occurs in other models with large representations. Consider for instance the symmetry group $G=\SU{N}\times\U{1}$ with a scalar field $\phi^{ijk}$ in the three-index symmetric representation of $\SU{N}$ and unit $\U{1}$ charge. Here $i,j,k=1,\dots,N$ are fundamental indices; the $\phi$ multiplet contains $N(N+1)(N+2)/6$ components. For $N=2$, one can check that the most general renormalisable scalar potential compatible with $G$ gives tree-level masses to all scalars, while accidents start appearing for $N\geq 3$. In the following, we give some details on the $N=3$ model, commenting on $N>3$ at the end of the section.

Let us thus consider $G=\SU{3}\times\U{1}$ with a scalar field $\phi\sim{\bf 10}_1$. Its components can be alternatively written as a vector with one index in the ten-dimensional representation, $\phi^I$ with $I=1,\dots,10$. The scalar potential for a ${\bf 10}$ representation of $\SU{3}$ has been partially analysed previously, e.g.~to derive discrete flavour symmetries \cite{Luhn:2011ip} or to stabilise a scalar \ac{DM} candidate \cite{Frigerio:2022kyu}. In the latter paper, the existence of tree-level flat directions was pointed out. Here, we investigate in more detail their nature and implications. 

The most general renormalisable potential invariant under  $\SU{3}\times\U{1}$  includes, besides the mass term, only two algebraically independent quartic invariants, 
\be\label{eq:V10}
V=-\mu^2\,S+\frac12\left(\lambda\, S^2+\delta\, A^a A^a\right)\,,
\ee
where the singlet and adjoint bilinears are defined by
\be
S=\phi^\dag\phi\,,\qquad\qquad A^a=\phi^\dag T^a\phi\qquad(a=1,\ldots 8)\,.
\ee
Here $T^a$ are the $\SU{3}$ generators in the $\bf{10}$ representation, satisfying $\tr(T^a T^b) =15/2\, \delta^{ab}$. The $\SO{20}$ global symmetry of a free theory of 20 mass-degenerate real scalars is respected by the $\lambda$ term, however it can be checked that $\delta$ breaks it explicitly to $\SU{3}\times\U{1}$, with no larger continuous symmetry surviving. We take $\mu^2>0$ to realise \ac{SSB}, while 
boundedness from below requires $\lambda>0$ and $\delta > -\lambda/3$. It so happens that accidentally massless scalars arise only for positive values of $\delta$, hence we focus on $\delta>0$.

Since only the $A^a A^a$ term is sensitive to the direction of the \ac{VEV} in field space, the potential is minimized in a direction where $A^a A^a$ is minimal, i.e.~where $\vev{A^a} =0$. The \ac{VEV} can then be rescaled to satisfy $\vev S=\mu^2/\lambda\equiv v^2/2$, in order to obtain a minimum of the full potential. One such direction can readily be identified as the common null eigenvector of the two Cartan generators of $\SU{3}$, conventionally taken to be $T_3$ and $T_8$, so that $\SU{3}$ breaks to $\U{1}_3\times \U{1}_8$. We call the corresponding direction in field space  point (I), with $\vev\phi=v_{\rm (I)}/\sqrt{2}$. In terms of the  $\phi^{ijk}$ components, $G$ transformations can be used to bring it to the form $\vev{\phi^{(123)}}=|v|/\sqrt{2}$, with all other components \ac{VEV}s vanishing. Here, the parentheses in $\phi^{(123)}$ stand for weighted symmetrisation. 

At point (I), out of the 20 real scalar fields, only seven are massive and 13 are massless at the tree level, despite the fact that only seven generators are spontaneously broken. The remaining six massless scalar fields are accidents. The diagonal form of the scalar mass matrix is
\be
M^2_{\rm(I)}=\diag\left(m_\lambda^2,\,6\times m_\delta^2\,,13\times 0\right)\,,
\ee
where the masses of the radial mode and the other massive modes are given by $m_\lambda^2=\lambda\,v^2$ and $m_\delta^2=\delta\,v^2$, respectively. When $\SU{3}\times\U{1}$ is gauged with couplings $g_3$ and $g_1$, the gauge boson masses are $m_V^2= g_3^2\,v^2~(\times\, 6)$, $m_{V_3}^2=m_{V_8}^2=0$ and $m_B^2=g_1^2\,v^2$. It can be checked that there are no points preserving a larger continuous symmetry than $\U{1}_3\times \U{1}_8$ on the vacuum manifold, and that the symmetry-preserving point is unique up to $G$-transformations.

Away from the special point (I), at a generic minimum defined by $\vev{A^a}=0$ and $\vev S=v^2/2$, we find that $G$ is completely broken, with nine \ac{NGB}s, nine massive modes, and two accidentally massless modes. To explicitly construct the associated flat directions, consider a second point (II) in field space, $\vev\phi=v_{\rm (II)}$, defined by $\vev{\phi^{111}}=\vev{\phi^{222}}=\vev{\phi^{333}}=|v|/\sqrt{6}$ and all other \ac{VEV}s vanishing. The points (I) and (II) are not equivalent under a $G$-transformation, as evident from  the different tree-level mass spectrum:
\be
M^2_{\rm (II)}=\diag\left(m_\lambda^2,\,6\times \frac{m_\delta^2}{2},\,2\times \frac{3m_\delta^2}{2},\,11\times 0\right)~.
\ee
Likewise, the $\SU{3}$ gauge boson masses at point (II) are $m^2_{V}/2~(\times 6)$ and $3 m_{V}^2/2~(\times 2)$.

In fact, any superposition of the \ac{VEV} directions defining points (I) and (II) gives rise to a minimum of the potential with $\vev{\phi}=v_{\rm(I)}\,\cos\alpha+v_{\rm(II)}\sin\alpha$, where $\alpha\in[0,\pi/2]$.
Moreover, it can be checked that the vacuum energy does not depend on the complex phases of the four \ac{VEV}s of $\phi^{(123)}$ and $\phi^{iii}$. Three of them can be rotated away using the $\U{1}_3\times \U{1}_8\times \U{1}$ generators. However, the fourth phase $\beta$ is physical, and so one obtains a two-parameter family of physically inequivalent vacua:
\be
\vev{\phi^{111}}=\vev{\phi^{222}}=\vev{\phi^{333}}=\frac{|v|}{\sqrt{6}}\sin\alpha\,,\qquad\vev{\phi^{(123)}}=\frac{|v|}{\sqrt{2}}\,\cos\alpha\;e^{i\beta}\,,\qquad\text{all other \ac{VEV}s }=0\,.
\label{vacuum}\ee

To summarize, the manifold of tree-level degenerate vacua is eleven-dimensional. At generic points in this manifold, $G$ is completely broken, and nine of the flat directions are Goldstone directions. When $G$ is gauged, gauge-equivalent points are identified, and nine \ac{NGB}s are absorbed by the gauge bosons. There remains a two-dimensional manifold of gauge-inequivalent vacua, corresponding to two accidentally massless scalars, parameterised by two angles $\alpha$ and $\beta$. At point (I) the vacuum manifold degenerates, there is a residual symmetry $\U{1}_3\times \U{1}_8$, and four additional tree-level accidents appear. Two of them correspond to the would-be \ac{NGB}s of restored symmetries, while the other two are of different nature; they are not associated to any of the 11 flat directions, but arise only at the special point.

The non-vanishing scalar masses along the vacuum manifold can be written as the roots of a certain cubic polynomial; their explicit expressions are not very illuminating. Nevertheless, they obey a simple sum rule, as it is easily checked directly from \eq{eq:V10}: $\tr M^2=\left(\lambda+C_2\,\delta\right)v^2$, where $C_2=6$ is the quadratic Casimir. Similarly, the sum of gauge boson masses squared is constant and equal to $(C_2\,g_3^2+g_1^2)v^2$; see Sec.~\ref{potT}.

Let us focus on the special point (I), where $\alpha=0$ and where $\beta$ becomes the Goldstone direction of the spontaneously broken $\U{1}$. The six non-Goldstone flat directions correspond to the real and imaginary parts of $\phi^{111}$, $\phi^{222}$ and $\phi^{333}$. The fate of these accidents beyond the tree level can be determined by computing the one-loop effective potential. In analogy with our computation of Sec.~\ref{sec:veff}, one finds that subtracting the tadpole term induced for the radial part of $\phi^{(123)}$ renders the one-loop masses finite. All six accidents receive a positive and equal one-loop mass given by
\be
m_{\text{acc}}^2=\frac{3}{16 \pi^2}\left( \delta\,m_{\delta}^2+3g_3^2\, m_V^2\right)\,.
\ee
The flat directions are therefore lifted, and the symmetry-enhanced point (I) becomes an isolated minimum of the effective potential, i.e.~the physical vacuum of this model. Point (II), on the other hand, is found to be destabilised by loops and becomes a saddle point of the effective potential.

The mass degeneracy of the six accidents (as well as the other mass degeneracies in the spectrum) are due to discrete symmetries preserved after \ac{SSB}. In particular, $G$ contains an $S_3$ subgroup acting by permutation on the $\phi^{ijk}$ indices, left unbroken in the vacuum of \eq{vacuum}. Note the remnant symmetries $\U{1}_3\times \U{1}_8$ and $S_3$ do not commute. For example, the accident components $\phi^{iii}$ carry different charges, yet they form a triplet under permutations. 
In principle, such remnant gauge symmetries could be radiatively broken by fermion loops, which might destabilise the special point, in the spirit of Sec.~\ref{subsec:abelianhiggs}; in this case finding the new minimum requires a multi-field effective potential computation.

Finally, let us comment on the analogous models obtained by replacing $\SU{3}$ with $\SU{N}$. The structure of the most general renormalisable potential $V$ is the same as in \eq{eq:V10}, for any $N$. For $\mu^2,\lambda,\delta>0$, the dimension of the tree-level vacuum manifold rapidly grows, yielding $N(N-1)(N-2)/3$ non-Goldstone flat directions.
As for $N=3$, there can be special points on the tree-level vacuum manifold where a subgroup of $G$ remains unbroken, and the number of accidents is enhanced. A systematic classification of these models and their vacua is left for future work.

\section{Possible applications}\label{sec:apps}

Could accidentally light scalars play a role in real-world particle physics or cosmology? Let us present a few possible applications of phenomenological interest.

\subsection{Accident dark matter}

    The toy models presented above could actually describe a realistic dark sector with accidents playing the role of \ac{DM} candidates. Assume the new scalar field $\phi$ to be a \ac{SM} singlet, and the \ac{SM} sector to be neutral with respect to the dark-sector symmetry $G$. The two sectors can communicate (and thermalise) through a Higgs portal interaction, $\lambda_{H\phi}(H^\dag H)(\phi^\dag \phi)$ with $H$ the \ac{SM} Higgs doublet. Let us consider the minimal model, with dark gauge symmetry $\SU{2}\times\U{1}$ spontaneously broken to $\U{1}'$. Indeed, we have shown that such symmetry-enhanced vacuum is selected, once the accident flat direction is lifted by radiative corrections (at least in the absence of dark fermions).

This remnant symmetry guarantees that the lightest charged particle is stable (without the need to assume an ad-hoc global symmetry). This is naturally the complex scalar accident, since it is charged under the unbroken $\U{1}'$ and its mass is generated only at loop level.\footnote{Let us remark that, even when fermion loops select a different vacuum, with no continuous unbroken symmetry, there are typically remnant discrete symmetries which also guarantee the accident stability. For example, when the minimum occurs at the second special point with $\alpha=\pi/6$, one can show that the accident is odd under a residual $Z_2$ symmetry, which also explains the mass degeneracies observed in Fig.~\ref{fig:TreeMass}. Therefore, accidents can be good \ac{DM} candidates even in the absence of $\U{1}$ factors.}

The unbroken gauge symmetry at the special point guarantees the presence of massless dark photons, into which the accidents can annihilate. The \ac{DM} phenomenology in the case of annihilation into dark, massless gauge bosons is discussed, e.g., in Sec.~5.2.1 of Ref.~\cite{Frigerio:2022kyu}. In our case, the observed relic density can be reproduced when these annihilations freeze out, as long as $Q_D^4 \alpha^2_{D} \simeq 2.2 \times 10^{-10} (m_{ \rm  DM}/\hbox{GeV})^2$, where $\alpha_{D} = g_D^2/(4\pi)$ with $g_{D}$ the 
$\U{1}'$ dark gauge coupling and $Q_D=2$ the accident charge. Alternatively, \ac{DM} annihilation through the Higgs portal can dominate. In this case direct detection constraints require $m_{\rm DM}\gtrsim 2$-$3$~TeV or $m_{\rm DM}$ within a very narrow window around the resonance $m_{\rm DM}\simeq m_h/2$, see, e.g., Ref.~\cite{Frigerio:2022kyu}. The \ac{DM} direct detection, indirect detection and collider constraints all depend on the size of the Higgs portal coupling, which can thus be tested.

A massless dark photon contributes to the  extra-radiation parameter $\Delta N_{\rm eff}$, which is constrained both by \ac{BBN} and \ac{CMB}. Therefore, it needs to decouple from the early-universe thermal bath sufficiently early (so that its contribution is sufficiently diluted by the \ac{SM} reheating, from the decoupling of the various \ac{SM} particles). 
For a single dark photon, one finds that the decoupling temperature should be above a few hundreds of MeV, which implies a \ac{DM} mass above a few GeV \cite{Frigerio:2022kyu}. Future \ac{CMB} observatories are expected to have the ability to rule out or establish an extra radiation component at the level of a single dark photon. Another relevant constraint comes from galactic-scale structure formation: the so-called ellipticity constraint gives an upper bound on the strength of the dark-photon long-range force, $Q_D^2\alpha_D\lesssim 0.4 \sqrt{10^{-11}(m_{\rm DM}/\hbox{GeV})^3}$ \cite{Feng:2009mn,Agrawal:2016quu, McDaniel:2021kpq}, which combined with the relic density constraint implies a \ac{DM} mass above $\sim 100$~GeV in our case. 

If the Higgs portal coupling is tiny, the \ac{SM} and dark sectors do not thermalise with each other in the early Universe, but still each sector can thermalise individually. Such a scenario with  two thermal baths leads to a different, but perfectly viable, \ac{DM} phenomenology \cite{Chu:2011be,Hambye:2019dwd}. In particular if the dark sector has a temperature $T'$ smaller than the one of the visible sector $T$, the dark photon contribution to $\Delta N_{\rm eff}$ is suppressed by a factor of $(T'/T)^3$, and becomes irrelevant.

An analogous analysis holds for the $\SU{3}\times \U{1}$ model: the natural \ac{DM} candidate is the multiplet formed by the six degenerate accidents, charged under the remnant gauge symmetry $\U{1}_3\times \U{1}_8$, corresponding to two dark photons
(see Ref.~\cite{Frigerio:2022kyu} for quantitative constraints).

\subsection{Cosmology along the accident potential}

As well known, any inflationary potential must be extremely flat along the inflaton scalar field direction. The possibility to invoke a shift symmetry to protect the flatness of the inflationary potential has been considered extensively since decades. This constitutes the ``natural inflation'' scenario \cite{Adams:1992bn} (also appearing in axion setups). In these scenarios the inflaton is the \ac{NGB} of a spontaneously broken continuous symmetry. Inflation requires that the shift symmetry is slightly broken for the potential not to be totally flat, and the inflaton is therefore a \ac{pNGB}.

It is interesting to note that the $\SU{2}\times \U{1}$ potential we obtain along the accident direction resembles the natural inflation one: $V = \Lambda^4[1+ a\, \cos (\varphi/v)]$, with $\varphi$ the accident field. This potential is known to be in tension with Planck data \cite{Planck:2018jri}. To be in agreement with Planck data one possibility is to let the inflaton slow roll down this potential and to end inflation by an extra waterfall scalar field \cite{Brummer:2024ejc}. A dedicated study of hybrid inflation realised along an accidentally flat direction is presented in Ch.~\ref{chap:AccInf}.

Note also that, as the accident oscillates around the bottom of its potential, it triggers a burst of (massless dark photon) particle production whenever it crosses the enhanced-symmetry point \cite{Kofman:1997yn, Kofman:1994rk}. This possibility of producing dark vector bosons could have interesting consequences for cosmology, see, e.g., Ref.~\cite{Dror:2018pdh} in a somewhat different context.

We conclude by commenting on the possibility to have a first-order cosmological \ac{PT} along the accident direction. We showed in Sec.~\ref{subsec:abelianhiggs} that the accident effective potential can develop a minimum away from the $\U{1}'$-preserving point at $\alpha=0$, so that $\U{1}'$ is spontaneously broken.
At sufficiently large temperatures, thermal corrections will dominate the effective potential, and tend to restore the symmetry with a minimum at $\alpha=0$. As the Universe expands, the temperature $T$ decreases, and the potential develops a second minimum which may be separated by a barrier from the one at the origin. This is typically the case if the tree-level potential at zero temperature is flat, and the flat directions are lifted radiatively by the one-loop effective potential. 
Then the \ac{PT} can be of the first order, proceeding via tunneling (this ends a period of supercooling, during which the scalar field lies in the false minimum and dominates the Universe energy density, see ,e.g., Refs.~\cite{Iso:2017uuu,vonHarling:2017yew,Hambye:2018qjv}).
As $T$ further decreases, the origin becomes a local maximum and, at $T=0$, we recover the potential shown in Fig.~\ref{fig:Veff}. \ac{FOPT}s have an interesting phenomenology, as mentioned in Sec.~\ref{sec:FOPTIntro}.

\subsection{The Higgs as an accident}

The only elementary scalar field in the \ac{SM} is the Higgs boson, which at $125$ GeV is much lighter than the scale of \ac{NP}. The absence of \ac{NP} up to the (multi-)TeV scale constitutes the ``little hierarchy problem''. Ultraviolet embeddings of the \ac{SM} predicting a loop-suppressed Higgs mass are therefore appealing, and have been widely explored. The accident mechanism may lead to an additional class of such models.

To address the little hierarchy problem with accidents, one would need to identify a model where, at some  special point in the tree-level vacuum manifold, the remnant symmetry contains the electroweak symmetry $\SU{2}_w\times \U{1}_Y$ (or even the entire custodial symmetry of the \ac{SM} Higgs potential), and the accidentally light scalars transform as an electroweak doublet. We have shown in Sec.~\ref{subsec:abelianhiggs} how radiative corrections may give a \ac{VEV} to the accidents, thus breaking the remnant symmetry at a scale parametrically smaller than the initial scale of \ac{SSB}. The accident models which we have identified so far feature only Abelian $\U{1}^n$ remnant symmetries, but this is likely a consequence of the simplest possible choices for the symmetry $G$ and for the representation of $\phi$. 

The phenomenology of an accidental Higgs would resemble that of composite \ac{pNGB} Higgs models (see Refs.~\cite{Bellazzini:2014yua,Panico:2015jxa} for reviews) or little-Higgs models (see Ref.~\cite{Schmaltz:2005ky} for a review). In the little-Higgs scenario, the Higgs mass is loop-suppressed because it is protected by a large global symmetry, explicitly broken only by the product of at least two independent couplings, typically gauge or Yukawa couplings. In the accident scenario, an enlarged symmetry is broken by a single scalar quartic coupling, nonetheless the accident mass is loop-suppressed thanks to the restrictive structure of the scalar potential. In composite Higgs models, it is typically assumed that the Goldstone-Higgs shift symmetry is exact within the composite sector, and broken only by external gauge and Yukawa couplings.

In our toy models, the effective theory of accidents has a simple ultraviolet completion, in terms of a weakly-coupled and renormalisable theory of an elementary scalar $\phi$ in a large representation of $G$. Notice that, in contrast with models where the Higgs is a \ac{NGB}, the full symmetry $G$ can be gauged, with no ad-hoc global symmetries assumed: the \ac{NGB}s are absorbed by the gauge bosons, while the Abelian Higgs is a tree-level massless accident. On the other hand, to address the ``big hierarchy problem'' associated with models of elementary scalars, it would be interesting to realise composite accident models, where $\phi$ emerges as a composite multiplet.

\section{Conclusions}

Given a generic renormalisable scalar potential with symmetry $G$ spontaneously broken to $H$, accidents are scalar fields which do not receive a tree-level mass, although they do not belong to the $G/H$ coset. There is no obvious way to infer from symmetry selection rules that the accident masses are suppressed.

We demonstrated that accidents appear in theories with a scalar multiplet in a large representation of $G$. This can be ascribed to the restrictive structure of the most general renormalisable $G$-invariant potential. Already in the minimal models -- the five-plet of $\SU{2}$ and ten-plet of $\SU{3}$ -- the vacuum manifold is non-trivial. It would be valuable to conduct a systematic analysis of the possible field-space geometries leading to accidents.

Accidents possess unsuppressed, tree-level, non-derivative couplings to other scalars. As one moves along the accidental tree-level flat directions, the tree-level mass spectrum of other scalars (and of vectors and fermions which obtain their masses from scalar \ac{VEV}s) changes, but the sum of the masses squared remains constant.  

When including loop corrections, the flatness of the potential is lifted and an isolated minimum appears, where a non-trivial symmetry $H$ remains unbroken and the number of accidents is enhanced. These models exhibit a one-loop hierarchy of scales, even if all dimensionless couplings are of the same order. It is an open question whether less minimal models exist where an accident mass arises only at two-loop or higher order.

An accidentally light Higgs may help to address the little hierarchy problem.
We built a toy model where an accident plays the role of Abelian Higgs with a small \ac{VEV}, motivating the quest for a less minimal model which could feature a $\SU{2}$ doublet of accidents.

The rolling of accidents along their tree-level flat directions provides a playground for natural inflation and/or resonant particle production. Accidentally flat directions lifted by loop corrections may also lead to cosmological \ac{FOPT}s. Finally, dark-sector accidents are excellent candidates for \ac{DM}, as they are naturally the lightest states charged under unbroken dark-sector symmetries.

\chapter{Accidental inflation}
\label{chap:AccInf}
\minitoc

We construct a hybrid-inflation model where the inflaton potential is  generated radiatively, as gauge symmetries guarantee it to be accidentally flat at tree level. The model can be regarded as a small-field version of \ac{NI}, with inflation ending when the mass of a second scalar, the waterfall field, turns tachyonic. 
This provides a minimal, robust realization of hybrid inflation, which
predicts specific correlations among \ac{CMB} observables.
Tachyonic preheating leads to the production of \ac{GW}s which,
for a low inflationary scale, might be detected by upcoming experiments.  Simple variations of the model can give rise to topological defects, such as unstable \ac{DW}s. 
Their dynamics produces a \ac{SGWB}, which
can be compatible with the recent detection  by \ac{PTA}. 

\section{Introduction}
Field-theoretical models of cosmic inflation can be roughly divided into two classes. In large-field models, the field excursion of the inflaton is transplanckian, while in small-field models it remains below the Planck scale. When allowing for large field excursions, very simple shapes of the potential can give rise to inflation (although many are now excluded by increasingly accurate \ac{CMB} measurements \cite{Planck:2018vyg, Planck:2018jri}). However, it is generally dubious if large-field models can be trustable  \ac{EFT}s, without being embedded in a complete quantum gravity framework. Small-field models, on the other hand, generally require more involved field-theoretic constructions. In particular, some mechanism should be included to protect the inflaton potential from radiative corrections. Once this is achieved, effects from transplanckian physics can be argued to be under control in the  \ac{EFT}. The only necessary fine-tuning left is that of the cosmological constant after inflation, and potentially of the initial conditions leading to an inflationary phase.

In small-field models, the flatness of the inflaton potential could be ensured, for example, by (spontaneously broken) supersymmetry, or by an (approximate) shift symmetry if the inflaton is a \ac{pNGB}. In the present chapter, we propose to employ a novel mechanism to this end: The inflaton is an accidentally massless scalar field, i.e., a scalar whose potential is constrained by symmetry to be flat up to loop corrections and higher-dimensional operators \cite{Brummer:2023znr}. Similarly to the \ac{pNGB} case, a small slope is generated radiatively. 

At first sight, our inflaton potential is very similar to that of \ac{NI}, where the inflaton is realized as a $\U{1}$ \ac{pNGB} \cite{Freese:1990rb}. Minimal \ac{NI}, however, is a large-field model, and, moreover, is excluded by Planck data \cite{Planck:2018jri}. Small-field \ac{NI} needs additional fields responsible for ending inflation, and such models of ``hybrid \ac{NI}'' turn out to be difficult to construct, since the couplings between the additional fields and the inflaton should not spoil the flatness of the potential (although this can be achieved with suitable discrete symmetries \cite{Ross:2009hg, Ross:2010fg}).\footnote{Alternatively, viable small-field \ac{NI} can also be realised in metric-affine models of gravity \cite{Racioppi:2024zva}.}

Our model, by contrast, is simple and economical; in the minimal case, the symmetry is simply a gauged $\SO{3}$ times a $\mathbb{Z}_2$ parity. Indeed, models with accidentally light scalars \cite{Brummer:2023znr} lend themselves particularly well to hybrid inflation model-building. This is because they feature an almost-flat inflaton direction by construction, yet the other scalar fields couple to the inflaton at the tree level, which allows inflation to naturally end by a second field becoming tachyonic.

We will compute the \ac{CMB} observables as a function of the model parameters, and study their correlations. We will also present naturalness arguments to identify the preferred region of parameters, and derive the corresponding predictions.

While naturalness would prefer an inflationary
scale not too far from the Planck scale, 
lower-scale scenarios are more interesting for the potential observation of \ac{GW} signals.
In hybrid models such as ours, inflation ends because a steep tachyonic ``waterfall'' direction appears in field space \cite{Linde:1993cn}, which leads to a transition from the inflationary slow-roll regime to a phase of tachyonic preheating \cite{Felder:2000hj, Felder:2001kt}. In this phase, \ac{GW}s can be produced, and may be within the reach of future detectors, if the inflationary scale is low.
In addition, \ac{SSB} in  the inflaton-waterfall potential may lead to the production of topological defects, which also source \ac{GW}s.
We will present slightly extended versions of the minimal model, which produce \ac{CS}s or \ac{DW}s after the end of inflation. These scenarios are phenomenologically interesting only if the inflationary scale is sufficiently low, as, in this case, the annihilations of topological defects produce \ac{GW}s which can be within 
experimental reach.

Cosmological \ac{GW} sources can potentially explain the recent observation of a stochastic background by the \ac{PTA} collaborations NANOGrav \cite{NANOGrav:2023gor}, EPTA and InPTA \cite{EPTA:2023fyk}, PPTA \cite{Reardon:2023gzh}, and CPTA \cite{Xu:2023wog}.  
We will compute the various \ac{GW} signals in our scenario, study their correlations with the \ac{CMB} observables, and compare them with present constraints from  \ac{LVK} \cite{KAGRA:2021kbb} and  NANOGrav \cite{NANOGrav:2023hvm}, as well as with the expected sensitivity of future \ac{GW} detectors.

Both large-field and small-field models are the subject of several swampland conjectures, which aim to characterize the set of  \ac{EFT}s that are supposedly incompatible with general principles of quantum gravity; see Ref.~\cite{Palti:2019pca} for a review. Specifically, models with transplanckian field excursions are concerned by the Distance Conjecture \cite{Ooguri:2006in}. This conjecture states roughly that, at large distances in field space, an infinite tower of fields becomes light, invalidating the  \ac{EFT}. On the other hand, slow-roll potentials with subplanckian field excursions and positive vacuum energy are claimed to be ruled out by various (arguably more speculative) de Sitter conjectures \cite{Obied:2018sgi,Garg:2018reu,Andriot:2018wzk}. While these conjectures are certainly interesting, none of them is firmly established. For this chapter, we, therefore, assume that single-field slow-roll inflation is not per se in the swampland, and that small-field models, in particular, can be viable  \ac{EFT}s of inflation.

\section{Hybrid inflation with an accidentally light scalar}\label{sec:AccInf}

Models with accidentally massless scalar fields were recently constructed in Ref.~\cite{Brummer:2023znr}, where the focus was on models with a single scalar in an irreducible representation and no ad-hoc discrete symmetries. To construct a model of hybrid inflation, however, we need scalar fields in at least two different multiplets, one of which will contain the inflaton and the other the waterfall field. While our model will share many features with the ones presented in Ref.~\cite{Brummer:2023znr}, the details of its field content and symmetry structure differ. 

The model is defined by a $G=\SO{3}$ symmetry with a real scalar $\phi$ transforming in the ${\bf 5}$ of $\SO{3}$, and a real scalar $\chi$ transforming in the ${\bf 3}$.
We also impose a $\mathbb{Z}_2$ symmetry under which $\phi$ is odd. The most general renormalizable potential is
\be\label{V5}
 V=-\frac{1}{2}\mu_\phi^2\phi^2-\frac{1}{2}\mu_\chi^2\chi^2+\frac{\lambda_\phi}{4}(\phi^2)^2+\frac{\lambda_\chi}{4}(\chi^2)^2+\frac{\varepsilon}{4}\phi^2\chi^2+
 \frac{\zeta}{4}\,T^a_{AC} T^b_{CB}\phi_{A} \phi_B \chi^{a}\chi^b\,.
\ee
Here, $T^a_{BC}$ are the $\SO{3}$ generators in the ${\bf 5}$-plet representation ($a=1,2,3$ and $A,B,C=1\ldots 5$). We refer the interested reader to App.~\ref{sec:AppendixDetails} for a detailed derivation of Eq.~\eqref{V5}, and a comprehensive analysis of the vacuum structure of this potential.

The continuous symmetry of $V$ is strictly $G=\SO{3}$: There is no larger custodial-type symmetry present. More precisely, the symmetry of the kinetic terms for $\phi_A$ and $\chi_a$ is $\SO{8}$, which is explicitly broken to $\SO{5}_\phi\times\SO{3}_\chi$ by the mass terms and the quartic couplings $\lambda_{\phi,\chi}$ and $\epsilon$. It is non-trivial to show that a single remaining coupling, $\zeta$, breaks
explicitly ten generators of this group, reducing the symmetry to the ``diagonal'' $\SO{3}$, which acts on both $\phi$ and $\chi$. The absence of a larger symmetry implies that, after \ac{SSB}, there will be at most three \ac{NGB}s; therefore, the appearance of additional, accidentally massless scalars is a non-trivial phenomenon.

It is instructive to consider first the case $\mu_\phi^2>0$ and $\mu_\chi^2<0$ with all quartic couplings positive. 
This parameter choice does not lead to hybrid inflation, but it is useful in order to understand how a slow-roll direction emerges. The potential is minimized at $\vev{\chi}=0$ and $|\vev{\phi}|=v$, where $v^2=\mu_\phi^2/\lambda_\phi$. 
Therefore, the vacuum manifold is a four-sphere, with $G$ completely broken at generic points. Three of the four flat directions are Nambu-Goldstone directions corresponding to the broken $G$ generators. If $G$ is gauged, the associated \ac{NGB}s are absorbed by the gauge bosons. The fourth flat direction corresponds to an accidentally (tree level-) massless scalar field. It is not associated to any symmetry of the scalar potential and as such is neither an \ac{NGB} nor, strictly speaking, a \ac{pNGB}. The flat direction will be lifted by loop corrections through both the scalar quartic coupling  $\zeta$ and the gauge coupling $g$. In the limit of small $\zeta$, it can be understood as a genuine \ac{pNGB} direction, as the scalar potential acquires a larger symmetry for $\zeta\to 0$.

At a special point along the flat direction, corresponding to a null eigenvector of one of the $G$ generators, an O$(2)=\SO{2}\rtimes\mathbb{Z}_2$ subgroup of $G$ is preserved, where the symbol $\rtimes$ denotes a semidirect product. At this point, there are only two \ac{NGB}s, while not one but two scalars remain accidentally massless at the tree level; the second one is the scalar which is absorbed by the $\SO{2}$ gauge boson outside the special point. This point turns out to be a minimum of the one-loop effective potential.

For concreteness, consider an explicit parametrization of the tree-level flat direction by writing the isospin-2 field $\phi$ as a $3\times 3$ traceless symmetric matrix $\Phi=\phi_A \lambda_A$. Here, $\lambda_A$ are the the symmetric Gell-Mann matrices $\lambda _{1,3,4,6,8}$. By an $\SO{3}$ transformation, $\Phi$ can be diagonalized, such that its  \ac{VEV} becomes
\be\label{eq:alpha}
\langle\Phi\rangle=v\left(\lambda _3\sin\alpha+\lambda _8\cos\alpha\right)\,,\qquad v^2=\mu_\phi^2/\lambda_\phi\,.
\ee
The mass of the radial mode in $\Phi$ is
\be
m_\rho^2=2\lambda_\phi v^2
\ee
and the $\chi$ masses are
\be\label{eq:mchi}
m_{\chi^{1,2}}^2=-\tilde\mu_\chi^2+2\,v^2\zeta\,\sin^2\left(\alpha\pm\frac{\pi}{3}\right)\,,\qquad
m_{\chi^3}^2=-\tilde\mu_\chi^2+2\,\zeta v^2\sin^2\alpha\,,
\ee
where we have defined
\be\label{eq:muchieff}
\tilde\mu_\chi^2=\mu_\chi^2-\frac{\varepsilon}{2}v^2\,.
\ee

Since, for the time being, we chose $\mu_\chi^2<0$ and $\epsilon$ and $\zeta$ positive, we have $\tilde\mu_\chi^2<0$ and so all masses are positive. The $\alpha$-dependence of the masses is a crucial feature of this model which will eventually allow us to realize hybrid inflation. 
As $G$ is gauged, the gauge bosons are also massive at the tree level.
The vacuum manifold of physically inequivalent points is parametrized by $\alpha\in[0,\,\pi/6]$, since points related by $\alpha\into -\alpha$ and $\alpha\into\alpha+\pi/3$ are identified under $G$. At $\alpha=0$, there is a symmetry-enhanced point with a massless $\SO{2}$ gauge field and one additional massless scalar field (which is otherwise absorbed by the third massive gauge boson).

Let us now consider the effect of one-loop corrections to the potential. The procedure to calculate the one-loop effective potential is detailed in App.~\ref{sec:AppendixDetails}. 
When including one-loop corrections, the vacuum degeneracy is lifted, the physical vacuum is at the point $\alpha=0$, and the two tree-level-massless scalars become massive. The only massless state is the gauge boson of the unbroken $\SO{2}$. However, there remains a one-loop mass hierarchy between the six heavy scalar modes having obtained their masses at the tree level, and the two accidentally light ones with one-loop masses. 
We emphasize that the latter are calculable and finite as functions of $\mu_\phi^2$ and $\mu_\chi^2$, and, therefore, they are quadratically sensitive to quantum corrections at the cutoff scale only at two loops (since contributions to the mass parameters themselves are quadratically divergent at one loop), much as in Little Higgs models (for a review of the latter, see, e.g., Ref.~\cite{Schmaltz:2005ky}).

Focusing on the one-loop effective potential along the tree-level flat direction, one finds that it is given by
\footnote{See App.~\ref{sec:AppendixDetails}
for the renormalisation prescriptions
employed to determine $V_{\rm eff}$,
and for the proof that it
can be expanded as  $\sum_n c_n \cos(na/f)$.}
\be\label{eq:CosPot}
V_{\rm eff}(a)=V_0-M^4\cos\frac{a}{f}+(\text{higher harmonics})~,
\ee
where $a=\alpha v$ is the canonically normalized ``accident'' field parametrizing the tree-level flat direction, $f=v/6$, and we have arbitrarily renormalized the vacuum energy to be $V_0-M^4$.
Moreover, we find
\be\label{eq:M4}
M^4 =\frac{1}{160\,\pi^2}\Bigg[
9\,g^4\,v^4
+2\,\frac{\tilde\mu_\chi^{10}}{\zeta^3\,v^6}\Bigg(F\left(\frac{\zeta v^2}{\tilde\mu_\chi^2}\right)
-T_{F,5}\left(\frac{\zeta v^2}{\tilde\mu_\chi^2}\right)\Bigg)\Bigg]~,
\ee
where $g$ is the $\SO{3}$ gauge coupling, $F(x)=(1-2\,x)^{5/2}$, and $T_{F,5}(x)=1-5\,x+\frac{15}{2}x^2-\frac{5}{2}x^3-\frac{5}{8}x^4-\frac{3}{8}x^5$ is the Taylor polynomial of $F(x)$ of degree 5.
Higher harmonics in the effective potential are suppressed relative to the first harmonic by small numerical factors.\footnote{For example, the coefficient of the second harmonic $\cos(2a/f)$ induced by $\zeta$ can be shown to always be smaller than its counterpart of Eq.~\eqref{eq:M4} by a factor $\gtrsim 10$ in the region of field space of interest. Likewise, the second harmonic induced by $g$ is always suppressed with respect to the first.} They will be neglected in the following.

The shape of the one-loop potential, therefore, closely resembles what one would obtain if $a$ were the \ac{pNGB} of an approximate global $\U{1}$ symmetry. Identifying $a$ with the inflaton and setting $V_0=M^4$, one obtains a realization of the famous \ac{NI} model \cite{Freese:1990rb}. However, vanilla \ac{NI} is excluded by Planck data \cite{Planck:2018jri}. Simple modifications of the minimal \ac{NI} model, such as a non-minimal coupling to gravity \cite{Ferreira:2018nav, Reyimuaji:2020goi,Salvio:2023cry} or allowing for some exotic phase of reheating \cite{Stein:2021uge}, can improve the fit, but in the preferred parameter region one finds $f\gtrsim \MP$.\footnote{We designate by $\MP$ the reduced Planck mass: $\MP \equiv (8\pi \GN)^{-1/2} = 2.4\times10^{18}\,{\rm GeV}$.} As already discussed, transplanckian field excursions, apart from being constrained by the Swampland Distance Conjecture, imply that the effects of higher-dimensional operators are not under control (although models have been constructed which address this issue \cite{Arkani-Hamed:2003xts,Kaplan:2003aj}).
To overcome these problems, it is desirable to identify a variant of the model where the field excursion is sub-Planckian, i.e., a small-field model of inflation.

In a small-field version of \ac{NI}, the inflationary potential should be dominated by the constant term $V_0$ in the vacuum energy. This implies that slow-roll is possible for $f<\MP$, but it also implies that slow-roll never ends on its own, because the slow-roll parameters never grow to become ${\cal O}(1)$, and that the vacuum energy density $V_0$ somehow needs to be converted into radiation eventually. To this end, one may extend the model into one of hybrid inflation \cite{Linde:1993cn}: add a second scalar field, the waterfall field, which is heavy during inflation but becomes tachyonic for some value of the inflaton, triggering the end of inflation and the start of preheating. 

Models of hybrid inflation with \ac{pNGB}s as inflatons were constructed, for example, in Refs.~\cite{Garcia-Bellido:1996mdl,Kaplan:2003aj}. In these models, the approximate symmetry protecting the inflaton is explicitly broken by the couplings to the waterfall field. These must be chosen ad hoc such as to preserve the flatness of the potential. Related models were subsequently built \cite{Ross:2009hg, Ross:2010fg}, where it was shown that the required structure of the potential can be protected by non-Abelian discrete symmetries. By contrast, we will now show that our model with an accidentally light scalar can furnish a natural model of hybrid inflation with the simple potential of Eq.~\eqref{V5}.

So far, the role of the field $\chi$ was solely to provide a source for loop corrections to the effective potential which lift the flatness of the inflaton potential. We will now identify $\chi$ with the waterfall field by choosing the model parameters such that the true vacuum of the model is no longer at $\vev{\chi}=0$ and $\vev{\phi}\neq 0$ but rather at $\vev{\phi}=0$ and $\vev{\chi}\neq 0$. Our previous discussion of the vacuum structure applies as long as the effective $\chi$ masses, given in Eq.~\eqref{eq:mchi}, are all positive, regardless of the sign of the mass parameter $\tilde\mu_\chi^2$ in Eq.~\eqref{eq:muchieff}. Even for $\tilde\mu_\chi^2>0$, the effective potential is still given by Eqs.~\eqref{eq:CosPot} and \eqref{eq:M4} upon replacing $M^4$ by its real part. It is only when a tree-level mass becomes tachyonic that our computation of $V_{\rm eff}$ can no longer be trusted. 

Now choosing parameters such that $\tilde\mu_\chi^2$ satisfies
\be\label{eq:mubound}
0 < \tilde\mu_\chi^2 <\frac{\zeta}{2} v^2\,,
\ee
one observes that $m^2_{\chi^3}$ is positive for $\alpha=\pi/6$, negative for $\alpha=0$, and crosses zero at
\be\label{eq:alphac}
\alpha_c=\arcsin\sqrt{\frac{\tilde\mu_\chi^2}{2\zeta\,v^2}}\,,
\ee
while all other masses remain strictly positive for $\alpha\geq\alpha_c$.
This can be achieved by choosing $\mu_\chi^2>0$ and suitable (positive) values for the quartics. 

\begin{figure*}[bt]
\begin{center}
      \includegraphics[width=13 cm]{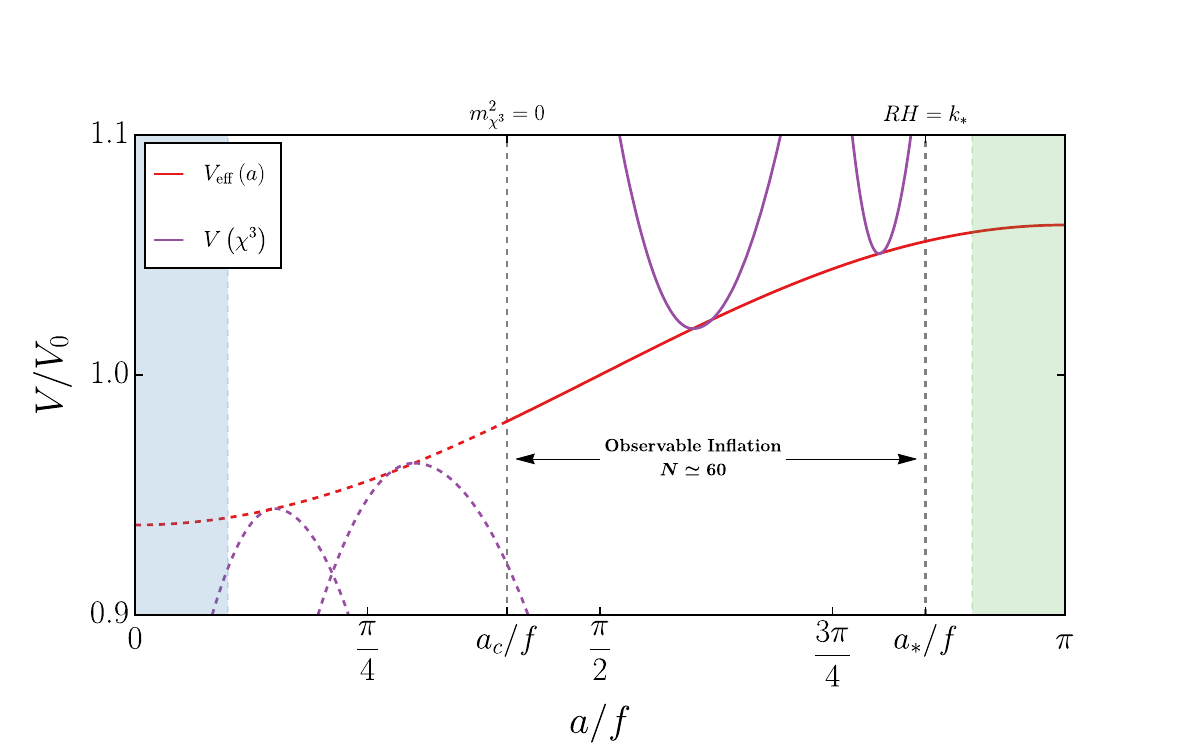}
    \caption{Pictorial representation of the inflationary potential, normalised to the inflationary scale $V_0$. Close to $a = \pi f$, where inflation begins, the potential along the $\chi^3$ direction (purple lines) has positive and large curvature, stabilizing the inflationary trajectory (red line). At the $a=a_*$ the pivot scale crosses the Hubble horizon, $R H = k_*$ (with $R$ the scale factor). As the inflaton slow-rolls down its potential, $m_{\chi^3}^2$ decreases until it becomes $0$ at the critical point $a_c$, where inflation ends. For $a<a_c$, the effective potential cannot be trusted, as the accidentally flat direction is no longer a minimum of the tree-level potential. A value of $a_*/f$ ($a_c/f$) in the green (blue) region would correspond to a tuning $\gtrsim 10\%$ on the initial (final) point of observable inflation, and, therefore, it is theoretically disfavoured.}
\label{fig:InflaSketch} 
\end{center}
\end{figure*}

In this regime, the inflaton $a$ starts rolling close to $\alpha=\pi/6$ or, equivalently, $a=\pi f$, where all scalars except $a$ have positive tree-level masses. In particular, $\chi$ is stabilized at $\chi=0$. For $\alpha>\alpha_c$, the one-loop effective potential Eq.~\eqref{eq:CosPot} can be trusted, and provides a small slope which the inflaton rolls down. As the inflaton crosses the point $\alpha_c$, the tree-level vacuum will no longer be given by $\chi=0$ and Eq.~\eqref{eq:alpha}, since a tree-level tachyon develops in the $\chi^3$ direction. Slow-roll ends as the tachyonic $\chi$ mass grows large, and $\chi$ modes are being excited. Eventually, the potential energy of the inflaton is converted into energy in the $\chi$ sector. In Fig.~\ref{fig:InflaSketch}, we provide a sketch of the potential along the $a$ (inflaton) and $\chi^3$ (waterfall) directions, for the sake of illustration.
Reheating should then proceed via some coupling, such as a Higgs portal coupling, between $\chi$ and the \ac{SM}; in this chapter, we will not explore reheating in detail.
It is easily checked that, for suitable values of the couplings, the true vacuum of the model is located at
\be
\chi^3=\sqrt{\frac{\mu^2_\chi}{\lambda_\chi}} \equiv v_\chi\,,\qquad
\text{all other fields} = 0
\ee
up to $G$-transformations. The residual symmetry is $\SO{2}\subset G$, and all fields except for the unbroken gauge boson pick up tree-level masses. Since the  true vacuum should be Minkowski up to a tiny cosmological constant, we can identify
\be\label{eq:V0}
V_0\simeq\frac{\mu_\chi^4}{4\,\lambda_\chi}-\frac{\mu_\phi^4}{4\,\lambda_\phi}\,,
\ee
which is the (tree-level) potential energy difference between the inflationary field configuration and the true vacuum. In the following, we will assume that the  second term in Eq.~\eqref{eq:V0}  is not only subdominant, but even negligible.

\section{Inflation parameters and CMB observables}\label{sec:CMB}

As inflation proceeds along the tree-level flat direction, the waterfall field $\chi$ is stabilized at $\chi=0$ by its effective mass; hence, the model qualifies as a single-field model. The evolution of scalar and tensor perturbations is then dictated by the potential in Eq.~\eqref{eq:CosPot}, with the slow-roll parameters defined by
\be
    \epsilon \equiv \frac{\MP^2}{2}\left(\frac{V_{\rm eff}'}{V_{\rm eff}}\right)^2\,,\qquad \eta \equiv \MP^2 \frac{V_{\rm eff}''}{V_{\rm eff}}\,,\qquad \xi \equiv \MP^4\frac{V_{\rm eff}'V_{\rm eff}'''}{V_{\rm eff}^2}\,,\qquad \sigma \equiv \MP^6 \frac{\left(V'_{\rm eff}\right)^2 V''''_{\rm eff}}{V_{\rm eff}^3}\,,
\ee
where primes denote derivatives with respect to the inflaton field $a$. 

At the leading order in the slow-roll parameters, the tensor-to-scalar ratio $r$, the amplitude of the curvature power spectrum $A_s$, the spectral index of the curvature power spectrum $n_s$, the running of the spectral index $n_r$, and the running of the running $n_{rr}$ can be computed, respectively, as
\be
\begin{aligned}
    &r = 16 \epsilon_*\,, \qquad A_s=\frac{H_*^2}{8\pi^2\MP^2}\frac{1}{\epsilon_*}\,, \qquad  n_s = 1 + 2\eta_* - 6\epsilon_*\,,\\ 
    &n_r = 16\epsilon_*\eta_*-24\epsilon_*^2-2\xi_*\,,
    \qquad n_{rr} = -192 \epsilon_*^3 + 192\epsilon_*^2\eta_* - 32\epsilon_*\eta_*^2 - 24\epsilon_*\xi_* + 2\eta_*\xi_* + 2\sigma_*\,,
\end{aligned}
\ee
where the asterisk means that all the quantities are evaluated at the moment when the pivot scale, $k_* = 0.05\;\rm{Mpc^{-1}}$, exits the Hubble horizon. Some of the above relations are derived in Sec.~\ref{sec:Infla}, where an introduction to cosmic inflation is also given. These quantities are constrained by \ac{CMB} observations. The most recent constraints on the scalar power spectrum are given by the Planck Collaboration \cite{Planck:2018vyg}: $\log \left(10^{10}A_s\right) = 3.047\,\pm\,0.014$, $n_s = 0.9647\,\pm\,0.0043$, $n_r = 0.0011\,\pm\,0.0099$, and $n_{rr} = 0.009\,\pm\,0.012$. The tensor-to-scalar ratio is constrained to be $r < 0.036$ by the BICEP/Keck Collaboration \cite{BICEP:2021xfz}.

The independent model parameters entering the calculation of the \ac{CMB} power spectrum can be taken as $V_0$, $M$, $f$, and $a_*$.
Considering that $V_0 \gg M^4$, the observables are
\be
\begin{aligned}\label{eq:CMBObs}
  & r \simeq 8\,\left( \frac{\MP}{f}\frac{M^4}{V_0}\right)^2\sin^2\frac{a_*}{f}\,,\qquad && A_s \simeq\frac{1}{12\pi^2} \frac{f^2\,V_0^3}{\MP^6\, M^8}\sin^{-2}\frac{a_*}{f}\,,\qquad  n_s \simeq 1+2\,\frac{\MP^2}{f^2}\frac{M^4}{V_0}\cos \frac{a_*}{f}\,,\\ & n_r \simeq 2\,\left(\frac{\MP^2}{f^2}\frac{M^4}{V_0}\right)^2\sin^2\frac{a_*}{f}\,,\qquad &&n_{rr} \simeq -4\,\left(\frac{\MP}{f}\right)^6\left(\frac{M^4}{V_0}\right)^3\sin^2\frac{a_*}{f}\cos\frac{a_*}{f}\,,
\end{aligned}
\ee
while the number of $e$-folds of observable inflation is given by
\be\label{eq:efolds}
    N\simeq \frac{1}{\MP^2}\int_{a_{\rm end}}^{a_*}\frac{V_{\rm{eff}}}{V'_{\rm{eff}}} da\simeq \frac{f^2}{\MP^2} \frac{V_0}{M^4}\left(\log\tan\frac{a_*}{2f}-\log\tan\frac{a_{\rm end}}{2f}\right)\,.
\ee
We identify the field value $a_{\rm end}$ where inflation ends with the critical point $a_c = 6 f \alpha_c$, where the waterfall field becomes tachyonic; see Eq.~(\ref{eq:alphac}).
We will comment on the assumptions underlying this identification at the end of this section. As the end of inflation is controlled by the coupling between the inflaton and the waterfall field, $a_{\rm end}$ is an independent parameter of our model, which can be freely adjusted: By suitably choosing $a_{\rm end}$, we are always able to accommodate enough $e$-folds of inflation. For concreteness, in what follows, we fix $N=60$ as a reference value.

A characteristic feature of the model is to predict positive $n_r$ and $n_{rr}$. A second feature concerns the running of the running of the spectral index, which for this model is not independent from $n_s$ and $n_r$, $n_{rr} \simeq (1 - n_s)n_r$, leading to the constraint $n_{rr} < 4 \times 10^{-4}$. This is more stringent than the direct experimental bound on $n_{rr}$. The upcoming satellite mission SPHEREx \cite{SPHEREx:2014bgr} will measure the running of the spectral index with a target sensitivity $\Delta n_r \sim 10^{-3}$, and, therefore, it could rule out the model in the near future.

The four parameters of the model are not all determined by the \ac{CMB} observables. Requiring that the model parameters do not suffer from fine-tuning problems will lead to further relations between them; see Sec.~\ref{sec:Naturalness}. To better understand how observation constrains them, we invert the relations given in Eq.~\eqref{eq:CMBObs} and obtain
\be
\begin{aligned}
    &V_0 = \frac{3\pi^2}{2}A_s r \MP^4\,,\qquad &&M^4 = \frac{3\pi^2}{16}\frac{A_s}{n_r}\sqrt{2 n_r+\left(n_s-1\right)^2}r^2 \MP^4\,,\\
    &f = \frac{1}{2}\sqrt{\frac{r}{n_r}} \MP\,,\qquad &&\frac{a_*}{f} = \arccos\left[\frac{n_s-1}{\sqrt{2n_r+(n_s-1)^2}}\right]\,.
\end{aligned}\label{eq:Par}
\ee

\begin{figure}[h!]
\begin{center}
\includegraphics[width=0.48 \textwidth]{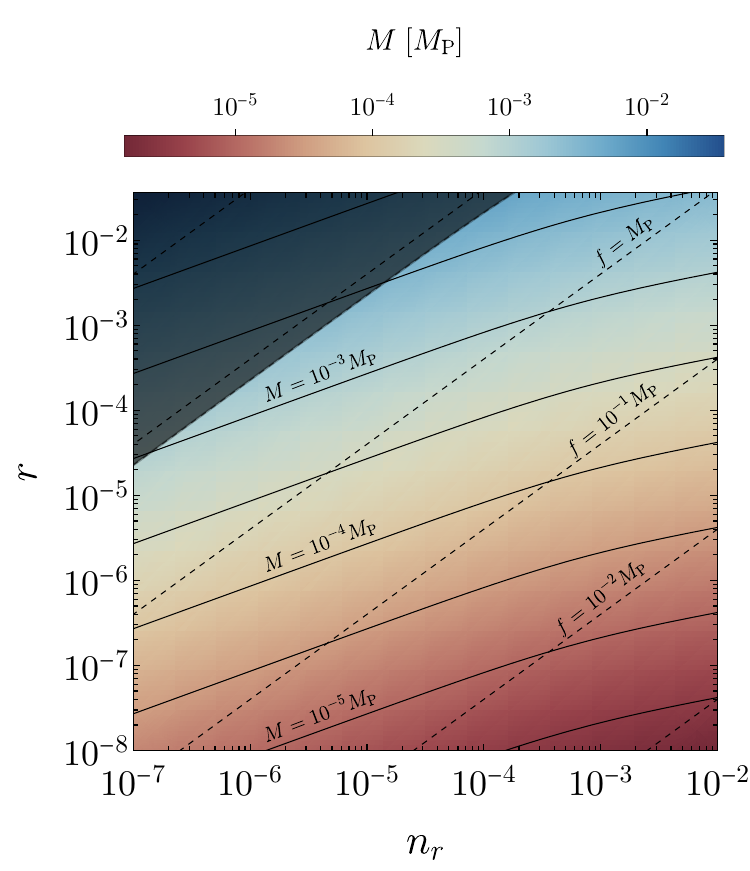}
\includegraphics[width=0.47\textwidth]{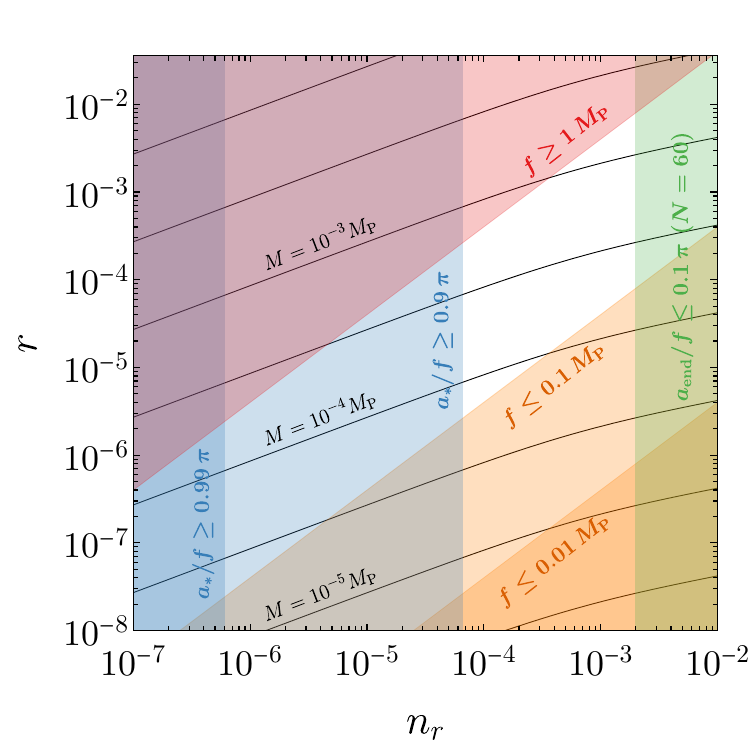}
\caption{
\textit{Left-hand panel.}
The values of $M$ and $f$ in units of $\MP$, as a function of $n_r$ and $r$,
while fixing $A_s = 2.105\times10^{-9}$ and $n_s = 0.965$. The largest displayed values of $n_r$ and $r$ correspond to the present experimental upper bounds. Solid (dashed) lines are contours of constant $M$ ($f$). The dark-gray region violates the slow-roll condition before the critical point.
\textit{Right-hand panel.} The white region is the one which complies with naturalness criteria, according to the discussion in Sec.~\ref{sec:Naturalness}. In the orange regions the value of $f$ is unnaturally small, while in the red region $f>\MP$. The parameter space covered by the blue and green regions requires a significant fine-tuning of the inflation initial and final conditions, respectively.}
\label{fig:M} 
\end{center}
\end{figure}

By imposing $A_s=2.105\times10^{-9}$, $n_s=0.965$, $n_r < 0.011$, and $r<0.036$, we can derive the inequalities
\be\label{eq:Mup}
    V_0^{1/4}< 6\times 10^{-3}\MP\,,\qquad \frac{M^4}{f^4}\simeq  3\sqrt{2}\pi^2 A_s n_r^{3/2}< 10^{-10}\,.
\ee
In the left-hand panel in Fig.~\ref{fig:M}, we chart the region allowed by observations in the $(n_r,\,r)$ plane, once $A_s$ and $n_s$ are fixed. All the above relations have been derived under the assumption that $V_0$ drives inflation, i.e., $V_0 \gg M^4$. Below, we quantify what separation is needed between the two scales, in order not to end inflation before the critical point $a=a_c$. In particular, we require $\epsilon < 1$ all along the inflationary trajectory:\footnote{For slow-roll dynamics to persist, we also have to check that $\lvert\eta\rvert<1$ for any value of $a$, but this leads to a constraint which is very similar to Eq.~\eqref{eq:MV0ratio}.}
\be
\epsilon = \frac{\MP^2}{2}\left(\frac{\frac{M^4}{f} \sin{\frac{a}{f}}}{V_0 - M^4\cos{\frac{a}{f}}}\right)^2 < 1\,, \quad \forall\;a\,.
\ee
This is equivalent to requiring
\be
    \frac{M^4}{V_0}<\sqrt{\frac{2f^2}{2f^2+\MP^2}} \,,
\ee
or, equivalently,
\be\label{eq:MV0ratio}
    r<n_r\left(-1+\sqrt{1+\frac{64}{2n_r+(n_s-1)^2}}\right)\,.
\ee
The region where this condition is not satisfied is shaded in dark gray in the left-hand panel in Fig.~\ref{fig:M}.

\begin{figure}[h!]
\begin{center}
 \includegraphics[width=0.48\textwidth]{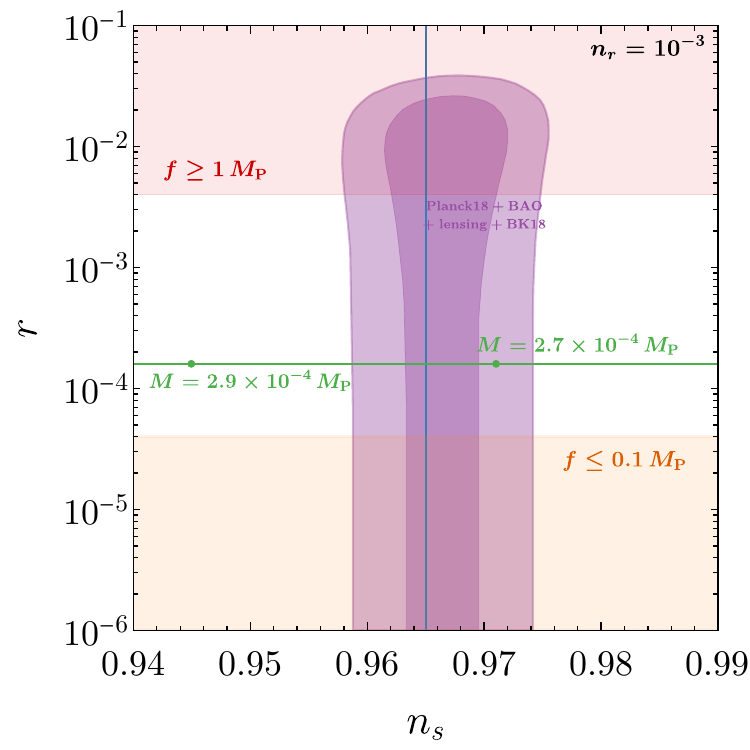} \includegraphics[width=0.48\textwidth]{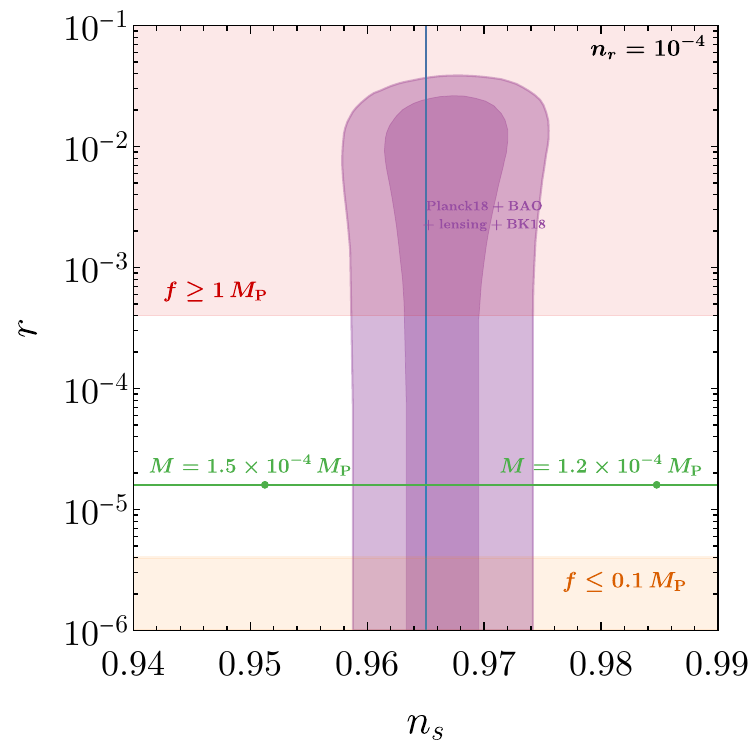}
\caption{The purple areas show the 95\% CL and 68 \% CL preferred regions from Planck 2018 + BAO + lensing + BICEP/Keck 2018 data \cite{BICEP:2021xfz}. We fixed the amplitude of the scalar power spectrum at $A_s = 2.105\times10^{-9}$. 
In the \textit{left-hand panel} we set the running of the spectral index at $n_r = 10^{-3}$, while in the \textit{right-hand panel} we took $n_r=10^{-4}$.
The horizontal, green lines are obtained by fixing $f=0.2\,\MP$ and varying $M$, while along the vertical, blue lines $M/f=3\times 10^{-3}$ is fixed and $f$ varies. The orange and the red regions are disfavoured by naturalness (see Sec.~\ref{sec:Naturalness}): Low values of the tensor-to-scalar ratio $r$ require unnaturally small $f$, while a large $r$ is associated to transplanckian $f$. The relation between $f$ and $r$ depends on the value assumed for $n_r$, according to Eq.~\eqref{eq:Par}.}
\label{fig:Planck} 
\end{center}
\end{figure}

In Fig.~\ref{fig:Planck}, we show the predictions of the model in the $(n_s,\,r)$ plane, together with the region allowed by the latest \ac{CMB} measurements \cite{Planck:2018jri, BICEP:2021xfz}. Our model is in agreement with observations for natural values of $f$, i.e., for $f$ smaller than, but not too far from $\MP$.

We require the end of inflation to occur within one Hubble time after the critical point has been reached by the accident. This allows us to identify $a_c = 6f\alpha_c$ with the point $a_{\rm end}$ in field space in which inflation ends. For this requirement to hold, the Hubble rate at the critical point, $H_c$, must be much lower than the value of the waterfall mass, after $\Delta t \sim H_c^{-1}$ has elapsed since $a = a_c$:
\be
    \lvert m_{\chi^3}^2\left(a_c + \Delta a\right)\rvert \gg H_c^2\,,
\ee
where $\Delta a$ is the accident excursion during $\Delta t$. 
This condition is also known as the ``waterfall condition'' \cite{Linde:1993cn}. Given that the energy density during inflation is dominated by the false vacuum, the Hubble rate can be approximated to $H_c \simeq V_0/3 \MP^2$. Moreover, the evolution of $a$ is governed, in the slow-roll approximation, by $3\,  H_c\,\dot{a} \simeq \partial V_{\rm eff}/\partial a$. With these considerations in mind, the above waterfall condition 
translates into a lower bound on the inflaton-waterfall quartic coupling:
\be \label{eq:WaterCond}
    \zeta \gg  \frac{V_0^2}{18\,M^4 \MP^4}\,\frac{1}{\sin\left( a_c/f\right) \sin\left( a_c/3f \right)}\,.
\ee

By making use of the relations in Eqs.~\eqref{eq:Par} and \eqref{eq:efolds}, we find that the right-hand side of the above condition is a monotonic, increasing function of $n_r$, once $A_s$, $n_s$, and $N$ are fixed. For $n_r$ taking its largest value allowed by observations, Eq.~\eqref{eq:WaterCond} becomes $\zeta \gg 10^{-6}$, meaning that we can always choose a perturbative value of $\zeta$ for which the waterfall condition holds.
We note that the requirement for inflation to end as soon the critical point is reached is not imposed by observations. In principle, one could also allow inflation to proceed along the waterfall direction for some 20 $-$ 40 $e$-folds as realized in the so called ``mild hybrid inflation'' scenario \cite{Clesse:2010iz, Clesse:2015wea,Kawasaki:2015ppx}.

\subsection{Naturalness considerations}\label{sec:Naturalness}

The aim of this section is to assess which portion of the parameter space allowed by observations complies with naturalness. An important motivation for considering this model in the first place is that the inflaton potential is nearly flat for generic values of the couplings, without requiring fine-tuning. To what extent does the model remain ``natural'', or minimally fine-tuned, once we impose that it should give rise to a quantitatively realistic phenomenology?

It turns out that there is a preferred range of natural values for the running of the spectral index $n_r$. A lower bound on $n_r$ is obtained as follows: From Eq.~\eqref{eq:CMBObs}, we have
\be
    n_r = \frac{1}{2}\left(n_s-1\right)^2 \tan^2 \frac{a_*}{f}\,.
\ee
Hence, fixing $n_s$ at its value preferred by \ac{CMB} measurements, $n_r$ could be arbitrarily small, but only at the cost of tuning the start of observable inflation $a_*$ arbitrarily close to $a_*=\pi f$. However, when remaining agnostic about any pre-inflationary physics, the inflaton could start rolling anywhere in field space. Taking, for instance, 10\% as an acceptable degree of fine-tuning of the initial condition,
\be
    \frac{a_*}{\pi f} \lesssim 0.9\,,
\ee
one obtains the naturalness bound $n_r\gtrsim 6.5\times10^{-5}$. Requiring the fine-tuning to be less than 1\% gives, instead, $n_r \gtrsim 6\times 10 ^{-7}$. Such lower bounds are shown as blue-shaded regions in the right-hand panel in Fig.~\ref{fig:M}.

An upper bound on $n_r$ is similarly given by the following argument: Fixing the number $N$ of observable $e$-folds to $N=60$, the parameter combination $f^2 V_0/(\MP^2 M^4)$ could, in principle, be made arbitrarily small by tuning the ending point of inflation $a_{\rm end}$ close to zero; see Eq.~\eqref{eq:efolds} (and by simultaneously tuning $a_*$ close to $a_*=\frac{\pi}{2} f$ to obtain the observed value of $n_s$ according to Eqs.~\eqref{eq:CMBObs}). This would imply, again by Eqs.~\eqref{eq:CMBObs}, that $n_r$ could be arbitrarily large. But $a_{\rm end}$ is given by a combination of independent model parameters which have no reason to conspire to produce $a_{\rm end}\simeq 0$. Imposing for concreteness
\be
\frac{a_{\rm end}}{\pi f} \gtrsim 0.1\,,
\ee
this yields $n_r\lesssim 2\times10^{-3}$, shown by green shading in the right-hand panel in Fig.~\ref{fig:M}. 

Moreover, naturalness requires $f$ to be not too far from the cutoff scale $\Lambda_{\rm UV}$ of the model. This is nothing but the well-known hierarchy problem of models with elementary scalar fields: Quadratically divergent loop corrections to the mass parameters $\mu_\phi^2$ and $\mu_\chi^2$ signal that they depend sensitively on the ultraviolet dynamics at the scale $\Lambda_{\rm UV}$. At this scale, the model might be embedded in a supersymmetric field theory, or in a strongly coupled field theory where $\phi$ and $\chi$ emerge as bound states. However, the most minimal assumption is that there is no field-theoretic embedding which cures the \ac{UV} sensitivity, in which case we should identify $\Lambda_{\rm UV}$ with the quantum gravity scale.

Quantum gravity corrections are, of course, greatly model-dependent. Identifying $\Lambda_{\rm UV}=\MP$ for definiteness, and assuming that \ac{UV} states have $\lesssim {\cal O}(1)$ couplings to $\phi$ and $\chi$, we can derive a tentative naturalness bound on the tensor-to-scalar ratio $r$. Cutting off the one-loop corrections to the $\mu_\phi^2$ parameter at the scale $\MP$ and requiring that they do not exceed the tree-level value leads to
\be \mu_\phi^2 \gtrsim\left(7\,\lambda_\phi + \frac{3}{2}\,\varepsilon + 3\,\zeta \right)\frac{\MP^2}{16\pi^2}\,.
\ee
Since $f^2=\mu_\phi^2/6\lambda_\phi$, this implies $f\gtrsim 0.1\,\MP$, under the technically natural assumption that $\varepsilon$ and $\zeta$ are at most of the order of $\lambda_\phi$. 
Therefore, the orange-shaded region in the right-hand panel in Fig.~\ref{fig:M} is disfavoured.
From Eqs.~\eqref{eq:CMBObs}, it follows that $r=4\,n_r\,(f/\MP)^2$ and, thus,
\be
    r \gtrsim (4\times 10^{-2})\,n_r\,,
\label{eq:NatBound}
\ee
shown in orange shading in Fig.~\ref{fig:Planck}. Depending on how much fine-tuning one is willing to accept, this bound can, of course, shift by several orders of magnitude, as also indicated in the right-hand panel in Fig.~\ref{fig:M}.

On the other hand, values of $f$ too close to $\Lambda_{\rm UV}$ imply that higher-dimensional operators, which may spoil the tree-level flatness of the inflaton potential, can no longer be neglected. In that sense, the ``least fine-tuned'' region of parameter space is also the one with the least theoretical control. In Figs.~\ref{fig:M} and \ref{fig:Planck}, we shaded in red the regions with $f\gtrsim \MP$, or, equivalently, $r\gtrsim 4n_r$.

Such naturalness constraints translate into some bounds on the model parameters. 
For instance, the inflationary scale is given by $V_0 \simeq \lambda_\chi v_\chi^4/4$, 
and the above bounds on $r$ and $n_r$ can be translated into bounds on $V_0$ once $A_s$ is fixed, according to Eq.~\eqref{eq:Par}. This, together with $0.1\,\MP \lesssim v_\chi \lesssim \MP$ (we require $v_\chi$ to be subplanckian, analogously to $f$), gives a natural range of values for the $\chi$ quartic coupling:
\be
   10^{-13} \lesssim \lambda_\chi \lesssim 10^{-5}
    \,.
\ee

\section{Gravitational waves from preheating}

As also described in Sec.~\ref{sec:Preh}, hybrid inflation is followed by a stage of tachyonic preheating \cite{Felder:2000hj, Felder:2001kt}. The tachyonic instability of the waterfall field exponentially enhances field inhomogeneities. Such inhomogeneous, bubble-like structures behave as scalar waves and scatter off each other until the Universe is homogenized and driven to local thermal equilibrium. A \ac{SGWB} is produced in the process  \cite{Garcia-Bellido:2007fiu,Garcia-Bellido:2007nns, Dufaux:2007pt, Dufaux:2008dn}. \ac{GW}s quickly decouple from the background dynamics and propagate (almost) undisturbed throughout cosmic history, hence carrying information about the process that sourced them. Given the energy scale of inflation, which is typically large, the frequency at which \ac{GW}s are produced during preheating lies well beyond the reach of current or even planned interferometers. However, it has been shown, e.g., in Ref.~\cite{Dufaux:2008dn}, that, by sufficiently lowering the scale of inflation, the signal might be brought within the projected sensitivity of the \ac{ET} \cite{Maggiore:2019uih}. It is also worth mentioning that there have been proposals of using optical atomic clocks to detect high-frequency \ac{GW}s; see, e.g., Ref.~\cite{Bringmann:2023gba} and references therein.

The spectrum of \ac{GW}s produced during preheating presents a peaked shape, with the frequency and amplitude at the peak given, respectively, by \cite{Dufaux:2008dn}
\be\label{eq:TachPre}
    f_\star \sim 4\times 10^{10}\,{\rm Hz} \frac{k_\star}{{\rho_c}^{1/4}}\,,\qquad h^2\Omega_\star \sim 10^{-6} \left(\frac{H_c}{k_\star}\right)^2\,,
\ee
where $k_\star$ is the typical momentum enhanced by the tachyonic instability, related to the characteristic size of inhomogeneities as $R_\star \sim 1/k_\star$.\footnote{Here $k_\star$ is unrelated to the pivot scale of \ac{CMB} observables $k_*$ in Sec.~\ref{sec:CMB}.} Here, $\rho_{c}\simeq V_0$ and $H_c^2\simeq V_0/(3\MP^2)$ are, respectively, the energy density and the Hubble rate at the end of inflation or, analogously, at the onset of the tachyonic instability. 
The value of $k_\star$ depends on the inflationary dynamics close to the waterfall transition.

From Eq.~\eqref{eq:TachPre}, one may already deduce that, for a signal of \ac{GW}s from preheating to be within the reach of the \ac{ET}, the inflation scale must lie in the range
\be
 3\times 10^5\,{\rm GeV}\lesssim V_0^{1/4}\lesssim 5\times 10^{11}\,{\rm GeV}\,,
\ee
which is associated to an unobservably small tensor-to-scalar ratio, according to Eq.~\eqref{eq:Par}. Should $r$ be measured in the future, along with a \ac{SGWB} at the \ac{ET} which can be traced back to tachyonic preheating, this would, therefore, rule out our model (and other models of hybrid inflation, where tachyonic preheating is an unavoidable feature).

In principle, two different regimes for \ac{GW} production from preheating can be identified  \cite{Dufaux:2008dn}, depending on the velocity $\dot{a}_c$ of the inflaton at the critical point $a=a_c$ where the waterfall field becomes tachyonic. In the first regime, $a$ is fast enough for the tachyonic instability to be driven by its classical dynamics. Another possibility is that $\dot a_c$ is sufficiently small, so that preheating is triggered by quantum fluctuations. In the present chapter, we take model parameters such that inflation ends as soon as the inflaton crosses the critical point; see Eq.~\eqref{eq:WaterCond}. In that case, classical rolling always dominates.

\begin{figure}[h!]
\begin{center}
\includegraphics[width=0.48\textwidth]{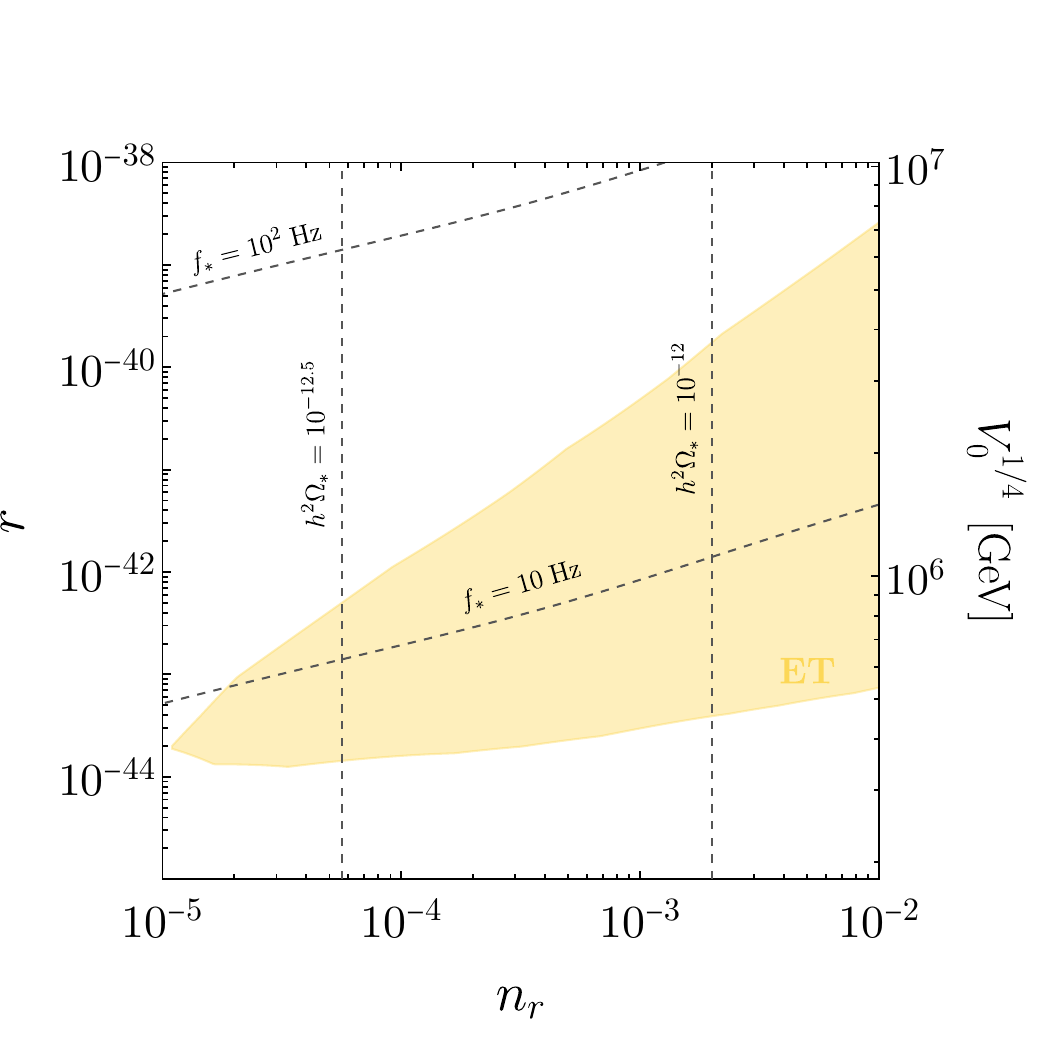}
 \includegraphics[width=0.48\textwidth]{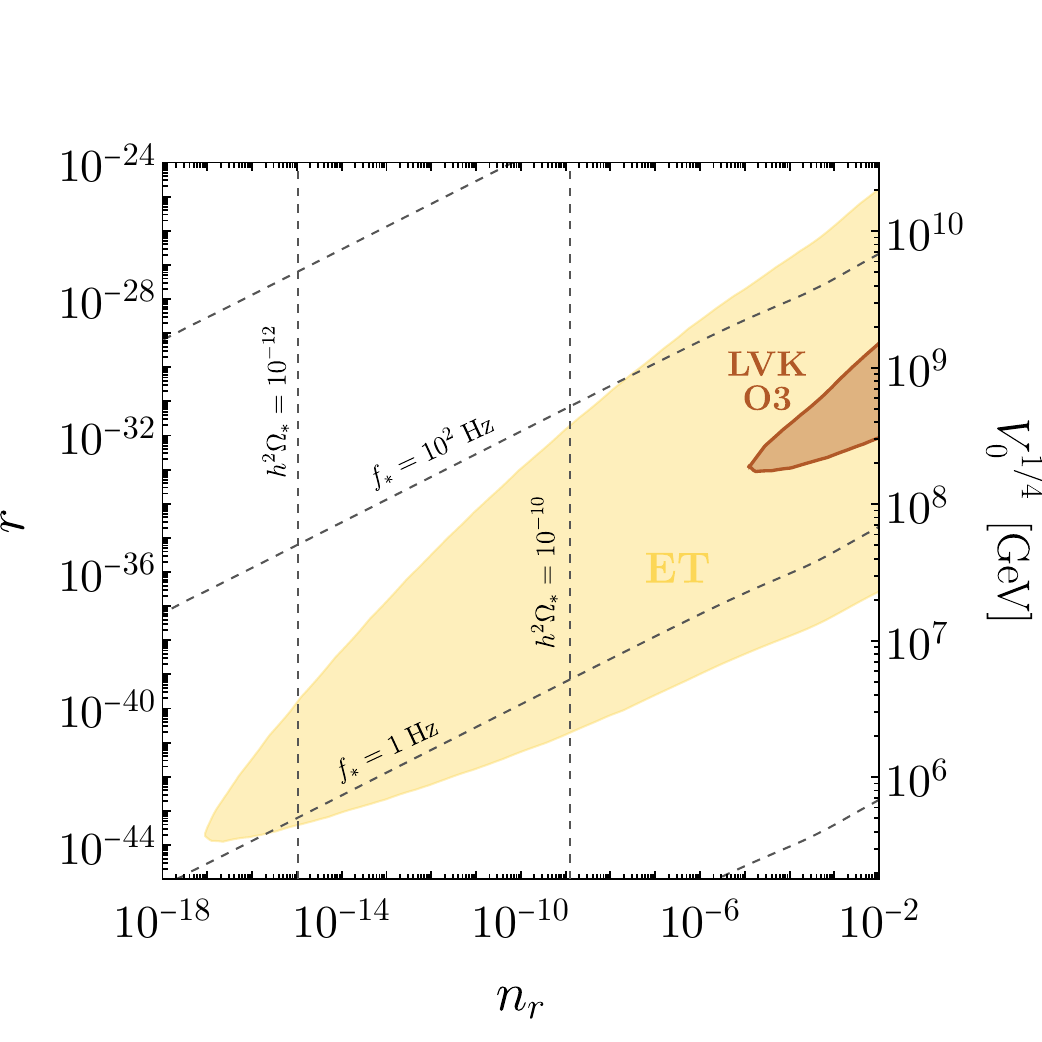}
    \caption{Sensitivity region (yellow) of the \ac{ET} \cite{Maggiore:2019uih} for the \ac{GW} spectrum produced during tachyonic preheating, as a function of $n_r$ and $r$ (or equivalently $V_0$).
    In the \textit{left-hand panel} we take $\tilde{\mu}_\chi^2/f^2 = 1$, while in the \textit{right-hand panel} we fix $\tilde{\mu}_\chi^2/f^2 = 10^{-6}$. 
    The projected sensitivity for the \ac{ET} is the one derived in Ref.~\cite{NANOGrav:2023hvm}, and it is shown in Fig.~\ref{fig:DWGW}.
    To guide the eye, we also plot lines of constant peak frequency and amplitude for the preheating \ac{GW} spectrum.
    The brown region shows the integrated sensitivity for the third observing run of the \ac{LVK} network \cite{KAGRA:2021kbb}.}
\label{fig:ET}  
\end{center}
\end{figure}

We are interested in an order-of-magnitude estimate of the amplitude and frequency of the \ac{GW}s signal at the peak. Because of the fact that tachyonic preheating is a non-linear process, a more precise computation of the \ac{GW} spectrum would require lattice simulations which are beyond the scope of this thesis. In the slow-roll approximation, $\ddot{a} \ll 3H\dot{a}$, the inflaton velocity at the critical point reads
\be
    \dot{a}_c \simeq \frac{\MP}{\sqrt{3}f}\frac{M^4}{V_0^{1/2}}
    \sin\frac{a_c}{f}\,.
\ee
If its velocity is large enough, the inflaton will keep rolling classically past the critical point. At $a=a_c$, the mass-squared of the waterfall field vanishes and then starts growing negative so that after an interval $\Delta t$ it becomes $m_{\chi^3}^2 \simeq 6 \, \zeta \, f \, \dot{a}_c\, \Delta t \,\sin \left(a_c/3f\right)$. The exponential growth of quantum fluctuations becomes efficient when the tachyonic mass of the waterfall field has grown large enough such that $\sqrt{|m_{\chi^3}^2|} \gtrsim \Delta t^{-1}$. This leads to the enhancement of the mode $k_\star \sim 1/\Delta t \lesssim \sqrt{|m_{\chi^3}^2|}$, so that
\be\label{eq:kCl}
    k_{\star,{\rm cl}}^3 \simeq 2\sqrt{3}\,\zeta\,\frac{M^4 \MP}{V_0
^{1/2}}\sin\frac{a_c}{f}\sin\frac{a_c}{3f}\,.
\ee
Fig.~\ref{fig:ET} shows the values of $r$ (or, analogously, of the inflationary scale $V_0$) and $n_r$ which correspond to a \ac{GW} signal within the \ac{ET} projected sensitivity for $\tilde{\mu}_{\chi}^2/f^2 = 1$ (left-hand panel) and $\tilde{\mu}_{\chi}^2/f^2 = 10^{-6}$ (right-hand panel), where the ratio $\tilde{\mu}_{\chi}^2/f^2$ measures the steepness of the waterfall direction.
 
We conclude this section by mentioning that, in principle, tachyonic instability could also lead to production of \ac{PBH}s.\footnote{As a side note, we touch on the possibility of producing \ac{PBH}s at the end of inflation thanks to an enhancement of curvature perturbations driven by a large running of the spectral index. It has been shown in Ref.~\cite{Kohri:2007qn} that this eventuality bounds $n_r \lesssim 10^{-2}$ in single-field, slow-roll inflation. However, larger values for $n_r$ are nowadays excluded by \ac{CMB} observations.} In our model, however, inflation ends almost immediately after reaching the critical point. This leads to enhancement of perturbations at very small scales and to a subsequent production of extremely light \ac{PBH}s which have completely evaporated by now. It is possible to envision a scenario in which inflation goes on for a certain number of $e$-folds along the waterfall direction, hence producing \ac{PBH}s which can be of phenomenological relevance \cite{Afzal:2024xci, Braglia:2022phb}.

\section{Topological defects}\label{sec:TopologicalDefects}

In Ch.~\ref{chap:TopoDef}, we have seen that models with spontaneously broken symmetries may give rise to topological defects such as magnetic monopoles, \ac{CS}s, and \ac{DW}s. 

Magnetic monopoles can appear if the second homotopy group of the  vacuum manifold is non-zero. As discussed in details in Sec.~\ref{sec:Mono}, when $G=\SO{3}$ is broken to $\SO{2}$ by a triplet \ac{VEV}, as is the case in our model, this gives rise to the celebrated 't Hooft-Polyakov monopole \cite{tHooft:1974kcl, Polyakov:1974ek} with $\pi_2(\SO{3}/\SO{2})= \pi_2[S^2]=\mathbb{Z}$. However, in our models, the relevant phase transition takes place before inflation, and, therefore, no monopoles are likely to remain in our Hubble patch. To be precise, symmetry breaking in our model proceeds in two steps: First, at the scale $v$, the symmetry group $G$ is broken completely by the 5-plet \ac{VEV}, while the triplet \ac{VEV} is zero in this phase. The vacuum manifold (or rather the scalar field manifold which minimizes the potential for all fields except the slowly-rolling inflaton) is $G$, and in different causally connected regions of the Universe, the scalar fields will take \ac{VEV}s in different domains of $G$. However, the subsequent phase of inflation stretches out these inhomogeneities, such that, at all scales relevant to us, the $5$-plet \ac{VEV} becomes effectively a single point in field space. Non-trivial field configurations which might evolve into monopoles are ``inflated away''. The second step of symmetry breaking happens at the waterfall transition at the end of inflation, when the $5$-plet \ac{VEV} relaxes to zero and a triplet \ac{VEV} is generated, thereby restoring an $\SO{2}$ symmetry. The vacuum manifold becomes $\SO{3}/\SO{2}\simeq S^2$; however, no new monopoles can be created, since this would require energies sufficiently high to restore the entire $\SO{3}$ symmetry. We conclude that, for better or worse, magnetic monopoles will not affect the cosmology of our model.

We showed in Sec.~\ref{sec:CS} that \ac{CS}s can appear if the scalar vacuum manifold is not simply connected. This is not the case in the minimal model, but it is interesting to consider a non-minimal extension to study the phenomenological consequences. Consider a version of the model in which the triplet $\chi$ is a complex scalar field transforming under $G=\SU{2}\times\U{1}_\chi$, with $\U{1}_\chi$ acting as a phase rotation. All or part of $G$ may be thought of as gauged. We can further extend the model by promoting also $\phi$ to a complex field and gauging $\U{1}_\phi$, which is appealing because then the model no longer relies on any ad-hoc discrete symmetries, the role of the $\mathbb{Z}_2$ in the original model now being played by $\U{1}_\phi$. The technical details of these model versions are worked out in App.~\ref{sec:AppendixComplexModel}.  These models do allow for additional quartic scalar couplings, but the qualitative features of the inflaton potential, as well as \ac{CMB} predictions, do not change: Crucially, there is still an accidentally light scalar field which can play the role of the inflaton, with a cosinusoidal loop-induced potential analogous to Eq.~\eqref{eq:CosPot}. Likewise, the effective mass parameter of the waterfall field $\chi$ depends on the value of the inflaton, and parameters can be chosen such that it becomes tachyonic at the end of inflation. The waterfall field is zero along the inflationary trajectory, so $\U{1}_\chi$ remains unbroken during inflation while $\SU{2}\times\U{1}_\phi$ is completely broken. The true vacuum is at $\vev{\phi}=0$ and $\vev{|\chi|}=\vev{|\chi^3|}=v_\chi\neq 0$, so the vacuum manifold is $S^2\times\U{1}_\chi$. 

As $\pi_1[U(1)]=\mathbb{Z}$, the \ac{SSB} phase transition results in the production of a network of stable, local \ac{CS}s with a typical tension \cite{Vilenkin:2000jqa}
\be
    \GN\mu \simeq 2\pi\, \GN\, v_{\chi}^2\,,
\ee 
where $v_{\chi}$ is the breaking scale of $\U{1}_{\chi}$. In Sec.~\ref{sec:CS} we argued that this network quickly reaches a scaling regime, where the fraction of energy density stored in the \ac{CS}s redshifts in the same way as the energy density of the background \cite{Hindmarsh:1994re,Vilenkin:2000jqa}. Strings can intercommute and produce loops, which oscillate under their own tension and radiate energy in the form of \ac{GW}s. Such a signal of \ac{GW}s appears to us as a \ac{SGWB} with a characteristic scale-invariant spectrum (if standard cosmology is assumed \cite{Gouttenoire:2019kij}). The peak and the amplitude of the \ac{GW}s signal for stable strings are both given by the string tension, which, in order to match the signal recently reported by \ac{PTA} collaborations, has to be $\GN\, \mu \simeq 10^{-10}$ \cite{NANOGrav:2023hvm,EPTA:2023xxk}, even though stable \ac{CS}s seem to be disfavoured with respect to other cosmological sources \cite{NANOGrav:2023hvm}. The reason is that, for stable \ac{CS}, both the amplitude and the shape of the \ac{GW} signal are controlled by $\GN\mu$ (see Fig.~\ref{fig:StringGW}); once we fix the string tension to match the amplitude of the reported signal, the \ac{CS} spectrum turns out to be flat in the nHz region, with data preferring a blue-tilted spectrum. $\GN\mu \simeq 10^{-10}$ represents a trade-off between a signal which is not too weak and one which is not too flat.  
However, as the computation of the expected signal of \ac{GW}s from a \ac{CS} network is subject to large uncertainties, we can take as upper bound on the string tension the one given by the \ac{CMB} constraint on the impact of \ac{CS}s on the \ac{CMB} power spectrum \cite{Planck:2015fie}, $\GN\mu \lesssim 10^{-7}$, which corresponds to an inflationary scale $V_0^{1/4}\lesssim \left(\lambda_\chi + \lambda_\chi'\right) \left(3\times 10^{14}\right)\,{\rm GeV}$.

Finally, \ac{DW}s can appear if the scalar vacuum manifold is not connected. This, again, is not the case in our minimal model. In the following, we exploit a non-minimal model with disconnected minima,
which is described in detail in App.~\ref{sec:AppendixZ4Model}.
In this case, the \ac{PTA} signal can be fitted with \ac{GW}s produced from annihilating \ac{DW}s, as explained in the next section. The waterfall field $\chi$ is taken to be a complex scalar triplet,
subject to a discrete symmetry $\mathbb{Z}_{4\chi}: \chi\into i\chi$.
This leads to disconnected minima which are degenerate, 
corresponding to stable \ac{DW}s.
However, the vacuum degeneracy can be easily lifted by
softly breaking the $Z_{4\chi}$ discrete symmetry, allowing for the \ac{DW}s to annihilate. Also in this model the inflationary predictions are the same as those for the minimal model.

\subsection{Gravitational waves from unstable domain walls}\label{secDW}

In the model described in App.~\ref{sec:AppendixZ4Model}, after the end of inflation the $\chi$ scalar potential has two disconnected, degenerate minima, so that \ac{DW}s are formed with a surface energy density \cite{Vilenkin:2000jqa}
\be\label{eq:DWTens}
    \sigma = \frac{2\sqrt{2}}{3}\lambda_{\chi}^{1/2}v_{\chi}^3\,,
\ee
where $\lambda_\chi$ is the $\chi$ quartic coupling and $v_\chi$ its \ac{VEV}. A more thorough discussion on \ac{DW}s can be found in Sec.~\ref{sec:DW}. The motion of \ac{DW}s is initially damped by their interaction with the surrounding plasma. Numerical simulations show that, after some time, friction becomes negligible and the \ac{DW} network reaches the ``scaling regime'' \cite{Saikawa:2017hiv} in which their energy density scales with time as $\rho_{\rm DW} = \sigma \mathcal{A}/t$, with $\mathcal{A} \simeq 0.8$ \cite{Hiramatsu:2013qaa}.
If the vacuum degeneracy were exact, the \ac{DW}s would be stable and they would quickly overclose the Universe, as,  in the scaling regime, $\rho_{\rm DW}$ redshifts more slowly than the energy density of the background. \ac{DW}s start dominating when $3H^2\MP^2 = \rho_{\rm DW}$, which for a previously radiation-dominated universe corresponds to the time
\be
    t_{\rm dom}^{\rm DW} = \frac{3\, \MP^2}{4\,\mathcal{A}\,\sigma}\,.
\ee
However, the discrete symmetry ensuring the vacuum degeneracy may be softly broken, which lifts the degeneracy and introduces a potential bias $\Delta V$; see Eq.~\eqref{eq:Vbias}. Provided that $\Delta V<0.795\,V_0$ \cite{Saikawa:2017hiv}, where $V_0$ is the potential energy at $\chi=0$, \ac{DW}s are formed, but they are unstable as long as $\Delta V\neq 0$. The different domains annihilate when the bias pressure $\Delta V$ dominates over the \ac{DW} surface tension at
\be\label{eq:tann}
    t_{\rm ann} \simeq C_d \mathcal{A} \frac{\sigma}{\Delta V}\,,
\ee
with $C_d \simeq 3$ \cite{Gouttenoire:2023ftk}, until all the observable Universe lies in the true vacuum. We require the \ac{DW} network to disappear before the bias energy density of the false-vacuum patches dominates the Universe:
\be
    t_{\rm dom}^{\Delta V} \simeq \frac{\sqrt{3}\MP}{2\sqrt{\Delta V}}\,.
\ee
The requirement $t_{\rm ann} < t_{\rm dom}^{\Delta V}$ leads to the constraint $\Delta V \gtrsim 8\,\sigma^2/\MP^2$. 

In an early phase, as far as we know, the Universe might well have been dominated by \ac{DW}s, provided that they annihilate before \ac{BBN}. However, the dynamics of a \ac{DW}-dominated Universe is not well understood, and its treatment requires dedicated numerical studies \cite{Saikawa:2017hiv}, so we restrict ourselves to a radiation-dominated scenario. By combining this requirement together with that of having a not too large $\Delta V$, in order to have \ac{DW}s appear in the first place, we obtain $v_\chi \lesssim 0.3\,\MP$, where we have made use of Eq.~\eqref{eq:DWTens} and $V_0 = \lambda_\chi v_\chi^4$.

As the \ac{DW} network evolves, the \ac{DW}s are driven to relativistic speed and radiate \ac{GW}s. The signal of \ac{GW}s, appearing to us as a \ac{SGWB}, has a peak frequency given by the Hubble parameter at the moment of annihilation, which after taking into account redshift becomes \cite{Saikawa:2017hiv}
\be\label{eq:fpDW}
    f_{p}\simeq \frac{R\left(t_{\rm ann}\right)}{R\left(t_0\right)}
    H_{\rm ann}\simeq  1.6 \times 10^{-7}\;{\rm Hz} \left(\frac{g_*(T_{\rm ann})}{100}\right)^{1/6}\frac{T_{\rm ann}}{ {\rm GeV}}\,.
\ee
Here $R\left(t_{\rm ann}\right)$ and $R\left(t_0\right)$ are the scale factors at the time of \ac{DW} annihilation and today, respectively,
$g_*$ is the number of effective degrees of freedom in relativistic energy, and $T_{\rm ann}$ is the temperature of the Universe at the moment of \ac{DW} annihilation:
\be\label{eq:Tann}
    T_{\rm ann} = \left[\frac{45}{2\pi^2g_*(T_{\rm ann})}\;\frac{\MP^2}{t_{\rm ann}^2}\right]^{1/4} \,.
\ee
The present energy density of radiated \ac{GW}s is \cite{Saikawa:2017hiv}
\be\label{eq:OmegaDW}
    h^2\Omega_{\rm GW}\left(f\right) \simeq 1.6\times10^{-5}\left(\frac{100}{g_*(T_{\rm ann})}\right)^{1/3}\, \frac{3}{32 \pi}\, \tilde{\epsilon}\, \alpha_{\rm ann}^2\,\mathcal{S}(f/f_p),
\ee
where $\tilde{\epsilon} =0.7$ is an efficiency parameter taken from simulations \cite{Hiramatsu:2013qaa},  $\alpha_{\rm ann}$ is the \ac{DW} network energy density fraction at the time of annihilation, 
\be\label{eq:AlphaAnn}
    \alpha_{\rm ann} \equiv \frac{\rho_{\rm DW}(t_{\rm ann})}{\rho_r(t_{\rm ann})} \simeq \frac{4}{3}\,C_d\,\mathcal{A}^2\frac{\sigma^2}{\MP^2\,\Delta V}\,,
\ee
and the spectral function $\mathcal{S}(x)$ can be modeled as \cite{NANOGrav:2023hvm}
\be
    \mathcal{S}(x) = \left(\frac{a+b}{b\,x^{-a/c}+a\, x^{b/c}}\right)^c\,.
\ee
Numerical simulations have shown that $b,c\simeq 1$ for $\mathbb{Z}_2$-type \ac{DW}s \cite{Hiramatsu:2013qaa}, and our model falls in this category, while causality fixes $a=3$ \cite{Hook:2020phx}.\footnote{In principle, the causality tail of the \ac{GW} spectrum is affected by the QCD phase transition and the subsequent change in the number of relativistic degrees of freedom. This leads to a deviation from the pure power-law behaviour with $a=3$. Including the effect of the QCD phase transition is beyond the scope of this work, and we refer to Ref.~\cite{Franciolini:2023wjm} for a detailed discussion on the topic.}  By plugging Eqs.~\eqref{eq:tann} and \eqref{eq:Tann} into Eq.~\eqref{eq:fpDW}, we observe that the \ac{GW} peak frequency scales with the \ac{DW} network parameters as $f_p\sim\sqrt{\Delta V/\sigma}$. Similarly, by plugging Eq.~\eqref{eq:AlphaAnn} into Eq.~\eqref{eq:OmegaDW}, we find that $h^2 \Omega_{\rm GW}(f_p) \sim (\sigma^2/\Delta V)^2$.

\begin{figure*}[tbp]
\begin{center}
      \includegraphics[width=16 cm]{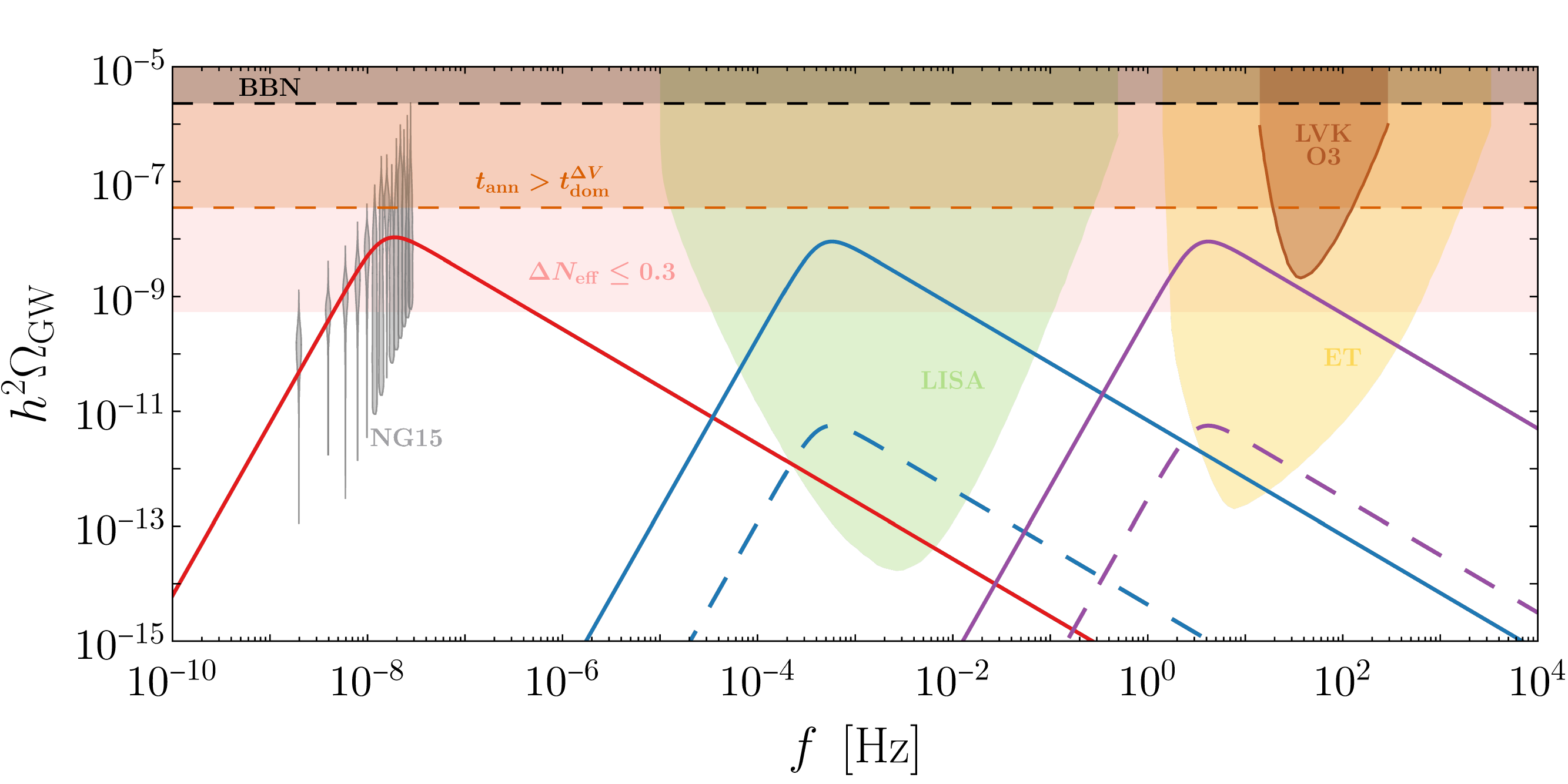}
    \caption{GW spectra produced by unstable \ac{DW}s for different benchmark points. The red curve is obtained with \ac{DW} tension $\sigma^{1/3} =3.2\times10^{5}\,{\rm GeV}$ and bias $\Delta V^{1/4} =0.23\,{\rm GeV}$. The blue solid curve corresponds to $\sigma^{1/3} = 3.2\times10^8\,{\rm GeV}$, and the purple solid one to $\sigma^{1/3} =1.2\times10^{11}\,{\rm GeV}$ ($\Delta V$ is varied accordingly to keep the amplitude of the signal fixed). The dashed spectra are obtained from the solid ones by decreasing $\sigma$ while keeping $f_p$ fixed: The dashed blue curve corresponds to $\sigma^{1/3} = 9.3\times 10^7\,{\rm GeV}$, while the dashed purple one to $\sigma^{1/3} = 3.5\times 10^{10}\,{\rm GeV}$. The gray violins show the \ac{SGWB} reported by the NANOGrav 15 yrs experiment \cite{NANOGrav:2023hvm}. The green and yellow shaded regions show the projected sensitivities of \ac{LISA} \cite{LISACosmologyWorkingGroup:2022jok} and \ac{ET} \cite{Maggiore:2019uih}, respectively, as derived in Ref.~\cite{NANOGrav:2023hvm}, assuming a signal-to-noise-ratio equal to unity and an observing time of one year. The integrated sensitivity of the  \ac{LVK} network during the third observing run \cite{KAGRA:2021kbb} is shown in brown. The region where the bias energy density dominates the Universe, leading to local departures from radiation domination, is shown in orange, while the black dashed line shows the bound on the amplitude of primordial \ac{GW}s given by \ac{BBN} \cite{LISACosmologyWorkingGroup:2022jok}. If the \ac{DW} network decays into dark radiation, the pink region is excluded by the \ac{BBN} bound on the effective number of neutrino species.
    }
\label{fig:DWGW} 
\end{center}
\end{figure*}

By taking the quartic couplings to be of the order of one, the \ac{DW} tension is related to the inflationary scale as $V_0 \simeq \sigma^{4/3}$. This allows us to relate the \ac{GW} signal generated by the \ac{DW} network to inflation. It has been shown \cite{NANOGrav:2023hvm,Gouttenoire:2023ftk,Ellis:2023oxs} that a network of unstable \ac{DW}s is a good candidate to explain the signal of a \ac{SGWB} recently detected by the NANOGrav Collaboration \cite{NANOGrav:2023hvm}. We find that a model with $V_0^{1/4} \simeq 2\times10^{5}\, {\rm GeV}$ and a suitable bias $\Delta V$ leads to a \ac{GW} signal which fits the NANOGrav detection. The peak of the \ac{GW} signal is within the reach of the upcoming space-borne \ac{LISA} \cite{LISACosmologyWorkingGroup:2022jok} for $3\times10^7\,{\rm GeV}\lesssim V_0^{1/4}\lesssim 2\times10^{10}\,{\rm GeV}$, while for $3\times10^{10}\,{\rm GeV} \lesssim V_0^{1/4}\lesssim 7\times 10^{12}\,{\rm GeV}$ the \ac{SGWB} might be detected by the \ac{ET} in the future. Fig.~\ref{fig:DWGW} shows several different benchmark \ac{GW} spectra together with the sensitivity curves of current and planned experiments.

The energy density released by \ac{DW} annihilation can potentially alter the expansion rate of the Universe. In order not to spoil \ac{BBN}, the products of \ac{DW} annihilation have to decay before the \ac{BBN} epoque, either into dark radiation or into \ac{SM} particles  \cite{Ferreira:2022zzo}. Indeed, $\chi$ can decay into the $\SO{2}$ dark photon, or into \ac{SM} degrees of freedom via, e.g., a Higgs-portal coupling. The dominant decay channel depends on the relative magnitude of the couplings and is related to the reheating mechanism, which we do not explore in this chapter. 
The usual assumption, which also leads to the weakest constraints, is that such decays are already active at $T_{\rm ann}$ \cite{Ferreira:2022zzo}. Assuming the \ac{DW} energy density is transferred entirely into dark radiation, 
the extra number of neutrinos species receives a correction \cite{Gouttenoire:2023gbn}
\be
    \Delta N_{\rm eff} = \frac{8}{7}\left(\frac{11}{4}\right)^{4/3}\frac{g_*(T_{\rm BBN})}{2}\alpha_{\rm ann}\,,
\ee
where the effective number of \ac{SM} relativistic degrees of freedom at the time of \ac{BBN} is $g_*(T_{\rm BBN}) \simeq 3.36$.
This quantity is constrained by \ac{BBN}, $\Delta N_{\rm eff} \leq 0.3$ \cite{Fields:2019pfx}, leading to an upper bound on the amplitude of \ac{GW}s generated by a network of \ac{DW}s decaying into dark radiation. The bound is shown in pink in Fig.~\ref{fig:DWGW}. On the other hand, if \ac{DW}s mainly decay into \ac{SM} degrees of freedom, such an upper bound is relevant only for networks decaying after the neutrino decoupling temperature, $T_{\rm ann} \lesssim 1\,{\rm MeV}$ \cite{Kawasaki:2000en, Bai:2021ibt}. Indeed, in this scenario, neutrinos produced by the \ac{DW} decay would not be in thermal equilibrium by the time of \ac{BBN}. Such a low annihilation temperature, however, is associated to much lower frequencies of the \ac{GW} signal.

\section{Conclusions}

Models with accidentally light scalars \cite{Brummer:2023znr} are particularly well suited for realizing hybrid inflation. They feature tree-level flat directions, lifted by loop corrections, along which slow-roll inflation takes place. They also feature tree-level couplings of the inflaton to other scalar fields, which can become tachyonic along the inflationary direction, thereby ending inflation by the waterfall mechanism.

In this chapter, we have presented a simple yet realistic example. 
It is defined by a symmetry group $G=\SO{3}$ and two scalar multiplets $\phi$ and $\chi$, transforming in the representation $\boldsymbol{5}$ and $\boldsymbol{3}$, respectively. The two fields may also carry an Abelian charge (discrete or continuous), depending on the variant of the model.
Gauging $\SO{3}$ and allowing for all renormalizable potential terms compatible with the symmetry, one degree of freedom $a$ in $\phi$ remains massless at tree level: This is what we call an
accidentally light field, which we identify with the inflaton. The inflationary potential arises at the one-loop level, induced by the gauge coupling and a quartic coupling between $\phi$ and $\chi$. It has a form similar to that of \ac{NI} \cite{Freese:1990rb}. One component of $\chi$ plays the role of the waterfall field: Its mass-squared starts out positive and decreases during inflation until, at some critical point along the inflationary direction, it turns tachyonic. This destabilizes the waterfall field, which terminates inflation by fast-rolling down its steep potential. 

The flatness of the inflationary potential is naturally protected: It is guaranteed by symmetry at the tree level, and it gets lifted by one-loop radiative corrections. Unlike pseudo-Goldstone inflatons, here the inflaton has tree-level non-derivative couplings to other scalars, which are necessary for an efficient waterfall. These couplings are typically a concern in earlier realizations of hybrid \ac{NI}, as they spoil the flatness, unless they are forbidden by some complicated additional symmetries or fine-tuned.

The predictions of the model are in agreement with the latest \ac{CMB} measurements \cite{Planck:2018vyg, Planck:2018jri} for a wide region of the parameter space. We identify a portion of parameter space for which the parameters are natural (assuming the \ac{UV} cutoff is identified with $\MP$), and fine-tuning of the initial and final points of inflation is less than 10\%. Inside this region, the predicted tensor-to-scalar ratio is $ \left(4\times 10^{-2}\right)n_r\lesssim r \lesssim 4\,n_r$, with $n_r$ natural in the range $6.5\times 10^{-5}\lesssim n_r \lesssim 2\times 10^{-3}$.

Inflation ends as the waterfall field becomes tachyonic, and the Universe is reheated with a reheating temperature of the order of the inflationary scale, $T_{\rm RH}\sim{\cal O}(V_0^{1/4})$. A negative curvature of the potential triggers tachyonic instability of the waterfall field quantum fluctuations, which are exponentially enhanced. This leads to production of \ac{GW}s, which reach us as a \ac{SGWB} signal. The peak frequency of the \ac{GW} spectrum is typically as large as $\mathcal{O}\left(10^{10}\right)\,{\rm Hz}$, but it can be lowered by sufficiently decreasing the inflationary scale. We find that the signal of \ac{GW}s produced during tachyonic instability is within the reach of the \ac{ET} for an inflationary scale in the range $3\times 10^5\,{\rm GeV}\lesssim V_0^{1/4}\lesssim 5\times 10^{11}{\rm GeV}$. 
Such a low inflationary scale, despite pointing to some \ac{UV} completion well below $\MP$, is perfectly viable and is associated to an unobservably small tensor-to-scalar ratio. This confirms that \ac{GW}s from a tachyonic instability can act as a possible probe of low-scale inflationary scenarios.

The inflationary phase may be followed by a production of topological defects, depending on the specific symmetries which undergo \ac{SSB}. 
In a model variation with $\phi$ and $\chi$ charged under an additional $U(1)$ gauge symmetry, the vacuum manifold is such that \ac{CS}s are produced during the waterfall transition. The tension of the string network is bounded by \ac{CMB}, and this constrains the inflationary scale to be $V_0^{1/4}\lesssim$ a few $\times 10^{14}\,{\rm GeV}$. 
Another model variation features a softly-broken $\mathbb{Z}_4$ symmetry acting on $\phi$ and $\chi$. In this case, the vacuum manifold is disconnected, and unstable \ac{DW}s are generated at the end of inflation. They produce a \ac{SGWB}, which can fit the signal recently detected by the NANOGrav Collaboration \cite{NANOGrav:2023hvm}, albeit for an extremely low inflationary scale, $V_0^{1/4} \simeq 2\times 10^{5}\,{\rm GeV}$,  assuming order one quartic couplings.
The \ac{DW} annihilation signal can also be relevant for future detection by \ac{LISA} (\ac{ET}) for $3\times10^7\,{\rm GeV}\lesssim V_0^{1/4}\lesssim 2\times10^{10}\,{\rm GeV}$  ($3\times10^{10}\,{\rm GeV} \lesssim V_0^{1/4}\lesssim 7\times 10^{12}\,{\rm GeV}$). 

Upcoming experiments will soon be able to further constrain or rule out our models, for example, by measuring the sign of $n_r$, or by a discovery of a \ac{SGWB} by the \ac{LVK} network. Concerning future directions, it would be interesting to couple our models to the \ac{SM} in order to study reheating in detail. Finally, the accident mechanism could be applied to other realizations of inflation.
\newpage

\begin{subappendices}

\section{Details on the vacuum structure of the model of Sec.~\ref{sec:AccInf}}
\label{sec:AppendixDetails}

In this appendix, we collect some technical details concerning the minimization of the scalar potential in Eq.~\eqref{V5}, as well as the computation of the associated one-loop effective potential.

\subsection{Tree-level analysis}

Eq.~\eqref{V5} is the most general polynomial of degree $\leq 4$ in $\phi$ and $\chi$ which is compatible with $\SO{3}\times\mathbb{Z}_2$ invariance (where $\mathbb{Z}_2$ acts as $\phi\into -\phi$). To see this, note that each term must contain an even number of $\phi$ factors, and that the symmetric products between two $\phi$'s and between two $\chi$'s decompose as $({\mathbf{5}}\otimes {\mathbf{5}})_{\rm sym}={\mathbf{1}}\oplus{\mathbf{5}}\oplus{\mathbf{9}}$ and $({\mathbf{3}}\otimes{\mathbf{3}})_{\rm sym}={\mathbf{1}}\oplus{\mathbf{5}}$, respectively. 
To form a singlet bilinear, the only possibilities are $(\phi\phi)_{\mathbf{1}}$ and $(\chi\chi)_{\mathbf{1}}$. Cubic $\SO{3}$-invariants can be constructed as $(\phi\phi)_{\mathbf{5}}\phi$ and $(\chi\chi)_{\mathbf{5}}\phi$, but these are forbidden by the $\mathbb{Z}_2$ symmetry.  Invariant quartic terms are $(\phi\phi)_\mathbf{1}(\phi\phi)_\mathbf{1}$, $(\phi\phi)_\mathbf{5}(\phi\phi)_\mathbf{5}$ and $(\phi\phi)_\mathbf{9}(\phi\phi)_\mathbf{9}$, the latter two of which can be checked to be proportional to the first; $(\chi\chi)_\mathbf{1}(\chi\chi)_\mathbf{1}$ and $(\chi\chi)_\mathbf{5}(\chi\chi)_\mathbf{5}$, which are likewise proportional to each other; $(\chi\chi)_\mathbf{1}(\phi\phi)_\mathbf{1}$ and $(\chi\chi)_\mathbf{5}(\phi\phi)_\mathbf{5}$. There are, therefore, four independent quartic invariants, which may be represented by the four quartic terms in Eq.~\eqref{V5}.

Moreover, the largest continuous symmetry under which Eq.~\eqref{V5} is invariant is indeed $\SO{3}$. We have checked this explicitly, by considering the full $\SO{8}$ symmetry which acts on the $\phi$ and $\chi$ components in the absence of any potential.
We verified that only three $\SO{8}$ generators are preserved by $V$, and they form an $\SO{3}$ algebra. This is a non-trivial statement; cfr., e.g., the electroweak sector of the \ac{SM}, where the most general renormalizable potential for the Higgs doublet which is invariant under $\SU{2}\times\U{1}$ gauge transformations turns out to preserve a larger, ``custodial'' $\SO{4}$ symmetry.

To find the extrema of the tree-level potential of Eq.~\eqref{V5}, we will limit ourselves to the case of positive quartic couplings, which is the one of interest for our inflationary model. For definiteness, we use the $5$-plet generators
\be
\begin{array}{c}
T^1={\scriptsize{\left(\begin{array}{ccccc}&&&-i& \\ &&&&-i \\ &&&\sqrt{3} i& \\ i&&-\sqrt{3} i&& \\ &i&&&\end{array}\right)}},\,
T^2={\scriptsize{\left(\begin{array}{ccccc}&-i&&& \\ i&&\sqrt{3}i&& \\ &-\sqrt{3}i&&& \\ &&&&i \\ &&&-i&\end{array}\right)}},\,
T^3={\scriptsize{\left(\begin{array}{ccccc}&&&&-2i \\ &&&-i& \\ &&0&& \\ &i&&& \\ 2i&&&&\end{array}\right)}}\,.
\end{array}\ee
The critical points are most easily identified after using
$\SO{3}$ invariance to rotate $\chi\into (0,\, 0,\, \chi_3)$. The potential becomes
\be
V=-\frac{1}{2}\mu_\phi^2\phi^2-\frac{1}{2}\mu_\chi^2\,\chi_3^2+\frac{\lambda_\phi}{4}(\phi^2)^2+\frac{\lambda_\chi}{4}\chi_3^4+\frac{\varepsilon}{4}\phi^2\chi_3^2+
 \frac{\zeta}{4}\chi_3^2\,\phi^{\rm T}(T^3)^2\phi\,,
\ee
with $(T^3)^2={\rm diag}\,(4,1,0,1,4)$. Hence, the quartic terms are all positive definite, which shows, in particular, that the potential is bounded from below. We have
\be
\begin{split}
\frac{\partial V}{\partial\phi_A}=&\;
\left[-\mu_\phi^2 +\lambda_\phi\,\phi^2+\frac{\varepsilon}{2}\chi_3^2+\frac{\zeta}{2}\chi_3^2(T^3)^2_{AA}\right]
\phi_A\,,\\
\frac{\partial V}{\partial\chi_3}=&\;\left[-\mu_\chi^2+\lambda_\chi\,\chi_3^2+\frac{\varepsilon}{2}\phi^2
+\frac{\zeta}{2}\phi^{\rm T}(T^3)^2\phi\right]\chi_3\,.
\end{split}
\ee
The critical points are, thus, seen to be as follows.
\begin{itemize}

 \item $\phi=0$, $\chi=0$.---This is a local minimum if $\mu_\phi^2<0$ and $\mu_\chi^2<0$, a local maximum if both are positive (which is the case in our inflationary model), and a saddle point otherwise.
 
 \item $\chi=0$ and  $\phi^2=\mu_\phi^2/\lambda_\phi\equiv v^2$
 (assuming $\mu_\phi^2>0$).--- Therefore, this set of critical points forms a four-sphere. 
 The mass matrix does not mix $\phi$ and $\chi$, and the $\phi$ block is easily seen to have eigenvalues $(2\lambda_\phi v^2,0,0,0,0)$. For a generic point on the vacuum manifold, there are three Nambu-Goldstone directions, as expected for a completely broken $\SO{3}$ symmetry, and one additional, accidentally flat
 direction. However, the ${\bf 5}$-plet generators each have a zero eigenvalue, and, therefore, annihilate a $\phi$ \ac{VEV} oriented in the direction of the corresponding eigenvector. These special points in the vacuum manifold preserve an O$(2)\subset\SO{3}$ symmetry, and, thus, feature two Nambu-Goldstone fields and two accidentally massless fields. All such points are related to one another by $\SO{3}$ transformations, and, therefore, correspond to a single physical point when $\SO{3}$ is gauged.
 
 The $\chi$ block of the mass matrix is 
 \be
 \left.\frac{\partial^2 V}{\partial\chi_a\partial\chi_b}\right|_{\chi=0,\,\phi^2=v^2}=\left(\frac{\varepsilon}{2}v^2-\mu_\chi^2\right)\delta^{ab}+\frac{\zeta}{2}\phi_A T^a_{AC} T^b_{CB}\phi_B\,.
 \ee
The last term is diagonalized by an $\SO{3}$ transformation which sends $\phi_{2,4,5}\into 0$, with eigenvalues $\frac{\zeta}{2}(\phi_1\pm\sqrt{3}\phi_3)^2$ and $2\zeta\phi_1^2$. The $\chi$ masses are, therefore, positive for $\mu_\chi^2<0$, but for $\mu_\chi^2>0$ the sign of the mass terms depends on the direction of the $\phi$ \ac{VEV}, as detailed in the main text. This is the field configuration realized on the inflationary trajectory.

\item $\phi=0$ and $\chi^2 = \mu_\chi^2/\lambda_\chi \equiv v_\chi^2 $ (assuming $\mu_\chi^2>0$).---The mass matrix in the $\chi$ sector has eigenvalues $(2\lambda_\chi v_\chi^2,0,0)$, with two Nambu-Goldstone directions as expected for $\SO{3}\into\SO{2}$ breaking by a triplet \ac{VEV}. The mass matrix in the $\phi$ sector is diagonal in a basis where $\chi_{1,2}=0$ and $\chi_3=v_\chi$:
 \be
  \left.\frac{\partial^2 V}{\partial\phi_A\partial\phi_B}\right|_{\chi=\begin{psmallmatrix} 0\\0\\v_\chi\end{psmallmatrix},\, \phi=0}=\bigg[\frac{\varepsilon}{2}v_\chi^2-\mu_\phi^2+\frac{\zeta}{2}v_\chi^2(T^3)^2_{AA}\bigg]\delta_{AB}\,.
 \ee
 Therefore, this critical point is a minimum provided that $\frac{\varepsilon}{2}v_\chi^2>\mu_\phi^2$. It is the true vacuum of our inflationary model, by a choice of parameter values.
 
 \item There are also critical points with both $\phi$ and $\chi$ non-zero. In a basis where $\chi=(0,\,0,\,\chi_3)$, one finds that this requires either $\phi_{1,2,4,5}=0$ or $\phi_{1,3,5}=0$ or $\phi_{2,3,4}=0$. Explicit expressions for the expectation values of the other fields can be derived, but we refrain from giving them here, since these critical points are saddle points for the parameter choices of interest.
\end{itemize}

\subsection{One-loop corrections}

At the one-loop level, the tree-level scalar potential is corrected by the well-known Coleman-Weinberg effective potential
\be
V_{\rm eff}=\frac{1}{64\pi^2}\,\tr\,\left({\cal M}^4\log\frac{{\cal M}^2}{\Lambda^2}\right)\,.
\ee
Here, ${\cal M}$ is the scalar-field-dependent mass matrix for gauge fields and scalars (with a factor 3 for the former, accounting for three polarizations for a massive vector field) and $\Lambda$ is the renormalization scale. 

We are interested, notably, in the loop-corrected effective potential for a field configuration with \ac{VEV}s $\phi^2=v^2$ and $\chi=0$, along the direction which remained accidentally flat at the tree level. This direction was parameterized by an angle $\alpha\in[0,\,\pi/6]$ in the main text; see Eq.~(\ref{eq:alpha}). Evaluating $V_{\rm eff}$ at some fixed value of $\alpha$, $\bar{\alpha}=\pi/6$, say, we find that there is a divergent tadpole term for the radial mode $\rho$, as well as a divergent contribution to the vacuum energy density. These divergences need to be subtracted by including counterterms for all parameters of the model and imposing suitable renormalization conditions. A particularly convenient set of conditions is to (i) renormalize the $\rho$ tadpole to be zero, such that $v^2\equiv m_\phi^2/\lambda_\phi$  remains the \ac{VEV} after including one-loop corrections (this is an implicit condition on the counterterms for $m^2_\phi$ and $\lambda_\phi$); and (ii) set the renormalized vacuum energy density to some finite constant. There are other divergences in the one-loop effective action, but these two conditions suffice to render the effective potential along the $\alpha$ direction finite. It is given by \cite{Brummer:2023znr}
\be\begin{array}{l}
V_{\rm eff}(\alpha)=\dfrac{1}{64\pi^2}\times
\\
\times\Biggl[\tr\left( M^4(\alpha)\log \dfrac{M^2(\alpha)}{\Lambda^2}-M^4(\bar{\alpha})\log \dfrac{M^2(\bar{\alpha})}{\Lambda^2}\right)\\
+3\,\tr \left( M_W^4(\alpha)\log \dfrac{M_W^2(\alpha)}{\Lambda^2}- M_W^4(\bar{\alpha})\log \dfrac{M_W^2(\bar{\alpha})}{\Lambda^2}\right)\Biggr]
\,,
\end{array}\ee
where $M$ and $M_W$ are the scalar and gauge boson mass matrices, respectively, and the $\Lambda$ dependence is, in fact, spurious and cancels out, $\tr M^4$ and $\tr M_W^4$ being $\alpha$-independent. Note that the choice of the subtraction point $\alpha=\bar{\alpha}=\pi/6$ is arbitrary: We could have chosen any other point provided that the tree-level mass matrix is positive definite, so that the effective potential is well-defined. The gauge symmetry leads to the identifications $\alpha\into -\alpha$ and $\alpha\into\alpha+\pi/3$; therefore, the effective potential is even and periodic. As such, $V_{\rm eff}(\alpha)$ can be expanded into a Fourier cosine series, $V_{\rm eff}(\alpha)=\sum_n c_n \cos(6n\alpha)$. The leading Fourier mode of $V_{\rm eff}$ is then found to be given by the expressions of Eqs.~\eqref{eq:CosPot} and \eqref{eq:M4}, and it provides a slow-roll potential for the inflaton field.

\section{A model with complex scalars, giving rise to cosmic strings}
\label{sec:AppendixComplexModel}

In Sec.~\ref{sec:AccInf}, we presented a minimal model with two real scalar $\SU{2}$ multiplets, a five-plet $\phi$ and a triplet $\chi$. Here, we take these fields to be complex, charged under additional \U{1} symmetries.

Let $\phi$ be a complex scalar transforming in the {\bf 5} of \SU{2} with unit $\U{1}_{\phi}$ charge. The most general renormalizable potential is
\be\label{V5Compl}
 V_\phi=-\mu_{\phi}^2\,\phi^\dag\phi+\lambda \,(\phi^\dag\phi)^2+ \kappa
 \left[(\phi^\dag\phi)^2-|\phi^T\phi|^2\right]+ \delta\, (\phi^\dag T^a\phi)^2\,,
\ee
with the generators $T^a$ defined below Eq.~\eqref{V5}.
A detailed study of this potential can be found in Ref.~\cite{Brummer:2023znr}.\footnote{The quartic couplings $\lambda,\kappa$, and $\delta$ here are normalized to be a factor 2 smaller than in Ref.~\cite{Brummer:2023znr}.} The quartic couplings $\lambda$, $\kappa$, and $\delta$ and the mass parameter $\mu_\phi^2$ are chosen positive. Notice that the $\kappa$ and $\delta$ terms vanish when $\phi$ is real.
Cubic terms are forbidden by $\U{1}_{\phi}$, with no need for ad-hoc discrete symmetries. The minimum of the potential is at $\vev{\phi^\dag\phi}=v^2/2$, where $v^2=\mu_\phi^2/\lambda$. When $\SU{2}\times\U{1}_\phi$ is gauged, all \ac{NGB}s are absorbed by the gauge bosons. However, there remains a continuous family of vacua at the tree level, parameterized by an angle $\alpha\in[0,\pi/6]$ as in Eq.~\eqref{eq:alpha}. 
Thus, $\alpha$ is an accidentally flat direction. For generic $\alpha$, all of the gauge symmetry is broken, but, at the special point $\alpha=0$, a subgroup $\U{1}'\subset\SU{2}$ remains unbroken. At this point, there is an additional massless scalar in the spectrum, besides the one corresponding to the flat direction.

In the limit $\delta \rightarrow 0$, the potential recovers an enlarged $\SO{5}\times \U{1}_{\phi}$ symmetry, spontaneously broken to $\SO{4}$. Switching on $\delta$ explicitly breaks the symmetry of the potential to $\SU{2}\times \U{1}_{\phi}$, so that the continuous symmetry of $V$ is strictly no larger than the gauge symmetry: Therefore, the accidentally light mode is not a \ac{pNGB} in the sense of Ref.~\cite{Weinberg:1972fn}.

Loop corrections lift the accidentally flat direction, analogously to the minimal model in Sec.~\ref{sec:AccInf}, with $\alpha=0$ being selected as the true vacuum. The one-loop effective potential along the tree-level flat direction takes the form $V_{\rm eff}\simeq V_0' -\left(M'\right)^4 \cos \left(a/f\right)$ (up to higher harmonics whose coefficients are numerically small for all values of the couplings of interest). The amplitude of the potential is given by 
\be\label{eq:M4Pr}
\left(M'\right)^4 =\frac{v^4}{160\,\pi^2}\Bigg[9\,g^4-\frac{\kappa^5}{\delta^3} \Bigg(F\left(-\frac{2\delta}{\kappa}\right)-T_{F,5}\left(-\frac{2\delta}{\kappa}\right)\Bigg)\Bigg]\,,
\ee
where $g_2$ is the $\SU{2}$ gauge coupling, and $F(x)$ and $T_{F,5}$ are defined below Eq.~\eqref{eq:M4}.
The tree-level massless scalar picks up a mass which is one-loop suppressed with respect to the masses of the other fields.

Now add to the model a complex scalar $\chi$ transforming in the ${\bf 3}$ of $\SU{2}$ with unit $\U{1}_\chi$ charge, which will play the role of the waterfall field. 
The most general renormalizable scalar potential now also includes the terms
\be
\begin{aligned}
V_\chi=&\;-\mu^2_\chi\chi^\dag\chi
+\lambda_\chi\,(\chi^\dag\chi)^2+\lambda'_\chi\,|\chi^T\chi|^2\,,
\\
V_{\phi\chi}=& \;\varepsilon\,\phi^\dag\phi\chi^\dag\chi 
+\zeta\,T^a_{AB} T^b_{BC}\phi^*_A \phi_C \chi^{*a}\chi^b+\vartheta\,T^a_{AB}(i\varepsilon^{abc})\phi^*_A \phi_B\chi^{*b}\chi^c\,.
\label{Vchi}
\end{aligned}
\ee
Notice that one can replace
the phase rotations $\U{1}_{\phi}\times \U{1}_\chi$ by a single gauged $\U{1}$, with charges $q_\phi$ and $q_\chi$ chosen to forbid additional mixed couplings, beside $\epsilon,\zeta$, and $\vartheta$.
In this case, by construction, the scalar potential has an accidental global $\U{1}_{\phi}\times\U{1}_\chi$
symmetry. As a consequence, when $\langle \chi \rangle \ne 0$ is generated after inflation, one produces \ac{CS}s with observable consequences, as discussed in Sec.~\ref{sec:TopologicalDefects}.

For $\mu^2_\chi<0$ and positive quartic couplings in Eq.~\eqref{Vchi}, the tree-level potential is easily seen to be minimized at $\chi=0$, with the tree-level vacuum structure for $\phi$ unchanged. In fact, when $\phi$ is set to its tree-level minimum of Eq.~\eqref{eq:alpha}, the terms $\mu^2_\chi$ and $\varepsilon$ give rise to a universal, $\alpha$-independent mass term for the three components of $\chi$, while the $\zeta$ term splits their masses in an $\alpha$-dependent fashion (and the $\vartheta$ term does not contribute to the mass matrix). 
The mass eigenvalues for the $\chi$ fields are the same as in Eq.~\eqref{eq:mchi}.
To obtain a model of hybrid inflation, we choose $\tilde\mu_\chi^2\equiv\mu_\chi^2-\frac{\varepsilon}{2}v^2>0$, and $\zeta v^2>2\tilde\mu_\chi^2$, such that the overall $\chi^3$ mass-squared is positive for $\alpha=\pi/6$ but negative for $\alpha=0$. At the intermediate value $\alpha=\alpha_c$ (given in Eq.~\eqref{eq:alphac}),
the effective $\chi^3$ mass crosses zero.

For suitable values of the couplings, the true vacuum of the model is located at
\be
\langle\lvert\chi^3\rvert\rangle=\sqrt{\frac{\mu^2_\chi}{2\left(\lambda_\chi+\lambda_\chi'\right)}}\,,\qquad
\text{all other fields} = 0
\ee
up to gauge transformations. The residual gauge symmetry is $\U{1}'\subset \SU{2}$, and all fields except the unbroken gauge boson pick up tree-level masses.
This holds for $\lambda_\chi'>0$, while for $-\lambda_\chi <\lambda_\chi'<0$, the \ac{VEV} of $\chi$ is aligned differently, and it breaks $\SU{2}$ entirely.

Loops of $\chi$ contribute to the accident effective potential so that Eq.~\eqref{eq:M4Pr} becomes
\be\label{eq:M4Przeta}
\begin{split}
\left(M'\right)^4 =\frac{v^4}{160\,\pi^2}\Bigg[&9\,g_2^4-\frac{\kappa^5}{\delta^3}\left(F\left(-\frac{2\delta}{\kappa}\right)-T_{F,5}\left(-\frac{2\delta}{\kappa}\right)\right)\\&+4\,\frac{\tilde\mu_\chi^{10}}{\zeta^3\,v^{10}}\left({\rm Re}\;F\left(\frac{\zeta v^2}{\tilde\mu_\chi^2}\right)-T_{F,5}\left(\frac{\zeta v^2}{\tilde\mu_\chi^2}\right)\right)\Bigg]\,.
\end{split}
\ee
Note that the ${\rm Re}\,F$ term in the second line vanishes for the parameter region of physical interest $\tilde\mu_\chi^2>0$.

The calculation of \ac{CMB} observables in this model is done in the same way as outlined in Sec.~\ref{sec:CMB}, upon replacing $M\into M'$ and $V_0\into V_0'$.

\section{A model with disconnected minima, giving rise to unstable domain walls}
\label{sec:AppendixZ4Model}

In this variant of the model, the potential has two physically distinct minima which can be chosen to be near-degenerate.
Let $\phi$ be a real five-plet of $\SO{3}$ which is odd under $\mathbb{Z}_{2\phi}:\phi\to-\phi$, and add to the model a pair of triplet waterfall fields $\chi_R$ and $\chi_I$; we represent them as a single complex field $\chi=(\chi_R+i\chi_I)/\sqrt{2}$, although no $\U{1}$ symmetry is implied. We do, however, impose a $\mathbb{Z}_{4\chi}$ symmetry acting as $\chi \rightarrow i \chi$. 
With these fields and symmetries, the most general renormalizable potential reads
\be
\begin{aligned}
 V=&-\frac{1}{2}\mu_\phi^2\phi^2+\frac{\lambda_\phi}{4}(\phi^2)^2-\mu_\chi^2\chi^* \chi +\lambda_\chi\left(\chi^*\chi\right)^2+\delta\,{\chi^*}^2\chi^2
 + \dfrac 12\left(\kappa\, \chi^2 \chi^2+{\rm h.c.}\right)\\
 &+\frac{\varepsilon}{2}\phi^2\left(\chi^*\chi\right)+
 \frac{\zeta}{2}\,T^a_{AC} T^b_{CB}\phi_{A} \phi_B {\chi^{ a}}^*\chi^{ b}\,,
\end{aligned}
\label{VZ4}\ee
where $T^a$ are the $\SO{3}$ five-plet generators, chosen as $5 \times 5$ imaginary and anti-symmetric matrices. For simplicity, we assume $\kappa$ real, which corresponds to a CP invariance $\chi\leftrightarrow\chi^*$, and all the quartic couplings to be positive. 

Let us first consider the case $\mu_\phi^2>0$ and $\mu_\chi^2<0$. The minimum is at $\langle \phi^2 \rangle =v^2\equiv\mu_\phi^2/\lambda_\phi$, $\langle \chi \rangle=0$, where the mass spectrum resembles the one of the minimal model of Sec.~\ref{sec:AccInf}. After the spontaneous breaking of $\SO{3}$, one observes the emergence of an accidentally flat direction parametrized by Eq.~\eqref{eq:alpha}. The masses of the complex $\chi$ components are given by Eq.~\eqref{eq:mchi}. 
One-loop corrections give rise to the usual potential along the accidentally flat direction $V_{\rm eff} = V_0'' - \left(M''\right)^4\cos\left(a/f \right)$, with $M''$ given by
\be
\left(M''\right)^4 
=\frac{1}{160\,\pi^2}\Bigg[9\,g^4\,v^4+4\,\frac{\tilde\mu_\chi^{10}}{\zeta^3\,v^6}
\Bigg({\rm Re}\;F\left(\frac{\zeta v^2}{\tilde\mu_\chi^2}\right)-T_{F,5}\left(\frac{\zeta v^2}{\tilde\mu_\chi^2}\right)\Bigg)\Bigg]\,,
\ee
and $F$ and $T_{F,5}$ defined below Eq.~\eqref{eq:M4}.

We now switch to the case interesting for inflation: $\mu_{\phi}^2>0$ and $\mu_{\chi}^2>0$. The $\chi_3$ mass is positive at the beginning of inflation, close to $\alpha = \pi/6$, and it crosses $0$ at $\alpha = \alpha_c$, given by Eq.~\eqref{eq:alphac}.
With this choice of couplings, the potential has a minimum at $\langle\phi\rangle=0$ and $\langle\chi\rangle\neq0$. The six equations to solve in order to find the extrema of the potential are 
\be
\begin{aligned}
    \chi_R^a\left[-\mu_{\chi}^2 + \lambda_\chi \left(\chi_R^2+\chi_I^2 \right)+(\kappa + \delta)(\chi_R^2 -\chi_I^2)\right]-2\chi_I^a \left(\chi_R\cdot\chi_I\right)\left(\kappa-\delta\right)&=0\,,\\
   \chi_I^a\left[-\mu_{\chi}^2 + \lambda_\chi \left(\chi_R^2+\chi_I^2 \right)-(\kappa + \delta)(\chi_R^2 -\chi_I^2)\right]-2\chi_R^a \left(\chi_R\cdot\chi_I\right)\left(\kappa-\delta\right)&=0\,,
\end{aligned}
\ee
for $a=1,2,3$. It is convenient to employ $\SO{3}$ invariance to set $\chi^1_R =\chi^2_R=0$ as well as 
$\chi^1_I=0$. Then, the system reduces to
\be
\begin{aligned}
    &\chi^2_I \chi^3_I \chi^3_R =0
    \,,\\
    &\chi_{I}^2\left[-\mu_\chi^2+\lambda_\chi
    [(\chi_R^3)^2+(\chi_I^2)^2+(\chi_I^3)^2]
    -(\kappa+\delta)
    [(\chi_R^3)^2-(\chi_I^2)^2-(\chi_I^3)^2]\right]=0
    \,,\\
    &\chi_{R}^3\left[-\mu_\chi^2+\lambda_\chi
    [(\chi_R^3)^2+(\chi_I^2)^2+(\chi_I^3)^2]
    +(\kappa+\delta)
    [(\chi_R^3)^2-(\chi_I^2)^2]
    - (3\kappa-\delta)(\chi_{I}^3)^2\right]=0
    \,,\\
     &\chi_{I}^3\left[-\mu_\chi^2+\lambda_\chi
     [(\chi_R^3)^2+(\chi_I^2)^2+(\chi_I^3)^2]
     +(\kappa+\delta)
     [(\chi_I^2)^2+(\chi_I^3)^2]
     - (3\kappa-\delta)(\chi_{R}^3)^2\right]=0\,.
\end{aligned}
\ee

For a certain hierarchy among the quartic couplings
($0<\delta<\kappa<\lambda_\chi+\delta < \varepsilon/2+\kappa$), the potential is minimized at
\be\label{eq:appBvacuum}
(\chi_R^3)^2 =(\chi_I^3)^2 = w^2 \equiv \frac{\mu_{\chi}^2}{2
\left(\lambda_\chi+\delta-\kappa\right)}\,,
\quad \chi_I^2=0\,.
\ee
In these minima, the symmetry is spontaneously broken as $\SO{3}\times \mathbb{Z}_{4\chi}
\to \SO{2}\times \mathbb{Z}_{2\chi}$. 
There are four solutions to Eq.~\eqref{eq:appBvacuum} but only two physically distinct vacua: one $(+)$ where the $\chi_R$ and $\chi_I$ \ac{VEV}s are parallel, $\chi_R^3 = \chi_I^3 = \pm w$, and the other $(-)$ where they are antiparallel, $\chi_R^3 =-\chi_I^3 = \pm w$. 
Thus, the vacuum manifold has two disconnected components, and \ac{DW}s can be produced during the \ac{SSB} phase transition, as discussed in Sec.~\ref{sec:TopologicalDefects}.

For a viable phenomenology, the \ac{DW} network needs to be unstable, as discussed in Sec.~\ref{secDW}. We, therefore, assume that $\mathbb{Z}_{4\chi}$ is softly broken by a term $V_{\rm soft}=i m_\chi^2 \chi\chi + {\rm h.c.}$, with $m_\chi^2$ real. This term preserves $\mathbb{Z}_{2\chi}:\chi\to-\chi$, while it breaks the CP symmetry $\chi\to\chi^*$. Consequently, it breaks the degeneracy between the two disconnected vacua ($+$) and ($-$), introducing a bias in the potential. We assume that soft $\mathbb{Z}_{4\chi}$ breaking is a small effect,  $m_\chi^2 \ll \mu_{\chi}^2$. Then the potential bias is, to leading order in $m_\chi^2$,
\be\label{eq:Vbias}
 \Delta V \equiv V_{(+)} - V_{(-)}= - \frac{2\,m_\chi ^2\mu_{\chi}^2}{\lambda_\chi+\delta-\kappa} + \mathcal{O}\left(m_\chi^4\right)\,,
\ee
where $V_{(+)}$ and $V_{(-)}$ are the potential energies evaluated at the minimum configuration in which $\chi_R$ and $\chi_I$ are parallel and anti-parallel, respectively.

\end{subappendices}

\chapter{Monopole dark matter}
\label{chap:MonoDM}
\minitoc

The magnetic monopole of a dark sector has been advocated as an appealing dark matter candidate. We revisit the computation of the monopole abundance $\Omega_M$, generated by a thermal phase transition in the minimal 't Hooft-Polyakov model.
We explore the three regimes where the phase transition is second order, weakly first order, or supercooled, identifying the parameter space regions where $\Omega_M$ can match the observed dark matter abundance.
However, the dark sector necessarily contains a stable electrically-charged particle, namely a massive vector boson, with a calculable abundance $\Omega_{W'}$.
We show that, 
under minimal assumptions,
$\Omega_{W'}$ is always far larger than $\Omega_M$:
dark monopoles cannot constitute a sizeable fraction of dark matter.

\section{Introduction}

A gauge interaction is said to belong to a dark sector if none of the \ac{SM} fields are charged under it. Dark-sector gauge symmetries may be broken by the Higgs mechanism, and can thus give rise to dark-sector magnetic monopoles. These are stable topological defects involving non-trivial gauge field and Higgs field configurations. At low energies, they effectively behave as massive classical particles. A detailed overview of monopoles is presented in Sec.~\ref{sec:Mono}.

Since monopoles are stable, they will contribute to the \ac{DM} of the Universe. Historically, this was regarded as a problem for \ac{GUT}s, as the breaking of the unified gauge group to the \ac{SM} gauge group would abundantly produce superheavy magnetic monopoles of ordinary electromagnetism, and thus overclose the Universe \cite{Preskill:1979zi}. This problem is famously solved by inflation after the grand-unified \ac{PT} \cite{Linde:1981mu}. If, however, the monopoles are part of a dark sector, the symmetry breaking scale can be much lower than the unification scale. Then the monopole abundance need not lead to overclosure, even if inflation takes place before the \ac{PT}. In fact, being stable massive particles, dark-sector monopoles might well account for all or part of the observed \ac{DM}.

Monopole \ac{DM} is an unusual scenario both conceptually, since the stabilizing symmetry is of topological origin, and technically, since the relevant production mechanisms are quite different from ordinary particle \ac{DM}. This makes it an interesting object for study. The possibility that all of the \ac{DM} consists of monopoles, produced via a thermal \ac{PT} in the early Universe, has been analysed in Refs.~\cite{Murayama:2009nj,GomezSanchez:2011orv,Baek:2013dwa,Khoze:2014woa,Kawasaki:2015lpf,Sato:2018nqy,Daido:2019tbm,Graesser:2020hiv,Yang:2022quy}. Monopoles produced during the stage of preheating have been studied in Ref.~\cite{Bai:2020ttp}.

The minimal dark sector model featuring monopoles with calculable properties has an $\SO{3}$ gauge group broken to $\SO{2}$ by a Higgs triplet \ac{VEV} \cite{Georgi:1972cj}, leading to 't Hooft-Polyakov monopoles \cite{tHooft:1974kcl, Polyakov:1974ek}. In this model there exists another stable \ac{DM} candidate, namely a massive $W'$ gauge boson which is the lightest state carrying electric charge under the unbroken symmetry. Indeed, in any model featuring dark sector monopoles, the lightest charged state will be stable. It is clearly an interesting question whether the \ac{DM} abundance can be dominated by monopoles, rather than by dark-sector charged elementary particles.

In this chapter, we will show that the answer to this question is negative for the minimal 't Hooft-Polyakov dark monopole, produced during a thermal \ac{PT}. With rather generic assumptions (essentially, we demand that the dark sector and the \ac{SM} were in thermal equilibrium at some point, due to a sizeable Higgs portal coupling) we show that the thermal $W'$ abundance always exceeds the monopole abundance, regardless of the order or the energy scale of the dark-sector \ac{PT}. This is in contrast with the results of previous studies \cite{Khoze:2014woa} which we extend and improve. To avoid this conclusion, one should consider extensions of the minimal model \cite{bigpaper}.

\section{Dark sector monopoles}\label{sec:Model}

\subsection{Model and mass spectrum}\label{ssec:Model}

Let $G=\SO{3}$ be a dark-sector gauge group and $\phi$ be a real scalar $\SO{3}$ triplet. The most general renormalizable potential for $\phi$ reads
\begin{equation}\label{eq:Pot}
    V(\phi) = -\frac{\mu^2}{2} \phi^2 + \frac{\lambda}{4} (\phi^2)^2+\frac{\lambda_{\phi H}}{2}\,\phi^2\,|H|^2\,.
\end{equation}
Here $H$ is the \ac{SM} Higgs doublet. For $\lambda>0$ and $\mu^2>0$, and neglecting the Higgs portal term for the time being, the potential is minimized at
\begin{equation}\label{eq:vev}
\vev{\phi^2}=\frac{\mu^2}{\lambda}\equiv\eta^2\,,
\end{equation}
and $\SO{3}$ is broken to $\SO{2}$. There is a massive scalar radial mode $\rho$ with $m_\rho^2 = 2\lambda \eta^2$. The two would-be \ac{NGB}s are absorbed by two of the gauge bosons $W'^\pm$, which become massive with $m_{W'}^2 = g^2 \eta^2$, where $g$ is the dark gauge coupling.  A third vector boson $\gamma'$, associated to the unbroken $\SO{2}$, remains massless.

The second homotopy group of the vacuum manifold is non-trivial, $\pi_2\left[\SO{3}/\SO{2}\right] = \mathbb{Z}$. The model therefore features stable monopole configurations, famously constructed in Refs.~\cite{tHooft:1974kcl, Polyakov:1974ek}. Monopoles are produced if the $\SO{3}$ symmetry is restored in the early Universe by thermal effects, and broken later as the Universe cools down.
The monopole $M$ with unit winding number has mass $m_M = c\times 4\pi \eta/g$, where $c$ depends on the ratio $\lambda/g^2$ with $1\leq c\lesssim 1.8$ \cite{Kirkman:1981ck}.
It has magnetic charge $q_M = 4\pi/g$, no electric charge (for simplicity, we assume a vanishing theta term for the dark gauge fields), and spin zero.  The monopole core radius $r_M$ scales as $r_M \sim (g \eta)^{-1}$.
For perturbatively small values of $g$,  $r_M$ is much larger than the monopole Compton wavelength $\lambda_M\equiv 1/m_M \le (g^2/4\pi) (g\eta)^{-1}$; therefore the monopole can be treated as a classical object.

The mass spectrum of the dark sector can be sketched as follows (for some arbitrary values of 
$\lambda$ and $g$):
{\small
\begin{align}\nonumber\label{eq:MassSketch}
	\begin{tikzpicture}
	 {\scalebox{1}{
    \draw[->,>=Latex][thick] (-7,0)--(7.,0);
    \draw[thick][thick] (-7,0.1)--(-7,-0.1);
    \node at (-7,-0.4) {\scalebox{1}{$0$}};
    \node at (-7,+0.4) {\scalebox{1.2}{$\gamma'$}};
    \draw[thick][thick] (-2,0.1)--(-2,-0.1);
    \node at (-2,+0.4) {\scalebox{1.2}{$\rho$}};
    \node at (-2,-0.4) {\scalebox{1}{$\sqrt{2\lambda}\,\eta$}};
    \draw[thick][thick] (0,0.1)--(0,-0.1);
    \node at (0,+0.4) {\scalebox{1.2}{$W'$}};
    \node at (0,-0.4) {\scalebox{1}{$g\eta$}};
    \draw[thick][thick] (5.5,0.1)--(5.5,-0.1);
    \node at (5.5,+0.4) {\scalebox{1.2}{$M$}};
    \node at (5.5,-0.4) {\scalebox{1}{$(4\pi/g)\eta$}};
    }}
	\end{tikzpicture}
\end{align}
}

This model contains two \ac{DM} candidates. The monopole is stable because it is the lightest object carrying a conserved topological charge, which can be identified with the $\SO{2}$ magnetic charge.  The massive dark gauge boson $W'$ is stable due to $\SO{2}$ electric charge conservation. Our goal is to study under which conditions monopoles can dominate the \ac{DM} abundance, assuming that they were created during a thermal \ac{PT}.

\subsection{Thermal effective potential and phase transitions}

We will assume that, in the early Universe, the dark sector and the \ac{SM} are thermalised via the Higgs portal coupling $\lambda_{\phi H}$, and that the reheating temperature is larger than the $\SO{3}$-symmetry breaking scale $\eta$. Thermal effects then correct the zero-temperature potential in Eq.~\eqref{eq:Pot} and lead to symmetry restoration, as discussed in Sec.~\ref{sec:TQFT}. As the Universe cools down, a symmetry-breaking \ac{PT} occurs at some critical temperature $T_c$. The Universe eventually settles in the symmetry-breaking vacuum, either smoothly in the case of a \ac{SOPT}, or via bubble nucleation in the case of a \ac{FOPT}. In either case, dark monopoles are produced.

If the quartic and gauge couplings are perturbatively small, which we assume throughout, one may attempt to compute the effective scalar potential in perturbation theory. However, it is well known that the loop expansion is useful only in a limited temperature range \cite{Dolan:1973qd}; in particular, $\lambda \ll g^2$ is required to trust the perturbative result close to $T_c$, and thus to make a reliable statement on the nature of the \ac{PT}. As discussed in Sec.~\ref{sec:VThermScal}, the one-loop thermal corrections to Eq.~\eqref{eq:Pot} are given by \cite{Dolan:1973qd}
\be\label{eq:ThermalPot}
    \Delta V_T(\phi) = \frac{T^4}{2\pi^2} \sum_i n_i \int_0^{\infty} dq\,q^2 \log\left[1-\exp\left(-\sqrt{q^2+\frac{m_i^2(\phi)}{T^2}}\right)\right]\,.
\ee
Here the sum runs over all fields with tree-level couplings to $\phi$, with $n_i$ degrees of freedom and effective squared mass $m_i^2(\phi)$ in the background of $\phi$.  Zero-temperature, one-loop corrections are given by the usual \ac{CW} formula
\be
    \Delta V_{\rm 1\text{-}loop}(\phi) = \sum_i \frac{n_i}{64\pi^2}\tr \bigg[m_i^4(\phi) \log \frac{m_i^2(\phi)}{\Lambda^2}\bigg]\,,
\ee
where $\Lambda$ is a suitable renormalisation scale.
The one-loop effective potential at zero temperature is reviewed in Sec.~\ref{sec:EffPot}. All in all, the potential in Eq.~\eqref{eq:Pot} becomes
\be\label{eq:FullPot}
    V(\phi,T) = V_{\rm tree}(\phi) + \Delta V_{\rm 1\text{-}loop}(\phi) + \Delta V_T(\phi)\,.
\ee

\begin{figure}[b!]
\begin{center}
\includegraphics[width=.7\textwidth]{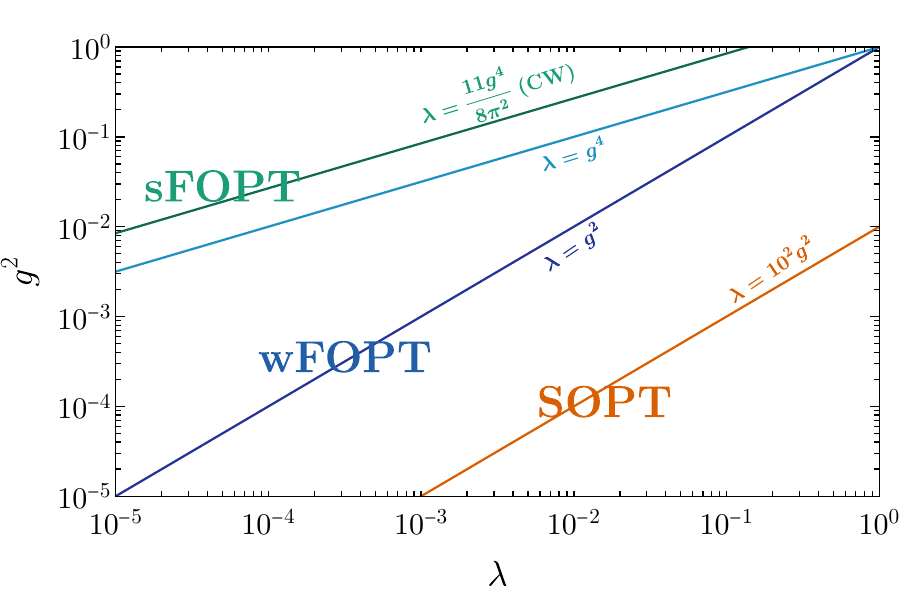}\quad 
    \caption{ The nature of the \ac{PT} depends on the choice of the couplings $\lambda$ and $g^2$. Above the dark-blue line $\lambda=g^2$, finite-temperature perturbation theory can be used to establish that the transition is of the first order \cite{Arnold:1992rz}. It is weakly first-order between the dark-blue and light-blue lines, which indicate the region amenable to a high-temperature expansion, and becomes progressively stronger towards smaller $\lambda/g^2$. Somewhere in the non-perturbative region $\lambda> g^2$, the \ac{FOPT} is expected to turn into a second-order one or a crossover. The green line indicates the \ac{CW} scenario, taken as an exemplary case for a \ac{FOPT} which is strong. The orange line indicates a benchmark scenario for a \ac{SOPT}.
    }
\label{fig:PhDiag}  
\end{center}
\end{figure}

The nature of the \ac{PT} depends on the relative size of $\lambda$ and $g$, as shown in Fig.~\ref{fig:PhDiag}. We now study three exemplary cases, and comment on the general case at the end.

\subsubsection{The high-temperature expansion and a weakly first-order phase transition}

For $\lambda\ll g^2$, finite-temperature perturbation theory can be used to establish that the \ac{PT} is of the first order \cite{Dolan:1973qd}. If, moreover, $\lambda\gg g^4$, then the one-loop thermal effective potential close to $T_c$, can be computed in the high-temperature expansion  \cite{Dolan:1973qd, Anderson:1991zb, Dine:1992wr, Arnold:1992rz}:
\be\label{eq:PotHighT}
 V(\phi,T)\simeq \frac{1}{2}\left(\frac{g^2}{2}\,T^2-\lambda\eta^2\right)\phi^2-\frac{g^3}{2\pi}T|\phi|^3+\frac{\lambda}{4}(\phi^2)^2\,.
\ee
Here, the zero-temperature \ac{VEV} $\eta$ is given by Eq.~\eqref{eq:vev}. We have neglected zero-temperature loop corrections and scalar loops. The latter are subdominant under the assumption $\lambda\ll g^2$. It is possible to resum the leading higher-loop ``daisy'' diagrams in order to obtain an improved expression for the effective potential \cite{Dolan:1973qd,Parwani:1991gq, Arnold:1992rz} but, for our purposes, including the finite-temperature corrections at the level of Eq.~\eqref{eq:PotHighT} is sufficient.

As the Universe cools, the symmetry-preserving point $\phi=0$ becomes metastable at a critical temperature 
\be
T_c \simeq T_0\left(1+\frac{g^4}{2\pi^2\lambda}\right)\,,
\ee
where
\be
T_0=\sqrt{\frac{2\,\lambda\eta^2}{g^2}}
\label{T0wFOPT}\ee
is the temperature where the point $\phi=0$ becomes classically unstable. 
The field expectation value in the symmetry-breaking vacuum is\footnote{Note that $\phi_c$ is not to be confused with the ``classical'' field appearing in the effective potential formalism presented in Sec.~\ref{sec:EffPot}.}
\be
\phi_c^2=\frac{2g^4}{\pi^2\lambda}\eta^2\,.
\ee

In the parameter region where the high-temperature approximation for the effective potential is valid (i.e.~for $g^4\ll \lambda$), the \ac{PT} is \emph{weakly} first-order, since bubbles start nucleating at a temperature $T_n$ immediately below $T_c$. Indeed, as we will show in more details in Sec.~\ref{sec:MonoFOPT}, one has $T_0 < T_n <T_c$, with $T_0 \simeq T_c$.

\subsubsection{The Coleman-Weinberg scenario: a strongly first-order phase transition}\label{sec:CW}
In the \ac{CW} scenario \cite{Coleman:1973jx}, at zero temperature, renormalisation conditions are chosen such that the second derivative of the one-loop effective potential vanishes at the symmetry-preserving point, $\phi=0$, and symmetry breaking is radiatively induced.  This requires that, in the model of Sec.~\ref{ssec:Model},
\be
\lambda=\frac{11\,g^4}{8\pi^2}\,,
\ee
where $\lambda$ and $g$ are understood to be the renormalised quartic couplings, defined at the symmetry-breaking minimum. 

At finite temperature, it is well known that the \ac{CW} scenario gives rise to a \ac{FOPT} \cite{Salvio:2024upo}. Defining the critical temperature $T_c$ to be the temperature where the true and the false vacua are degenerate in energy, and $\phi_c$ to be the field expectation value in the false vacuum at $T=T_c$, one finds
\be\label{eq:TcFOPT}
    T_c \simeq 0.31\,g\, \eta\,,\quad \phi_c \simeq 0.95\,\eta\,,
\ee
where $\eta$ is the renormalised zero-temperature \ac{VEV}. 

As we will review in Sec.~\ref{sec:MonoFOPT}, the \ac{PT} in the \ac{CW} scenario is \emph{strongly} first-order, in the sense that the tunneling rate to the true vacuum at $T\lesssim T_c$ is still highly suppressed. Bubbles of true vacuum will only form at a much lower nucleation temperature, $T_n\ll T_c$. Such \ac{PT}s are often referred to as “supercooled'' in the literature.

\subsubsection{The global limit: a second-order phase transition}
\label{sec:SOPT}

In the limit $g\into 0$ of the $\SO{3}$ symmetry becoming global, perturbation theory breaks down in the vicinity of the critical temperature and does not permit to conclude about the nature of the \ac{PT} (see, e.g., Ref.~\cite{Arnold:1992rz}). It can be argued that the \ac{PT} is of the second order at $g=0$, since by dimensional reduction \cite{Ginsparg:1980ef} the theory can be mapped to the ${\rm O}(3)$ model in three Euclidean dimensions, which is a classic system in the study of critical phenomena via the $\epsilon$ expansion \cite{Zinn-Justin:2002ecy}. By continuity, the \ac{PT} should also be of the second order ``for all practical purposes'' for non-zero but sufficiently small $g$. This is the physically relevant case for our study, since in the global limit $g\into 0$, monopoles become infinitely massive and their core radius infinitely large (see also Sec.~\ref{ssec:generalcase} for further discussion).

The critical temperature for a \ac{SOPT} is given by the temperature where the symmetry-preserving point at $\phi=0$ becomes classically unstable, $T_c=T_0$. At the one-loop level, $\phi$ obtains a thermal mass due to its self-interaction,
\be\label{eq:thermalmass}
V(\phi,T)=\frac{\lambda}{2}\left(\frac{5}{12}T^2-\eta^2\right)\phi^2+\ldots
\ee
and so one would be led to conclude that
\be\label{eq:T0SOPT}
T_0=\sqrt{\frac{12}{5}}\,\eta\,.
\ee
This is the leading-order result in perturbation theory; however, as we have mentioned, perturbation theory is well known to be unreliable in the ``Ginzburg region'' of temperatures around $T_0$. More precisely, the expansion parameter for the perturbative calculation of the effective potential, after daisy resummation, is $\delta\equiv(3\,\lambda T)/m_{\rm eff}(\phi^2)$. Here $m_{\rm eff}^2(\phi^2)=\lambda\left(\phi^2+5 T^2 / 12-\eta^2\right)$ is the field-dependent scalar mass including its leading-order thermal correction. Since the critical temperature $T_0$ of a \ac{SOPT} is characterized precisely by the condition that $m_{\rm eff}^2(0)\into 0$, $\delta$ becomes large in the vicinity of $T_0$, and perturbation theory is unreliable. However, for estimating $T_0$ it is sufficient to bound the Ginzburg region by the criterion $\delta<1$, which leads to
\be
\left(\frac{5}{12}-9\,\lambda\right)^{-1/2}\lesssim\frac{T_0}{\eta}\lesssim \left(\frac{5}{12}+9\,\lambda\right)^{-1/2}\,.
\ee
This amounts to at most a $\lesssim$ 10\% deviation from Eq.~\eqref{eq:T0SOPT} if $\lambda\lesssim 10^{-2}$. In fact, an argument from dimensional reduction and \ac{EFT} indicates that the leading-order result of Eq.~\eqref{eq:T0SOPT} should be accurate for even larger $\lambda$. The effective potential of Eq.~\eqref{eq:ThermalPot}, as well as its extensions to higher loop order and its (partially) resummed variants, are obtained by summing over the quantum corrections due to an infinite tower of Matsubara modes. In dimensional reduction, instead, one defines a three-dimensional Euclidean \ac{EFT} by integrating out these modes. The effective couplings are obtained by matching to the original four-dimensional theory at some fixed temperature, using perturbation theory (which is justified so long as the quartic coupling is perturbatively small). The \ac{EFT} itself is reliable if there is a large hierarchy of scales between $T$ and the masses of the light dynamical fields, which is indeed the case for $T\simeq T_0$.

\subsubsection{The general case}\label{ssec:generalcase}
As we have reviewed, perturbation theory predicts a \ac{FOPT} in the perturbative region $g^2\gg\lambda$, which for fixed $\lambda$ becomes progressively weaker as $g$ decreases (see Fig.~\ref{fig:PhDiag}). The nature of the \ac{PT} in the non-perturbative region $g^2<\lambda$ can only be reliably assessed on the lattice. Lattice studies of the closely related electroweak \ac{PT} in the \ac{SM} \cite{Karsch:1996yh,Gurtler:1997hr,Rummukainen:1998as} and of the Abelian Higgs model \cite{Kajantie:1997hn, Hove:2000rzx} indicate that it turns into a crossover at some critical ratio $\left(g^2/\lambda\right)_{\rm crit}\sim{\cal O}(0.1)$ in the non-perturbative region. We can plausibly expect that the model we are studying will behave similarly. Between the critical point and the second-order point $g=0$, the broken and the unbroken phase should then be connected by a crossover, with the field correlation length enhanced but not strictly divergent around the critical temperature.

\section{Monopole relic abundance}
\label{sec:MonoAbun}

As the \ac{PT} takes place, the order parameter  changes from $\langle \phi \rangle = 0$ in the symmetry-preserving phase to a non-zero value in the symmetry-breaking one. While the absolute value of $\langle \phi \rangle$ is unequivocally determined by energy minimisation, its orientation in field space is not; on the contrary, during the \ac{PT}, the field takes random orientations in the vacuum manifold on scales larger than the field correlation length $\xi$. Therefore, the Universe is fragmented into domains of typical length $\xi$, each characterised by a different orientation of $\langle\phi\rangle$. At the intersection of these domains, a monopole forms with a probability $p$ which depends on the topology of the vacuum manifold. Therefore, the monopole number density can be estimated as $n_M \simeq p\,\xi^{-3}$.
In our case, the vacuum manifold is $S^2$ and $p=1/8$ \cite{Kibble:1976sj}. We define the comoving monopole number density by
\be
    Y_M = \frac{n_M}{s} = p \left( \xi^3s\right)^{-1}\,,
\ee
where the entropy density is $s = 4\gamma_* T^3$, and $\gamma_*\equiv \pi^2 g_*/90$. The relic density of monopoles is then given by
\be
    h^2\Omega_M = 0.12\left( \frac{Y_M}{4.4\times 10^{-13}}\right)\left(\frac{m_M}{{\rm TeV}}\right)\,,
\ee
as derived in Eq.~\eqref{eq:OmegaDM}.

The computation of the correlation length $\xi$ strongly depends on the details of the \ac{PT}, as we show in Sec.~\ref{sec:MonoProd}. After being produced, the monopole number density can be reduced by monopole-antimonopole annihilation \cite{Zeldovich:1978wj,Preskill:1979zi}, as discussed in Sec.~\ref{sec:MonoAnn}, and summarised in Sec.~\ref{sec:MonoAnnSummary} below.

\subsection{Monopole production}
\label{sec:MonoProd}

\subsubsection{Second-order phase transitions}\label{sec:MonoSOPT}

After a \ac{SOPT}, monopoles are formed by the \ac{KZ} mechanism \cite{Kibble:1976sj,Zurek:1985qw}; see, e.g., Refs.~\cite{Zurek:1996sj, delCampo:2013nla} for reviews. As the temperature approaches the critical temperature $T_0$, both the correlation length $\xi$ and the relaxation time $\tau$ diverge as
\be\label{eq:xitau}
    \xi(t) = \xi_0 \lvert\epsilon(t) \rvert^{-\nu},\qquad \tau(t) = \tau_0 \lvert\epsilon(t) \rvert^{-\mu}\,.
\ee
Here $\epsilon(t) \equiv \left(T(t) - T_0 \right)/T_0$ is the quench, and $\mu$ and $\nu$ are critical exponents. Since our model is in the Heisenberg universality class of the O(3) model in 3 dimensions, we take $\nu=0.705\pm0.003$ \cite{Zinn-Justin:2002ecy}, and since its dispersion relation is relativistic, we have $\mu\approx\nu$ \cite{Zurek:1996sj, Murayama:2009nj}.  The quench can be linearised around the time $t_0$ when $T=T_0$: $\epsilon(t) \simeq \left(t_0 - t\right)/\tau_Q$, where $\tau_Q$ thus defined is the quenching timescale. In a cosmological context, the decrease of $T$ is due to the expansion of the Universe with $dT/dt = -HT$, and therefore $\tau_Q \simeq H^{-1}$ for $t \simeq t_0$. As time evolves, the relaxation time grows according to Eq.~\eqref{eq:xitau}, until it becomes equal to the time left before reaching the critical point $t_0$. We call this moment in time $t_*$:
\be
t_0 - t_*  = \tau\left(t_*\right).
\ee 
 For $t>t_*$, the system is no longer able to respond to external changes and fluctuations freeze, and therefore $\xi$ remains constant. It is only after a time interval $\lvert t_* - t_0 \rvert$ has elapsed since the critical time that the system is able to equilibrate again. This happens on spatial scales given by
\be
\label{eq:correlation_lgth}
   \xi(t_*)\simeq H^{-1}(T_0)\, \left[H(T_0)\xi_0\right]^\frac{1}{1+\nu}\equiv \xi_{\rm KZ}. 
\ee

To determine $\xi_0$, we write the effective potential in the Ginzburg-Landau approximation at a temperature $T$ close to $T_0$, just before critical behaviour sets in:
\be
    V_{\rm GL} = a_0 \frac{(T-T_0)}{T_0}\phi^2 + b_0 \phi^4 \,.
\ee
Then $\xi_0^2\simeq 1/(2\,a_0)$. By comparing with Eq.~\eqref{eq:thermalmass} we find
\be
\xi_0^2=\frac{6}{5\lambda\,T_0^2}=\frac{1}{2\lambda\eta^2}\,.
\ee
Therefore, using  $n_M \simeq \xi^{-3}(t_*)/8$, the monopole yield produced by the \ac{SOPT} in the limit $g^2\ll\lambda$ reads
\be
\begin{aligned}\label{eq:MonoSOPT}
Y_M \approx \frac{1}{32} \left(\frac{5}{6}\lambda\right)^{3/(2+2\nu)} \gamma_*{}^{(\nu-2)/(2+2\nu)}\left(\frac{T_0}{\MP}\right)^{3\nu/(1+\nu)}\,.
\end{aligned}
\ee
This scaling with $T_0$ matches the generic prediction  \cite{Murayama:2009nj} for monopoles from a \ac{SOPT}, while the prefactor is specific to the 't Hooft-Polyakov model studied here.

The above discussion is valid as long as the monopoles are effectively point-like. However, in the global limit $g\into 0$, the monopole mass and radius diverge. For $r_M>\xi_{\rm KZ}$, the \ac{KZ} estimate of the number density must clearly break down. Analytical studies and numerical simulations have shown that global monopoles enter a scaling regime with an ${\cal O}(1)$ number $\zeta$ of monopoles per Hubble volume, $n_M=\zeta H^3$ \cite{Barriola:1989hx, Rhie:1990kc,Yamaguchi:2001xn,Martins:2008zz}. In radiation domination, Ref.~\cite{Yamaguchi:2001xn} finds $\zeta= 3.44\pm 0.56$. This global limit is a good approximation as long as $r_M>1/H$.
To roughly estimate the present-day monopole number density, 
we assume that the scaling regime ends once $r_M\lesssim 1/H$, and that afterwards the monopoles redshift as matter. This leads to a comoving number density 
\be\label{eq:Monoscaling}
Y_M\approx \frac{\zeta}{4}(r_M\MP)^{-3/2}\gamma_*^{-1/4}\,.
\ee
In the intermediate case $\xi_{\rm KZ} <r_M<H(T_c)^{-1}$ , the field configuration after the phase transition still consists of many overlapping monopoles which will efficiently annihilate. Assuming that this takes place on a short timescale, the effective correlation length right after the phase transition is given by the monopole radius, $\xi\approx r_M$, and the comoving number density becomes
\begin{equation}\label{eq:Monoann}
Y_M\approx \frac{1}{4} {\gamma}_{*}^{-1}(r_M T_c)^{-3}\,,
\end{equation}
which interpolates between the \ac{KZ} and global limits. In all cases the abundance is further reduced by monopole annihilations, see Sec.~\ref{sec:MonoAnn}.

\subsubsection{First-order phase transitions}\label{sec:MonoFOPT}

The hallmark of a \ac{FOPT} is the presence, at the critical temperature, of a barrier separating the two degenerate minima of the thermal effective potential. The scalar field remains trapped in the false vacuum as the Universe cools down, until quantum tunneling or thermal fluctuations are efficient enough for the \ac{PT} to take place. Therefore, the \ac{PT} proceeds via nucleation of bubbles of true vacuum, in a background of metastable phase, as we reviewed in Sec.~\ref{sec:Tunnel}. Neglecting the effect of quantum tunneling, the rate per unit volume at which bubble nucleation proceeds is given by \cite{Linde:1981zj}
\be\label{eq:Gammav}
    \Gamma(T) \simeq T^4 e^{-S_3/T},
\ee
where $S_3$ is the euclidean action of a bubble, evaluated along the $\rm{O}(3)$-symmetric bounce solution describing finite-temperature transitions.

The \textit{nucleation temperature}, $T_n$, is defined as the temperature at which there is, on average, one bubble per Hubble volume\footnote{In principle, one should integrate the probability of nucleation per Hubble volume over time. However, the integral is dominated by the nucleation time $t_n$ and Eq.~\eqref{eq:Tn} provides a good estimate for $T_n$.}
\be\label{eq:Tn}
    T_n: \qquad \Gamma(T_n) \simeq H(T_n)^4\,,
\ee
and is often taken as the reference temperature signaling the onset of the \ac{PT}. Once nucleated, bubbles of the new phase expand due to the gradient of potential energy across the bubble wall, $\Delta V(T)$. The latent heat parameter $\alpha$ is defined as the ratio between vacuum and radiation energy densities at the moment of nucleation, when the \ac{PT} starts:
\be\label{eq:alpha}
    \alpha \equiv \frac{\Delta V(T_n)}{\rho_r(T_n)} = \frac{30}{g_* \pi^2}\frac{\Delta V(T_n)}{T_n^4}\,.
\ee
Notice that the energy difference between the two vacua is a temperature dependent quantity: it vanishes at the critical temperature, and then it grows to its zero-temperature value, as the Universe expands and cools down. If the separation between the critical temperature and the nucleation temperature is large, $T_n \ll T_c$, the vacuum energy density eventually comes to dominate before the first bubbles start nucleating. This happens at $T=T_{\rm eq}$, with 
\be
    T_{\rm eq} \equiv \left(\frac{30}{g_* \pi^2}\Delta V(T_{\rm eq}) \right)^{1/4}\,,
\ee
Approximating $\Delta V(T_{\rm eq})\simeq \Delta V(T_n)\simeq \Delta V$, with $\Delta V$ the vacuum energy difference at zero temperature, gives $\alpha \simeq (T_{\rm eq}/T_n)^4$. If $\alpha >1$, bubbles nucleate in a vacuum-dominated Universe and a \ac{sFOPT} is realised. On the contrary, if $\alpha < 1$, the \ac{PT} takes place during radiation domination. This is the case for a \ac{wFOPT}. We will see below that these two scenarios lead to qualitatively different phenomenologies.

The progress of the \ac{PT} is described by the probability, at a given time (or, analogously, at a given temperature), of finding one point in space in the false vacuum \cite{Guth:1979bh}
\be\label{eq:PfT}
    \mathcal{P}_{f}(T) = \exp\left[-\frac{4\pi}{3}\int_{T}^{T_c} dT' \frac{\Gamma(T')}{T'^4 H(T')}\left(\int_{T}^{T'} v_b \frac{dT''}{H(T'')}\right)^3\right]\equiv e^{-I(T)}\,,
\ee
where expanding bubbles are expected to quickly reach their terminal velocity and, therefore, the wall velocity, $v_b$, is assumed to be constant. The above expression for $\mathcal{P}_f$ is derived under the further assumption that the initial radius, at which bubbles are nucleated, provides a subleading contribution to the bubble size at the end of the \ac{PT}. That is, the size of a bubble at the completion of the \ac{PT} is mostly given by its expansion.

The transition completes at a temperature $T_e$. Since its determination is subtle and depends on the details of the \ac{PT} \cite{Levi:2022bzt,Athron:2023xlk}, we will postpone it till we discuss \ac{wFOPT}s and \ac{sFOPT}s separately. Since, for the kind of \ac{PT}s of interest here, the nucleation rate decays exponentially with time, we can write it as $\Gamma(t) \simeq C \exp\left[\beta (t-t_e)\right]$, where $C$ is a constant and we have introduced the completion rate $\beta$, defined as
\be\label{eq:beta}
    \beta \equiv \frac{d\log \Gamma}{d t}\bigg\rvert_{t_e} = -H(T_e) T_e \frac{d \log \Gamma}{d T}\bigg\rvert_{T_e}\,,
\ee
with $t_e$ the instant of cosmic time at which $T=T_e$. In the second equality, we have made use of the fact that, as long as entropy is conserved, we can exchange cosmic time $t$ and temperature $T$ through the relation $dT = - H(T) T dt$. In the following, we will dub as \textit{``fast''} those \ac{PT}s for which $\beta/H(T_e) \gg 1$.

Monopoles form  with probability $p=1/8$ when bubbles of true minimum collide, around $T=T_e$. Assuming that the scalar field is correlated within each nucleated bubble, the relevant correlation length $\xi$ is given by the average bubble radius  $R_e$ at the completion of the \ac{PT}. By $n_M=p\,\xi^{-3}$, we obtain for the comoving monopole number density
\be\label{eq:nMFOPT}
Y_M\approx \frac{1}{32} (\gamma_*^{\rm reh})^{-1}\left( R_p\, T_{\rm reh}\right)^{-3}\,,
\ee
where $T_{\rm reh}$ is the temperature at which the Universe gets reheated by bubble collisions. 

Let us now estimate $R_e$. The average bubble radius at a given temperature can be inferred from the bubble number density distribution \cite{Turner:1992tz}
\be\label{eq:dndR}
    \frac{d n_b(t)}{dR} = \frac{\Gamma[t_i(t,R)]\mathcal{P}_{f}[t_i(t,R)]}{v_b}\left[\frac{a[t_i(t,R)]}{a(t)}\right]^4\,,
\ee
where $a(t)$ is the scale factor of the Universe and $t_i(t,R)$  is the time of nucleation of a bubble whose radius at a later time $t$ is $R$. The quantity $t_i$ can be found by inverting the relation
\be\label{eq:R}
    R(t_i;t) = \int_{t_i}^t dt' v_b \frac{a(t')}{a(t_i)}\,,
\ee
which is valid, again, under the assumption that the initial radius of a bubble can be neglected.
By integrating Eq.~\eqref{eq:dndR}, we get the number density of bubbles
\be\label{eq:nt}
    n_b(t) = \int_{0}^{\infty} dR \frac{d n_b(t)}{dR}\,.
\ee
Finally, the correlation length is the mean bubble size at the end of the \ac{PT}, i.e.~the typical distance between the centers of two adjacent bubbles at $t=t_e$: $\xi\simeq R_e \simeq n_b(t_e)^{-1/3}$.

Our discussion so far, with the caveats we mentioned, is general and applies to any \ac{FOPT}. However, \ac{wFOPT}s and \ac{sFOPT}s are quite different and, as we shall see in the remainder of this section, lead to different phenomenology.

\begin{figure}[h!]
\begin{center}
\includegraphics[width=.47\textwidth]{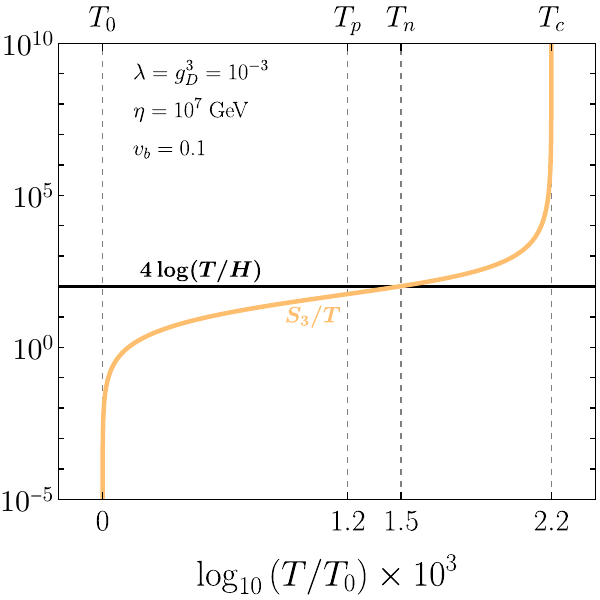}\qquad\includegraphics[width=.47 \textwidth]{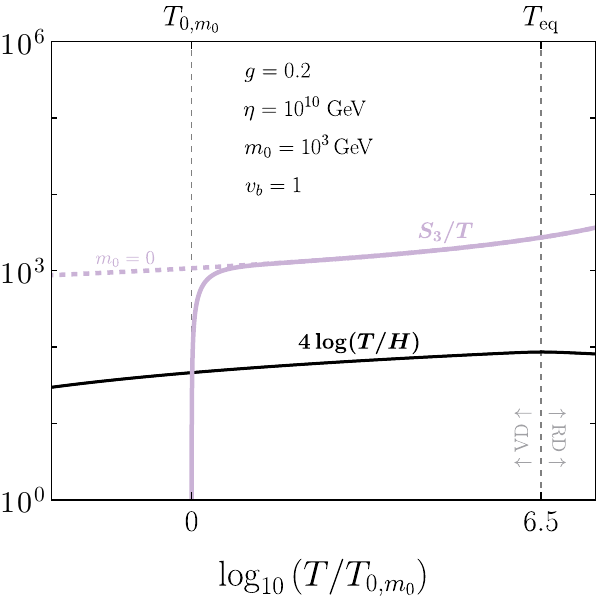}
    \caption{\textit{Left-hand panel.} In orange, evolution of the bounce action described by Eq.~\eqref{eq:S3Tw}, as a function of the temperature, in the scenario of a \ac{wFOPT}. The nucleation temperature, $T_n$, defined as the temperature at which the nucleation rate per Hubble volume matches the Hubble rate, as per Eq.~\eqref{eq:Tn}, corresponds to the crossing of the orange and black lines. The \ac{PT} ends at $T=T_p$, where bubbles of the true vacuum percolate. The temperature $T_0$, defined in Eq.~\eqref{T0wFOPT}, is the temperature at which the symmetry-preserving point ceases to be a local minimum, and the thermal barrier vanishes. \textit{Right-hand panel.} Same as in the left-hand plot but for a \ac{sFOPT} in the \ac{CW} limit ($\lambda = 11 g^4/8\pi^2$). The bounce action is given by Eq.~\eqref{eq:S3s}. As shown by the solid purple line,  the \ac{PT} tales place at $T\simeq T_{0,m_0}$, defined in Eq.~\eqref{eq:T0m0}. Bubbles nucleate and percolate around this temperature: $T_n\simeq T_p \simeq T_{0,m_0}$, due to the presence of a small bare mass term $m_0$. In the pure \ac{CW} scenario ($m_0 = 0$), the Universe remains stuck in the false vacuum (dashed purple line).}
\label{fig:FOPT}  
\end{center} 
\end{figure}

\underline{\textbf{Weakly first-order phase transition.}} We focus here on $g^4 \ll \lambda \ll g^2$, i.e.~the region delimited by the blue lines in Fig.~\ref{fig:PhDiag}, where the effective potential is well approximated by Eq.~\eqref{eq:PotHighT}. In this case, the bounce action is numerically fitted by \cite{Levi:2022bzt}
\be\label{eq:S3Tw}
   S_3^{(w)}(T) \simeq
 \frac{m^3(T)}{\delta^2(T)} \frac{2\pi}{3(\kappa - 2/9)^2} F(\kappa)\,,
\ee
where 
\be
\label{eq:potential_param}
    m^2(T) \equiv \frac{g^2}{2}T^2 - \lambda \eta^2\,,\qquad \delta(T) \equiv \frac{3g^3}{2\pi}T\,,\qquad\kappa\equiv \lambda\frac{m^2(T)}{\delta^2(T)}\,,
\ee
and
\be
\begin{aligned}
    F(\kappa) = \frac{16}{243}\Bigg [& 1-38.23\left(\kappa-\frac{2}{9}\right)+115.26\left(\kappa-\frac{2}{9}\right)^2\\
    &+58.07 \sqrt{\kappa}\left(\kappa-\frac{2}{9}\right)^2+229.07 \kappa\left(\kappa-\frac{2}{9}\right)^2\Bigg ]\,.
\end{aligned}
\ee
The bounce action, evaluated for a benchmark point in parameter space, is plotted in the left-hand panel in Fig.~\ref{fig:FOPT}. 

We choose, as the moment when the \ac{PT} completes, the \textit{percolation temperature} $T_e=T_p$, defined as the temperature at which 29\% of the Universe is occupied by bubbles: $\mathcal{P}_{f}(T_p)\simeq 0.71$ and $I(T_p) \simeq 0.34$, according to the definition in Eq.~\eqref{eq:PfT}. A \ac{wFOPT} is generally fast, in the sense we have defined below Eq.~\eqref{eq:beta}. Therefore, we can regard the Hubble parameter as constant over its duration. This allows us to write $I(t)$, defined by Eq.~\eqref{eq:PfT}, as
\be\label{eq:I(t)}
     I^{(w)}(t) \simeq \frac{4\pi}{3}v_b^3\int_{t_c}^{t}dt'\Gamma(t')(t-t')^3 \simeq 8\pi v_b^3\frac{\Gamma(t)}{\beta^4}\,.
\ee
Here we have used $\Gamma(t')\propto e^{\beta t'}$ so that the integral is exponentially dominated by its upper boundary.  The above expression for $I(t)$ allows for a numerical estimate of  $T_p$. 

Neglecting the expansion of the Universe altogether between bubble nucleation and the end of the \ac{PT}, Eq.~\eqref{eq:R} can be trivially solved to find $t_i(t,R) = t-R/v_b$. Eqs.~\eqref{eq:nt} and \eqref{eq:dndR} then lead to
\be\label{eq:nbwFOPT}
     n_b^{(w)}(t) = \frac{\Gamma(t)}{\beta I(t)}\left[1-\mathcal{P}_f(t)\right]\,.
\ee
Therefore, the mean bubble size at percolation, for a \ac{wFOPT}, is 
\be\label{eq:Rp}
    R_p^{(w)} = \left[\frac{8\pi v_b^3}{1-\mathcal{P}_f(t_p)}\right]^{1/3}\beta^{-1}\,,
\ee
and the comoving monopole number density becomes
\be\label{eq:YMw}
    Y_M^{(w)} = \frac{1}{32}(\gamma_*^p)^{-1}\frac{0.29}{8\pi v_b}\frac{\beta^3}{T_p^3}\,.
\ee
In the above estimate we have taken $T_{\rm reh} = T_p$, since for a \ac{wFOPT} the Universe remains radiation dominated and its evolution remains approximately adiabatic. The bubble velocity is treated as a free parameter here. On general grounds, we expect the radiation bath to exert a non-negligible friction on bubble walls, so that $v_b\ll 1$.

\underline{\textbf{Strongly first-order phase transition.}} A \ac{sFOPT} is realised, e.g., in scale-invariant potentials \cite{Salvio:2023qgb}, like the one presented in Sec.~\ref{sec:CW}  (green line in Fig.~\ref{fig:PhDiag}). In this limit, the high-temperature expansion is not a good approximation of the potential around the critical temperature \cite{Jaeckel:2016jlh}. However, as long as $T_n \ll \eta$, a \textit{supercool expansion} can still be used for modelling the tunneling process \cite{Salvio:2023qgb,Salvio:2023ynn}. Neglecting the effect of the Higgs-portal coupling, the thermal effective potential can be approximated by
\be\label{eq:VsFOPT}
\begin{aligned}
    V(\phi,T) &\simeq \frac{g^2}{4}T^2\phi^2 - \frac{g^3}{2\pi}T\phi^3 - \frac{3 g^4}{16\pi^2}\log \left(\frac{\eta}{T}\right)\phi^4\\
    &\equiv \frac{\tilde{m}^2(T)}{2}\phi^2 - \frac{\tilde{\delta}(T)}{3}\phi^3 -\frac{\tilde{\lambda}(T)}{4}\phi^4\,,
\end{aligned}
\ee
and the bounce action is \cite{Levi:2022bzt}
\be\label{eq:S3s}
    S_3^{(s)}(T) \simeq
\frac{\tilde{m}^3(T)}{\tilde{\delta}^2(T)} \frac{27\pi}{2} 
\left[ \frac{1 + \exp\left(-1/\sqrt{|\tilde{\kappa}|}\right)}
{1 + 9|\tilde{\kappa}|/2} \right]\,, \qquad {\rm with}\qquad \tilde{\kappa} \equiv \tilde{\lambda}(T)\frac{\tilde{m}^2(T)}{\tilde{\delta}^2(T)}\,. 
\ee

The potential in Eq.~\eqref{eq:VsFOPT} always features a local minimum at the origin, as long as $T \neq 0$. In other words, contrary to what happens for a \ac{wFOPT}, a barrier between the false and the true vacua persists down to $T=0$. It is then possible for the Universe to remain stuck in the false vacuum for long enough to enter a regime of vacuum domination, with $\alpha >1$. Since this is the scenario we are interested in, in what follows we will approximate $H \simeq \Delta V/3\MP^2$. 
Under the assumption of a perfect de Sitter metric, $a(t) = a_0 e^{Ht}$, we obtain from Eq.~\eqref{eq:PfT}
\be
    I^{(s)}(T) \simeq \frac{4\pi}{3}v_b^3 \frac{\Gamma(T)}{\beta^4}\left[1+6\,\frac{H}{\beta}+11\left(\frac{H}{\beta}\right)^2+6 \left(\frac{H}{\beta}\right)^3\right]^{-1}\,,
\ee
which allows for a numerical computation of the percolation temperature $T_p$. 
However, for bubbles growing during vacuum domination, the spacetime accelerated expansion may prevent the transition from completing, even though the percolation condition, $I = 0.29$, is fulfilled \cite{Athron:2023xlk}. This happens, in particular, if, below the percolation temperature $T_p$, the fraction of the Universe in the unbroken phase keeps increasing. In this scenario, the transition can be said to be complete as soon as the false vacuum fraction starts decreasing,
\be
    \frac{d}{dt}\left[a(t)^3 \mathcal{P}_f(t)\right]<0\,.
\ee
It is easy to show that, for exponential nucleation, this condition is equivalent to $I(T) > 3 H/\beta$. We define $T_d$ as the temperature at which the physical volume in the false vacuum starts decreasing,
\be\label{eq:Td}
    T_d:\qquad I^{(s)}(T_d) = 3 \frac{H}{\beta}\,.
\ee
However, for fast-enough \ac{PT}s with $\beta \gtrsim 9$, which is the scenario of interest for us, we can take $T_e = T_p$ \cite{Levi:2022bzt}.

The computation of the average bubble radius at transition completion is analogous to the one we carried out for a \ac{wFOPT}, with the only difference that, in  a de Sitter Universe, a bubble that at a given time $t$ has a radius $R$ was nucleated at
initial time
\be
    t_i(t,R) = t -\frac{1}{H}\log \left(1+\frac{R H}{v_b}\right)\,,
\ee
according to Eq.~\eqref{eq:R}. Hence, using Eqs.~\eqref{eq:dndR} and \eqref{eq:nt}, the bubble number density is
\be\label{eq:nbsFOPT}
    n_b^{(s)}(T) = \frac{\Gamma(T)}{\beta}I(T)^{-1-3\frac{H}{\beta}}\left[\tilde{\Gamma}\left(1+3\frac{H}{\beta},0\right)-\tilde{\Gamma}\left(1+3\frac{H}{\beta},I(T)\right)\right]\,,
\ee
where $\tilde{\Gamma}$ is the incomplete gamma function. For fast \ac{PT}s ($\beta/H \gg 1$), in which the Universe expansion can be neglected, the above bubble number density matches the one obtained in a \ac{wFOPT}, cfr.~Eq.~\eqref{eq:nbwFOPT}. The correlation length is, once again, given by $R_e^{(s)} = [n_b^{(s)}(T_e)]^{-1/3}$:
\be\label{eq:Rps}
    R_p^{(s)} = \left[\frac{8\pi v_b^3}{1-\mathcal{P}_f(T_p)}\right]^{1/3}\beta^{-1}\,.
\ee
After the \ac{PT} completes, the Universe is reheated at a temperature $g_*(T_{\rm reh})T_{\rm reh}^4 = g_*(T_p) T_p^4  + g_*(T_{\rm eq})T_{\rm eq}^4$, as it is given by the conservation of $H$ across the different steps of the transition, upon assuming instantaneous reheating. A large amount of supercooling implies that $T_p\ll T_{\rm eq}$, and $T_{\rm reh} \simeq T_{\rm eq}$.

In supercool scenarios, the accelerated expansion of the Universe dilutes any pre-existing thermal bath of particles, and bubbles expand in an empty Universe. Due to the absence of any plasma-induced friction, bubble walls enter a runaway regime, in which they quickly reach ultrarelativistic velocities. Therefore, whenever \ac{sFOPT}s are concerned, we will assume $v_b = 1$.

Due to the presence, in the thermal effective potential in Eq.~\eqref{eq:VsFOPT}, of a thermal barrier down to $T=0$, Eq.~\eqref{eq:Tn} defining the nucleation temperature does not always admit a solution, as shown by the dashed, purple line in the right-hand panel of Fig.~\ref{fig:FOPT}. When this is the case, the Universe is never able to thermally tunnel to the true vacuum, and enters a permanent stage of inflation, unless some other mechanism, which we have ignored in our analysis, aids the transition in completing. This can happen, for example, because of quantum fluctuations becoming relevant at low temperatures and assisting the nucleation of bubbles \cite{DelleRose:2019pgi,Lewicki:2021xku}. 

Another possibility is that a \ac{PT} into some other sector catalyses bubble nucleation. At the electroweak \ac{PT} temperature, $T_{\rm EW} \simeq 160\,{\rm GeV}$ \cite{DOnofrio:2015gop}, the \ac{SM} Higgs acquires a \ac{VEV}, $\langle h \rangle \simeq 246\,{\rm GeV}$. The non-zero Higgs \ac{VEV} induces, in turn, a mass for the dark scalar below $T_{\rm EW}$: $\tilde m^2(T)\rightarrow \tilde m^2(T) +  \lambda_{\phi H} \langle h\rangle^2/2$, for $T<T_{\rm EW}$. Therefore, by choosing $\lambda_{\phi H}<0$, $\phi=0$ ceases to be a local minimum of the dark sector potential at a temperature $T_{0,h} = \min\left(T_{\rm EW},\sqrt{\lvert\lambda_{\phi H}\rvert}\langle h\rangle/g\right)$, and the dark \ac{PT} is triggered just before this moment. Note  that the dark \ac{VEV}, $\eta$, back-reacts onto the Higgs potential mass parameter, $\mu_H^2 \to \hat\mu_H^2\equiv \mu_H^2+ \lambda_{\phi H} \eta^2/2$. However, $\mu_H^2$ needs to be negative by itself, for the electroweak symmetry breaking to happen before the dark one. Therefore, the back-reaction cannot exceed the measured value of $\hat\mu^2_H  = -\lambda_H\langle h \rangle^2 =-m_h^2/2 \simeq - (88\,{\rm GeV})^2$. We then assume that the Higgs portal coupling  satisfies $|\lambda_{\phi H}|\ll 2\lambda_H \langle h \rangle^2/\eta^2$, where $\lambda_H \simeq 0.1$ is the Higgs quartic coupling \cite{ParticleDataGroup:2024cfk}. However, as we will show in Sec.~\ref{sec:DMVecvsMono}, monopoles acquire a sizeable relic abundance for $\eta \gg \langle h \rangle$. For such values of the symmetry breaking scale, the portal coupling has to be tiny, not allowing thermal equilibrium to establish between the dark and the \ac{SM} sectors. Therefore, in such scenario the temperatures of the two sectors could differ from each other.

An alternative, allowing to ensure thermalisation via a sizeable portal coupling, is to slightly deform the scale-invariant potential discussed in Sec.~\ref{sec:CW}. We modify the choice of the renormalisation conditions, so that $\phi$ gets a small, zero-temperature negative mass, $m_0^2$:
\be
    \tilde m ^2(T) \quad \longrightarrow \quad \tilde m ^2(T) - m_0^2\,.
\ee
We want it to be a small perturbation to the \ac{CW} limit, $m_0^2/\lambda \sim m_0^2/g^4 \ll \eta^2$. At temperatures around
\be\label{eq:T0m0}
    T_{0,m_0}^2 = \frac{2}{g^2} m_0^2\,,
\ee
the symmetry-preserving point ceases to be a minimum and the dark-sector \ac{PT} occurs, as shown in the right-hand plot in Fig.~\ref{fig:FOPT}. We tune the Higgs mass parameter, $\mu_{H}^2$, against the dark-sector induced contribution, 
\be\label{eq:mh2}
    m_h^2 =-2\hat\mu_H^2 = -2\mu_H^2 - \lambda_{\phi H}\eta^2\simeq (125\,{\rm GeV})^2\,,
\ee
so that large values of $\lambda_{\phi H}$ are possible. In the following, we study the scenarios of positive and negative $\lambda_{\phi H}$ separately, since they lead to different cosmic histories. 

\underline{$\boldsymbol{\lambda_{\phi H} < 0.}$} 
Eq.~\eqref{eq:mh2} implies that the Higgs quadratic coefficient, $\mu_H^2$, is large and positive to compensate for the dark-sector contribution. This implies that the electroweak \ac{PT} always happens after the dark sector one. At $T\simeq T_{0,m_0}$, the dark sector undergoes a \ac{sFOPT} and the \ac{VEV}, $\langle\phi\rangle = \eta$, induces a negative contribution to $\hat\mu_H^2$. The temperature at which the electroweak \ac{PT} occurs depends on the choice of parameters. For a given value of $m_0$, if $g$ is small enough, then $T_{0,m_0} > T_{\rm EW}$, cfr.~Eq.~\eqref{eq:T0m0}. In this scenario, at $T\simeq T_{0,m_0}$ the Higgs quadratic coefficient $\mu_H^2$ jumps to its measured value $\hat\mu^2_H$. However, thermal fluctuations are still large enough to stabilise the origin of the Higgs potential. The electroweak crossover \ac{PT} takes place, as in the \ac{SM}, at $T_{\rm EW} = 160\,{\rm GeV}$ \cite{DOnofrio:2015gop}. In the opposite scenario, in which $g$ is large, we have $T_{0,m_0}< T_{\rm EW}$. In this case, the temperature at the dark \ac{PT} is low enough for the dark \ac{VEV} to trigger the electroweak \ac{PT} right away at $T \simeq T_{0,m_0}$. In this scenario the interplay between the dark transition and electroweak one is non-trivial. In particular, one cannot assume anymore that the tunneling process takes place along the $\phi$-direction only. 
The study of this case is beyond the scope of the present work and is left for future investigation. Nonetheless, notice that, independently from the hierarchy between $T_{0,m_0}$ and $T_{\rm EW}$, the Universe is reheated by bubble collisions at a temperature $T_{\rm reh}\simeq T_{\rm eq}$, which is always smaller than the dark critical temperature, $T_c$, but large enough to restore the electroweak symmetry. After reheating, the Higgs potential is exactly the \ac{SM} one, modulo radiatively-small corrections from the portal coupling, and it will undergo a \ac{PT} a second time, at $T_{\rm EW} = 160\,{\rm GeV}$.

\underline{$\boldsymbol{\lambda_{\phi H} > 0.}$} If we take a positive portal coupling, Eq.~\eqref{eq:mh2} implies $-\mu_H^2 \simeq \lambda_{\phi H}\eta^2/2 \gg m_h^2$ (remember that $\eta \gg \langle h \rangle$ and $\lambda_{\phi H}$ large enough to thermalise the two sectors). Therefore, the electroweak \ac{PT} happens at larger temperatures, as soon as $\mu_H^2$ overcomes the thermal Higgs mass.\footnote{The dominant thermal correction to the Higgs mass comes from top-quark loops, $m_h^2(T)\sim \frac{y_t^2}{4} T^2$, where $y_t \sim 1$ is the top-quark Yukawa coupling \cite{CMS:2020djy}.}
This leads to a temperature $T_{\rm EW}' \simeq \eta\sqrt{2\lambda_{\phi H}}$, and a Higgs \ac{VEV} $\left(\langle h \rangle'\right)^2 \simeq \lambda_{\phi H}\eta^2/(2 \lambda_H)$. Here, we use a prime to differentiate this scenario from the one in which the electroweak \ac{PT} happens in a \ac{SM}-like way, with $T_{\rm EW} = 160\,{\rm GeV}$ and $\langle h \rangle = 246\,{\rm GeV}$. The effect of the Higgs \ac{VEV} on the dark potential is that of stabilising the symmetry-preserving point 
\be
\begin{aligned}
    \tilde m^2(T<T_{\rm EW}') &= -m_0^2 + \frac{g^2 T^2}{2} + \frac{\lambda_{\phi H}}{2}(\langle h\rangle')^2\\
    &= -m_0^2 + \frac{g^2 T^2}{2} + \frac{\lambda_{\phi H}^2}{4 \lambda_H}\eta^2\,.
\end{aligned}
\ee
One can, then, tune $-m_0^2$ against the Higgs-induced contribution, so that their sum is small and negative, in order to drive a \ac{sFOPT}. This amounts to choosing $m_0^2 \simeq \lambda_{\phi H}^2 \eta^2/(4 \lambda_H)$. In this way, after the electroweak \ac{PT}, the dark sector potential becomes approximately scale-invariant in the zero-temperature limit, up to a small quadratic deformation, $(m_{0}')^2 \equiv m_0^2 - \lambda_{\phi H}^2 \eta^2/(4 \lambda_H)$, and it undergoes a \ac{sFOPT} at $T\simeq T_{0,m_0'} = \sqrt{2} m_0'/g$. In order for the above discussion to be consistent, $T_{0,m_0'}< T_{\rm EW}'$ is needed, which implies $(m_0')^2 < \lambda_{\phi H} g^2 \eta^2$. To summarise, at large temperatures $T \gg T'_{\rm EW}, T_{0,m_0'}$, both the \ac{SM} and the dark potential feature large and negative temperature-independent, quadratic coefficients. At $T\simeq T_{\rm EW}'$, the electroweak \ac{PT} occurs, the Higgs acquires a large \ac{VEV} $\langle h \rangle'$, which backreacts on the potential for $\phi$ making it almost scale invariant, featuring a small negative, quadratic deformation, $-(m_0')^2$. The temperature is still large enough to preserve the dark $\SO{3}$ symmetry. Finally, at $T \simeq T_{0,m_0'}$, a dark \ac{sFOPT} takes place. However, the \ac{PT} happens most likely in multi-field space, and a dedicated analysis is required. 

For concreteness, hereafter, we will take $\lambda_{\phi H}$ negative and sizeable, focusing, in particular, on the case $T_{0,m_0} > T_{\rm EW}$, in which the transitions in the two sectors can be studied independently from one another. The dark sector \ac{PT} realised in this scenario is extremely fast, being characterised by large values of $\beta/H$, as it can be seen from the right-hand panel in Fig.~\ref{fig:FOPT} (remember that $\beta \propto d(S_3/T)/dT$, as per Eq.~\eqref{eq:beta}). Since the bubble radius at percolation scales as $R_p^{(s)} \propto 1/\beta$, cfr.~Eq.~\eqref{eq:Rps}, in this scenario bubbles are extremely small at percolation. Intuitively, this is due to the fact that many small bubbles are nucleated at the same time, and soon after percolate, with little time to expand. For bubbles nucleating at $T_n \simeq T_{0,m_0}$, we find that $R_p^{(s)}$ is smaller than the initial, critical radius
\be\label{eq:Rc}
    R_c^2=\frac{3\tilde\lambda(T_n)}{3\tilde\lambda(T_n)\tilde m^2(T_n)-\tilde\delta^2(T_n)+\tilde\delta(T_n)\sqrt{\tilde\delta^2(T_n)-4\tilde\lambda(T_n)m^2(T_n)}}\,,
\ee
which we have computed in the thick-wall approximation as discussed in Sec.~\ref{sec:thick-wall}.
Having, $R_p^{(s)} < R_c$ is obviously nonsensical. This outcome is due to the fact that the formalism we have reviewed at the beginning of this section relies on the assumption that most of the bubble size at percolation comes from their expansion (see discussion below Eq.~\eqref{eq:PfT}). This assumption obviously breaks down for $R_c >R_p^{(s)}$, rendering our computation for the radius at percolation non-consistent. Nonetheless, when this happens, we can estimate a lower bound on the correlation length of the scalar field $\phi$ by taking it to be $\xi \simeq R_c$. This overestimates the number density of monopoles produced at the \ac{PT}, with a more refined computation of the bubble radii that would only result in a smaller abundance of monopoles. As it is shown in Sec.~\ref{sec:DMVecvsMono}, such a refinement would not change our conclusions. Therefore, whenever a \ac{sFOPT} is discussed we will consider a correlation length
\be
    \xi \simeq R_p^{(s)} + R_c\,.
\ee

All the discussion present in this section is based on the assumption that thermal fluctuations are the dominant process driving the transition to the stable vacuum, neglecting quantum fluctuations and gravitational effects. While the former can be shown to be negligible for $g \lesssim 2$ \cite{Levi:2022bzt}, the latter becomes relevant for $T_n \lesssim H$. For such low nucleation temperatures, the bounce action should be modified to take into account the spacetime curvature \cite{Coleman:1980aw,Hawking:1981fz}.

\subsection{Monopole annihilation}\label{sec:MonoAnnSummary}

After being produced, the monopole number density may be reduced by monopole-antimonopole annihilation (see Sec.~\ref{sec:MonoAnn} for details). Monopoles can dissipate their energy by moving through a plasma of relativistic charged particles, in our case the $W'$ gauge bosons.\footnote{Energy loss from bremsstrahlung can be shown to be negligible for sufficiently heavy monopoles.} This allows the formation of monopole-antimonopole bound states, which will eventually annihilate.

 The diffusive capture process is efficient as long as the mean free path of the monopole in the plasma is smaller than the capture radius. It stops once the comoving monopole number density is sufficiently diluted, or when the $W'$ bosons become non-relativistic, whichever happens first. It may also happen that the monopole density at production was never large enough to allow for efficient annihilation in the first place.

In the case that it is monopole-antimonopole annihilation which determines the final monopole yield, the latter is given by
\be\label{eq:YAnn}
   Y_M = \frac{\mathcal{B} g^2}{16\pi} \gamma_*^{-1}\frac{T_f}{\MP}\,.
\ee
Here $T_f$ is the temperature where annihilation ceases to be effective, and $\mathcal{B}$ is a constant depending on the plasma properties, with $\mathcal{B}=2\zeta(3)/\pi^2$ for $W'$ bosons. The relic abundance of monopoles is given by the smallest quantity between the comoving number density at production, and Eq.~\eqref{eq:YAnn}.

\section{Dark matter relic abundance: vector bosons vs monopoles} 
\label{sec:DMVecvsMono}

Having established the relevant mechanisms which give the monopole number density, we can now study under which conditions dark monopoles can constitute most of the observed \ac{DM}. $W'$ vector bosons, as the lightest (and only) particles carrying dark electric charge, are also stable \ac{DM} candidates; we will investigate whether their relic abundance can be subdominant. For an early study of $W'$ \ac{DM} phenomenology, see Ref.~\cite{Baek:2013dwa}.

Contrarily to monopoles, the $W'$ abundance can be computed rather straightforwardly using the well-established formalism of thermal freeze-out, which we have introduced in Sec.~\ref{sec:Freeze-out}. We start by observing that the abundance of transversely polarized $W'$ bosons and dark photons may be parametrically small in the limit of small $g$, where the $\SO{3}$ symmetry becomes approximately global. However, by assumption, $\phi$ is thermalised with the \ac{SM} via the Higgs portal, and therefore there is always a thermal abundance of dark Higgs bosons,  and of longitudinally polarized $W'$ dark gauge bosons, which are \ac{NGB}s in the global limit.  From the thermally averaged cross-section for (massless) $\phi-H$ scattering, 
\be
   \langle \sigma v\rangle_{\phi\phi \to HH} = \frac{\lambda_{\phi H}^2}{128\pi T^2}\,,
\ee
one deduces that the two sectors are thermalised at the time of the dark \ac{PT}, $ n^{\rm eq}_\phi \langle\sigma v\rangle >H$, provided that
\be
T_c\lesssim \lambda_{\phi H}^2\times 10^{14}\,{\rm GeV}\,,
\ee
where $n_\phi^{\rm eq}$ is the equilibrium number density of relativistic $\phi$ particles. For a larger critical temperature, an additional thermalisation mechanism should be invoked. Such $W'$ thermal bath is present at the time of monopole production, at least in the absence of supercooling, 
thus it can catalyse monopole annihilation. If $g$ is large enough, then the transverse $W'$s and the $\gamma'$ will also be thermalised. The relevant process here is $W'-\phi$ scattering before the dark sector phase transition, with
\begin{equation}
    \langle \sigma v \rangle_{W'W'\rightarrow \phi\phi}=\frac{41 g^4}{4608\pi T^2} \,.
\end{equation}
The transverse polarisations of the three dark gauge bosons are thermalised at the phase transition if $n_{W'}^{\rm eq}\langle \sigma v\rangle > H$, which requires  $T_c\lesssim g^4 \left(1.2\times10^{15}\,{\rm GeV}\right)$. 

The present $W'$ abundance is determined by  the type of  freeze-out.  If
$W'$s freeze-out when they are non-relativistic ($z_f\equiv m_{W'}/T_f \gg 1$, where $T_f$ is the freeze-out temperature), 
today's abundance is given by 
\be\label{eq:Wfo}
    \Omega_{W'}h^2 = 1.1\times 10^{-9}\left(\gamma_*^f\right)^{-1/2}\left(\frac{x_f}{20}\right)\left[\langle\sigma v\rangle\;{\rm GeV}^2\right]^{-1}\,.
\ee
where $g_*(z_f) \simeq 100$, $\langle \sigma v\rangle$ depends on the specific process keeping the dark gauge bosons in equilibrium, and
\be\label{eq:zf}
    z_f \simeq \log\left[0.076 \frac{g_{W'}}{\sqrt{g_*(z_f)}}  m_{W'}\langle \sigma v\rangle\MP\right]-\frac{1}{2}\log\log \left[0.076 \frac{g_{W'}}{\sqrt{g_*(z_f)}}  m_{W'}\langle \sigma v\rangle\MP\right]\,.
\ee
The relevant internal degrees of freedom of the dark gauge bosons are $g_{W'} = 6$ if all the three polarisations are thermalised, but $g_{W'}=2$ if only longitudinal $W'$s are in the bath. Since the value of $z_f$ depends only logarithmically on the cross-section, non-relativistic freeze-out typically occurs at $z_f\sim 20$. Alternatively, $W'$s may freeze-out when they are relativistic ($z_f\ll 1$), and in this case their abundance is independent from the annihilation cross-section, according to 
\begin{equation}
\label{eq:hotDM}
    \Omega_{W'} h^2 = 0.12
    \dfrac{g_{W'}}{6}\left(\gamma_*^f\right)^{-1}\dfrac{m_{W'}}{2.39\ {\rm eV}}\,.
\end{equation}
Then, the relic abundance of dark gauge bosons can be suppressed by taking $g\lll1$. These dark gauge bosons would overclose the Universe for masses above $\sim 100$ eV, therefore they cannot constitute cold DM. Sub-KeV $W'$s are subject to strong constraints on hot DM, as well as to bounds on dark radiation.

The dark gauge bosons can annihilate into dark sector particles, namely $\rho$ and $\gamma'$, as well as into the \ac{SM} Higgs, via a $\rho$ mediator. The relevant process determining the final energy density of $W'$s depends on the hierarchies among the couplings $\lambda$, $g$ and $\lambda_{\phi H}$. The rest of our discussion is organised as follows: we divide the parameter space in regions depending on the type of \ac{PT} the dark sector undergoes (second-order, weakly first-order, or strongly first-order); for each case we explore different hierarchies among the couplings, leading to different freeze-out processes for the dark gauge bosons. Throughout the whole discussion, we assume that the relevant contributions to the effective potential for $\phi$ come from dark sector interactions: i.e.~$\lambda_{\phi H}$ is always subleading with respect to at least one between $\lambda$ and $g$. At the same time, $\lambda_{\phi H}$ is sizeable to ensure that the dark sector and the visible one share a common temperature at the dark \ac{PT}. 

\subsection{A second-order phase transition: \texorpdfstring{$\lambda \gg g^2$}{\lambda \gg g^2}}

For $\lambda \gg g^2$, the dark $\rho$ is much heavier than the $W'$ and, therefore, it decouples from the thermal bath soon after the \ac{PT}. The dominant annihilation channel for the dark gauge bosons depends on the relative size of $\lambda_{\phi H}$ and $g^2$.

\begin{figure}[h!]
\begin{center}
\includegraphics[width=.48\textwidth]{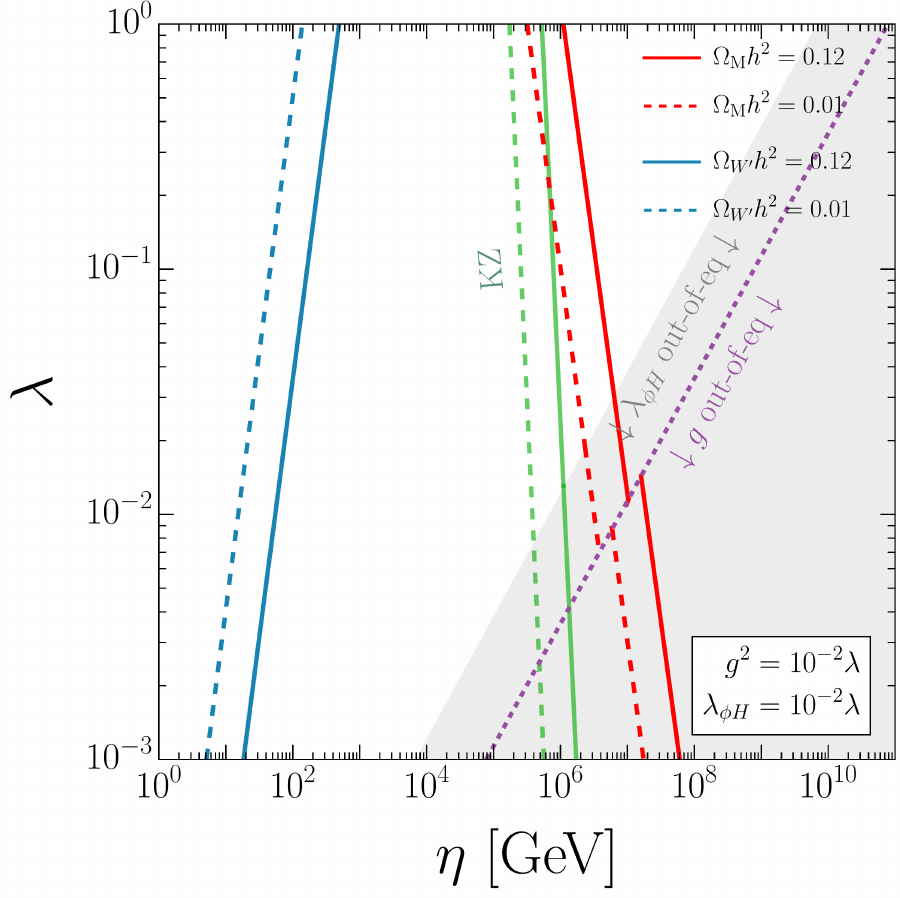}\quad \includegraphics[width=.48\textwidth]{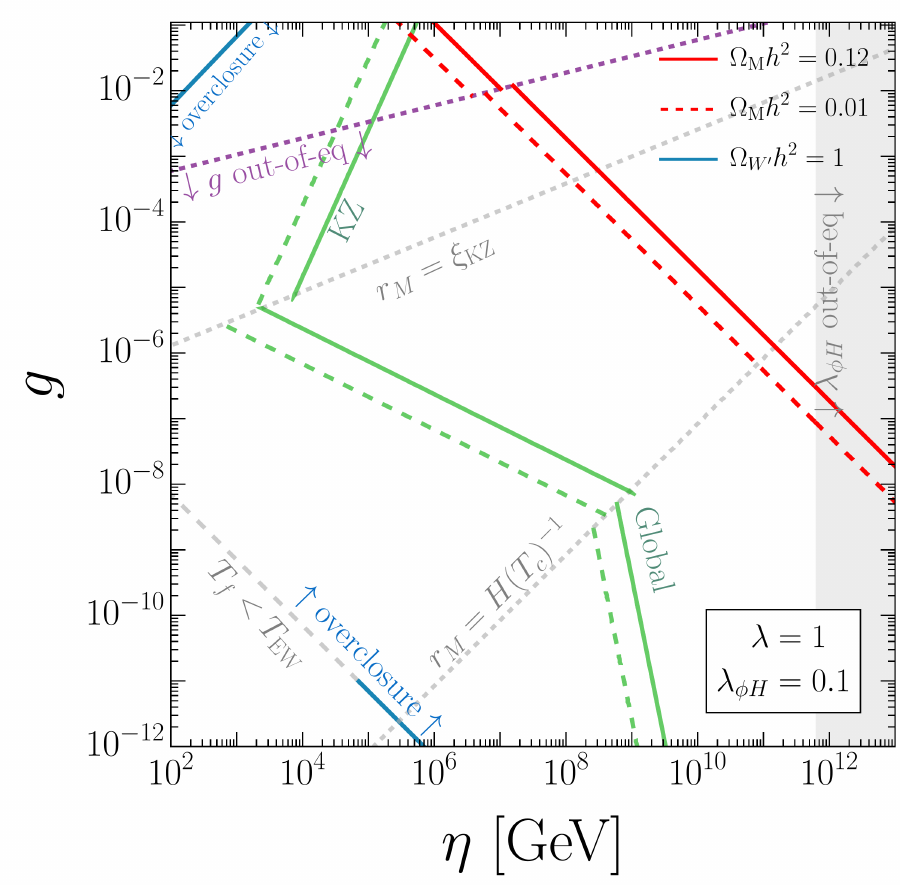}
    \caption{Relic density of the vectors $W'$ (blue) and of the monopoles $M$ (red) after a \ac{SOPT}.
    In the grey region the whole dark sector is not thermalised with the \ac{SM}; below the purple line, the transverse dark gauge bosons are not thermalised with the rest of the dark sector. 
    For the choice of parameters in the \textit{left panel}, 
    the $W'$s freeze-out when they are non-relativistic, 
    while the monopoles are produced by the \ac{KZ} mechanism (green) and later undergo annihilations (red).
    In the \textit{right panel}, 
    the $W'$ overclosure abundance is determined by non-relativistic freeze-out in the top-left corner, and by relativistic freeze-out in the bottom-left corner (we neglected electroweak symmetry breaking, which would further enhance $\Omega_{W'}$ for $T_f < T_{\text{EW}}$). For $r_M< \xi_{\rm KZ}$, the monopoles are produced by the \ac{KZ} mechanism, for $r_M>1/H(T_c)$ they are produced as global monopoles, while for the intermediate region we took  $\xi\approx r_M$; 
    in all the three cases, later monopole annihilations control the final abundance.
    }
\label{fig:GlobalAbun}  
\end{center}
\end{figure}

\underline{$\boldsymbol{\lambda \gg g^2 \gtrsim \lambda_{\phi H}}$}. Annihilation into the \ac{SM} Higgs is suppressed both by a small portal coupling, and by the heavy mediator exchanged. Therefore, the dominant channel is  $W'^+W'^- \rightarrow \gamma'\gamma'$, whose cross section is
\be\label{eq:2W2gamma}
   \langle \sigma v \rangle_{W'^+W'^-\to \gamma'\gamma'} \simeq \frac{19 g^4}{72\pi m_{W'}^2}\,,
\ee
where we have retained only the $s$-wave component. The \ac{PT} is of the second order, and monopole abundance at production is given by the \ac{KZ} estimate presented in Sec.~\ref{sec:MonoSOPT}. The relative abundances of the two relics are shown in the left-hand panel in Fig.~\ref{fig:GlobalAbun}, for $\lambda = 10^2 g^2 = 10^2 \lambda_{\phi H}$. 

\underline{$\boldsymbol{\lambda  \gg \lambda_{\phi H} \gg g^2}$}. Dark gauge bosons annihilate most efficiently into Higgs bosons via $W'^+ W'^- \rightarrow HH$. Due, again, to the heavy-propagator suppression, this process freezes out early, when the temperature of the Universe is much larger than the $W'$ mass: $T_f \gg m_{W'}$. Since the dark gauge bosons freeze-out when non-relativistic, their relic abundance is set by Eq.~\eqref{eq:hotDM}, which does not depend on the annihilation cross section. The relic abundance of dark gauge bosons can, then, be suppressed by taking a tiny $g^2$. In this limit, the monopole radius $r_M \sim 1/(g\eta)$, becomes larger than the dark Higgs correlation length in Eq.~\eqref{eq:correlation_lgth}, and the \ac{KZ} argument does not hold anymore. The monopole number density at production is, instead, given by Eq.~\eqref{eq:Monoscaling} and Eq.~\eqref{eq:Monoann}. The total relic density of dark gauge bosons and monopoles is shown in the right-hand panel in Fig.~\ref{fig:GlobalAbun}. 

\subsection{A weakly first-order phase transition: \texorpdfstring{$g^4 \ll \lambda \ll g^2$}{g^4 \ll \lambda \ll g^2}}

\begin{figure}[tb!]
\begin{center}
\includegraphics[width=.5\textwidth]{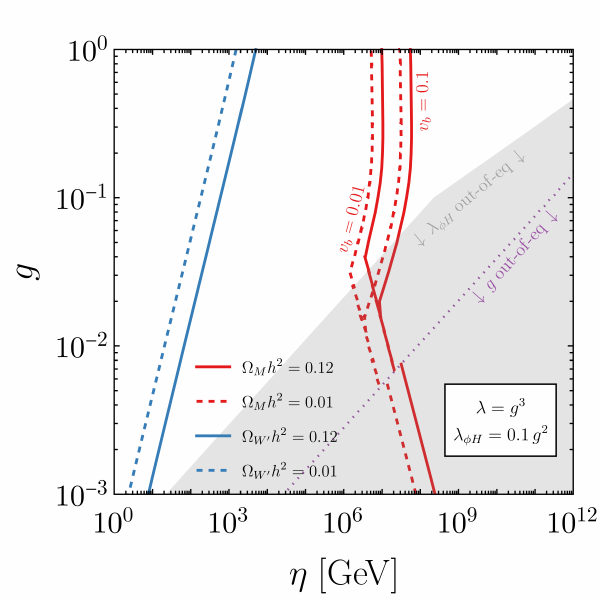}
    \caption{The monopole abundance (red) after a \ac{wFOPT}, taking into account later annihilations which become relevant for $g\lesssim 2\times 10^{-2}$.
    The corresponding $W'$ abundance (blue) 
    is determined by standard freeze-out, mostly driven by annihilations into $\gamma'$s and $\rho$s. 
    In the grey region, the two sectors do not reach thermal equilibrium for $T>T_c$; 
    below the purple dotted line, the transverse polarisations of the dark gauge bosons are not thermalised with the rest of the dark sector.
   }
\label{fig:wFOPTAbun}  
\end{center}
\end{figure}

For $\lambda \ll g^2$, the dark gauge bosons annihilate into $\rho$, $\gamma'$ and $H$, with the total $s$-wave cross section given by
\be
\label{eq:sigmavtot}
    \langle \sigma v \rangle_{\rm tot} \simeq \frac{49 g^4}{144\pi m_{W'}^2}-\frac{14g^2\lambda}{192\pi m_{W'}^2}+\frac{31\lambda^2}{2304\pi m_{W'}^2} + \frac{\lambda_{\phi H}^2}{384\pi m_{W'}^2}\simeq \frac{49 g^4}{144\pi m_{W'}^2}\,.
\ee
In the last equality we have retained only the dominant contribution. The \ac{PT} is weakly of the first order and the monopole abundance at production is given by Eq.~\eqref{eq:YMw}. For concreteness, we fix $\lambda=g^3$, and we show the relic abundances of gauge bosons and monopoles in Fig.~\ref{fig:wFOPTAbun}, for two values of the bubble wall velocity at percolation $v_b$.

\subsection{A strongly first-order phase transition: \texorpdfstring{$\lambda \ll g^4$}{\lambda \ll g^4}}

In the \ac{CW} scenario, we have $\lambda = 11 g^4/(8\pi^2)\ll g^4$, and the \ac{PT}, which is a strongly first-order one, proceeds as discussed in Sec.~\ref{sec:MonoFOPT}. At large temperatures, the Universe is in the unbroken phase and the dark gauge bosons are massless and in  equilibrium: their number density is $ n_{W'} = 4\zeta(3)
T^3/\pi^2$. At $T_{\rm eq}$, if the transition to the broken phase has not taken place yet, a stage of thermal inflation begins,  during which the temperature of the Universe drops by many orders of magnitude. Bubble collisions, which denote the end of the transition, reheat the Universe, now lying in the global minimum, to a temperature $T_{\rm reh}\simeq T_{\rm eq}$. In the new phase, the dark gauge bosons, among other particles, acquire a mass, $m_{W'} = g \eta$. If the reheating temperature is larger than their freeze-out temperature, $T_f = z_f\,m_{W'}$, with $z_f$ defined in Eq.~\eqref{eq:zf}, the $W'$s re-establish thermal equilibrium and their relic density is given by the freeze-out mechanism, cfr.~Eqs.~\eqref{eq:Wfo} and \eqref{eq:sigmavtot}. If, instead, $T_{\rm reh} < T_f$, the dark gauge bosons are not able to reach thermal equilibrium again, and their number density remains the one they had at the end of thermal inflation, when $T=T_e$. However, their relative fraction of energy density is diluted by the large entropy injection following reheating \cite{Hambye:2018qjv}
\be\label{eq:WabunSupercool}
    h^2\Omega_{W'} = 1.6\times 10^8 \frac{m_{W'}}{\rm GeV}\frac{\zeta(3)}{\pi^2} \frac{g_{W'}}{g_*(T_{\rm reh})} \frac{T_{0,m_0}^3}{T_{\rm eq}^3}\,,
\ee 
where $g_*(T_{\rm reh})=2+106.75$ represents the relativistic degrees of freedom of particles that are in thermal equilibrium after reheating. Having in mind the scenario in which the \ac{PT} completes thanks to the small deformation of the otherwise-scale-invariant potential, as discussed in Sec.~\ref{sec:MonoFOPT}, we have set $T_e\simeq T_{0,m_0}$. The number of $e$-folds of thermal inflation is $N = \log T_{\rm eq}/T_{0,m_0}$, so that $h^2\Omega_{W'} \propto e^{-3N}$. Therefore, a large amount of supercooling, $N\gg 1$ or, analogously, $T_{0,m_0} \ll T_{\rm eq}$, can efficiently dilute the relic abundance of dark gauge bosons.\footnote{The intermediate stage of thermal inflation has nothing to do with the primordial inflation responsible for the generation of \ac{CMB} scales. On the contrary, \ac{CMB} observations constrain $N\lesssim 50$, if we assume \ac{GUT}-scale inflation followed by an instantaneous reheating \cite{Lewicki:2021xku}.} However, even for $T_{\rm reh}\lesssim T_f$, a significant subthermal $W'$ population can be generated. Assuming instantaneous reheating, this latter contribution is given by \cite{Hambye:2018qjv}
\be
    \Omega_{W'}h^2 = 3.3\times 10^{23}\left(\gamma_*^{\rm reh}\right)^{-3/2} \left(\frac{g_{W'}}{2}\right)^2 m_{W'}^2 \langle \sigma v\rangle \left(1+2 \frac{m_{W'}}{T_{\rm reh}}\right)e^{-2 m_{W'}/T_{\rm reh}}.
\ee
This effect is particularly relevant in our minimal model, where the ratio $m_{W'}/T_{\rm reh}$ is fixed and so the exponential suppression never leads to a parametrically small $\Omega_{W'}h^2$.  

\begin{figure}[h!]
\begin{center}
\includegraphics[width=.48\textwidth]{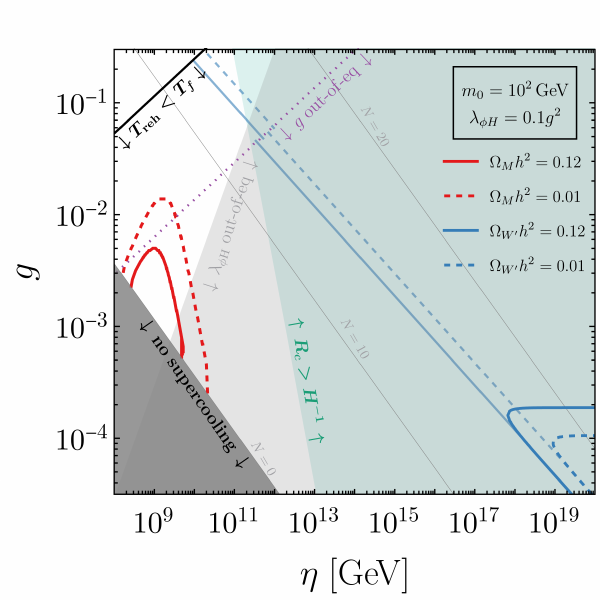}\quad\includegraphics[width=.48\textwidth]{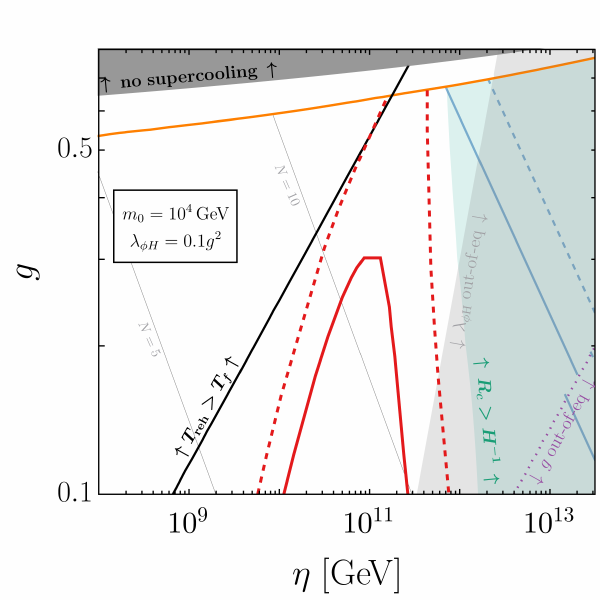}
    \caption{Relic abundance of monopoles (red) and dark gauge bosons (thick blue), for a supercooled \ac{sFOPT}, followed by instantaneous reheating. The thin blue lines show the $W'$ abundance at the end of supercooling, neglecting the sub-thermal population which could be generated after reheating. 
    Grey lines show isocontours of the number $N$ of $e$-folds of thermal inflation, which dilute the $W'$ abundance. Above the orange line, nucleation occurs independently of $m_0$ at much larger temperatures so that $N$ sharply decreases.
    Above the black line, the universe is reheated to a temperature large enough to restore $W'$ thermal equilibrium. 
    Inside the dark grey region the phase transition completes before the universe enters vacuum domination.
In green shading, the region where the effect of gravity  on the tunneling rate should be taken into account.}
\label{fig:sFOPTAbun}  
\end{center}
\end{figure}

The left panel of Fig.~\ref{fig:sFOPTAbun} shows the total $W'$ abundance (thick blue lines) compared with the abundance left after supercooling (thin blue lines), which would be relevant in a scenario where the reheating temperature can be made parametrically small and the subthermal population from reheating is subdominant. In such a scenario, the monopole abundance could be closer to (but still smaller than) the $W'$ abundance in some corners of parameter space (right panel); in the minimal model with instantaneous reheating, however, the whole plane of the right panel is ruled out as the subthermal $W'$ population would overclose the Universe.

We conclude that a \ac{sFOPT}, which is triggered by a small zero-temperature mass term, leads to a subleading monopole relic density compared to the one of $W'$, regardless of the large dilution of the $W'$ relic abundance achieved thanks to supercooling. Properly taking into account a non-negligible critical radius for the bubbles at the level of Eq.~\eqref{eq:PfT} can only result into a larger correlation length for the scalar field, thus not changing our conclusions.

\ac{sFOPT}s may also lead to the production of another stable relic: 
\ac{PBH}s. Due to its stochastic nature, bubble nucleation does not take place in all the Hubble patches at the same time. If the \ac{PT} is supercooled, some regions of the Universe may be trapped in the false vacuum for a longer time than the average patch. Since radiation energy density gets diluted by the expansion of the Universe, while vacuum energy density remains constant, late tunneling results in an overdensity in the corresponding patch. Such overdensities were previously thought to be so large to collapse and form PBHs \cite{Kodama:1982sf,Liu:2021svg,Hashino:2021qoq,Kawana:2022olo,Gouttenoire:2023naa,Lewicki:2023ioy,Baldes:2023rqv,Banerjee:2024cwv}. However, recent developments in the field have shown that the \ac{PBH} abundance resulting from a supercooled \ac{PT} has been greatly overestimated by the previous literature, due to a non-consistent gauge choice \cite{Franciolini:2025ztf}.\footnote{Notice that the published version of Ref.~\cite{Banerjee:2024cwv} presents a gauge-invariant treatment, also accounting for non-linear effects. Their results differ from those of Ref.~\cite{Franciolini:2025ztf}.} In light of this recent result, we find the abundance of produced \ac{PBH}s to be phenomenologically irrelevant in the model under study.

\section{Conclusions }

The model defined by an $\SO{3}$ symmetry spontaneously broken to $\SO{2}$ by the \ac{VEV} of a triplet scalar field is the simplest one featuring the presence of monopoles. Monopoles carry a non-zero magnetic charge under the residual $\SO{2}$. Such a magnetic charge is associated with the non-trivial topology of the vacuum manifold. Therefore, monopoles are stable, and contribute to the total relic abundance of \ac{DM} observed in the Universe. Alongside monopoles, the dark gauge bosons that acquire a mass from the triplet \ac{VEV} also contribute to the relic abundance, as they are the only particle electrically charged under $\SO{2}$.  

We studied a dark sector of the type just described, in which interaction with the \ac{SM} is achieved via a sizeable Higgs portal coupling, ensuring thermalisation. In the early Universe, the $\SO{3}$ symmetry is restored by thermal effects. When the temperature becomes of the order of the symmetry-breaking scale, a cosmological \ac{PT} takes place, with the triplet field acquiring a non-zero \ac{VEV}. When this happens, dark monopoles are produced, with an initial number density that strongly depends on the details on the \ac{PT}. After their production, monopoles annihilate, due to the presence of a plasma of charged vector bosons, and their abundance is reduced. The relic abundance of dark gauge bosons is computed through the freeze-out mechanism. They can annihilate both into dark sector particles and \ac{SM} Higgs bosons. 

The dark \ac{PT} can be either of the second order, if the triplet field adjusts to the symmetry breaking \ac{VEV} smoothly, or of the first order, if it tunnels to the new minimum through a barrier. In the latter scenario the \ac{PT} proceeds through the nucleation of bubbles, and completes once such bubbles percolate. Under some circumstances, a \ac{FOPT} implies an intermediate period of vacuum domination. This is the case, for example, for scale-invariant potentials of the \ac{CW} type. The order of the \ac{PT} depends on the values of the parameters under consideration. We have explored the different corners of parameter space, carefully studying the type of cosmological \ac{PT} undergone by the dark sector and, therefore, computing the number density of produced monopoles.

Both for a \ac{SOPT} and a \ac{wFOPT}, we find that the relic abundance of dark monopoles is greatly diminished by the effect of monopole-antimonopole annihilation. As a result, the massive gauge bosons always dominate the total \ac{DM} density. In a \ac{sFOPT}, the Universe experiences a stage of supercooling before the \ac{PT} completes. Such a period of thermal inflation dilutes the pre-existing population of gauge bosons, avoiding monopole annihilation. Nonetheless, we find that, in this scenario, monopole abundance at production is negligible compared to the one of gauge bosons. Our conclusion is that, in the minimal realisation of the model, the total relic abundance is always dominated by massive dark gauge bosons, with monopoles always comprising a negligible fraction of \ac{DM}.

Our analysis suggests possible ways to achieve monopole domination, which are worth being explored. In our discussion we have always assumed that the dominant contribution to the dark scalar effective potential came from the dark sector itself, taking $\lambda_{\phi H}$ smaller than $\lambda$ and/or $g^2$. However, the hierarchy $\lambda_{\phi H}\gg g^2\gg \lambda$ is interesting to suppress the $W'$ abundance: the cross-section for $W'^+W'^- \rightarrow HH$ (last term in Eq.~\eqref{eq:sigmavtot}) can be parametrically large, as the annihilation proceeds through a large coupling $\lambda_{\phi H}$, while the $W'$ mass can be taken small.
In this case the monopoles could dominate, but to determine their precise abundance one has to study the \ac{PT} for a two-scalar-field system, as now the $\phi$ and $H$ dynamics are tightly coupled by the large portal. Moreover, the monopole solution could in principle differ from the 't Hooft-Polyakov one. Analogously, in our discussion about the \ac{sFOPT} scenario in Sec.~\ref{sec:MonoFOPT}, we have focused on the eventuality in which the dark sector \ac{PT} can be studied independently from the electroweak one. This need not be the case and the scenario of a \ac{sFOPT} taking place in multi-field space could possibly lead to a larger abundance of dark monopoles, as well as to interesting phenomenology including a first-order electroweak \ac{PT}. Finally, the model discussed in this chapter could be extended to include light fermions charged under the residual $\SO{2}$. In this way, the dark gauge bosons are unstable (pretty much like what happens in the \ac{SM}), and the relic abundance of the dark fermions can be suppressed by taking them sufficiently light. We explore this scenario in Ref.~\cite{bigpaper}.

\chapter{Conclusions and future prospects}
\label{chap:Concl}
\minitoc

In this thesis, we studied the cosmological implications of spontaneous symmetry breaking in models in which the structure of the vacuum manifold is non-trivial, either due to the presence of extra flat directions that are not explained by symmetries, or due to a non-trivial topology that allows for the presence of topological defects in the spectrum.  

We have focused on accidentally light scalar fields: scalar fields that do not receive a tree-level mass even though they do not belong to the coset of the theory. In other words, accidents are neither Nambu-Goldstone bosons nor pseudo-Nambu-Goldstone bosons, and yet they are massless at the lowest order in perturbation theory. Their presence is associated with extra flat directions appearing in the vacuum manifold. We studied, in particular, the most minimal model featuring the presence of accidents: an $\SU{2}\times \U{1}$ symmetry broken by a scalar field transforming in the $\boldsymbol{5_1}$. Points lying along the accident direction in the vacuum manifold are not physically equivalent to each other, for there is no symmetry associated with this direction. In fact, we find that $\SU{2}\times \U{1}$ is generally completely broken, with the presence of one accident in the spectrum, except for an enhanced-symmetry point in field space. Here, the symmetry breaking pattern is $\SU{2}\times \U{1}\rightarrow \U{1}'$, and another accident appears. We studied the effect of radiative corrections on the accident direction, showing how bosonic interactions tend to stabilise the $\U{1}'$-preserving point, while the presence of fermions tends to break such $\U{1}'$ symmetry. An analogous analysis has been carried out considering the $\SU{3}$ ten-plet, as another example of a model featuring the presence of accidents. We then proceeded to discuss possible interesting applications. In the five-plet model, we showed how, by tuning fermionic and bosonic contributions against each other, we are able to construct an Abelian Higgs effective field theory, with the accident taking a non-zero vacuum expectation value due to the presence of radiative corrections, and breaking the $\U{1}'$ symmetry restored at the distinguished point in field space. The Abelian Higgs mass is naturally one-loop suppressed with respect to the ultraviolet scale, thus mitigating the hierarchy problem. Moreover, we showed how accidents prove to be a viable dark matter candidate, given that they are the lightest particles in the model, charged under the $\U{1}'$ symmetry. As a future prospect, it would be interesting to further investigate the accident phenomenon, with the idea of developing a systematic way of analysing different potentials and classifying their accidentally flat directions. The final goal is to find a model in which an $\SU{2}$ symmetry is restored at some enhanced-symmetry point along the accident direction. Provided that the accident transforms as a doublet under such $\SU{2}$ symmetry, it could be identified with the Standard Model Higgs boson. This would provide a novel way of addressing the electroweak hierarchy problem.

In chapter \ref{chap:AccInf}, we constructed a model of hybrid inflation in which the inflaton is an accidentally light scalar field. Small-field models of inflation require a flat potential to satisfy the slow-roll conditions. This is not always easy to achieve since the inflationary potential usually receives large radiative corrections. In our model, however, the flatness of the inflaton potential is explained via the accident mechanism. The simplest scenario we considered features a gauged $\SO{3}$ symmetry group and two scalar fields: $\phi$ transforming in the $\boldsymbol{5}$ and $\chi$ transforming in the $\boldsymbol{3}$. After spontaneous symmetry breaking, one of the components of $\phi$ remains massless at tree level. We identify this component with the inflaton field. One of the components of $\chi$ plays the role of the waterfall field, ending inflation when its mass becomes tachyonic. In spite of the tree-level coupling between $\phi$ and $\chi$, required for the waterfall mechanism to work, the flatness of the inflationary potential is protected by the fact that the inflaton is an accident. At the end of inflation, a stage of tachyonic preheating occurs, during which large overdensities are produced. We studied the stochastic gravitational wave background produced during this stage, showing that it could be possibly detected by the Einstein Telescope in the near future, provided that the inflationary scale is low enough. We have also explored minimal modifications of the model described above, in which cosmic strings and biased domain walls are produced at the end of inflation. Unstable domain walls generate a signal of gravitational waves which could explain the recent Pulsar Timing Array detection, or which could be detected in the near future by the Laser Interferometer Space Antenna or the Einstein Telescope, depending on the inflationary scale. 

In chapter \ref{chap:MonoDM}, we have analysed a different class of models, which posses a topologically non-trivial vacuum manifold. We have focused, in particular, on an $\SO{3}$ symmetry broken by a triplet as  $\SO{3}\rightarrow\SO{2}$. This model is known to feature the presence of 't Hooft-Polyakov monopoles. Two stable objects are present in the spectrum: the massive gauge bosons are the only particle electrically charged  under the residual $\SO{2}$, while monopoles are the only object possessing a magnetic (topological) charge. An Higgs-portal interaction ensures thermalisation between the dark sector and the Standard Model species. Depending on the values of the couplings, the dark sector can undergo either a second-order phase transition or a first-order one. We explored both possibilities, computing the scalar field correlation length: in a second-order phase transition this is roughly given by the mass of the dark Higgs, while in a first-order phase transition it is the average radius of bubbles at the moment of percolation. The presence of a thermal bath of charged, massive gauge bosons causes the formation of monopole-antimonopole pairs, that quickly annihilate. This efficiently reduces the relic abundance of monopoles after they have been produced. We studied the scenario of a weakly first order phase transition, as well as that of a strongly first order phase transition. In the latter case, the Universe experiences an intermediate period of thermal inflation which could potentially dilute the relic density of gauge bosons, thus preventing monopoles from annihilation. The relic abundance of gauge bosons has been computed by the freeze-out mechanism, as, in the early Universe, they efficiently annihilate into dark sector particles, as well as Standard Model Higgs bosons. We have identified the relevant annihilation channel case by case. We found that the dark monopoles always provide a negligible relic abundance, compare to that of the massive vector bosons, in all the scenarios discussed. In fact, monopole annihilation prevents them from acquiring a sizeable number density. Moreover, in the supercooled scenario, in which annihilation does not take place, the number density of monopoles at production is subleading compared to the one of gauge bosons. Therefore, we exclude the possibility of monopole dark matter, in the minimal model we have considered. Our analysis motivates the search for ways around gauge boson dominance. In this direction, it would be interesting to study scenarios in which the dark sector phase transition and the electroweak one are intertwined. Besides possibly providing a sizeable relic density for monopoles, this could have interesting implications for the electroweak phase transition itself, including the production of gravitational waves detectable by future experiments. Another possibility, which is currently under investigation, is that of extending the minimal model we have presented in this dissertation. The addition of charged fermions renders the gauge bosons unstable. At the same time, the relic density of dark fermions can be suppressed if they are sufficiently light.

\chapter{Résumé en Français}
\label{chap:Resumee}
\minitoc

Le Modèle Standard de la physique des particules et le modèle cosmologique $\Lambda$CDM offrent une description réussie, mais incomplète, de notre Univers. Ces deux modèles rencontrent des défis majeurs, notamment l’explication de la nature de la matière noire, l’origine des conditions initiales de l’Univers, et l’ajustement fin de la masse du boson de Higgs. Cette thèse étudie les implications cosmologiques de la brisure spontanée de symétrie afin de traiter certains de ces problèmes, en se concentrant sur des théories dans lesquelles la variété de vide possède une structure non triviale.
Nous introduisons une nouvelle classe de scalaires élémentaires appelés « accidents », qui apparaissent comme des directions accidentellement plates dans la variété de vide : contrairement aux directions des bosons de Nambu-Goldstone, les directions accidentelles ne sont associées à aucune symétrie. Des corrections radiatives induisent une masse pour ces accidents, qui reste cependant supprimée à une boucle par rapport aux attentes naïves : ces champs scalaires sont donc naturellement légers. Nous proposons que les accidents puissent constituer des candidats viables à la matière noire, ainsi que le champ d’inflaton à l’origine de l’inflation cosmique. Nous construisons un modèle d’inflation hybride dans lequel le potentiel de l’inflaton est une direction accidentelle et est donc naturellement plat. Un signal d’ondes gravitationnelles est produit durant la phase de préchauffage tachyonique. Nous explorons également la possibilité de produire des cordes cosmiques et des murs de domaine après la fin de l’inflation, dans des modèles d’inflation accidentelle où la variété de vide possède une topologie non triviale. Les réseaux de ces défauts topologiques génèrent un fond stochastique d’ondes gravitationnelles.

Par ailleurs, nous étudions la formation et l’impact cosmologique des monopôles magnétiques, qui résultent de la brisure spontanée de symétrie dans des théories à topologie de vide non triviale. Nous examinons en particulier les monopôles de 't Hooft-Polyakov, issus du schéma de brisure $\SO{3}\rightarrow \SO{2}$. Ce modèle inclut également des bosons de jauge massifs stables en tant que candidats à la matière noire. Nous explorons à la fois les transitions de phase du second ordre et du premier ordre, et évaluons avec soin si les monopôles peuvent représenter une fraction significative de l’abondance relique mesurée de matière noire. Nous concluons que ce n’est jamais le cas, excluant ainsi la possibilité de matière noire constituée de monopôles dans le modèle minimal.


\cleardoublepage


\bibliographystyle{hieeetr}
\bibliography{Thesis}


\cleardoublepage
\begin{vcenterpage}
\noindent\rule[2pt]{\textwidth}{0.5pt}
\begin{center}
{\large\textbf{Cosmological consequences of spontaneous symmetry breaking\\}}
\end{center}
{\large\textbf{Abstract:}}
The Standard Model of particle physics and the $\Lambda$CDM model of cosmology provide a successful, yet incomplete, description of our Universe. Both models face significant challenges, including explaining the nature of dark matter, the origin of the Universe's initial conditions, and the fine-tuning of the Higgs boson mass. This thesis investigates the cosmological implications of spontaneous symmetry breaking to address some of these issues, focusing on theories in which the vacuum manifold has a non-trivial structure.

We introduce a novel class of elementary scalars called ``accidents'', which emerge as accidentally flat directions in the vacuum manifold: unlike Nambu-Goldstone boson directions, accident directions are not related to any symmetry. Radiative corrections induce a mass for the accidents that is one-loop suppressed with respect to naive expectations, making such scalar fields naturally light. We propose that accidents can act as viable dark matter candidates, and as the inflaton field driving cosmic inflation. We construct a model of hybrid inflation in which the inflaton potential is an accident direction and, hence, is naturally flat. A signal of gravitational waves is produced during tachyonic preheating. We also explore the possibility of producing cosmic strings and domain walls after the end of inflation, in models of accident inflation where the vacuum manifold has a non-trivial topology. Such topological defects generate a stochastic background of gravitational waves.

Finally, we investigate the formation and cosmological impact of dark magnetic monopoles, arising from spontaneous symmetry breaking in theories with non-trivial vacuum topology. Focusing on 't Hooft-Polyakov monopoles from $\SO{3}\rightarrow \SO{2}$ symmetry breaking, we explore both second-order and first-order phase transitions, and we carefully identify the regions of parameter space where monopoles comprise a sizeable fraction of the measured dark matter relic abundance. This model also features stable massive dark gauge bosons.  We find that the relic density of dark gauge bosons is always much larger than the one of monopoles. We conclude that dark monopoles cannot constitute a sizeable fraction of dark matter.
\\

{\large\textbf{Keywords:}}
Accidentally Light Scalars, Monopole Dark Matter, Hybrid Inflation, Gravitational Waves

\noindent\rule[2pt]{\textwidth}{0.5pt}
\end{vcenterpage}

\end{document}